\documentclass[11pt]{article}
\pdfoutput=1

\usepackage[utf8]{inputenc}

\usepackage[hidelinks]{hyperref}

\usepackage{xcolor}

\usepackage{caption}

\usepackage{enumerate}

\newcommand{\operator}[1]{\hat{#1}}

\newcommand{\R}{\mathbb{R}}

\newcommand{\T}{\mathbb{T}}

\newcommand{\iless}[1]{\bigg\lfloor #1 \bigg\rfloor}

\newcommand{\pauli}[1]{
    \ifnum#1=1
        \operator{\sigma}_{x}
    \else
        \ifnum#1=2
           \operator{\sigma}_{y}
        \else
            \ifnum#1=3
                \operator{\sigma}_{z}
            \else
                \errmessage{Incorrect number given to pauli}
            \fi
        \fi
    \fi
}

\usepackage{geometry}
\usepackage{comment}

\usepackage{soul}
\usepackage{subcaption}

\usepackage{hyperref}
\usepackage{amsmath,amssymb,amsthm}
\usepackage{bm}
\usepackage{slashed}
\usepackage{graphicx}
\usepackage[T1]{fontenc}
\usepackage{fix-cm}

\makeatletter
\newcommand*\bigcdot{\mathpalette\bigcdot@{.5}}
\newcommand*\bigcdot@[2]{\mathbin{\vcenter{\hbox{\scalebox{#2}{$\m@th#1\bullet$}}}}}
\makeatother

\newlength{\dummysp}
\usepackage{mwe}

\def\R{{\mathbb R}}

\def\T{{\mathbb T}}

\def\tr{\,{\rm tr}\,}

\usepackage{authblk}
\usepackage{hyperref}
\newcommand*{\email}[1]{%
    \normalsize\href{mailto:#1}{#1}\par
    }
 \title{\bf Notes on Confinement on $\mathbf{R^3 \times S^1}$: From
 Yang-Mills, super-Yang-Mills, and QCD(adj) to QCD(F)\footnote{An expanded and updated version of lectures originally given at the 22$^{\rm nd}$ W. E. Heraues ``Saalburg'' Summer School in Wolfersdorf, Germany, in September 2016.  
Prepared for the 2021 Special Issue of the journal ``Symmetry'' on  ``New Applications of Symmetry in Lattice Field Theory.''

 {\it This 2024 update  includes  references, in footnote \ref{added2024}, to important  work from the early 1990s missed in the published version. }}  }

\date{}

\author{\bf  Erich Poppitz}

\affil{Department of Physics, University of Toronto\\ 60 St George St, Toronto, ON M5S 1A7, Canada\\ \email{poppitz@physics.utoronto.ca}}

\begin{document}

\maketitle

\thispagestyle{empty}

\vspace{1.4cm}
\begin{abstract} 

This is a pedagogical introduction to the physics of confinement on $R^3 \times S^1$, using  $SU(2)$ Yang-Mills with    massive or massless adjoint fermions  as the prime example; at the  end, we also add fundamental flavours. The small-$S^1$ limit  is remarkable, allowing for controlled semiclassical determination of the nonperturbative physics in these, mostly non-supersymmetric, theories. We begin by reviewing  the Polyakov confinement mechanism on $R^3$. Moving on to $R^3 \times S^1$, we show how introducing adjoint fermions stabilizes center symmetry, leading to abelianization and semiclassical calculability.~We explain how  monopole-instantons and twisted monopole-instantons arise.~We describe the role of various novel topological excitations in extending Polyakov's confinement to the locally four-dimensional case, discuss the nature of the confining string, and the $\theta$-angle dependence.~We study the global symmetry realization and, when available, present evidence for the absence of phase transitions as a function of the $S^1$ size. As our aim is  not to cover all work on the subject, but to prepare the interested reader for its study, we also include brief descriptions of  topics not covered in detail: the necessity for analytic continuation of path integrals, the study of more general theories, and the 't Hooft anomalies involving higher-form symmetries. 
\end{abstract}

%%%%%%%%%%%%%%%%%%%%%%%%%%%%%%%%%%%%%%%%%%
\setcounter{page}{0}
\setcounter{section}{0} %% Remove this when starting to work on the template.

%\end{paracol}
\newpage

\tableofcontents

\section{Introduction.}

I have overheard graduate students, after having taken a standard quantum field theory course, say that ``confinement occurs because the beta function is negative and the coupling becomes strong at long distances.'' Loosely, I translate this  to ``it's complicated and I won't think about it.'' That the students' explanation is insufficient is underscored by the fact that   we now know, due to insight from supersymmetry \cite{Seiberg:1994bp}, of gauge theories where the coupling becomes strong, but which become  nontrivial conformal field theories at long distances. Conversely, while confinement is a nonperturbative phenomenon, as evidenced by the framework described in these notes, it  does not always require strong coupling.

 Nonetheless, the students' attitude  is not unreasonable: in  real-world QCD the long-distance physics of chiral symmetry breaking and the emergence of hadrons occurs at strong  coupling, making it difficult to handle analytically. Hence, many are happy to leave\footnote{While others stick to the already mentioned supersymmetric world (a full disclosure requires us to state that these notes will also  not remain immune to the charms of supersymmetry).} the study of strong-coupling phenomena to experiments, performed either in the lab or on the computer. Indeed, the numerical approach to lattice QCD has been very successful in relating the short-distance description in terms of quark and gluon fields to the hadron physics emerging at long distances.

In summary,\footnote{The quotation marks indicate the need to more carefully define what we mean. See Sections \ref{sec:polyakov_pass4}, \ref{sec:holonomyandcenter}, \ref{sec:digression}.}  ``confinement'' is 45+ years old news and we have gotten used to it. It is believed to be a property of pure Yang-Mills theory, but we have no proof.\footnote{For extra motivation, see the Clay Institute website \url{https://www.claymath.org/millennium-problems}.} Even a more modest goal, a physicist's analytical understanding in continuum asymptotically-free pure Yang-Mills theory on $R^4$ is lacking. 
We do have analytical understanding via the strong-coupling expansion on the lattice, but this is far from the continuum limit. There is also the overwhelming numerical evidence from  lattice simulations. 

At this point, let me make a disclaimer and a recommendation. The disclaimer is that I can not possibly review all existing analytical---theoretically controlled or otherwise---and numerical approaches to confinement. The recommendation is, for an overview of the existing ideas about confinement,  to consult  the comprehensive and refreshingly (self-) critical monograph  \cite{Greensite:2011zz} (a minor warning  is that its relatively small size is both a blessing and a curse). 

After this preamble, we now  turn to the topic of these notes.

 \subsection{What are these lectures  about?}
\label{sec:whatabout}

These lectures are about an approach to the study of confinement that emerged within the past decade or so.
This is an analytical approach to confinement, within asymptotically free QFT in four dimensions (thus, not using AdS/CFT and other string-inspired tools \cite{Aharony:1999ti}) that is under theoretical control. In the case at hand, a weak-coupling semiclassical expansion using objects defined in the UV theory is valid and the physics is weakly coupled at all scales, all the way from the UV to the IR.
This  makes the $R^3 \times S^1$ setup distinct from the few analytical approaches on $R^4$ that do not involve uncontrolled approximations, notably  Seiberg-Witten theory \cite{Seiberg:1994rs}. This is Yang-Mills theory with extended  ${\cal{N}}=2$ supersymmetry, with a soft mass term preserving minimal ${\cal{N}}=1$  supersymmetry. Here, owing to supersymmetric dualities, the IR physics has a weakly coupled description and confinement can be shown to be due to the condensation of monopole or dyon particles, in a kind of dual Higgs mechanism.
 Other examples, also in theories with (extended) supersymmetry, describe a type of confinement  dual to the confinement of electric charges:  the confinement of monopoles via nonabelian strings, see e.g. \cite{Gorsky:2007ip} and references therein.

The scope of these lectures is as follows: we  study the $R^3 \times S^1$ approach to  nonperturbative physics on the example of four-dimensional $SU(2)$ nonabelian gauge theory with a number of Weyl fermions, denoted by $n_f$, in the adjoint (or vector) representation of the gauge group. The theory is asymptotically free for $n_f \le 5$. The fermions can be taken massless, or be given gauge invariant Majorana masses. The masses of the different flavours can be taken different, but for our discussions here we shall assume them to be of the same order and  denote their    overall scale  by $m$. As  mentioned before, the class of theories with calculable nonperturbative dynamics is larger, but we stick to $SU(2)$ for pedagogical reasons.\footnote{While there are many interesting observations to be made upon replacing $SU(2)$ with $SU(N)$, there is already more than enough material to cover; thus, introducing more group theory is   left for future self-study.}

When formulated on $R^4$, analytical approaches to the nonperturbative physics determining the vacuum structure and symmetry realization are not available, due to the strong-coupling infrared problem. Here, we shall study the above class of theories, but formulated on $R^3 \times S^1$, with periodic boundary conditions for the fermions. Thus, our $S^1$  is not a  thermal circle (which would require anti-periodic boundary conditions for the fermions) but   a  compact spatial direction\footnote{A purist would then insist that we refer to $R^{1,2} \times S^1$ instead, but since most of our studies will be Euclidean, for brevity we  stick with   $R^3 \times S^1$.} of circumference $L$.
As we shall explain, calculability is ensured by taking the size of the $S^1$ circle to be smaller than the inverse of the strong coupling scale of the four-dimensional theory $\Lambda$, often called the ``confinement scale.''

Before we continue, we note that the idea of the ``femto-universe,'' where Yang-Mills theory is considered in a small volume, was first put forward by Bjorken long ago \cite{Bjorken:1979hv}.\footnote{Incidentally, Bjorken is also the one rumoured to have coined the phrase ``voodoo QCD,'' presumably to characterize approaches to nonperturbative physics whose validity is not a priori clear and is hard to justify (except for agreement with a set of data, often judiciously chosen). Jokes aside, while modelling the nonperturbative dynamics has its raison d'\^ etre, our emphasis here is on  studying gauge theories from first principles and without uncontrolled approximations, hence we stay far from such approaches. } He envisaged, however, taking all three directions of space   smaller than the confinement scale, of order $\sim 1$ Fermi, hence the name. Significant effort has gone into studying the physics of Yang-Mills theories on small spatial three-tori   since \cite{Luscher:1982ma}, see the review \cite{vanBaal:2000zc}. It turns out that formulating the perturbative and semiclassical expansion on a small $T^3$ is quite complex and difficult to handle. In addition there is a center-symmetry breaking phase transition separating the small-$T^3$ theory from the infinite volume one. However, it also turns out that taking two of the dimensions of the $T^3$ be infinite, thus considering  the limit $R^1 \times T^3 \rightarrow R^{1,2} \times S^1$, brings, in a large class of theories, great advantages with respect to calculability, at least to leading order in the semiclassical expansion.\footnote{Due to important insight from the mid to late 1990's, see \cite{Lee:1997vp,Kraan:1998pm} and Section \ref{sec:from3to3x1}.} A  center-symmetry breaking phase transition upon compactification can also be avoided.\footnote{\label{added2024}{A  remark regarding earlier developments  not mentioned in the published version of this review is due.  It was shown in the early 1990s that imposing 't Hooft twisted boundary conditions ensures a smooth transition between small and large $\T^3$  in an $\R \times \T^3$ compactification. In addition, at small-$\T^3$, the theory is semiclassical and an area law for Wilson loops can be shown as being due to fractional instantons: while an explicit analytic form for these $\R \times \T^3$ solutions is not known, it was shown in \cite{RTN:1993ilw}, by a combination of numerical and analytic studies, that a dilute gas of such objects disorders Wilson loops (wrapped around circles in $\T^3$ and separated along $\R$) and  leads to an area law. Furthermore, continuity of the string tensions to the infinite volume limit was demonstrated in \cite{Gonzalez-Arroyo:1995ynx}. A recent review of these developments, including also  references to more recent work on the subject, is in \cite{Gonzalez-Arroyo:2023kqv}. I thank Antonio Gonz\' alez-Arroyo for discussions on this point.}}

It was first realized by \" Unsal \cite{Unsal:2007jx,Unsal:2007vu} that the nonperturbative physics of these theories at  $L \Lambda \ll  \pi$ can be studied  in a weak-coupling semiclassical expansion. One can show that a nonperturbative mass gap for the gauge fluctuations is generated,  a confining string can be seen to form when charged probes are inserted in the vacuum, and, in the theories with  chiral symmetries, one can study their realization in the ground state and show their spontaneous breakdown.  In addition, a center-symmetry breaking phase transition, typically expected after compactification, can be avoided and one can argue for a smooth connection to the $R^4$ theory of ultimate interest.

The space of $SU(2)$ theories with adjoint fermions, defined by the parameters $\Lambda, n_f, m$, and the $S^1$ size $L$, can be roughly split into three classes, each with distinct calculable dynamics at $L \Lambda \ll  \pi$. We enumerate them below in the order we  study them in these lectures. We introduce our jargon (dYM, QCD(adj), SYM) and also briefly advertise their main properties that shall be explained below:

 \begin{enumerate}
\item {\bf dYM}, or deformed Yang-Mills theory \cite{Unsal:2008ch,Shifman:2008ja}: for our purpose, this theory is defined in a UV complete manner as Yang-Mills theory with two or more massive adjoint fermions, whose mass is  taken $m \sim 1/L$. The fermions decouple from the physics at energy scales $\ll 1/L$, but leave an important imprint: center stability and abelianization, ensuring calculability of the long-distance physics. Confinement and the $\theta$-angle  dependence can be analytically studied. 
Notice that dYM is in the universality class of pure Yang-Mills theory. It is believed to be continuously (i.e. without phase transition) connected to the $R^4$ pure Yang-Mills theory. The agreement with lattice data, particularly regarding topological properties is quite remarkable.
\item {\bf QCD(adj)}, or adjoint QCD, is the theory with $2 \le n_f \le 5$ massless adjoint Weyl fermions. The long distance physics of confinement and chiral symmetry breaking at $L \Lambda \ll  \pi$ is calculable \cite{Unsal:2007jx}. The study of this theory highlights the importance of novel topological excitations in the dynamics of confinement and chiral symmetry breaking, the so-called ``magnetic bions.'' The structure of confining strings is markedly different from that in dYM, reflecting certain recently found ``generalized'' 't Hooft anomalies.\footnote{``Generalized 't Hooft anomalies'' is a topic that we can not cover in any detail here, see \cite{Gaiotto:2014kfa,Gaiotto:2017yup,Gaiotto:2017tne}.}
\item {\bf SYM}, or supersymmetric-Yang-Mills, is the same as QCD(adj), but with $n_f = 1$.  This theory stands out because it automatically has ${\cal{N}}=1$ supersymmetry (for $m=0$ only). Thus, it can be studied using the powerful tools of holomorphy introduced by Seiberg \cite{Seiberg:1994bp}. In addition,  just like QCD(adj), it also becomes semiclassical at small $L$, allowing for a calculation of the mass gap and confinement properties, not accessible to the ``power of holomorphy.''
 It turns out that the structure of confining strings in SYM is like that in QCD(adj), for similar anomaly-related reasons.  
 Most tantalizingly,  in the semiclassical limit one finds    a novel kind of topological excitation \cite{Poppitz:2011wy,Argyres:2012ka,Unsal:2012zj,Poppitz:2012sw} the so-called ``neutral bions,'' ensuring center-stability and abelianization.\footnote{These topological excitations highlight the need for  analytic continuation of path integrals and are, ultimately, relevant for the idea of ``resurgence,'' also outside the topic of these lectures (see Section \ref{sec:neutralbions} for motivation and brief discussion).}
\end{enumerate}
Obviously, by varying the mass parameters of the fermions (i.e.  taking some masses to be   $\gg 1/L$ allows for the decoupling of flavours) one can arrange for interesting renormalization group flows between the above theories. In some cases, one can also make conjectures about the nature of the thermal deconfinement transition in pure Yang-Mills theory which can be confronted by lattice data (we shall consider an example in these notes). 

\hfill\begin{minipage}{0.85\linewidth}
\textcolor{red}{{\flushleft{\bf Summary of \ref{sec:whatabout}:}} {The answer to the question posed in the title of this section is: we shall explain in some detail how the nonperturbative properties listed above arise in the $L \Lambda \ll \pi$ limit. }}
   \end{minipage}
\bigskip

\subsection{Why   study small-$L$ theories? In lieu of conclusion.}
\label{sec:whybother}

Skeptical students (and colleagues) often ask: ``You are going to tell us how to study gauge theories on a small circle. To boot, you study theories with unphysical---adjoint, not fundamental as in the standard model---fermion representations. This is definitely not the real world. Why bother?'' This is a fair question: if your interest is mainly in a calculation of hadron spectra that can be directly compared with experiment, stop reading; if you are interested in the inner workings of QFT, stay on.  As my goal is  not  to (over-) sell you anything,   I will simply tell you why I think this topic is interesting.

First, as we shall see, the stories I will tell about confinement, chiral symmetry breaking, and the thermal deconfinement transition are rather elegant. It is a rare luxury to be able to make statements about the nonperturbative phase structure of a locally four dimensional quantum field theory. To a theoretical physicist, this alone is quite satisfying. It makes the study intrinsically worthwhile and fun.

Second, it can be expected that, while honestly studying a theoretically controlled regime, one may encounter surprising new features that are more generally valid. The recent resurgence of interest, reviewed in \cite{Dunne:2016nmc}, in ``resurgence in QFT'' (studying the nature of the divergent perturbative series in QFT and their resummation) did, in fact, arise from these  small-$L$ studies. This was due to, among others, the peculiar nature of the magnetic and neutral bion topological molecules. Other unusual features observed in the calculable regime, in addition to the many novel and strange topological excitations (Sections \ref{sec:bionstructure} and \ref{sec:neutralbions}),  include the appearance of doubly-exponential nonperturbative effects  (Sections \ref{sec:dymvacua} and \ref{sec:symvacuum}) and the emergence of latticized dimensions in the abelian large-$N$ limit (Section \ref{sec:summary} and \cite{Cherman:2016jtu}).

Third, the theoretically-controlled study of the symmetry realization at small $L$ has led to the conjectured existence of novel possible phases of various theories on $R^4$ (Sections \ref{sec:chiralqcdadjphasesj} and \ref{sec:summary}). Admittedly, one needs dedicated numerical studies to confirm or refute these conjectured phases. Further, we can also use SYM with soft-breaking mass on $R^3 \times S^1$ to  make conjectures about the nature of the thermal phase transition in  pure YM theory. These are borne out by available numerical simulations (Section \ref{sec:thermalsym}).

Fourth, the small-$L$ theories offer an interesting arena to study the realization, at various energy scales  of the recently discovered 't Hooft anomalies involving  traditional, or ``0-form,'' global symmetries as well as higher-form symmetries; their implications are discussed in Sections \ref{sec:doublestringdYM}, \ref{sec:chiralqcdadj}, \ref{sec:symvacuum}, and \ref{sec:summary}. One's hope, then, is that this will lead to better understanding of these anomalies and their implications.

Finally, I will also mention that theories with fundamental fermions have been incorporated in an interesting way. They exhibit various nonperturbative properties of real QCD in a calculable setup. At small $L$, chiral symmetry is broken by the expectation value of a monopole operator and one can argue that a chiral  phase transition between the small and large $L$ theories is absent. We  explain this in detail, for $SU(2)$ QCD(F)  (Section \ref{sec:fundamental}).

\hfill\begin{minipage}{0.85\linewidth}
\textcolor{red}{{\flushleft{\bf Summary of \ref{sec:whybother}:}} The calculable nonperturbative dynamics on $R^3 \times S^1$ offers a rare theoretical opportunity to analytically study nonperturbative phenomena in 4d gauge theories. My answer to the question posed in the title of this Section is that this alone makes these explorations worthwhile, as they extend our understanding of the nonperturbative properties of quantum field theories in ways that are not always obvious from the start. } \end{minipage}

\bigskip

{\flushleft{\bf In lieu of conclusion:} Our hope is that these notes give the necessary background, collected all in one place, to help the interested reader through the literature on $R^3\times S^1$ compactifications. The main emphasis is on introducing the ideas and techniques  used to study the small-$L$ calculable regime in a range of different theories, focusing on  IR energy scales $\mu \ll 1/L$ and exhibiting the physics of confinement and (chiral) symmetry breaking. On a few occasions,\footnote{As in Sections \ref{sec:cfc13} and  \ref{sec:dymvacua}. We stress that despite the fact that the small-$L$ theory is weakly coupled at all scales, there are situations (as mentioned in Section \ref{sec:doublestringdYM}) where accounting for this backreaction is a nontrivial open problem.}  the backreaction of the IR physics on the UV modes of mass $1/L$ will be also mentioned. As the reader will no doubt notice, the continuity of the symmetry realization towards large $L$ is largely conjectural, with evidence  based on  comparisons with lattice data or on other expectations, such as consistency with anomalies. Notably a theoretical proof of (the absence of) large-$L$/small-$L$ continuity in most of the theories we discuss, apart from SYM, is not known. Clearly, there is room to advance our understanding.}

\subsection{Philosophy and a reader's guide.}

These notes assume some background knowledge. This includes the basics of non-abelian gauge theory and asymptotic freedom as well as some familiarity with 't Hooft-Polyakov monopoles, chiral anomalies, instantons, and the dilute instanton gas approximation, in quantum mechanics and on $R^4$ (references covering some of this material in more depth are provided throughout). 
A few general remarks regarding the philosophy behind these notes are due:
\begin{enumerate}
\item As the notes are rather long, every subsection ends with a paragraph summarizing  the main results (highlighted in \textcolor{red}{red}). This should help the returning reader while allowing those familiar with the subject of a given Section to quickly review its content.
\item We have tried to balance hand-waving explanations and careful derivations, often leaning towards the former. We feel that keeping in mind the order of magnitude and the leading parametric dependence of the physical quantities is more important for a qualitative understanding than the precise numerical factors.\footnote{We did, however, make every effort to have the correct numerical coefficients when they really matter---in discussions of charge quantization, monodromies, and other topological features.} This attitude will be pervasive on many occasions in these notes and various factors we omit can be found in the literature.

\item Throughout the notes, some straightforward technical derivations are relegated to exercises, intended to improve the reader's appreciation of the topics. These are especially recommended to those encountering them for the first time. However,  the notes can be read  without solving the exercises, by simply understanding (and trusting) the statements they make. 
\end{enumerate}
To help the readers---who might have vastly different backgrounds and interests---through these long notes, we now include some guidance of how one may go through the various Sections. 
We stress that to understand the post-2010 developments, familiarity with background discussed in two of the eight Sections of these notes, Sect.~\ref{sec:polyakov} and \ref{sec:from3to3x1},   is an absolute necessity. Due to their importance they take about half of the space.

The first is Section \ref{sec:polyakov}, where we discuss in  detail Polyakov's confinement on $R^3$, known since the 1970's. 
 
 The second is  Section \ref{sec:from3to3x1}, devoted to notions specific to $R^3\times S^1$. These include the holonomy, center symmetry, its relation to confinement/deconfinement, as well as the holonomy (or ``GPY'') potential, familiar from the 1980's. These are discussed in Sections \ref{sec:holonomyandcenter} and \ref{sec:perturbative3x1}.  A newer crucial development on $R^3 \times S^1$ is the  ``dissociation'' of an instanton into its monopole-instanton constituents, the ``M'' and ``KK'' monopole-instantons (for $SU(2)$),  discovered in the 1990's and explained in Section \ref{sec:bpsandkkmonopoles}. The final crucial set of ideas, from the 2000's, is the perturbative center-symmetry stabilization due to massive or massless adjoints, the subject of Section \ref{sec:adjointgpy} and Appendices \ref{appx:gpy} and \ref{appx:gpymass}.  
 
Readers familiar with the subjects of Sections \ref{sec:polyakov} and \ref{sec:from3to3x1} can proceed directly to the Sections devoted to studying the nonperturbative dynamics of the   classes of theories they may be interested in:
 
 Section  \ref{sec:dym} discusses deformed Yang-Mills theory, adjoint QCD is the subject of Section \ref{sec:adjsym}, and super-Yang-Mills is in Section \ref{sec:softthermal}. The many topics discussed for each class of theories can be found in the table of contents.
  
Section \ref{sec:fundamental} is devoted to a detailed description of $SU(2)$
  QCD with fundamental quarks in the context of colour-flavour-center symmetric compactifications. 
  
 Section \ref{sec:summary}  contains much briefer descriptions and references involving all classes of theories discussed here, but for $SU(N)$ gauge groups with $N>2$, as well as other classes of theories and gauge groups.

 \section{Flashback to the 1970's: Polyakov confinement   on $\mathbf{R^3}$.}
 \label{sec:polyakov} 
 
 We begin our journey towards  confinement on $R^3 \times S^1$ by studying a theory in $R^3$. This is the so-called ``Polyakov model''  \cite{Polyakov:1976fu,Polyakov:1987ez}. More recent treatments can be found  in Witten's lectures in vol. 2 of \cite{Deligne:1999qp} and, for example, in the textbooks by Shifman \cite{Shifman:2012zz} and Banks \cite{Banks:2014twn}. As this is not material taught in standard QFT courses,  we include an extended discussion,  intended to be self-contained. (The familiar reader can skip this Section and move to Section \ref{sec:from3to3x1} discussing $R^3 \times S^1$.)
 
 The Polyakov model is an $SU(2)$ gauge theory with a real adjoint scalar field in three dimensional spacetime.\footnote{On $R^4$, this theory is known as the bosonic sector of the ``Georgi-Glashow model,'' a pre-cursor of modern electro-weak  theory, which lacks the $Z$-boson mediated neutral currents.}  
 The gauge field is $A_\mu^a$.\footnote{See Appendix \ref{appx:notation} for a summary of our notation and a few warnings about possible confusions. } The real scalar field in the adjoint representation of $SU(2)$ shall be labeled $A_4^a$, with some hindsight aimed towards our future $R^3 \times S^1$ study. The theory has the Euclidean Lagrangian:
 \begin{equation}\label{polyakovlagrangian}
 L = {1 \over 4 g_3^2} F_{\mu\nu}^a F^{\mu\nu \; a} + {1 \over 2 g_3^2} (D_\mu A_4)^a (D^\mu A_4)^a + {\lambda \over g_3^2}(A_4^a A_4^a - v^2)^2~,
 \end{equation}
 where the 3d gauge coupling $g_3^2$ has dimension of mass and the gauge field $A_\mu^a$ and scalar $A_4^a$ both have mass dimension one. We have normalized the entire Lagrangian so that the action $S = \int\limits d^3 x L$ has an overall factor $1/g_3^2$. As $g_3^2$ is dimensionful, the role of $\hbar$, the semiclassical expansion parameter is played by a ratio $E/g_3^2$, where $E$ is a relevant energy scale, the nature of which shall be revealed below. Finally, $\lambda$ is a dimensionless coupling, which shall be assumed to be ${\cal{O}}(1)$ or smaller (also see further below).
 
 \subsection{Perturbative analysis of the IR theory.}
\label{sec:pert_polyakov}
 
 The potential for the adjoint (triplet, or vector) scalar $A_4^a$ is similar to that found for the Higgs field in the standard model. To minimize the energy, we let $A_4$ have an expectation value (vev), which we take along the 3rd ``isospin'' direction: 
 \begin{equation}\label{polyakovvev}
 \langle A_4^a \rangle = \delta^{a3} v~.
 \end{equation}
  Expanding around the vev for $A_4$ it is easy to see that the gauge symmetry is broken, $SU(2) \rightarrow U(1)$. For the choice of vev we have made, the massless gauge boson is $A_\mu^3$, while $A_\mu^1 \pm i A_\mu^2$ are massive $W^\pm$ bosons. Their mass, as can be seen   from (\ref{polyakovlagrangian}) is of order $v$. The $A_4^1$ and $A_4^2$ components of the triplet scalar are ``eaten'' by the massive gauge bosons, while the fluctuation of $A_4^3$ around the vev (\ref{polyakovvev}), the radial component of the higgs field, has mass which depends on the parameter $\lambda$ and is of order $\sqrt{\lambda} v$. For $\lambda \sim 1$, this is of the same order as the mass of the $W$-bosons, while taking $\lambda \ll 1$ makes for a hierarchy in the massive spectrum.

Next, observe that in this model, there are two dimensionful parameters: the scale of the vev (\ref{polyakovvev}), $v$, and the scale set by the 3d gauge coupling,  $g_3^2$. In our classical analysis, the scale $v$ sets the mass of the heavy $W$-bosons and radial higgs mode $A_4^3$. We shall call the vev $v$  the scale of ``abelianization,'' i.e. the scale below which the nonabelian degrees of freedom decouple and the only relevant (here: massless) degree of freedom is that of the  $U(1)$ ``photon'' $A_\mu^3$.

If we had set the  $v^2$ to zero (or taken it negative), we wouldn't have had a ``Mexican hat'' potential for the triplet scalar and the theory would remain  nonabelian  at all scales. What, then, is the IR dynamics expected of such an $SU(2)$ theory (+ adjoint scalar) in 3d? Based solely on dimensional analysis, one expects that at scales of order $g_3^2$, the 3d theory becomes strongly coupled. Such behaviour is typical in lower dimensional theories, where the couplings have positive mass dimension. These represent relevant parameters whose importance grows in the IR. Formally, as in any theory with dimensionful coupling constant, one can define a dimensionless coupling  using an appropriate power of the relevant energy scale $\mu$. In our case, as $g_3^2$ has dimension of mass (and   is, by standard power-counting,  a relevant coupling) the dimensionless combination is  $g_3^2 
\over \mu$,  showing that this coupling grows in the IR and becomes large when $\mu \sim g_3^2$. Those  not familiar with these sort of  arguments should follow the  more explicit discussion between eqns.~(\ref{polyakoveffective}) and (\ref{polyakovcoupling}) below. 

 However, if the scale $v$ is taken to be $v \gg g_3^2$, the abelianization of the theory stops  the running of the coupling towards large values in the IR. At energies less than $v$ the theory is free, as there is only the $A_\mu^3$ photon and there are no $U(1)$-charged particles that are light.\footnote{Notice that this is crucial in 3d where a $U(1)$ theory with light charged particles becomes strongly coupled in the IR.} 
The point we shall make more explicit below is that our analysis of the spectrum above is sensible only in the $v \gg g_3^2$ limit. Thus, the dimensionless expansion parameter in our theory is expected to be ${g_3^2 \over v} \ll 1$. Its smallness guarantees the validity of the (essentially classical) analysis above.

 \begin{figure}[ht]
 \centerline{
\includegraphics[width=12cm]{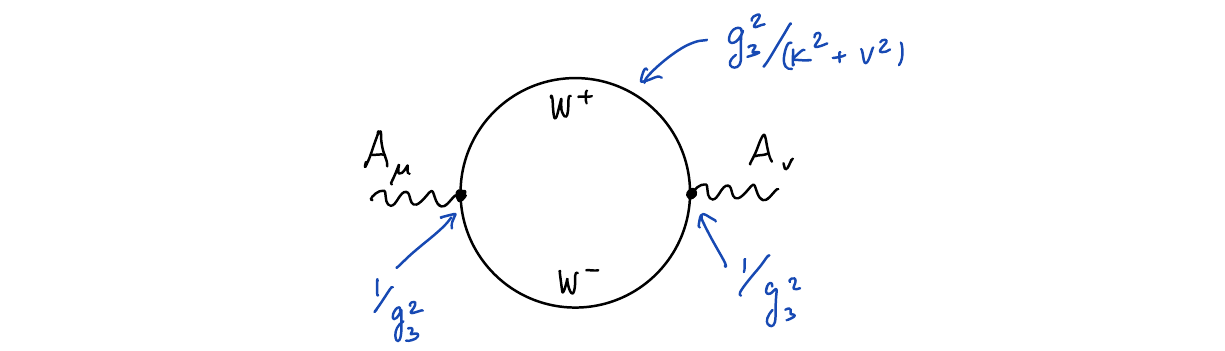}}
\caption{The one-loop heavy $W$-boson contribution to the photon kinetic term. From (\ref{polyakovlagrangian}), all vertices come with a factor ${1\over g_3^2}$ and the $W$-boson propagator is $\sim {g_3^2 \over {k^2 + v^2}}$. Thus, the loop integral scales as $  \int {d^3 k \over (k^2 + v^2)^2} \sim {1\over v}$, producing the   second term in  $L_{eff}$ of (\ref{polyakoveffective}), with a calculable coefficient $C$. (For this estimate, the momenta at the interaction vertices can  be taken to act on the external $A_\mu^3$ to produce the field strength in (\ref{polyakoveffective})). \label{fig1}}
\end{figure} 

Let us elaborate on the above two paragraphs. To this end, we  write an effective Lagrangian governing the physics at $\mu \ll {\rm min} (v, \sqrt{\lambda} v)$. We claim that it has the schematic form:
\begin{equation}\label{polyakoveffective}
L_{eff} = {1 \over 4 g_3^2}(F_{\mu\nu}^3)^2 \left[1 + C {g_3^2 \over v}\right] + \ldots~,
\end{equation}
where $C$ is a numerical coefficient.
The overall term multiplying the square brackets   is self-explanatory:  we simply kept the kinetic term for the massless photon $A_\mu^3$ from (\ref{polyakovlagrangian}) and ignored the terms involving massive fields (this is equivalent to the ``tree-level'' integrating out the heavy fields; here, it amounts to crossing them out). The second term in the brackets can be seen as the result of the calculation of the  one-loop graph shown on Fig. \ref{fig1} (naturally, there are higher loop corrections not shown),
and the dots in (\ref{polyakoveffective}) denote higher-dimensional terms suppressed by powers of the $W$-boson mass $v$ (e.g. $\sim (F_{\mu\nu}^3)^3$, not showing the Lorentz index contractions), which  are irrelevant at $\mu \ll v$.\footnote{This is similar to the Heisenberg-Euler lagrangian in QED describing the photon self-interactions at energies below  the mass of the electron.} 
 The precise calculation of the second term in the brackets requires some care. However, simple  dimensional analysis suffices to guess the parametric scaling of the operator given in the term in $L_{eff}$  multiplied by $C$. The normalization of our lagrangian (\ref{polyakovlagrangian}) determines the scaling of vertices and propagators and thus the final scaling of the one-loop result, as explained in the caption of Fig.~\ref{fig1}.

 To conclude, the dimensional analysis explained in Figure \ref{fig1} determines the form of the term in square brackets in (\ref{polyakoveffective}). It tells us that the $U(1)$ gauge coupling in the effective theory of the massless modes is of the form 
\begin{equation}\label{polyakovcoupling}
{1 \over g_{eff}^2} = {1\over g_3^2} + {C \over  v} + \ldots~,
\end{equation}
 given by the sum of a tree-level term and the loop corrections. It is clear from (\ref{polyakovcoupling}) that perturbation theory breaks down whenever the second term becomes bigger than the first,\footnote{The precise calculation reveals the true behaviour: the sign of   $C$   is negative. Taken at face value, it implies that the coupling of the massless photon blows up at a finite value of $v$, indicating a breakdown of the abelian effective field theory. This is another way to see the relevant nature of the gauge coupling in the 3d Yang-Mills theory. See \cite{Smilga:2004zr} for a calculation in a supersymmetric 3d theory. For a nonsupersymmetric theory, this behaviour can  be inferred from the $R^3 \times S^1$ calculation in QCD(adj)  \cite{Anber:2014sda}, after throwing out the contributions of adjoint fermions and Kaluza-Klein modes on the circle.} i.e. when $v \ll g_3^2$. Conversely, as advertised earlier, when $v \gg g_3^2$ the one-loop correction is small and the semiclassical (for now: perturbative) approximation works, at the usual physicist level of rigour. 
As already declared, this is the limit we shall be exploring to study the  physics of the Polyakov model semiclassically.

\hfill\begin{minipage}{0.85\linewidth}
\textcolor{red}{{\flushleft{\bf Summary of \ref{sec:pert_polyakov}:}} The upshot of our analysis here is that the perturbative IR physics of the Polyakov model (\ref{polyakovlagrangian}), taken with $v \gg g_3^2$, is rather boring: there is a free massless photon, which has irrelevant self interactions due to the ``$\ldots$'' terms in (\ref{polyakoveffective}). However, as we shall see in the following sections, nonperturbative effects due to (\ref{polyakovlagrangian}), still calculable  in the semiclassical limit, completely change the IR behaviour. A mass gap  for the perturbatively free IR $U(1)$ theory is generated nonperturbatively, giving rise to confinement of probe electric charges.}
\end{minipage}

\bigskip

 \subsection{Finite action monopole-instantons in the Polyakov model.}
\label{sec:instantons_polyakov}

The striking fact about the Polyakov model is that the simple free-photon lagrangian (\ref{polyakoveffective}) does not give a correct account of the IR physics. Nonperturbative effects, not accounted for by any finite number of  loop diagrams, like that on Fig.~\ref{fig1}, completely change the long distance physics. The goal of this section is to discuss the objects responsible for this change of IR behaviour:  the ``monopole-instantons'' of the Polyakov model (\ref{polyakovlagrangian}).  It is their proliferation in the vacuum that ultimately causes confinement of fundamental charges.

The monopole-instantons are finite Euclidean action solutions of the equations of motion of the Polyakov model (\ref{polyakovlagrangian}), akin to the famous Belavin-Polyakov-Schwarz-Tyupkin (BPST) instantons in $R^4$. In order to exhibit them in a  straightforward manner, we shall work in the $\lambda= 0$ limit, where $A_4^a$ is a scalar field without a potential and thus $v$ is arbitrary. Studying this limit is, strictly speaking, not necessary in the purely 3d theory. However, in our $R^3 \times S^1$ study, the physics of the compactified 4d theory will force us into a  regime where the scalar potential is small, i.e.  $\lambda \ll 1$.  The $\lambda= 0$ limit is known as the ``Bogomolny-Prasad-Sommerfield,'' or BPS, limit, known to simplify the study of monopole-instantons.\footnote{As stated earlier, some familiarity of the reader with 't Hooft-Polyakov monopoles is assumed here and our discussion may appear terse. For a review of BPS monopoles, see  Harvey's lectures \cite{Harvey:1996ur} (our monopole-instantons are related to the finite-energy solitonic monopole  solutions on $R^{1,3}$ discussed there by forgetting the time direction). A pedagogical reference that also goes beyond the BPS limit is E. Weinberg's book \cite{Weinberg:2012pjx}.} (As an aside, this limit is natural in supersymmetric theories, where a vanishing potential, $\lambda=0$, is perturbatively stable.) 

In the $\lambda = 0$ limit, we can simplify matters further if we write (\ref{polyakovlagrangian}) in a ``4d'' form as follows. Let us introduce indices $M,N=1,2,3,4$ and extend the Euclidean metric to the usual 4d one. Then, observe that
\begin{equation} \label{bps3dlagrangian}
L = {1 \over 4 g_3^2} F_{\mu\nu}^a F^{\mu\nu \;a} + {1 \over 2 g_3^2} (D_\mu A_4)^a (D^\mu A_4)^a =
{1 \over 4 g_3^2} F_{MN}^a F^{MN \; a}~,
\end{equation}
where $F_{MN}^a = \partial_M A_N^a - \partial_N A_M^a + i ([A_M, A_N])^a$, and $F_{\mu 4}^a = (D_\mu A_4)^a$, i.e. $\partial_4 \equiv 0$, the dependence on the fictitious (for now) 4th dimension is neglected. Again, I stress that this 4d terminology is simply convenient for the study of the $R^3$ theory, where no $\partial_4$ appears, and will only become indispensable once we transition to $R^3 \times S^1$. Further, let us also introduce the components of the field-strength tensor:
\begin{equation} \label{eandb}
B_\mu^a = {1 \over 2} \epsilon_{\mu\nu\lambda} F^{\nu \lambda \; a}~,~~ E_\mu^a = (D_\mu A_4)^a = F_{\mu4}^a~,
\end{equation}
and call these fields ``magnetic'' and ``electric,'' respectively ($\epsilon_{123} = 1$).  Notice that the names we attach to the above $B$ and $E$ are a convenient choice of words and should not be taken literally, as these are Euclidean $R^3$ fields. If we go to Minkowski space, $R^{1,2}$, by say, declaring $x^1$ to be the Euclidean version of time, $B_1$ will represent the true magnetic field (a pseudoscalar $F_{23}$ in 3d) while $B_2$ and $B_3$ are actually electric fields, related to $F_{12}$ and $F_{13}$ by (\ref{eandb}). On the other hand, $E_\mu$ is simply the covariant derivative of a scalar.\footnote{There exists confusing nomenclature in the literature related to taking the ``electric'' and ``magnetic'' labels attached to $E$ and $B$ seriously. The reason for why these names stuck will become clear shortly. It would be nice to banish them, but they have by now been deeply ingrained, so we use them but remember the context. } Now, using (\ref{eandb}) we can rewrite (\ref{bps3dlagrangian}) (for brevity, we  do not distinguish upper and lower indices, all are assumed to be lowered/raised by the unit metric) as follows
\begin{equation}
L = {1 \over 2 g_3^2}\left[ (B_\mu^a)^2 + (E_\mu^a)^2 \right] =  {1 \over 2 g_3^2}\left[ (E_\mu^a \mp B_\mu^a)^2 \pm 2 E_\mu^a B_\mu^a \right]~.
\end{equation}
The above equation, due to the positivity of $ (E_\mu^a \pm B_\mu^a)^2$, implies that the (positive definite) Euclidean action obeys an inequality known as the ``BPS bound'' 
\begin{equation}\label{bpsbound}
S = \int d^3 x L \ge \pm {1 \over g_3^2} \int d^3 x E_\mu^a B_\mu^a  \equiv S_{BPS}~.
\end{equation}
What is interesting about these manipulations? Suppose we are looking for finite Euclidean action solutions of the equations of motion of (\ref{bps3dlagrangian}), i.e. for instantons. Finiteness of the action requires that $E^2$ and $B^2$,  and hence, $(E \pm B)^2$ and $E \cdot B$ are all integrable on $R^3$. Then, (\ref{bpsbound}) implies that 
the action will reach its minimum value $S_{BPS} = {1 \over g_3^2} \int d^3 x |E_\mu^a B_\mu^a|$ whenever $E = \pm B$. Solutions of the equations of motion of (\ref{bps3dlagrangian}) which obey $E_\mu^a = B_\mu^a$ are called ``self-dual,'' while the ones where $E_\mu^a = - B_\mu^a$ are the ``anti-self-dual'' ones. Further, we notice that $E_\mu^a B_\mu^a$ is a total divergence, hence (taking the positive sign for definiteness and denoting by $d^2 s^\mu$ the surface element)
\begin{equation} \label{bpsaction}
S_{BPS} = {1 \over g_3^2} \int d^3 x E_\mu^a B_\mu^a = {1 \over g_3^2} \int d^3 x \partial_\mu ( A_4^a B_\mu^a) = {1 \over g_3^2} \oint\limits_{S^2} d^2 s^\mu A_4^a B_\mu^a,
\end{equation}
showing that $S_{BPS}$ can be written as an integral over an $S^2$ at Euclidean space-time infinity. Thus, the minimum value of the action is determined by boundary conditions at infinity, determined by the behaviour of the Higgs field vev $v$ and ``magnetic'' field $B_\mu^a$ at infinity.

Let us concentrate on the simplest self-dual Euclidean solutions of finite action, the spherically symmetric ones. We shall now give their form and in later Sections discuss their relation to the static 't Hooft-Polyakov monopole solutions on $R^{1,3}$ (it is, in fact, this relation that has made the name ``monopole-instantons'' stick to the instanton solutions of the 3d Polyakov model). 
The solutions are known as the ``BPS monopoles.'' The solution centered at the origin $r=0$ of $R^3$ is
\begin{equation}\label{bpsmonopole1}
A_4^a = - n_a v P(vr)~, ~~A_\mu^a = \epsilon_{a \mu\nu} n_\nu {1- A(vr) \over r}~.
\end{equation}
Here, $n_a = {r^a \over r}$ is a unit vector in $R^3$, and you should imagine that all indices are lifted and lowered with   Kronecker delta's. The functions $P(x) = \coth x - {1 \over x} \rightarrow 1 - {1 \over x}$, $A(x) = {x\over \sinh x} \rightarrow {\cal{O}}(x e^{-x})$, where both limits are as $x \rightarrow \infty$. 
For the BPS self-dual monopole-instanton (\ref{bpsmonopole1}), the field strength can be straightforwardly calculated. It is given in terms of two functions
$F_1(v,r) = {v^2 \over \sinh vr} \left( {1 \over vr} - \coth vr \right) \rightarrow v^2 {\cal{O}}(e^{- vr})$ and $F_2 (v,r) =  {v^2 \over \sinh^2 vr} - {1 \over r^2} \rightarrow -{1 \over r^2}$, so that 
\begin{equation}\label{bpshedgehogstrength}
B_\mu^a = E_\mu^a = (\delta_{\mu a} - n_\mu n_a) F_1(v,r) + n_\mu n_a F_2(v,r) \rightarrow -  { n_\mu n_a \over r^2}~ {\rm as} ~ r \rightarrow \infty~.
\end{equation}
Notice that the solution has a characteristic core size, of order $v^{-1}$, inside which the field configuration is nonabelian and quite complicated. However, as $r \gg v^{-1}$, the fields drastically simplify, as shown above.
The solution given above is everywhere regular and is known as the ``hedgehog'' gauge solution. This is due to the fact that the isospin orientation of the $B$ field at spatial infinity varies as a function of direction.
The action of the BPS monopole is easily computed from (\ref{bpsmonopole1},\ref{bpshedgehogstrength}) and (\ref{bpsaction}) as an integral over spatial infinity
\begin{equation}\label{bpsaction2}
S_{0} = {1 \over g_3^2} \oint\limits_{S^2} d^2 s^\mu A_4^a B_\mu^a = {v \over g_3^2} \oint\limits_{S^2} d^2 s^\mu {n_\mu \over r^2} = {4 \pi v \over g_3^2}~.
\end{equation}
Notice that the action is large in the semiclassical limit $v \gg g_3^2$. The anti-BPS solution is given by the same equations (\ref{bpsmonopole1}, \ref{bpshedgehogstrength}), but with $A_\mu^a$  and $B_\mu^a$ taken with opposite sign. 

\begin{quote}
{{\bf Exercise 1}}: Verify that (\ref{bpsmonopole1}) solve the equations of motion of (\ref{bps3dlagrangian}) and that the field strength is given by (\ref{bpshedgehogstrength}).

\end{quote}

\begin{figure}[h]
\centerline{
\includegraphics[width=6.0 cm]{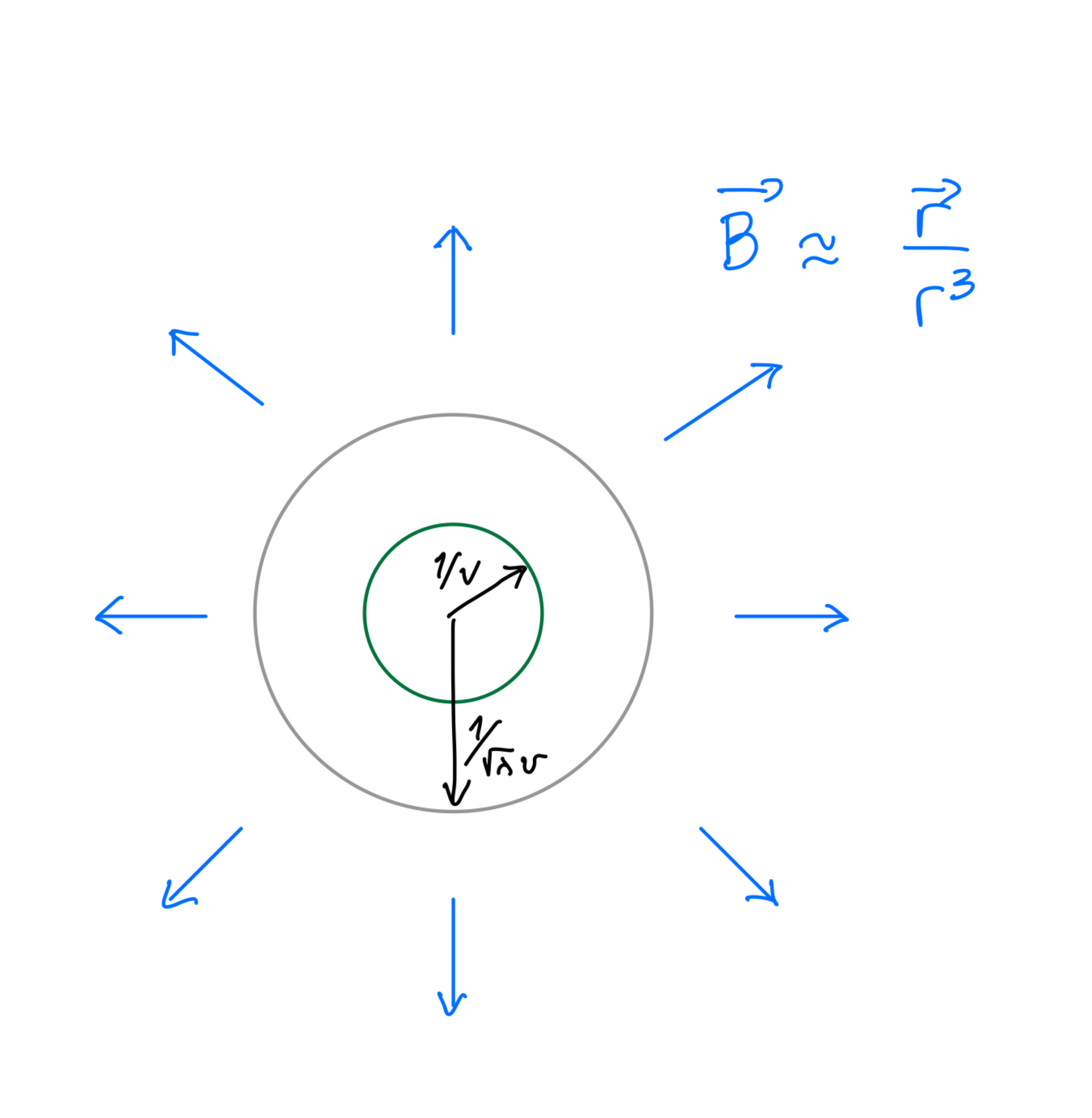}}
\caption{A representation of a monopole-instanton solution in $R^3$, showing the scales characterizing the $\lambda < 1$ spherically-symmetric monopole-instanton: the nonabelian core of size $v^{-1}$, the outside core region of size $(\sqrt{\lambda} v)^{-1}$ with abelian $E_\mu$ field, and the ``long-range''  region $r >  (\sqrt{\lambda} v)^{-1}$ where only an abelian $\vec B$ field in the unbroken-$U(1)$ is present. For our applications, only the long-range region will be relevant, as the monopole-instantons are well separated in the dilute-gas approximation. The action of the $\lambda \ll 1$ solution is approximately the BPS action (\ref{bpsaction}), $S_0 = {4\pi v\over g_3^2}$, and the magnetic charge (\ref{magneticcharge}) is unity. \label{fig:monopole1}}
\end{figure} 

Before we continue, let us also give the large-$r$ asymptotics of the BPS monopole-instanton solution in a gauge where the hedgehog has been combed, the so-called ``string'' gauge.\footnote{The hedgehog can not be combed everywhere without a singularity. One needs to cover $R^3$ by at least two coordinate charts  to describe the string-gauge solution for all $r, \theta, \phi$. We shall not dwell on this as it will not be relevant for us. The formulae of the string gauge solution covering also all $r < v$ in the two coordinate patches can be found in e.g. \cite{Anber:2011de}, along with the gauge transformations relating them to the regular hedgehog solution.}  In this gauge, the $B$ (and $E$) field far away from the core of the monopole-instanton is rotated to point solely in the 3rd isospin direction, so it really resembles the field of a pointlike charge under the unbroken $U(1)$. The $A_4$ and $B$ field, given now as a matrix in the $SU(2)$ Lie algebra, approach, as  $r \gg v$, 
\begin{equation}\label{stringgauge}
A_4(r) \rightarrow{ v \sigma^3 \over 2} ,  ~~B_r \rightarrow {1 \over r^2} {\sigma^3 \over 2},
\end{equation} with exponentially small angular-direction components $B_{\theta, \phi} \rightarrow v^2 {\cal{O}}(e^{- vr})$ (we defined the polar-coordinate components of $B$ via $B_\mu dx^\mu = B_r dr + B_\theta d\theta + B_\phi d \phi$). In this gauge, we can write the action  (\ref{bpsaction}) as 
$S_{0} = {v \over g_3^2} \oint\limits_{S^2} d^2 \vec{s} \cdot \vec{B}^{3} \equiv {4\pi v \over g_3^2} \; Q_m~$, with $Q_m = 1, $
where we defined the ``magnetic charge'' in the usual way, as the integral of the ``magnetic field'' $\vec{B}^{3}$ at spatial infinity
\begin{equation}\label{magneticcharge}
~ Q_m = {1 \over 4 \pi} \oint\limits_{S^2} d^2 \vec s \cdot \vec{B}^{3} ~.
\end{equation}
The minimal action (anti-) BPS solutions have $Q_m= \pm 1$. The $Q_m=1$ solutions given here are also the lowest action ones. Outside of the core (of order $v^{-1}$), the field configuration is determined by the  essentially abelian $B^3_\mu$ (and $E^3_\mu$) field. This is schematically shown on Figure \ref{fig:monopole1}. We stress that the long-range abelian field, the charge, and the action of the monopole-instanton are the main features relevant for our discussion, while the details of the core are inessential.

Here, we presented   these solutions in the BPS limit $\lambda = 0$, where no potential for the $A_4$  field is present. The  BPS monopole-instantons also carry long-range ``electric'' fields $E_\mu^3 \sim n_\mu/r^2$. In the nonsupersymmetric 3d Polyakov model and in most of our future $R^3\times S^1$ applications, $\lambda \ll 1$ is finite, causing $A_4$ to become massive. Denoting the $A_4$ mass by $m_4 \sim \sqrt{\lambda} v$, the corresponding (now non-BPS) monopole-instantons have an $E$ field that dies exponentially away from the core, $E_\mu^3 \sim n_\mu{e^{- m_4 r}/r^2}$, thus they carry no long-range ``electric'' fields. We stress, however, the most relevant feature: the ``magnetic'' field remains long-range, $B_\mu^3 \sim n_\mu/r^2$, also for $\lambda \ne 0$. The action is, as before, given by $S_0$ of (\ref{bpsaction}), with small, for $\lambda \ll 1$, corrections.\footnote{When $\lambda \ne 0$, the monopole-instanton solutions are no longer (anti-) self-dual. Exact analytic forms of the solutions do not exist and they have to be studied numerically or by employing matched asymptotic expansions to find approximate solutions in different regions of space. See \cite{Anber:2011de} for some expressions and references. }  A cartoon, useful for our future applications, of the field configuration at large distances is given in Figure \ref{fig:monopole1}.

\hfill\begin{minipage}{0.85\linewidth}

\textcolor{red}{
{\flushleft{\bf Summary of \ref{sec:instantons_polyakov}:}} The (anti-) BPS monopole-instanton solutions (\ref{bpsmonopole1}) are finite action (\ref{bpsaction}) Euclidean solutions of the classical equations of motion of the $\lambda = 0$ Polyakov model (\ref{polyakovlagrangian}). Most importantly for our application, even for nonzero $\lambda \ll 1$, these instantons carry long-range ``magnetic'' fields $B_\mu^3 \sim n_\mu/r^2$ and thus have ``magnetic'' charges (\ref{magneticcharge}). Their action  $S_0$ is given by (\ref{bpsaction}), up to small-$\lambda$  corrections.}

\end{minipage}

\bigskip

 \subsection{Monopole-instantons and the IR: I. first pass.}
\label{sec:polyakov_1stpass}

Armed with the knowledge of finite action instantons in the Polyakov model we now return to study the IR physics in  the semiclassical $v \gg g_3^2$ regime. Recall from Section \ref{sec:pert_polyakov}
that, perturbatively, the IR theory is the rather boring one of a massless photon and nothing else.

 In the semiclassical limit the theory is weakly coupled and we expect, using our intuition from quantum mechanics, that summing up the contributions of the trivial and nontrivial saddle points of the path integral will give a good guidance to the physics.\footnote{To use quantum mechanics as a motivation, recall the example of the double well potential  \cite{Polyakov:1976fu,Vainshtein:1981wh,Coleman:1985rnk,Polyakov:1987ez,Schafer:1996wv}. The energy of the ground state receives perturbative contributions calculated using an expansion around a minimum. At any finite order, these corrections are blind to tunneling and the existence of the second minimum. There are also nontrivial saddle points: instantons, anti-instantons, instanton-anti-instantons, etc. When summed over, their contributions imply that the true ground state is the symmetric linear combination of  states built around the two minima. The energy splitting between the symmetric and anti-symmetric combinations is also determined. There is a lot of structure here, which in many cases can be made quite precise, see the review \cite{Dunne:2016nmc}, relevant to the fascinating and difficult topic of ``resurgence'' theory (that we can't go into). }  Saddle points are classical solutions extremizing the action functional. Ordinary perturbation theory, as in Section \ref{sec:pert_polyakov}, is an expansion in small fluctuations around the trivial saddle point of the path integral, i.e. around the classical solution $A_4  = v$, $A_\mu = 0$ with action $S=0$.
However,  in  Section \ref{sec:instantons_polyakov} we found that there are other  saddle points of the path integral, the monopole-instantons with action (\ref{bpsaction}) $S_0= {4 \pi v \over g_3^2} \gg 1$. The natural question that arises, then, is whether and how these other saddles affect the IR physics? 

We shall begin answering this question using the Euclidean path integral framework. We  found the nontrivial solutions of the equations of motion, the monopole-instantons with action  $S_0 = {4 \pi v \over g_3^2} \gg 1$. Next, we recall   the equivalence between Euclidean field theory (here obtained from a $d=2+1$ quantum field theory) and classical statistical mechanics (in $d=3$). It implies that the Euclidean path integral can be given a probability interpretation: any field configuration will occur with probability proportional to its ``Boltzmann'' factor $e^{- S}$. Thus, we expect that in the Euclidean path integral of the Polyakov model, such as the one representing the vacuum to vacuum amplitude, monopole-instanton field configurations can contribute, with probability governed by their action, $\sim e^{- S_0}$. A monopole-instanton is a field configuration in the $R^3$ spacetime, which is characterized by its location (the position of its center) and characteristic core size $ \sim v^{-1}$, or core volume $v^{-3}$. Such a fluctuation can spontaneously appear anywhere in $R^3$. If we take a box in $R^3$ of   volume $V$, the classical statistical mechanics interpretation leads one to expect  that  a monopole-instanton fluctuation in the path integral can appear with probability $ {V v^3} e^{- S_0}$, where $Vv^3$ represents the ``entropy'' enhancement and $e^{-S_0}$ the ``energy'' suppression. Thus, the probability of a monopole-instanton fluctuation per unit spacetime volume is $\sim v^3 e^{-S_0}$.

To get an idea about the effect of this nonperturbative fluctuation, let us study the gauge-invariant two-point field-strength correlation function, say  $\langle B_\mu^3(\vec x) B_\mu^3(\vec y) \rangle$. The perturbative contribution is easily evaluated, as per Section \ref{sec:pert_polyakov}, 
\begin{equation}\label{b3perturbative}
\langle B_\mu^3(\vec x) B_\mu^3(\vec y) \rangle\big\vert_{pert.} \sim {g_3^2 \over |\vec x -\vec y|^3}~.
\end{equation}
This  is simply the contribution of the massless photon and the r.h.s. can be guessed by dimensional analysis (once again, the coefficient is calculable but inessential). As per our discussion above, a monopole-instanton fluctuation can also appear and contribute to this two-point function. Suppose now that such a fluctuation appears at some point $\vec{r} \in R^3$, with (per unit-volume) probability $\sim v^3 e^{- S_0}$.  As in the quantum mechanics instanton calculus, we sum over $d^3 r$, the possible positions of the instanton (as the fluctuation is equally likely to appear anywhere, at least in this leading approximation). We also recall that   for  an instanton at $\vec r$, the long-range ``magnetic'' field is $\vec B^3(\vec x)\vert_{1-inst.\; at \; \vec{r}} \sim  {\vec x - \vec r \over |\vec x - \vec r|^3}$. Thus, we find that the 1-instanton contribution to the two-point function  is given by
\begin{eqnarray}\label{b3instanton}
\langle \vec B^3(x) \cdot \vec B^3(y) \rangle\big\vert_{1-instanton} &\sim& v^3 e^{- S_0} \int d^3 r\; \vec B^3(\vec x)\vert_{1-inst.\; at \; \vec{r}} \;\cdot  \vec B^3(\vec y)\vert_{1-inst.\; at \; \vec{r}} \nonumber \\
 &\sim& {v^3 e^{- S_0} \int d^3 r}~ {\vec x - \vec r \over |\vec x - \vec r|^3} \cdot  {\vec y - \vec r \over |\vec y - \vec r|^3}~\sim ~ {v^3 e^{- S_0} \over  |\vec x - \vec y|}~.
\end{eqnarray}
To get the last equality, one can calculate the integral over $d^3 r$, or simply recall from electrostatics that it represents the interaction energy of two electric charges at $\vec x$ and $\vec y$. 

Let us now add the perturbative (\ref{b3perturbative}) and 1-instanton (\ref{b3instanton}) contribution to the two-point function
\begin{equation}\label{b3both}
\langle B_\mu^3(\vec x) B_\mu^3(\vec y) \rangle\big\vert_{pert.+1-inst.} \sim g_3^2\left( {1 \over |\vec x -\vec y|^3} +  C e^{- {4 \pi v \over g_3^2}} \;{v^2 \over |\vec x -\vec y|}\right)~,
\end{equation}
where $C$ is a dimensionless constant (dependent on $g_3^2/v$) that also incorporates a proper integration over the collective coordinates of the monopole-instanton (this detail shall not concern us here). 

Most important for what follows is the structure of (\ref{b3both}).  In the second term, we restored the explicit small-coupling dependence of $S_0$. This emphasizes the fact that the two terms have very different behaviour as $g_3^2/v \rightarrow 0$: the second term in (\ref{b3both})  is nonperturbative, i.e. nonanalytic in the small dimensionless parameter. The most  important conclusion that one can draw from the two-point correlator (\ref{b3both}) is that, despite its exponential smallness in the ${g_3^2 \over v }\ll 1$ limit, the nonperturbative term dominates the long-distance behaviour of the two-point function, as it becomes more important for exponentially large separations, $|\vec x -\vec y|^2 \gg v^{-2} e^{ 4 \pi v \over g_3^2}$. Thus, the effect of nonperturbative fluctuations is to significantly alter the IR physics. The behaviour of (\ref{b3both}) also indicates that a nonperturbative mass scale   appears in the IR theory, given by the inverse of the square root of the crossover distance indicated above
\begin{equation}
\label{mfirst}
m_{IR}^2 \sim v^2 e^{- S_0} = v^2 e^{- {4 \pi v \over g_3^2}}~.
\end{equation}
The appearance of an IR mass scale also makes one suspect that the true behaviour of the correlator (\ref{b3both}) at large distances may have an exponential rather than a powerlaw falloff.

In what follows, we shall argue that the above guess  is essentially correct. We shall show that the  IR physics of the Polyakov model is even ``more boring'' than the free massless photon of Section \ref{sec:pert_polyakov}---it is, instead, gapped, with a nonperturbative mass gap given by $m_{IR}$. Perhaps some of the boredom will disappear when we show that, upon insertion of fundamental probe electric charges, the same mechanism that gave rise to a nonvanishing mass gap, $m_{IR}$, is also responsible for the confinement of these charged probes.

However, before we get there, we must address the   deficiencies in our ``derivation'' of (\ref{b3both}). Let's first recall the relevant lessons we learned. Think of the monopole-instantons as particles of size $1/v^3$ and  a ``Boltzmann'' suppression factor $e^{- S_0}$. In a box of 3-volume $V$, a monopole-instanton can appear with probability ${V \over (1/v)^3} e^{- S_0}$, indicating the  ``energy'' suppression $e^{-S_0}$ and ``entropy'' enhancement $V v^3$. Note that for large $V v^3$ the entropy contribution wins over the Boltzmann suppression and one has to account for the appearance of multiple monopole-instanton fluctuations. This is one of the  main points we need to address to improve upon (\ref{b3both}). In the quantum mechanics double-well problem, this is addressed by summing up multiple instanton contributions using the dilute instanton gas approximation, and we shall do a similar summation here. 

In the model at hand, 
one encounters another issue not present in quantum mechanics, due to the long-range instanton interactions. One  expects the probability that two such objects appear to scale as  $v^6 e^{- 2 S_0} e^{- S_{inter.}}$. Here, $S_{inter.}$ is due to the fact that the monopole-instantons have a long-range ``magnetic'' field and hence non-negligible interactions at long distances (as opposed to the double-well instantons in quantum mechanics which have exponentially suppressed long-distance interactions). Thus, the action of a pair of such objects taken some distance apart will be  larger or smaller than $2 S_0$, depending on whether they attract or repel. The long-range interactions between two monopole-instantons is due to their long-range $B_\mu^3$ tail and we expect them to have Coulomb-like interactions at large separations (recall  the end of Section \ref{sec:instantons_polyakov}, where we argued that at $\lambda \ne 0$, the $E_\mu^3$ field is exponentially damped away from the instanton core, as pictured on Figure \ref{fig:monopole1}). 

Thus, in order to properly calculate correlation functions like (\ref{b3both}), we have to account for the contribution of multiple monopole-instantons (as in quantum mechanics) and for their  long-range interactions (a feature not present in quantum mechanics). The picture that will emerge is that the sum over saddle points of the Euclidean path integral of the Polyakov model can be recast in the form of  the partition function of a classical gas of monopole-instantons and anti-monopole-instantons, with pairwise Coulomb interactions at large separations. This classical gas is ``grand-canonical,'' i.e. the number of either monopole-instantons and anti-monopole-instantons is not fixed. What is fixed is the ``fugacity'' $e^{- S_0}$ that each object's contribution to the partition function is weighted by. If the gas is sufficiently dilute, accounting for only the long-range interactions should suffice to describe the physics, as the various monopole-instantons never significantly overlap. The semiclassical expansion parameter controlling the diluteness of the gas is now $e^{- S_0} \ll 1$. 

\hfill\begin{minipage}{0.85\linewidth}

\textcolor{red}{
{\flushleft{\bf Summary of \ref{sec:polyakov_1stpass}:}} In this Section, we saw the first indication that  monopole-instanton fluctuations drastically alter the perturbative IR physics of the Polyakov model, leading to the appearance of a new infrared scale (\ref{mfirst}). However, to proceed, we need to develop more technology to properly study the effect of multiple instantons and their interactions.}

\end{minipage}

\bigskip

 \subsection{Monopole-instantons and the IR: II. duality, 't Hooft vertices, and monopole operators.}
\label{sec:polyakov_pass2}

The IR physics of the 3d Polyakov model is most conveniently described using a dual language.\footnote{Recently, 3d dualities more general than the one we use in these notes have received some attention. For a pedagogical introduction and review, see  \cite{Turner:2019wnh}.} The dual description employs the fact that in 3d, a photon has one polarization and is thus equivalent to a scalar field. The duality that we shall perform is thus valid in the IR theory (\ref{polyakoveffective}) of the UV  $SU(2)$ model (\ref{polyakovlagrangian}). Recall that to obtain the IR theory, the $W$-bosons are integrated out, and  the only light degree of freedom is $A_\mu^3$ (and possibly the  $A_4^3$ neutral scalar, which remains massless for $\lambda = 0$).  The long distance theory is defined by the action (\ref{polyakoveffective}), 
\begin{equation}
\label{dual1}
L_{eff.} = {1 \over 4 g_3^2} F_{\mu\nu}^2,
\end{equation} and a path integral over the $A_\mu$ field modulo gauge transformations. For brevity, we shall  henceforth omit the isospin index on the fields $A_\mu$, $F_{\mu\nu}$, with the understanding that these are the massless modes describing the unbroken $U(1)$ subgroup of $SU(2)$. There are no charged degrees of freedom in this theory and the path integral over $A_\mu$ can be rewritten as a path integral over $F_{\mu\nu}$, with an additional constraint imposed to ensure that the field strength obeys the Bianchi identity $\epsilon^{\mu\nu\lambda} \partial_\mu F_{\nu\lambda}=0$ (which guarantees that locally $F$ is the curl of a vector). The constraint can be imposed via a Lagrange multiplier scalar field $\sigma$. To avoid the appearance of factors of $i$ in the action, we shall now go to Minkowski space with metric $(+,-,-)$ and $\epsilon^{012} = +1$. The action is 
\begin{equation}
\label{dual2}
S_{Mink.}[F_{\mu\nu}, \sigma] = - {1 \over 4 g_3^2} \int d^3 x F_{\mu\nu} F^{\mu\nu} - {1 \over 8 \pi} \int d^3 x \;\partial_\mu \sigma \; F_{\nu\lambda} \epsilon^{\mu\nu\lambda}
\end{equation} and the path integral is over $F_{\mu\nu}$ and $\sigma$. Integrating out $\sigma$ imposes the Bianchi identity for $F_{\mu\nu}$, i.e. gives back the Minkowski space version of the original theory (\ref{dual1}). 
On the other hand, integrating out $F_{\mu\nu}$, and  substituting back into (\ref{dual2}) its saddle point value 
\begin{equation}
\label{dualityrelation}
F^{\mu\nu} = - {g_3^2 \over 4 \pi} \epsilon^{\mu\nu\lambda} \partial_\lambda \sigma,
\end{equation}
 we obtain the ``dual photon'' action 
\begin{equation}
\label{dualaction}
S_{Mink.}[\sigma] =  {1 \over 2} { g_3^2 \over (4 \pi)^2} \int d^3 x \partial_\lambda \sigma \partial^\lambda \sigma~.
\end{equation}
The field $\sigma$ will be called, from now on, the ``dual photon.'' The description in terms of the dual photon is often called the ``magnetic description,'' in contrast with the electric description in terms of a $U(1)$ gauge field.

The duality relation (\ref{dualityrelation}) shows how to map local gauge invariant operators between 
the electric description (\ref{dual1}) and the magnetic one (\ref{dualaction}). For future use, we note that the spatial gradients of $\sigma$ give the $F^{0i}$, $i=1,2$, components of the field strength, i.e. they represent the true electric field on the $R^{1,2}$ Minkowski space. The time derivative of $\sigma$, on the other hand, represents the $F^{12}$ component, i.e. the only magnetic field component on $R^{1,2}$. 

There are, however, other operators that will be of interest to us, that may be slightly less familiar. Begin, in the electric theory (\ref{dual1}), where one can define line operators representing the insertion of static nondynamical charge probes, such as
\begin{equation}
\label{wilsonline}
W_{Q_e}(\vec x_*) = e^{i Q_e \int dx^0 A_0(\vec x_*, x^0)},
\end{equation}
the Wilson line representing a static electric charge $Q_e$ located at $\vec x_* \in R^2$. Such operators are meant to be inserted in the path integral and are important to the study of confinement. Thus, it will be of interest to us to learn how Wilson lines are represented in the dual theory (\ref{dualaction}). 

Conversely, in the magnetic theory (\ref{dualaction}), one can define operators that create ``fluxons.'' To justify their introduction, notice that the dual photon theory (\ref{dualaction}) has a global symmetry under which $\sigma$ shifts by a constant. The corresponding current is $j_\mu = {g_3^2 \over 16 \pi^2} \partial_\mu \sigma$, which is mapped, by (\ref{dualityrelation}) to $j_\mu = - {1 \over 8 \pi} \epsilon_{\mu\nu\rho} F^{\nu\rho}$. Thus, the shift symmetry of the dual photon maps to the ``0-form'' magnetic symmetry of the electric 3d $U(1)$ theory, whose current is conserved due to the Bianchi identity.  We note that this magnetic symmetry is an ``emergent'' symmetry of the IR $U(1)$ theory, and is not present in the $SU(2)$ UV theory of the Polyakov model.  In fact, its breaking by  the UV $SU(2)$ theory is crucial for the physics of confinement.\footnote{This is also not a symmetry   in the compact-$U(1)$ lattice theory which also exhibits confinement  \cite{Polyakov:1975rs,Polyakov:1987ez}. For a  study of the compact $U(1)$ theory, including in the Hamiltonian formalism, see \cite{Kogan:2002au}.}

The charge corresponding to the $\sigma$-shift symmetry is
\begin{equation}
\label{qoperator}
Q = {g_3^2 \over 16 \pi^2} \int d^2 x \partial_0 \sigma  \leftrightarrow -{1 \over 4 \pi} \int d^2 x F_{12},
\end{equation}
and maps to the integral over $R^2$ of the magnetic field $F_{12}$. In a canonical quantization of the $\sigma$ theory (\ref{dualaction}), the state
\begin{equation}
\label{fluxonoperator}
e^{i \sigma(\vec x_*)} | 0 \rangle
\end{equation}
is an eigenstate of $Q$ with eigenvalue $1$, as you will show in the following
\begin{quote}
{{\bf Exercise 2}}: Canonically quantize the dual theory (\ref{dualaction}) (for brevity, omitting hats over operators). Show that the canonical commutation relations imply that $Q e^{\pm i \sigma(\vec x_*)} | 0 \rangle = \pm e^{\pm i \sigma(\vec x_*)} | 0 \rangle$, i.e. the operator $e^{\pm i \sigma(\vec x_*)}$ creates a unit magnetic flux, a ``fluxon'' at $\vec x_* \in R^2$.
\end{quote}
Notice that in the electric description (\ref{dual1}), the operator $e^{i \sigma}$, creating a pointlike fluxon with $-{1 \over 4 \pi} \int d^2 x F_{12} = 1$ does not have a simple description. This is an example of a ``disorder'' operator which does not have a local expression in terms of the electric theory fields. 

\begin{figure}[h]
\centerline{
\includegraphics[width=9.5cm]{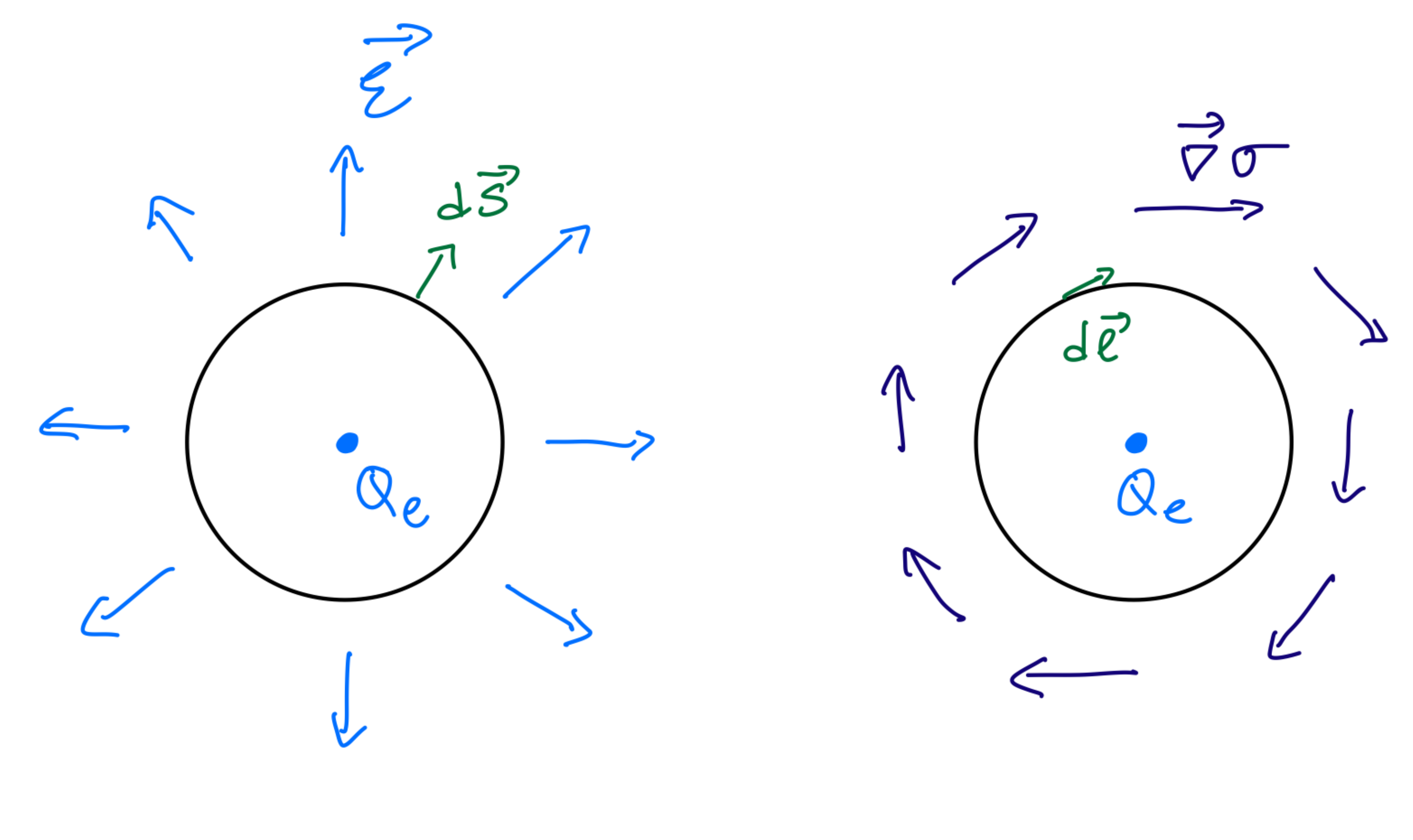}}
\caption{ The photon-dual photon duality (\ref{dualityrelation}) maps an electric charge $Q_e$ into a vortex of the dual photon field, with monodromy determined by the charge. As shown in Exercise 3, the monodromy equals $\pm 2\pi$ for ``quarks'' in the fundamental of $SU(2)$. A fundamental Wilson loop is thus mapped to a disorder operator for the $\sigma$ field, defined by imposing $2\pi$ monodromy of the dual photon around the loop.\label{fig:duality1}}
\end{figure} 

Conversely, the Wilson loop operator (\ref{wilsonline}) is simple in the electric theory, but not in the magnetic theory (\ref{dualaction}), where it is represented by a disorder operator. As Wilson loops are important for the study of confinement, let us now flesh out the details. As already discussed $W_{Q_e}(\vec x_*)$ describes the insertion at $\vec x_*$  of a static electric probe charge $Q_e$. A static charge at the origin of $R^2$ creates electric field $\vec{\cal{E}}(\vec{r}) \sim Q_e {\vec{r} \over r^2}$ (for the proportionality constant, consult Exercise 3 below).
By Gauss' law, a line integral of the electric field over a loop $C \in R^2$ enclosing the charge can be used to find the charge, $Q_e \sim \oint\limits_C \vec{\cal{E}} \cdot d \vec{s}$, with $\vec{s}$ normal to $C$. 
From the duality relation (\ref{dualityrelation}), we have that $Q_e \sim \oint\limits_C \vec{\cal{E}} \cdot d \vec{s} \sim \oint\limits_C d \vec{l} \cdot \vec \nabla \sigma$, where $\vec l$ is tangent to $C$.  In other words, we have shown that the $\sigma$ field has nonvanishing monodromy,\footnote{A nonzero monodromy simply means that  $\sigma$ is not single-valued around electric charges: taking $C$ to be a circle parametrized by a polar angle, we have $\oint_C d \vec{l} \cdot \vec \nabla \sigma = \sigma(2 \pi) - \sigma(0) \sim Q_e$.} proportional to $Q_e$, around loops surrounding static electric charges, see Figure \ref{fig:duality1}. Thus, the Wilson loop operator $W_{Q_e}(\vec x_*)$ should be defined by the following prescription: when inserted in the path integral of the dual photon theory, impose boundary conditions on the $\sigma$ field in the path integral, requiring it to have monodromy $\sim Q_e$ along the line $\vec x = \vec x_*$ in $R^3$. To determine the coefficient, consider the following

\begin{quote}
{{\bf Exercise 3.1}}: Our electric theory (\ref{dual1}), which descends from an $SU(2)$ gauge theory (\ref{polyakovlagrangian}), is an example of a compact $U(1)$ theory---electric charge under the unbroken $U(1)$ is quantized, with the smallest value corresponding to sources in the fundamental representation of $SU(2)$.\footnote{\label{Wloopfootnote}To do this problem, we use the fact that a static charge  in the fundamental representation of $SU(2)$ can be introduced by inserting a fundamental Wilson line, $W(\vec x_*) = \text{tr}\;{\cal{P}} e^{i \int dx^0 A_0^a(\vec x_*, x^0) T^a}$ in the path integral, see the discussion around (\ref{wilsonloop2}). This leads, in the unbroken-$U(1)$ theory, to the insertion of the sum of two operators like (\ref{wilsonline}).  Here, ${\cal{P}}$ denotes path ordering, which can be ignored when projecting  (ignoring the massive components of $A_\mu$) to the  $U(1)$ sector. Thus, introducing a static fundamental charge corresponds to adding a term $\int d x^0 A_0^3 (\vec x_*, x^0) (T^3)_{ii}$, with $i=1$ or $i=2$, to the Minkowski space version of the IR theory action (\ref{dual1}). Here $(T^3)_{ij}$ denotes the $ij$-th entry of the Cartan generator of $SU(2)$. } Notice that the insertion of a static charge in the fundamental representation of $SU(2)$ at $\vec{x}_* \in R^2$ corresponds to adding  $\int d x^0 A_0^3 (\vec x_*, x^0) (T^3)_{ii}$, with $i=1$ or $i=2$, to the Minkowski-space unbroken-$U(1)$ theory action. Following the above discussion, be mindful of the coefficients and show that the monodromy of the $\sigma$ field around a static charge in the fundamental of $SU(2)$ is $\pm 2 \pi$.

{{\bf Exercise 3.2}}: The result of this  Exercise can be used to further the study of the deconfinement transition briefly described in Section \ref{sec:dymtemperature}.  We include it here because it may help  solidify the understanding of the  charge-vortex duality of Figure \ref{fig:duality1}. Consider the following static $\sigma$-field configuration:
\begin{equation}
\label{sigmavortices}
\sigma(\vec{r}) = q_1 \theta(\vec{r}-\vec{r}_1) + q_2 \theta(\vec{r}-\vec{r_2})~,
\end{equation}
where $\theta(\vec{r})$ is the angle the vector $\vec{r} \in R^2$ makes with, say, the positive-$x$ axis. This ``two-vortex'' configuration has  $\sigma$ monodromy $2 \pi q_1$ around $\vec{r}_1$ and $2 \pi q_2$ around $\vec{r}_2$. Thus, according to the duality, it represents two electric charges. As per Exercise 3.1, if $|q_{1,2}|=1$, these are fundamental charges. While the angle $\theta(\vec{r})$ is not defined at $\vec{r}=0$, showing that a UV definition is needed (like a lattice cutoff), this does not affect the calculation of the interaction energy between the charges. Use the dual photon action (\ref{dualaction}) to calculate the interaction energy between the charges represented by (\ref{sigmavortices}). Letting $\vec{R} = \vec{r}_1 - \vec{r}_2$, show that the interaction energy is
\begin{equation}
\label{chargesinteraction}
E(R) = - q_1 q_2 \;{g_3^2 \over 8\pi}\; \log {R \over a}~,
\end{equation}
such that e.g. like charges logarithmically repel (here $a$ is some irrelevant short distance cutoff needed to define the vortex configurations). As a sanity check, also convince yourself that for $|q_1|=|q_2|=1$, the same  interaction energy between static fundamental charges can be obtained from the electric theory. (There are various ways to do this problem, but the relation $\partial_i \theta(\vec{r}) = \epsilon_{ij} \partial_j \log |\vec{r}|$ and judicial integration by parts helps.)
\end{quote}

The conclusion from Exercise 3 is that, to permit the insertion of static fundamental charge probes, which are natural in the UV $SU(2)$ theory, $2\pi$ monodromies of the $\sigma$ field should be allowed. These only make sense if the dual photon field  is regarded as  a compact scalar field, $\sigma \equiv \sigma + 2 \pi$, i.e. the field $\sigma$ is a map from the $R^3$ spacetime to the $S^1$ field space.
We also notice that the $2 \pi$ periodicity of $\sigma$ is consistent with the $e^{i \sigma}$ being a well-defined operator (while a $\pi$ periodicity would not be consistent).

 In Exercise 2, you showed that in the canonical picture the operator $e^{i \sigma}$ creates ``fluxons,'' pointlike excitations for which  $-{1 \over 4 \pi} \int d^2 x F_{12}$ $=$ $1$. 
Notice that the unit magnetic flux  of the ``fluxon'' just given is equal to the magnetic charge of the monopole-instanton (\ref{magneticcharge}) with $|Q_m|= 1$ (recall that $B_3^3 = F_{12}$). This suggests that magnetic monopole-instantons of unit charge (\ref{magneticcharge}) can be interpreted as tunneling events between states differing by one flux quantum. This can be argued by deforming the integral over $S^2$ in (\ref{magneticcharge}) to equal the difference between two $R^2$-integrals like (\ref{qoperator}),  specifying the number of fluxons in the states at the initial and final time $\pm T$ (as stressed in \cite{Banks:2014twn}). That the number of fluxons can change reflects the fact that (\ref{qoperator}) is not conserved in the full theory.

In fact, we shall now argue an important relation valid for the Polyakov model and also useful for our future $R^3 \times S^1$ studies: in the Euclidean path integral of the dual-photon theory, insertions of  $e^{\pm i \sigma (x)}$, with $x \in R^3$, represent the appearance, at $x$, of a monopole-instanton of minimal ``magnetic'' charge (\ref{magneticcharge}), discussed in Section \ref{sec:instantons_polyakov}.\footnote{To avoid confusion, the object $e^{\pm i \sigma}$ represents the insertion of a pointlike object, i.e. the structure of the monopole-instanton inside its core ($\sim v^{-1}$) is ignored. This suffices in the dilute gas approximation where only the long-distance monopole-instanton interactions are important.} 
To begin, define the following object (usually called generating functional)
\begin{equation}\label{gas1}
\langle e^{i \int d^3 x \rho(x) \sigma(x)} \rangle \equiv \zeta^{-1} \int {\cal{D}} \sigma \;e^{- {\kappa\over 2} \int d^3 x (\partial_\mu \sigma)^2 + i \int d^3 x \rho(x) \sigma(x)}~~, ~\text{where}\; \kappa \equiv {g_3^2 \over (4 \pi)^2},
\end{equation}
and $\zeta$ is a normalization factor ensuring that the l.h.s. equals unity when $\rho=0$. It is a standard result that
$
\langle e^{i \int d^3 x \rho(x) \sigma(x)} \rangle = e^{- {1 \over 8 \pi \kappa} \int d^3 x d^3 x' \rho(x) {1 \over |x-x'|} \rho(x')}$.
We next concentrate on a particular form of $\rho(x)=\rho_N(x)$:
\begin{equation}\label{source}
\rho_N(x) = \sum_{a = 1}^N q_a \delta^{(3)} (x - x_a),
\end{equation}
 where $q_a$ are integers. Notice that $\rho_N(x)$ can be interpreted as the charge density at $x$ of a gas of $N$ pointlike charges $q_a$ located at $x_a \in R^3$.
\begin{quote}
{{\bf Exercise 4}}: Show that,\footnote{\label{footnoteneutral}For future use, note that integration over the zero mode of  $\sigma$ gives $\int\limits_0^{2 \pi} d \sigma_0 e^{i \sigma_0 \sum\limits_{a=1}^N q_a}$. For integer $q_a$ this is only nonzero provided $\sum\limits_{a=1}^N q_a = 0$.} with $\rho(x) = \rho_N(x)$, 
\begin{equation}
\label{gas3}
\langle e^{i \int d^3 x \rho(x) \sigma(x)} \rangle = e^{-{1 \over 4 \pi \kappa} \sum\limits_{a>b} {q_a q_b \over |x_a - x_b|}}~,
\end{equation}
where divergent terms with $a=b$ have been omitted (these divergences are to be absorbed in the UV definition of the monopole-instanton fugacities, see further below). As a corollary, show that with the same omission, 
\begin{equation}
\label{interaction1}
\langle e^{i q_1 \sigma(x_1) + i q_2 \sigma(x_2)} \rangle = e^{ - {4 \pi \over g_3^2} {q_1 q_2 \over |x_1 - x_2|}}.
\end{equation}
\end{quote}
To interpret the last relation, we now do the following
\begin{quote}
{{\bf Exercise 5}}:  Consider two monopole-instantons of Section \ref{sec:instantons_polyakov}, one at $x_1 \in R^3$ and the other at $x_2 \in R^3$.  According to the discussion there, for $\lambda \ll 1$ we ignore their $E_\mu^3$ fields.\footnote{These extra interactions present for $\lambda=0$  have to be accounted for in SYM. We shall do so when we discuss the supersymmetric case. Notice that the calculation of the ``electric'' long-range interactions present when $\lambda = 0$ is significantly more involved than Exercise 5 (see  the discussion in Section 2.3 of \cite{Poppitz:2017ivi} and references therein). However, the result can be stated simply.} Let the monopole-instantons have magnetic charges $q_i = \pm 1$, $i=1,2$. Let also $|x_1 - x_2| \gg v^{-1}$, so that the cores of the monopole-instantons do not overlap. Outside their cores the field reduces to the sum of the abelian magnetic monopole terms,  $B_\mu^3(y) \simeq q_1 B_\mu^3(y-x_1) + q_2 B_\mu^3(y-x_2)$ (with $B_r^3 = 1/r^2$ around each monopole-instanton). Compute the interaction action of the two monopole-instantons by calculating the contribution to the action $S = {1 \over 2 g_3^2} \int d^3 y (B_\mu^3(y))^2$ from the region outside the cores, extract the interaction term, and show that 
\begin{equation}\label{interaction2}
e^{- S_{inter.}} =  e^{ - {4 \pi \over g_3^2} {q_1 q_2 \over |x_1 - x_2|}}, 
\end{equation}
exactly reproducing the r.h.s. of (\ref{interaction1}). 

Notice that the above result has a simple intuitive explanation: two magnetic monopoles, considered as static particles in $R^3$, interact via a magnetic version of the Coulomb law. If $q_1$ and $q_2$ are of the same sign, the probability to find the two charges close to each other, controlled by $e^{- S_{inter.}}$,  is vanishingly small, corresponding to repulsion. Conversely,  if the charges have opposite signs, the probability grows with decreasing separation, showing that the charges attract.
\end{quote}
The conclusion we draw from comparing (\ref{interaction1}) with (\ref{interaction2}) is that, in the long-distance abelian IR theory, in the dilute gas regime where monopole-instantons are sufficiently far away so that their cores do not overlap, the appearance of a monopole-instanton  of magnetic charge $q$, at $x \in R^3$,  can be represented by inserting $v^3 e^{- S_0} e^{ i q \sigma(x)}$ in the path integral of the dual-photon theory. The $v^3 e^{- S_0}$ fugacity factor was already discussed in Section \ref{sec:polyakov_1stpass}. The novelty here is that the $e^{ i q \sigma(x)}$ insertions correctly account for the long-distance interactions of the monopole-instantons of magnetic charge $q \;(= \pm 1)$.

We can now introduce the following mnemonic accounting for nonperturbative fluctuations in the Euclidean path integral of the dual photon theory (again, we leave SYM for later). The semiclassical saddles of lowest action are (approximately-) BPS and anti-BPS monopole instantons of minimal charge $\pm 1$. We denote them by $M$ and $M^*$, respectively, and associate with them the following ``'t Hooft vertices'' \cite{tHooft:1976snw},  or ``monopole operators:''
\begin{eqnarray}\label{thooftvertices}
M \;\text{at} \; x&:& v^3 e^{- S_0} e^{ i \sigma(x)}~, \\
M^*\; \text{at} \; x&:& v^3 e^{- S_0} e^{- i \sigma(x)}~. \nonumber
\end{eqnarray}
In the classical 3d statistical mechanics picture, the  $v^3 e^{- S_0}$ factor can be interpreted as the fugacity of $M$ or $M^*$.\footnote{
Similar vertices can also be written in the electroweak sector of the 4d standard model, where the instantons generate exponentially suppressed $B+L$ violating interactions, as first done by 
't Hooft  \cite{tHooft:1976snw,tHooft:1976rip}.
 The main difference with the present setup is that the 4d instantons have no Coulomb-like
  long-range interactions, so no analogue of the $e^{\pm i \sigma}$ factors are present. (Nonetheless, one can account to the interactions between instantons due to gauge field exchange by modifying the 't Hooft vertex to include the interactions of instantons with gauge fields, see the description in \cite{Shifman:2012zz}.)}
  
  A slight subtlety that we shall ignore is that there is additional dependence on the dimensionless coupling constant $g_3^2/v$, which multiplies both $M$ and $M^*$ above. These factors arise upon taking into account  the integration over collective coordinates and  the determinants in the monopole-instanton backgrounds and give a power-law dependence on the dimensionless coupling of the pre-exponential factor in the 't Hooft vertices. As this power-law dependence can not compete (at small $g_3^2/v$) with the exponential $e^{- S_0} = e^{- {4 \pi v\over g_3^2}}$, we shall ignore it. We shall often refer to the neglect of these pre-exponential terms in (\ref{thooftvertices}) as  working with ``exponential-only accuracy.''\footnote{For the interested reader, the computation of these pre-exponential terms has been performed (to one-loop order) in the greatest detail in SYM, see \cite{Poppitz:2012sw,Anber:2014lba}. Accounting for collective coordinate integrations is discussed in  many textbooks, e.g. \cite{Vainshtein:1981wh,Coleman:1985rnk,Schafer:1996wv,Shifman:2012zz}. }

\hfill\begin{minipage}{0.85\linewidth}

\textcolor{red}{
{\flushleft{\bf Summary of \ref{sec:polyakov_pass2}:}} In this Section, we introduced a duality transformation in the abelian IR theory of the Polyakov model, leading to the dual-photon description. We discussed the mapping of operators between the electric and magnetic description. We showed that the insertion of a fundamental Wilson loop corresponds to the requirement of a $2 \pi$ monodromy  of the dual photon field around the loop. We also showed that the dual photon picture
allows one to incorporate the effect of monopole-instantons in the dilute gas approximation, accounting for their long-range interactions.
In the following Sections, we shall sum over all possible insertions of ``monopole operators'' (\ref{thooftvertices}) in the partition function of the dual theory  to find the effect of the monopole-instanton fluctuations in the dilute gas approximation and to study the physics of confinement.}

\end{minipage}

\bigskip

 \subsection{Monopole-instantons and the IR: III. dilute gas and mass gap.}
\label{sec:polyakov_pass3}

Now that we have done all preparatory work, we are ready to  start enjoying the fruits of our labour. Our IR theory is, at the perturbative level, defined via a path integral of the dual photon theory with action (\ref{dualaction}). We argued in Section \ref{sec:polyakov_1stpass} that including nonperturbative monopole-instanton fluctuations  changes the IR physics. We shall now include the fluctuations of arbitrary numbers of monopole-instantons and anti-monopole-instantons. 

Consider the probability of a fluctuation of $n$ monopole-instantons $M$ and $m$ anti-monopole-instantons $M^*$ in the Euclidean path integral. Per our discussion above (\ref{thooftvertices}), we expect that such a fluctuation corresponds to the insertion in the $\sigma$-theory Euclidean path integral of the following object
\begin{equation}\label{nMandmMbar}
P_{n,m} = {\left(v^3 e^{- S_0}\right)^n \over n!}{\left(v^3 e^{- S_0}\right)^m \over m!} \int d^{3(n+m)} r \;
e^{i \sum\limits_{a =1}^{n+m} q_a \sigma(r_a)}~=  {\left(v^3 e^{- S_0}\right)^n \over n!}{\left(v^3 e^{- S_0}\right)^m \over m!} \int d^{3(n+m)} r \;
e^{i \int d^3 x \sigma(x) \rho_{n+m}(x)}~,
\end{equation}
where $\rho_{n+m}(x)$ is defined in (\ref{source}) via the $n+m$ $q_a$ and $r_a$.

As the notation is somewhat condensed, let us elaborate. Here the index $a$ runs from $1$ to $n+m$, i.e. runs over both $M$ and $M^*$. $q_a$ denote the corresponding charges, which take values $+1$ or $-1$  (we only take into account $q = \pm 1$, as all other monopole-instantons of higher charges have higher action and their effect will be exponentially suppressed compared to $e^{- S_0}$). Likewise, $r_a \in R^3$ are the corresponding positions of $M$ or $M^*$. The measure $d^{3(n+m)} r$ should be understood to mean  the product over all $n+m$ $d r_a$'s.
The $v^3 e^{-S_0}$ factors are the fugacities of these objects, while the term $e^{i \sum\limits_{a =1}^{n+m} q_a \sigma(r_a)}$ (written also in terms of the charge density $\rho_{n+m}$ in the second equality above) accounts for the long-distance Coulomb interactions of the $n+m$ $M$ and $M^*$, as discussed in the previous Section. The factors of $n!$ and $m!$ take into account the fact that these fluctuations are indistinguishable: a fluctuation of a monopole-instanton $M$ at $r_1$ and another one, $M'$ at $r_2$ is the same as $M'$ at $r_1$ and $M$ at $r_2$.

We now insert $P_{n,m}$ into the Euclidean path integral of $\sigma$, to obtain the partition function of the IR theory accounting for the appearance of the $n,m$ fluctuation:
\begin{eqnarray}\label{nmcontribution}
Z_{n,m}=&& \nonumber\\
= \zeta^{-1}&&\int {\cal{D}} \sigma \;e^{- {\kappa\over 2} \int d^3 x (\partial_\mu \sigma)^2 } P_{n,m} = 
\zeta^{-1}  {\left(v^3 e^{- S_0}\right)^n \over n!}{\left(v^3 e^{- S_0}\right)^m \over m!} \int d^{3(n+m)} r \int {\cal{D}} \sigma \;e^{- {\kappa\over 2} \int d^3 x (\partial_\mu \sigma)^2 }\; e^{i \sum\limits_{a =1}^{n+m} q_a \sigma(r_a)} \nonumber \\
= \zeta^{-1}&&\int {\cal{D}} \sigma \;e^{- {\kappa\over 2} \int d^3 x (\partial_\mu \sigma)^2 } {\left(\int d^3 r v^3 e^{-S_0} e^{i  \sigma(r)}\right)^n \over n!} {\left(\int d^3 r v^3 e^{-S_0} e^{-i  \sigma(r)}\right)^m \over m!}
\end{eqnarray}
The second line is simply a rearrangement of the first, taking into account that all $n$ $M$ terms have identical form, as do all $m$ $M^*$ terms. Notice that if we integrate over $\sigma$, we find the $n+m$-particle contribution to the grand-canonical partition function of a classical nonrelativistic $3d$ Coulomb gas. (In fact, the treatment that we give it here corresponds to the Debye-Hueckel approximation in the theory of charged gases  \cite{Polyakov:1976fu,Polyakov:1987ez}.) 

Next we want to sum over all values of $n$, $m$, i.e. over all possible $M$ and $M^*$ fluctuations. Recall now that in (\ref{gas1}) we introduced the $\zeta^{-1}$ factor simply to make the normalization convenient and that,  with $\zeta^{-1}$ present, the IR theory partition function (\ref{nmcontribution}) with $n=m=0$ would be trivial. Thus, to obtain our IR theory partition function, when summing over $n$ and $m$ we drop this factor. The partition function of our IR theory with all monopole-instantons summed over becomes
\begin{eqnarray}\label{sumnmcontribution}
Z  &=& \sum\limits_{n,m=0}^\infty \zeta Z_{n,m} =  \int {\cal{D}} \sigma \;e^{- {\kappa\over 2} \int d^3 x (\partial_\mu \sigma)^2 } \left( \sum\limits_{n=0}^\infty{\left(\int d^3 r v^3 e^{-S_0} e^{i  \sigma(r)}\right)^n \over n!}\right) \left(\sum\limits_{m=0}^\infty {\left(\int d^3 r v^3 e^{-S_0} e^{-i  \sigma(r)}\right)^m \over m!} \right)\nonumber \\
&=&\int {\cal{D}} \sigma \;e^{- \int d^3 x {\kappa\over 2}  (\partial_\mu \sigma)^2} \; e^{ \int d^3 x v^3 e^{-S_0} e^{i  \sigma(x)}}  \; e^{ \int d^3 x v^3 e^{-S_0} e^{-i  \sigma(x)} } ~,
\end{eqnarray}
where, on the last line we performed the sums over $n$ and $m$.

Combining everything, we find that the IR theory partition function, accounting for arbitrary (anti-) monopole-instantons in the dilute gas approximation, is\footnote{\label{footnotenegative}For future use, let us note an important implication of the above summation of the dilute instanton gas contributions, which is quite generally valid. It shows that semiclassical objects of positive fugacity (as our monopole instantons) contribute   a negative term to the ground state energy: evaluated at $\sigma=0$, the minimum of the potential in (\ref{sumnmcontribution1}), their effect is to give a negative contribution to the ground state energy (this is well known in quantum mechanics of the double-well potential, where tunneling lowers the ground state energy). In the future, we shall call objects of positive fugacity ``real saddles'' (see Sections \ref{sec:symfields} and \ref{sec:neutralbions}).}
\begin{eqnarray}\label{sumnmcontribution1}
Z  &=& \int {\cal{D}} \sigma \;e^{- \int d^3 x \left[ {\kappa\over 2}  (\partial_\mu \sigma)^2 - 2 v^3 e^{-S_0} \cos \sigma \right]} ~,
\end{eqnarray}
hence, the Euclidean action of the IR dual-photon theory is, after shifting the potential by an inessential constant:\footnote{A remark we elaborate on in Sections \ref{sec:dym} and \ref{sec:firstbions} is due here as well. The effective action given below contains only the leading terms in a combined perturbative expansion in powers of $g_3^2/v$ and a semiclassical expansion in powers $e^{- S_0}$. The higher-order terms in the semiclassical expansion in the Polyakov model have not received much scrutiny on their own, but we expect that they share many features with the ones briefly discussed for $R^3\times S^1$ theories in Section \ref{sec:neutralbions}. (N.B. As these notes were being finalized, ref. \cite{Pazarbasi:2021ifb} appeared, discussing these issues.)}
\begin{equation}\label{iraction1}
S_{IR} = \int d^3 x {g_3^2 \over (4 \pi)^2} \left[ {1 \over 2} \partial_\mu \sigma \partial^\mu \sigma + {2 v^3 \over \kappa} e^{- {4\pi v\over g_3^2}}(1- \cos \sigma) \right]~, ~ \kappa = {g_3^2 \over (4\pi)^2}.
\end{equation}

Notice that the potential term generated by summing over $M$ and $M^*$ simply looks like adding to the dual photon action the 't Hooft vertices of $M$ and $M^*$ (\ref{thooftvertices}), a result due to the dilute-gas approximation.

We can now take stock and discuss what the result (\ref{iraction1}) represents.
We immediately notice that the dual photon field acquires a mass. As written, the potential in (\ref{iraction1}) has a unique minimum at $\sigma=0$ (recall that $\sigma$ has a $2\pi$ periodicity). Expanding the potential around the minimum, we see that the mass of the dual photon is 
\begin{equation}\label{msigma}
m_\sigma^2 = {2 v^3 \over \kappa} e^{- {4 \pi v \over g_3^2}},
\end{equation} in agreement with our earlier guess (\ref{mfirst}), and with exponential-only accuracy (in fact, a more  precise calculation gives an extra factor of $v^2/g_3^4$ due to monopole-instanton collective coordinate integrations; in what follows, we simply absorb these into the definition of $m_\sigma$). 
Thus, at weak coupling, we find that an exponential scale hierarchy is generated:  
\begin{equation}
\label{hierarchy}
v \gg g_3^2 \gg m_\sigma \sim v e^{- 4 \pi v/g_3^2}.
\end{equation}

It is the separation of the various scales due to the $v \gg g_3^2$ weak-coupling condition (recall Section \ref{sec:pert_polyakov}) which makes our effective field theory treatment useful. 
In contrast with instanton calculations in 4d QCD \cite{Schafer:1996wv}, the weak-coupling (due to Higgsing) leads to the  important difference that the instanton size is fixed and the problem of large-size instantons does not arise.

Let us also discuss the scales characterizing the  dilute monopole-instanton gas. The probability (per unit volume) of a monopole-instanton fluctuation  implies that the number of instantons per unit volume is $v^3 e^{- S_0}$. Thus, the typical separation between instantons is $\bar \ell \sim v^{-1} e^{S_0/3}$. Notice that this is exponentially larger than the $W$-boson Compton wavelength $v^{-1}$. The inverse dual-photon mass defines another length scale $m_\sigma^{-1} \sim \ell_D$, called the Debye screening length (this terminology is borrowed from the charged nonrelativistic plasma). From (\ref{hierarchy}), we have that $\ell_D \sim v^{-1} e^{S_0/2} \gg \bar\ell$, i.e. the Debye length is exponentially larger than the typical distance $\bar\ell$ between $M$ and $M^*$. Thus, the validity of the dilute gas treatment can be rephrased by requiring that there be a large number of $M$ and $M^*$ in a Debye volume.  

\begin{quote}
{\flushleft{\bf{Exercise 6:}}} Further properties of the system can be studied, using, instead of (\ref{sumnmcontribution1}), the following 
\begin{eqnarray}\label{sumnmcontribution2}
Z[\eta]  &\equiv& \int {\cal{D}} \sigma \;e^{- \int d^3 x \left[ {\kappa\over 2}  (\partial_\mu (\sigma - \eta))^2+ m_\sigma^2 (1-\cos \sigma) \right]} ~.
\end{eqnarray}
Reverse the arguments leading  to (\ref{sumnmcontribution1}) to show that a nonzero $\eta$ corresponds to the insertion of $e^{i \int d^3 x \eta(x) \rho_{n+m}(x)}$ in the sum over monopole instantons. In other words, argue that $Z[\eta]$ is a generating functional for correlation functions of the monopole-instanton  magnetic charge density. Show that the average charge density vanishes, $\langle \rho(x) \rangle = - i {\delta Z[\eta]\over \delta \eta(x)}\big\vert_{\eta = 0}= 0$ and find the scale of exponential fall-off of the two-point function $\langle \rho(x) \rho(y) \rangle$.
\end{quote}

\begin{figure}[h]
\centerline{
\includegraphics[width=9.5 cm]{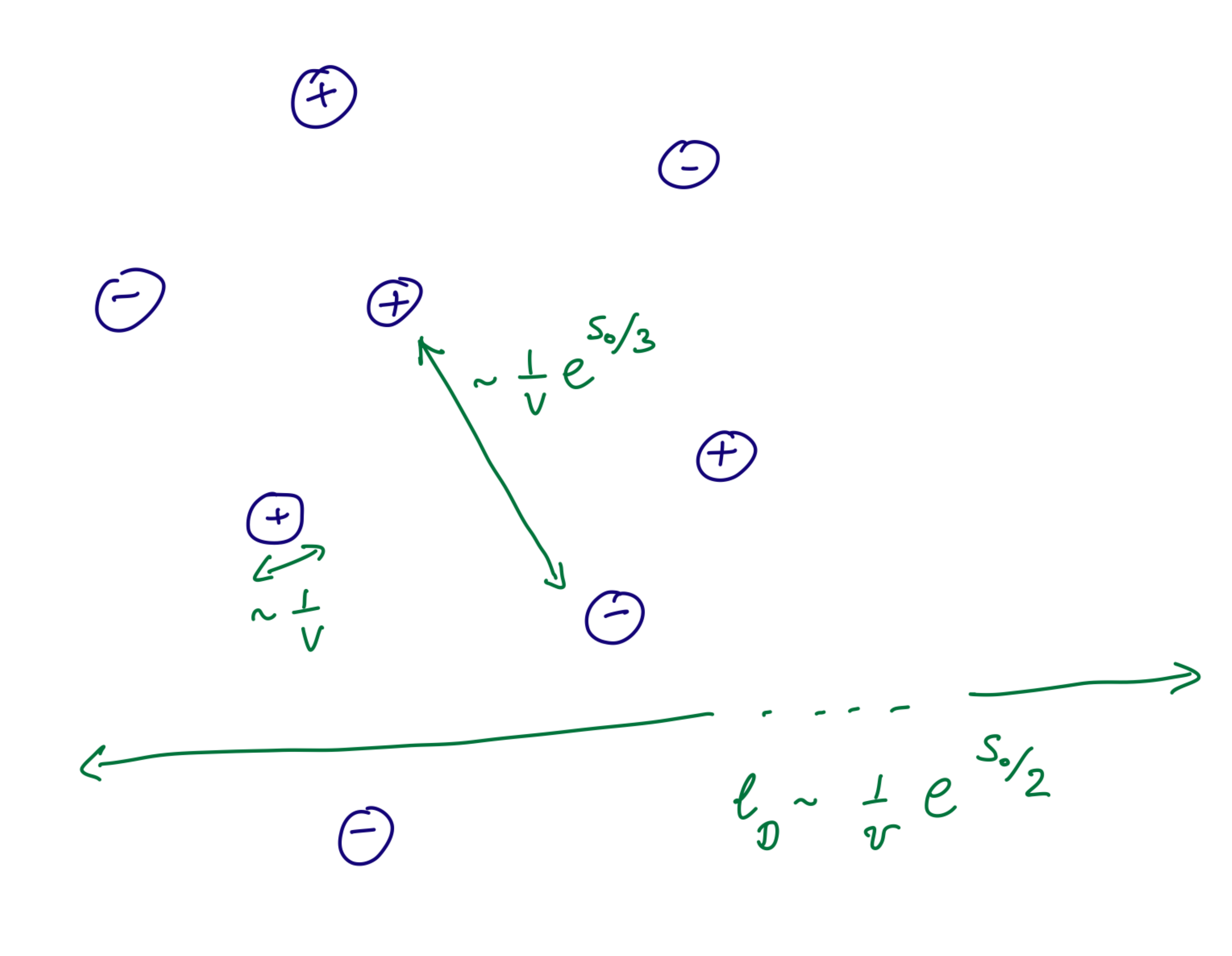}}
\caption{ The hierarchy, controlled by the exponentially large $e^{S_0}$, between the various length scales characterizing the the dilute $M$-$M^*$ monopole-instanton gas. The dual photon Compton wavelength (the Debye screening length) is the largest length scale. A Debye volume contains a large number of $M$ and $M^*$ fluctuations. \label{fig:MMbargas}}
\end{figure} 
A final comment, before we continue to discuss confinement, relates to terminology. One often hears that confinement is due to ``magnetic monopole condensation,'' in a manner dual to the usual Higgs mechanism.  While this is demonstratively true in the Seiberg-Witten theory, where magnetic monopoles are particle-like physical excitations, in the Polyakov model, as well as in its $R^3 \times S^1$ generalizations, this terminology should be taken with a grain of salt. The ``monopoles'' discussed here,   leading to the mass gap generation, are instantons. They are pseudoparticles (in 't Hooft's terminology \cite{tHooft:1976snw})  localized in spacetime, thus existing only ``for an instant.'' It is the proliferation of these Euclidean monopole-instanton fluctuations in the vacuum that leads to the mass gap and, as we show next, to confinement.\footnote{The relation of the monopole-instantons on $R^3 \times S^1$   to physical monopole particles on $R^4$ whose condensation (in Seiberg-Witten theory) leads to a dual superconductivity is, at best, far from straightforward, see the discussion in \cite{Poppitz:2011wy}.}

To end this Section, we rewrite (\ref{iraction1}) in terms of the dual photon mass $m_\sigma$, absorbing the inessential (for the present discussion)  coefficient into its definition,  a form that we shall use from now on:
\begin{equation}\label{iraction2}
S_{IR} = \int d^3 x {g_3^2 \over (4 \pi)^2} \left[ {1 \over 2} \partial_\mu \sigma \partial^\mu \sigma + m_\sigma^2 (1- \cos \sigma) \right]~.
\end{equation}

\hfill\begin{minipage}{0.85\linewidth}

\textcolor{red}{
{\flushleft{\bf Summary of \ref{sec:polyakov_pass3}:}} The major  result obtained here is the IR effective action for the dual photon field (\ref{iraction2}). It shows that calculable nonperturbative effects at weak coupling generate an exponentially small mass gap $m_\sigma$ (\ref{msigma}). The exponential hierarchy of scales (\ref{hierarchy}) arises due to  $v \gg g_3^2$ and is responsible for calculability and for the utility of the effective field theory approach to describing the long-distance physics. }

\end{minipage}

\bigskip

 \subsection{Monopole-instantons and the IR: IV. confinement  and the string tension.}
\label{sec:polyakov_pass4} 

We have come close to the end of our discussion of the Polyakov model. To proceed, as indicated in the Introduction, we first need to define what we mean by ``confinement.'' The most precise definition of confinement applies to theories without fundamental-representation dynamical fields, such as pure Yang-Mills theory, or Yang-Mills theory with adjoint-representation fields, in any dimension. The Polyakov model falls in the latter class, as do most theories we discuss in these notes. The modern-day way to phrase this distinction is to note the presence of a ``1-form''\footnote{This symmetry shall be more carefully  defined in Section~\ref{sec:holonomyandcenter}, in a way sufficient for our discussion. Here we note that the so-called ``generalized global symmetries,'' of which 1-form symmetries are an example, act not on local fields (operators) but on extended objects, such as Wilson line operators.  The most familiar symmetries usually discussed in QFT courses---those acting on local fields and operators---are called ``0-form symmetries'' in this generalized framework.  We  refer the reader  to \cite{Gaiotto:2014kfa} for a detailed  definition of higher-form symmetries in the continuum and to \cite{Greensite:2011zz} for a lattice description of center symmetry. We note that both the lattice and continuum ways of thinking about the 1-form center symmetry are quite useful.} $Z_2$ (for $SU(2)$ gauge group) center symmetry in theories without dynamical fundamental fields. Center symmetry will play an important role in our $R^3\times S^1$ discussion and we shall introduce the relevant aspects in what follows. For now, we proceed with a somewhat hand-waving discussion of confinement. 

Confinement is probed for by studying the potential energy between probe static ``quarks.'' We shall take these to be nondynamical colour sources in the fundamental representation of the gauge group (one can think of them as the limit of heavy $m \rightarrow \infty$ dynamical fields). As already mentioned, recall Exercise 3 and footnote \ref{Wloopfootnote}, to study the way the dynamics reacts to nondynamical fundamental probes, one  inserts a fundamental Wilson loop in the path integral of the theory and studies its expectation value.
\begin{figure}[h]
\centerline{
\includegraphics[width=5.5 cm]{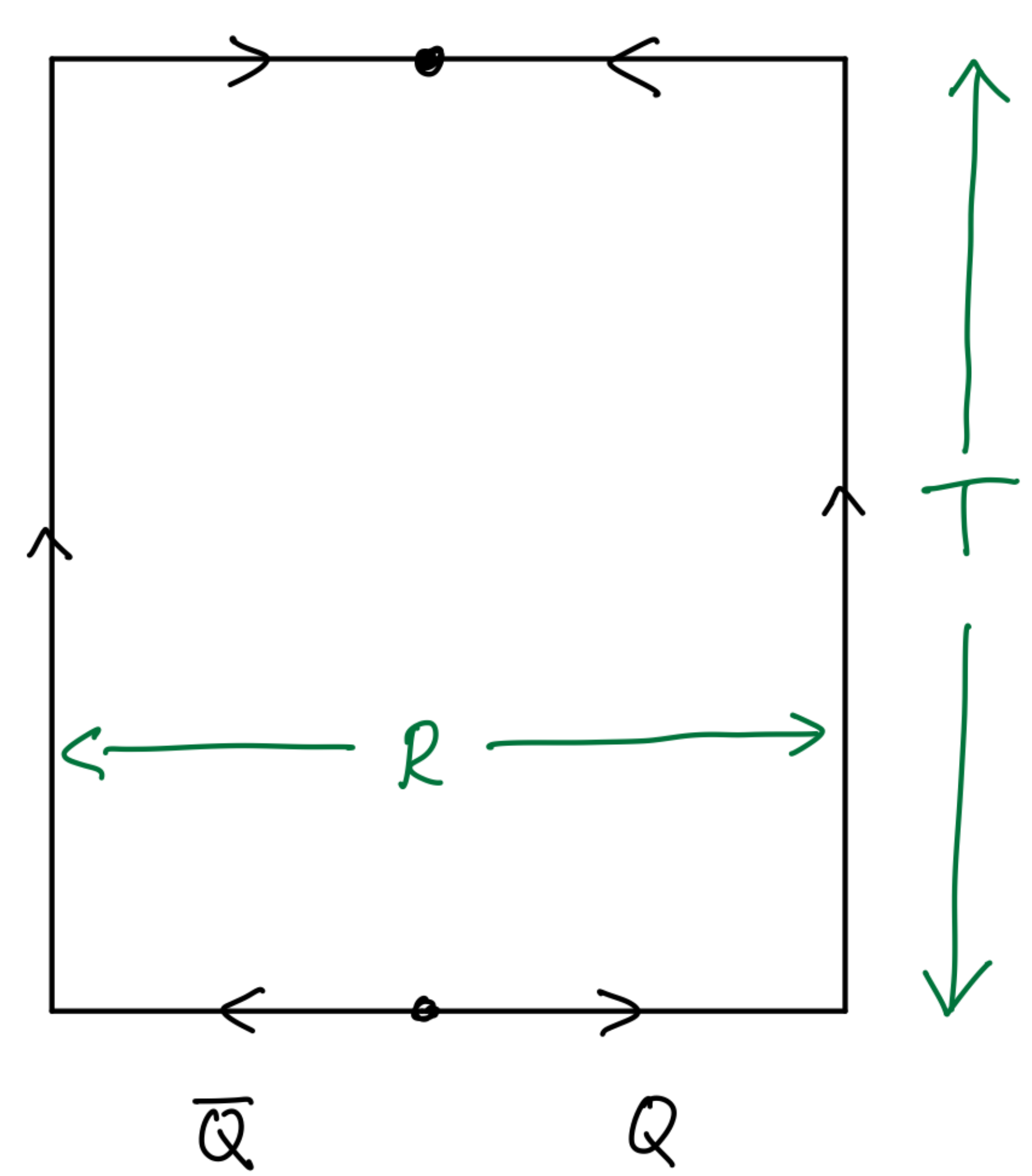}}
\caption{ The rectangular Wilson loop used to study the potential between probe fundamental sources. \label{fig2}}
\end{figure}  

The Wilson loop along a closed spacetime contour $C$ is defined as
\begin{equation}
\label{wilsonloop2}
W[C] = \text{tr} \; {\cal{P}} e^{\; i \oint\limits_C A_\mu^a T^a d x^\mu}~,
\end{equation}
where the trace is in the fundamental representation (and so are the generators $T^a$) and ${\cal{P}}$ denotes path ordering along the contour $C$. That $W[C]$ is gauge invariant follows from the following
\begin{quote}
{\flushleft{\bf Exercise 7:}} Consider the open Wilson line defined as ${\cal{P}} e^{\; i \int\limits_{x_1}^{x^2} dx^\mu A_\mu^a(x) T^a}$, where the integral is over any path from $x_1$ to $x_2$. Notice that path ordering is  defined similar to the usual time ordering appearing in the Dyson formula, only with respect to a parameter $\tau$ parameterizing the spacetime contour $x^\mu(\tau)$. Show\footnote{There are various ways to proceed: it helps to first consider $x_2$ and $x_1$ infinitesimally close and then exponentiate the result, or one might write a differential equation w.r.t. one of the end points, or  consult a textbook proof, say \cite{Peskin:1995ev}.
} that under gauge transformations, $A_\mu \rightarrow g(A_\mu - i \partial_\mu)g^\dagger$,
\begin{equation}
\label{wilsonlinetransform}
{\cal{P}} e^{\; i \int\limits_{x_1}^{x^2} dx^\mu A_\mu^a(x) T^a} \rightarrow g(x_1) {\cal{P}} e^{\; i \int\limits_{x_1}^{x^2} dx^\mu A_\mu^a(x) T^a} g(x_2)^\dagger~.
\end{equation}
Then argue that if the path $C$ is closed, so that $g(x_1)=g(x_2)$, $W[C]$ is invariant.
\end{quote}
 Taking the contour $C$ to be the one shown on Figure \ref{fig2}, we   interpret the insertion of $W[C]$  as the creation of a quark-antiquark pair, which we let separate a distance $R$, then propagate for some time $T$, and then annihilate it. The expectation value of this Wilson loop operator, taking $R$ and $T$ to infinity, can then be interpreted as giving the potential energy $V(R)$ between the static sources:\footnote{\label{footnotewilsonrenormalization}For brevity, the coefficient multiplying the exponent here is set to unity. In reality, the Wilson loop operator suffers renormalization, which affects precisely this prefactor (as divergences are local, they effect the operator defined on the loop $C$, but  not the potential we are interested in). This is already seen in the calculation of $W[C]$ in QED.  }
\begin{equation}\label{wilsonpotential}
\langle W[C]\rangle = e^{- V(R) T} ~.
\end{equation}
We say that the theory is confining if, in the infinite $T$ and infinite $R$ limit, $V(R) = \Sigma R$ with $\Sigma$ a proportionality  constant of mass dimension two known as the string tension.\footnote{A general   result following from reflection positivity  is that the static potential  can not grow faster than $R$  \cite{Bachas:1985xs}.} The expectation value of $W[C]$, for  large loops $C$,  is then said to obey the ``area law'' $\langle W[C]\rangle = e^{- \Sigma \text{Area}\;[C]}$.\footnote{Running somewhat ahead,  we shall see below that for theories with dynamical fundamental matter fields, an area law does not hold due to string breaking. Thus such theories are not ``confining'' in the sense discussed here  \cite{Fradkin:1978dv}, see also \cite{Greensite:2011zz} .}

 For confining theories at strong coupling, the area law has been demonstrated numerically, using lattice gauge theory, or via the analytic strong-coupling expansion on the lattice. The presence of a linear confining potential $V(R) = \Sigma R$ can be intuitively explained by the presence of a string, or a colour-field flux tube whose tension  or energy per unit length is $\Sigma$. The world sheet of the string spans the area of the Wilson loop on Figure \ref{fig2}, leading to the area law.
 
 In  free Maxwell theories, the dependence of the expectation value of $W[C]$ on $C$, for large contours,  can be easily calculated. One finds that $V(R)$ is the Coulomb potential in the relevant dimension. This result applies to our perturbatively-free IR  theory (\ref{dual1}), where $V(R)$ is  logarithmic, as appropriate in 3d.

Our goal in this Section is to show that in the Polyakov model, after taking into account the monopole-instanton gas, $W[C]$ obeys the area law. To this end, imagine that we insert a Wilson loop operator $W[C]$ in the path integral of the $SU(2)$ theory. As we want to study large loops, with size larger than $v^{-1}$ (as well as $1/m_\sigma$), it makes sense to use the IR effective field theory (\ref{iraction1}), but to do this, we have to reduce $W[C]$ from (\ref{wilsonloop2}) to a form using the relevant IR variables. We do this in two steps: first we reduce $W[C]$ to the Cartan-subalgebra variables: $W[C]$ of eqn.~(\ref{wilsonloop2}) becomes $W_{IR}[C] =  e^{\; {i \over 2} \oint\limits_C A_\mu^3  d x^\mu} + e^{\; -{i \over 2} \oint\limits_C A_\mu^3  d x^\mu}$, i.e. the sum of two $U(1)$ Wilson loop operators of charges $\pm 1/2$.  This is expected, as an $SU(2)$-doublet quark has two components that have opposite charges under the unbroken $U(1)$. As our long-distance theory is abelian, we may as well consider separate insertions of the charges $\pm 1/2$ Wilson loops in the long-distance theory partition function. Notice, however, that in the $UV$ theory these contribution come as a sum, as indicated above. 

Thus, let us introduce the $U(1)$ Wilson loop of quantized integer charge $q$, where we dropped the isospin index, 
\begin{equation}
\label{U1loop}
W_q[C] =   e^{\; i {q \over 2} \oint\limits_C A_\mu d x^\mu}~.
\end{equation}
Here, $q = \pm 1$ corresponds to fundamental representation probes, while, e.g. $q= 2$ would be one of the three component of the adjoint (vector) representation, which has Cartan-$U(1)$ charges $\pm 2$ and $0$. The definition of $W_q[C]$ given on the r.h.s. of (\ref{U1loop}) is in terms of the electric variables appropriate to the electric version of the IR theory (\ref{dual1}). As discussed in Section \ref{sec:polyakov_pass2} (recall Exercise 3), in the dual-photon theory, the insertion of the operator $W_q[C]$ in the path integral is defined by the prescription to integrate over  $\sigma$-field configurations that have a $2\pi q$ monodromy around the contour $C$.\footnote{This can also   be expressed in a way that we shall not utilize  and only mention for completeness. It is, however, very  useful for  numerical calculations of string tensions in more complicated models, for details see \cite{Cox:2019aji,Bub:2020mff}. Integrate by parts  in the exponent of  (\ref{U1loop}) to rewrite $W_q[C]= \exp( i {q \over 2} \int_{S, \partial S=C} d^2 s^\mu {1 \over 2} \epsilon_{\mu\nu\lambda}  F^{\nu\lambda})$, where $s^\mu$ is normal to the surface $S$ spanning $C$. As this form involves only the gauge invariant field strength, it can be written via  $\partial_\mu \sigma$ using  the duality relation (\ref{dualityrelation}). Then insert  $W_q$ into the path integral of the dual photon theory to find its expectation value. At the semiclassical level, one calculates this integral by solving the classical equation of motion for $\sigma$. One finds that the contribution of the $W_q$ insertion to the equation of motion requires $2\pi q$ jump of $\sigma$ upon crossing $S$,  ensuring  correct monodromy around $C$. We also note that Polyakov used the instanton gas picture to justify a  calculation essentially identical to the one we describe \cite{Polyakov:1976fu,Polyakov:1987ez}, but using the results of  Exercise 6 along the way. Another approach to calculating string tensions, allowing to smoothly interpolate between the logarithmic Coulomb behaviour of the quark-antiquark potential at short distances and the confining behaviour at distances greater than $\ell_D$,  was developed in \cite{Anber:2013xfa}.}

\begin{figure}[h]
\centerline{\includegraphics[width=7.5 cm]{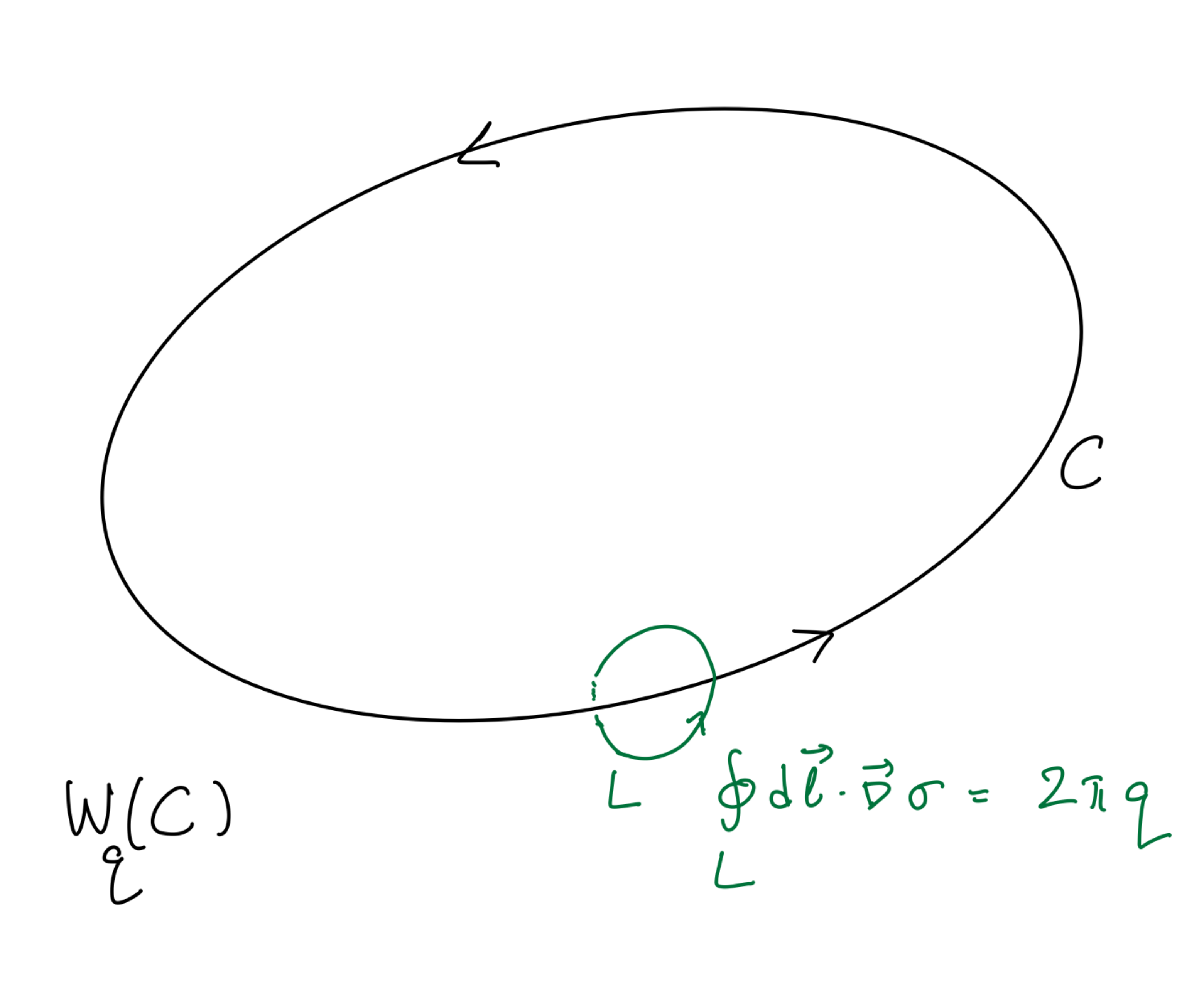}}
\caption{ The charge-$q$ Wilson loop $W_q[C]$ requires a $2\pi q$ monodromy of $\sigma$  around any loop $L$ linked with $C$. To obtain an analytic expression, we imagine that $C$ is in the $xy$-plane and take it to be infinitely large, running around the perimeter of the plane. 
 \label{fig:Wilsonmonodromy}}
\end{figure}

We can now proceed and calculate the expectation value of the Wilson loop in the theory (\ref{iraction2}). All we need to do is compute
\begin{equation}
\label{wilsonsigma}
\langle W_q[C] \rangle = Z^{-1} \int {\cal{D}} \sigma \; \text{``}W_q[C]{\text{''}}\; e^{\; - {g_3^2 \over (4\pi)^2}\int d^3 x\left[{1\over 2} (\partial_\mu\sigma)^2 + m_\sigma^2 (1 - \cos \sigma)\right]}~,
\end{equation}
where  the insertion of $\text{``}W_q[C]{\text{''}}$ indicates, as shown on Figure \ref{fig:Wilsonmonodromy}, that the path integral has to be taken over $\sigma$-field configurations with $2\pi q$ monodromy around any loop $L$ linked with $C$. Calculating this path integral precisely is beyond our current ability. However, we shall now show that  a classical field configuration, i.e. a saddle point  of the path integral (\ref{wilsonsigma}), with the correct monodromy around $C$  exists. Furthermore, we shall argue that the action of this classical field configuration is proportional to the area of $C$, for sufficiently large $C$. Thus, evaluating the path integral (\ref{wilsonsigma}) in the saddle point approximation gives rise to the area law 
\begin{equation}
\label{wilsonsigma1}
\langle W_q[C] \rangle \sim e^{- S_{class}[\sigma_q[C]]}~= e^{ - \Sigma_q \text{Area}(C)}~.
\end{equation}
As per footnote \ref{footnotewilsonrenormalization}, calculating the prefactor is both difficult and not of particular interest to us. 
The quantity $\Sigma_q$ is the string tension of the confining string, labelled by the charge $q$. In what follows we shall  consider the minimal charge $q=1$, corresponding to a fundamental $SU(2)$ colour source, but shall make some comments about adjoint charges later.
\begin{figure}[h]
\centerline{
\includegraphics[width=8.5 cm]{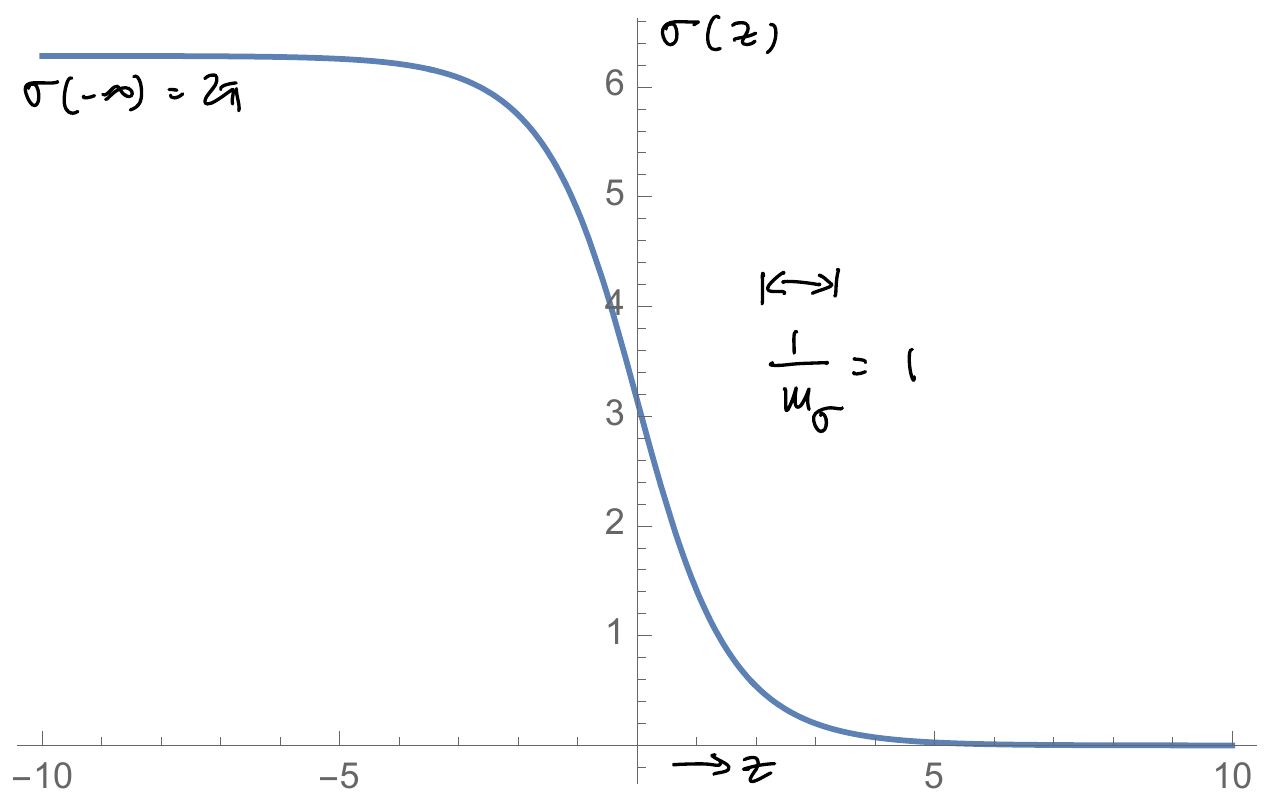}}
\caption{ The domain wall solution (\ref{domainwall1}) plotted for $m_\sigma = 1$. It is clear that the thickness of the wall is a few times the Compton wavelength of the dual photon, as the naive estimate shows.\label{fig:dw1}}
\end{figure}

\begin{figure}[h]
\centerline{
\includegraphics[width=9.5 cm]{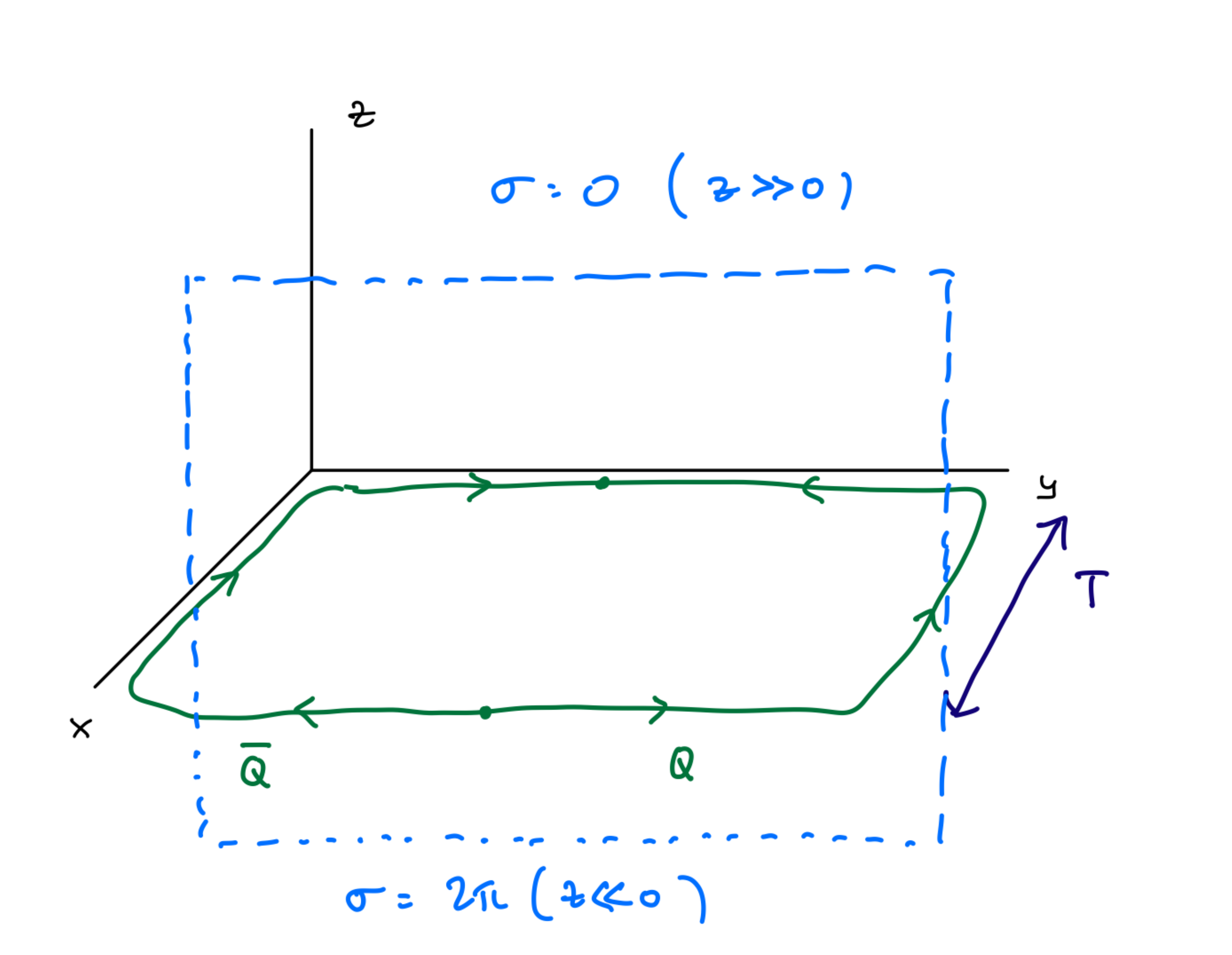}}
\caption{ Continuing the Euclidean picture to Minkowski space. Taking $x$ to be the Euclidean time, we intersect the Wilson loop with a fixed-time plane (shown by the dotted line) and plot the resulting field configuration on Figure \ref{fig:conf2}. \label{fig:conf1}}
\end{figure}

It should be clear that even the saddle point evaluation of (\ref{wilsonsigma}) is not easy, as one needs  to analytically find classical $\sigma$-field configurations that extremize the action in (\ref{wilsonsigma}) and have the right monodromy, for general contours $C$.  To simplify matters and get an analytic handle, we imagine that the Wilson loop $C$ is a planar one, encompassing the entire $xy$-plane (i.e., we take the loop from  Figure \ref{fig:Wilsonmonodromy} to be like the one on Figure \ref{fig:conf1}). In this limit having a $2\pi$ monodromy around $C$ implies that as one crosses the $z=0$ plane from $z = -\infty$ to $z=+\infty$, the $\sigma$ field should exhibit a $2\pi$ jump. Further, in the limit when $C$ is very (infinitely) large, we expect that the classical solution for $\sigma$ will not depend on $x$ and $y$. Thus, we have reduced the problem to a rather manageable one, namely to find a solution of the classical equation of motion $- \partial_z^2 \sigma + m_\sigma^2 \sin\sigma = 0$, such that $\sigma(-\infty) = 2\pi$ and $\sigma(+\infty) = 0$.  Needless to say, this solution is easy to find analytically, as the equation is easily integrable. It is known as the domain wall,  or (better) kink,  solution of the sin-Gordon model.  The analytic form of the domain wall solution is easy to find, see Figure \ref{fig:dw1}, but it is also instructive to use simple reasoning to find an estimate for its action. 

\begin{figure}[h]
\centerline{
\includegraphics[width=8cm]{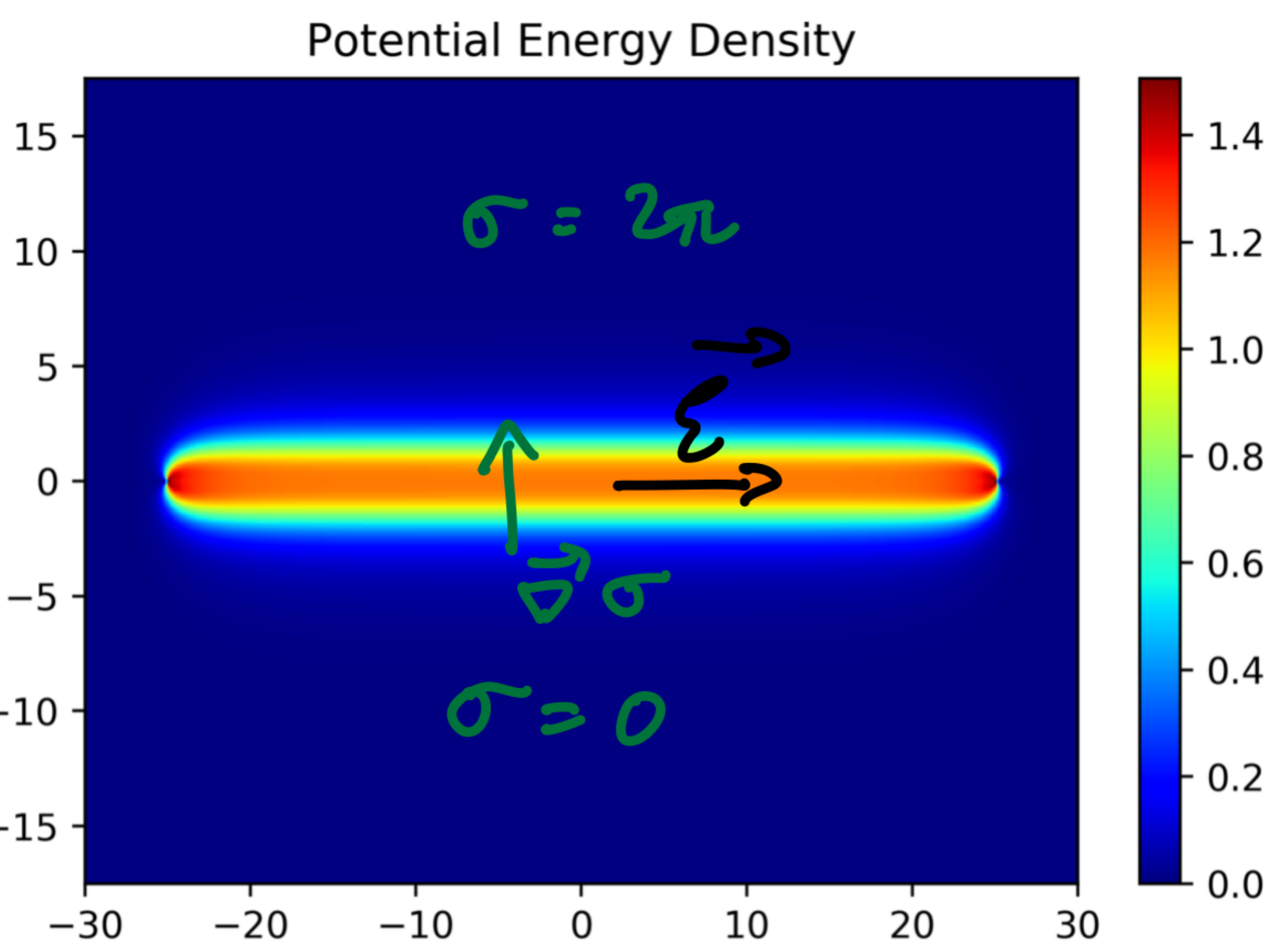}}
\caption{ The spatial configuration of  static quark/antiquark sources taken at a finite separation,  obtained as explained in Figure \ref{fig:conf1}, after taking $T$ to infinity. The potential energy density of the $\sigma$ field is shown. Far from the sources, one can check that the $\sigma$ field profile approaches that of the analytic solution from Figure \ref{fig:dw1}. The gradient of the $\sigma$ field is nonzero only in a flux tube region of width $1/m_\sigma$ connecting the static sources. In terms of the electric variables, recall  the duality relation (\ref{dualityrelation}), the $\vec{\cal{E}}$-field flux of the sources is collimated in a flux tube (line in $R^2$), the semiclassical confining string of the Polyakov model. (This Figure as well as Figures \ref{fig:doublestring1} and \ref{fig:decwall1} are taken from \cite{Cox:2019aji,Bub:2020mff}  with the sole purpose to qualitatively illustrate the physics, as the Figures are not always for  $SU(N=2)$. The numerical methods used to obtain them are explained in these references.) \label{fig:conf2}}
\end{figure}

Recall that the $\sigma =0$ and $\sigma=2\pi$ ground states are identified (thus, the $\sigma$-field configuration we are looking for  is not a true domain wall, which requires distinct vacuum states on the two sides, but we shall continue to use this name). First, the domain wall is $x,y$-independent, so the domain wall action $S_{class.} = {g_3^2 \over (4\pi)^2}\int d^3 x\left[{1\over 2} (\partial_\mu\sigma)^2 + m_\sigma^2 (1 - \cos \sigma)\right]$ will be proportional to the area of the $xy$-plane, which we denote as $A_{xy}$.\footnote{One can numerically solve for contours of arbitrary shape and confirm this for very large $C$, see \cite{Cox:2019aji,Bub:2020mff} and  Figure \ref{fig:conf2}. } Further, as $z \rightarrow \pm \infty$, the solution ends up in a state where the potential vanishes. As the $\sigma$ field is massive, these vacuum values at infinity will be approached exponentially. Thus, we expect that the change $\delta \sigma = 2\pi$ across the domain wall occurs over some finite distance $\delta z$. Then, an estimate of the action is $g_3^2 A_{xy}( \delta z ({2 \pi \over \delta z})^2 + m_\sigma^2 \delta z)$, dropping  numerical constants. To find the minimal action, extremize w.r.t. $\delta z$ to find $\delta z \sim m_\sigma^{-1}$ and find the action:
\begin{equation}
\label{actionstring1}
S_{class.} = A_{xy} \; {1 \over 2 \pi^2} \; g_3^2 m_\sigma~.
\end{equation}
Thus, the Wilson loop path integral, evaluated at the saddle point (\ref{wilsonsigma1}), yields for the fundamental Wilson loop
\begin{equation}\label{wilsonsigma2}
  W_1[C_{xy}] \sim e^{- A_{xy} \Sigma_1}, ~\text{and string tension}~ \Sigma_1 = {1 \over 2 \pi^2} \; g_3^2 m_\sigma~.
\end{equation}
 The precise numerical coefficient in (\ref{actionstring1}) can be found in the following
 \begin{quote}
{\flushleft{\bf{Exercise 8:}}} Show that the $z$-dependent domain wall solution with boundary conditions as shown on Fig.~\ref{fig:dw1} is 
\begin{equation}\label{domainwall1}
\sigma(z) = 4 \arctan e^{- m_\sigma z}.
\end{equation}
 Then, compute its action per unit area and show that it is as given in (\ref{actionstring1}). 
\end{quote}
 
Equation (\ref{actionstring1}) is the fruit of our long labour through the many steps of this rather long Section \ref{sec:polyakov}. The upshot is that we have shown that large fundamental Wilson loops in the Polyakov model obey an area law. We have obtained this result by calculating the path integral (\ref{wilsonsigma}) of the dual-photon theory in the saddle point approximation (whose validity  requires a large action, and is, a posteriori, justified in the limit of large area). We have found that the fundamental string tension is $\Sigma_1 \sim g_3^2 m_\sigma$ and is determined by the dual photon mass, due to the proliferation of monopole-instantons.

In addition to having calculated the string tension (\ref{wilsonsigma2}), we can now also elaborate on the emergence of the confining string by interpreting the result of the calculation in Minkowski spacetime. Begin by considering (see Figure \ref{fig:conf1}) the Wilson loop surrounding  the $xy$-plane whose expectation value (\ref{wilsonsigma2}) we computed. The profile of the $\sigma$ field as a function of $z$ is as shown on Figure \ref{fig:dw1}. Of course, for a finite-size loop, this $z$-only dependence should hold far away from the edges. This can be verified numerically, see Figure \ref{fig:conf2}. A detailed numerical analysis shows that   the value of the string tension  (\ref{actionstring1}) for infinite separation of the sources (computed using (\ref{domainwall1})) is quickly achieved by the string configuration on the figure, already for quark-antiquark separations of a few (${\cal{O}}(10)$) dual-photon Compton wavelengths.

Now we imagine   taking the $x$ direction to be the Euclidean time and take a cross section of the picture across the dotted plane (along $yz$) on Figure \ref{fig:conf1}. The result is shown on   Figure \ref{fig:conf2}. The intersections of the dotted plane with the Wilson loop are  the locations of the static quark sources. The $\sigma$-field configuration is such that $\sigma$ is constant above and below the $z=0$ plane (the plane of the Wilson loop), and the gradient is only nonzero in the region indicated by the dotted lines. The width of this region is proportional to the dual photon Compton wavelength. Now recall that from the duality (\ref{dualityrelation}), the gradient of $\sigma$ is a $90$-degree rotation of the electric field $\vec{\cal{E}}$. Thus, we conclude that the electric field of the quark-antiquark pair is zero everywhere except for a flux tube region of width $1/m_\sigma$ connecting the sources. This flux tube is the confining string,  which, of course, is a line, as our space is $R^2$. 

We find that confinement of colour caused the electric field of the static sources, instead of representing the dipole picture found in Maxwell electrodynamics, to become collimated in the flux tube (line) between the two quarks. The relevant scales we found are the string tension, $\sim g_3^2 m_\sigma$, and the width of the flux tube, $\sim 1/m_\sigma$.

We are now almost ready to move to $R^3 \times S^1$. But before doing this, let us go back to the beginning of this Section \ref{sec:polyakov_pass4} and the 1-form $Z_2$ center symmetry mentioned there. It was stated there that it is the theory without fundamental dynamical quarks that enjoys the $Z_2$ center symmetry. Using the picture of confinement that we developed, we can now physically explain the distinction between theories with $Z_2$ center symmetry (the ones without dynamical fundamental representation quarks) and those without $Z_2$ center symmetry (the ones with finite-mass fundamental representation quarks). 

We used the IR theory of the Polyakov model to find the string tension confining $q=1$ quark sources, i.e. fundamental representation quarks. But what about higher charges, for example $q=2$? As explained above, this includes adjoint representation colour sources. In the IR theory, one can find, similar to what we did here, configurations composed of two domain walls that create a $2\pi \times 2$ monodromy of the $\sigma$ field and thus carrying the necessary flux to confine $q=2$ sources (see Figure \ref{fig:doublestring1} in Section \ref{sec:doublestringdYM}). Thus, one would conclude that $q=2$ colour sources are also confined. However,  $q=2$ is the charge of adjoint fields (the $W$-bosons) under the unbroken $U(1)$, in our normalization where fundamentals have $q=1$. In the IR theory,  the $W$ bosons were integrated out, as they had mass $m_W \sim v$. Now imagine having a  $q=2$ confining flux tube, whose energy grows linearly as $\Sigma_2 R$, where $R$ is the separation between the $q=2$ sources. Clearly, there exists an $R$ sufficiently large so that $\Sigma_2 R > 2 m_W$. But then, energetics suggests that it would be advantageous for a $W^\pm$ pair to be created out of the vacuum, as this would lower the energy: the $W^+$ would combine with the $q=-2$ source into an uncharged object, as would the $W^-$ with the $q=2$ source. In other words, the pair creation of $W^\pm$ would break the $q=2$ flux tube. The linear potential turns into a constant at large $R$, as the screened sources do not interact. Thus, confinement, the area law, and the linear potential have disappeared.

{\flushleft{
We}} learn two lessons from this discussion:

\begin{enumerate}
\item First, we learn that in the Polyakov model, there is a breakdown of the IR effective field theory in a $q=2$ flux tube background for asymptotically large $R$, due to  the breaking of the confining flux tubes. This is a qualitative argument, but there is no reason to doubt its correctness, despite the fact that calculating the $W^\pm$ pair-creation probability in the flux tube background is a difficult problem. 
\item
Another lesson one can draw from the above is the following. We could repeat the  argument in the theory with dynamical fundamental quarks, no matter how heavy: a $q=1$ flux tube with tension $\Sigma_1$ will, upon increasing $R$, find it energetically advantageous to break by the pair creation of a heavy quark-antiquark pair, which will then screen the fundamental sources. Thus, in the theory with  dynamical fundamental charges, the Wilson loop can not have an area law, as the string connecting fundamental sources is unstable to pair creation.\footnote{It is clear, then, that there is no clear-cut notion of confinement in the standard model, which has dynamical fundamental quarks. In fact, there is no symmetry distinction \cite{Fradkin:1978dv} between the confined and Higgs phases,  see also \cite{Greensite:2011zz} and, for recent developments, \cite{Cherman:2020hbe} and \cite{Greensite:2021fyi}.}
\end{enumerate}

 It would seem, then, that there is some property that distinguishes the theory with or without dynamical fundamental fields, and that this property is related to the stability of confining strings. 
In fact, there is a global symmetry that acts on flux tubes (which are lines in space, hence the name ``1-form'' symmetry, meaning that the symmetry acts on line operators). This 1-form center symmetry ensures the stability of flux tubes. In our $SU(2)$ theory, it is a $Z_2$ symmetry. This means that a single flux tube ($q=1$) is stable, but two flux tubes (combined, two $q=1$ tubes correspond to $q=2$) are unstable, as explained above. For more formal definitions, see \cite{Greensite:2011zz} for a lattice perspective and \cite{Gaiotto:2014kfa} for a modern ``generalized symmetry'' perspective.
We shall use and explain in more detail a restricted notion of center symmetry in Section \ref{sec:holonomyandcenter}. As explained in Section \ref{sec:digression}, it will play an important role in our construction.

\hfill\begin{minipage}{0.85\linewidth}

\textcolor{red}{
{\flushleft{\bf Summary of \ref{sec:polyakov_pass4}}}: We showed that the fundamental representation Wilson loop in the Polyakov model obeys an area law. We found that there is a linear confining potential between fundamental representation static colour sources. We determined the string tension (\ref{wilsonsigma2}) in terms of the fundamental parameters of the UV $SU(2)$ theory.  
We also explained the semiclassical nature of the confining flux tube, see Figure \ref{fig:conf2}, and estimated its parameters. Finally, we qualitatively discussed the role of the ``1-form'' $Z_2$ center symmetry. This symmetry distinguishes $SU(2)$ theories with or without dynamical fundamental representation fields and its presence ensures the stability of confining flux tubes. 
}

\end{minipage}

\bigskip

 \section{From $\mathbf{R^3}$ to $\mathbf{R^3 \times S^1}$: generalities.}
  \label{sec:from3to3x1}
  
  We are now ready to move from $R^3$ towards $R^3 \times S^1$. However, before studying confinement,  we have to come to grips with some developments dating back to the 1980's and 1990's. The first is the so-called ``GPY''-potential, found in studies of high-temperature gauge theories \cite{Gross:1980br}. The second concerns the properties of monopole-instanton solutions in circle compactified theories discovered  simultaneously (and independently) in studies of quantum field theory \cite{Kraan:1998pm}  and string-theory D-branes \cite{Lee:1997vp,Lee:1998bb}. 
  
\subsection{Holonomy,  Polyakov loop, center symmetry, and the Weyl chamber.}
\label{sec:holonomyandcenter}

The first thing we shall do is to ``dispose'' of the Higgs field $A_4^a$ of the Polyakov model (\ref{polyakovlagrangian}) as an entity distinct from the gauge field. More precisely, we shall incorporate it into the theory as an intrinsic part of the 4d $SU(2)$ theory: the component of the gauge connection in the $S^1$ direction.

 In other words, we shall now reverse the logic around (\ref{bps3dlagrangian}), where we used the 4th dimension as  a formal tool to study the BPS limit of monopole-instanton solutions. Explicitly, we now consider a 4d gauge theory with  one coordinate compactified on a circle, i.e. $x^4 \equiv x^4 + L$, where $L$ is the $S^1$ circumference. We use $M,N$ to denote $R^4$ indices, while $\mu,\nu$ are $R^3$ indices;  $M=4$ denotes the $S^1$ component. The 4d  Euclidean Lagrangian is
\begin{equation} \label{4dlagrangian}
L_{4d} =
{1 \over 4 g_4^2} F_{MN}^a F^{MN \; a}~ - i \theta q,
\end{equation}
and $g_4$ denotes the dimensionless 4d gauge coupling taken at some UV scale to be specified later, $\theta$ is the theta-angle and $q$ is the topological charge density.\footnote{See eqn~(\ref{topologicalcharge}) and Exercise 11. The topological term will not be important until the next Section and so we omit it in what follows. We assume familiarity with the standard construction of $\theta$ vacua in Yang-Mills theory, see \cite{Shifman:2012zz}.}

Without loss of generality,\footnote{This may require a (necessarily short) explanation. On a general compact manifold, one needs to introduce coordinate patches and relate the gauge fields on different patches via transition functions (gauge group elements).  $S^1$  can  be considered as an interval $[0,L]$ with  $A(0)$ and $A(L)$  related  by a transition function. One can show that there exists a gauge transformation making the transition function unity, implying $A(0)=A(L)$. } the gauge fields can be taken periodic on the $S^1$: $A_M^a (x_4 + L) = A_M^a (x_4)$. Expanding $A_M$ in Fourier (also called Kaluza-Klein, KK, or Matsubara) modes on the $S^1$, it is easy to see that modes with $x^4$-dependence carry momentum quantized in units of  $1/L$. 

Our studies of the nonperturbative properties of the $R^3 \times S^1$ gauge theory will, as in the Polyakov model, use a tower of effective theories. As alluded to in the Introduction, and will be discussed in more detail below, to ensure calculability, we shall take the highest energy scale in the problem to be the compactification scale $1/L$. To reduce the 4d theory to one where excitations of mass $\ge 1/L$ are integrated out, we  proceed, at tree-level, to
 neglect all $x^4$ dependence, taking $\partial_4 =0$ in (\ref{4dlagrangian}).
 Then, we  obtain 
  an effective IR theory valid on $R^3$, at energy scales $\mu \ll {1\over L}$ by integrating the 4d lagrangian $L_{4d}$ over $x^4$. The resulting 3d Lagrangian is: 
\begin{equation} \label{3dlagrangian}
L_{3d} = {L \over 4 g_4^2} F_{\mu\nu}^a F^{\mu\nu \;a} + {L \over 2 g_4^2} (D_\mu A_4)^a (D^\mu A_4)^a + \ldots.
\end{equation}

This lagrangian is quite similar to (\ref{bps3dlagrangian}). 
One difference is the replacement $1/g_3^2 \rightarrow L/g_4^2$. This is simply the way the 3d (dimensionful) and 4d (dimensionless) couplings are related upon compactification, not taking into account quantum corrections. The factor of $L$ in the numerator occurs  because we integrated (\ref{4dlagrangian}) over the $S^1$ coordinate to obtain (\ref{3dlagrangian}). This is a correct procedure  at tree level, but the question of how the quantum loop corrections affect $L_{3d}$ needs to be addressed. In particular, in a 4d theory, the coupling $g_4^2$ runs logarithmically and one needs to indicate what scale this coupling is taken at in (\ref{3dlagrangian}). We shall make this more precise in what follows (although we shall not delve into all detail, as some of  it is inessential for our purposes). At this point, we appeal to common sense: the natural expectation is that the relevant scale should be of order $1/L$, since neglecting the Kaluza-Klein modes means that all excitations heavier than $1/L$ have been integrated out. 

The other difference between (\ref{bps3dlagrangian}) and  $L_{3d}$ of (\ref{3dlagrangian}) are the  ``$\ldots$'' terms. These indicate that quantum corrections can generate terms of a form  different from the ones already shown in $L_{3d}$. We can appeal to symmetries and dimensional analysis to guess what additional terms may occur. The terms in the $\mu\ll 1/L$ effective action should be gauge invariant, with respect to 3d gauge transformations, and they should respect the global symmetries of the 4d theory (as usual, we assume that the theory can be regulated in a way respecting the symmetries so that the effective action  preserves them).
Some of the terms that are allowed to occur in $L_{3d}$ are traces of higher powers of the 4d gauge invariant field strengths $F_{\mu\nu}$ and $F_{\mu 4}$, with spacetime indices appropriately contracted. These terms should be suppressed by inverse powers of $1/L$, the lightest KK-mode mass scale, and are expected to be irrelevant at $\mu \ll 1/L$ (in our weak-coupling set up).

Other allowed terms may appear less obvious and have the form of a potential for the $A_4$ component.  Notice that $A_4$ is part of the $SU(2)$ connection and is not invariant under $SU(2)$ gauge transformations, but transforms as $A_4 \rightarrow g(A_4 - i \partial_4) g^\dagger$, where $g$ is an $SU(2)$ group element. A simple, but not quite precise, argument in favour of the possible appearance of a potential term for $A_4$ is that, from the perspective of the $\mu \ll 1/L$ theory, $A_4$ is an adjoint scalar. This is because the inhomogeneous term in the gauge transformation of $A_4$ vanishes if $g$ is taken independent on $x^4$.

A better argument, bringing extra insight, is to notice the following. We have compactified the 4d theory on an $S^1$ by integrating out all modes of energy $\ge 1/L$. Thus, the resulting 3d theory is expected to be local in 3d, but not  in the $S^1$-coordinate. In particular, there exist an object that is not local on the $S^1$ but is local in $R^3$, the Wilson loop winding around the $S^1$. It is also known as the Polyakov loop; we shall interchangeably call $\Omega$ a Polyakov loop or a winding Wilson loop. The fundamental representation Polyakov loop is defined as
\begin{equation}
\label{polyakovloop1}
\Omega(\vec{x}) = {\cal{P}} e^{\; i \int\limits_{0}^L dx^4 A_4^a(\vec{x}, x_4) T^a}~, ~~ \text{det} \Omega=1, ~~ \Omega^\dagger \Omega = 1~,
\end{equation}
where we stress that $\Omega$ is an $SU(2)$ group element. Sometimes, $\Omega$ is also called the ``gauge holonomy'' around $S^1$.
Here ${\cal{P}}$ denotes path ordering, already discussed near (\ref{wilsonloop2}) and $T^a$ are the fundamental-representation generators of $SU(2)$. Recalling Exercise 7, from eqn.~(\ref{wilsonlinetransform}) there we know that under gauge transformations on $R^3 \times S^1$, $g(\vec{x}, x_4) \in SU(2)$,
\begin{equation}
\label{omegatransform}
\Omega(\vec{x}) \rightarrow g(\vec{x}, 0) \Omega(\vec{x}) g(\vec{x}, L)^{\dagger}~.
\end{equation}
This makes it clear that  $\Omega(\vec{x})$ transforms as a unitary adjoint field under $x_4$-independent gauge transformations. Further, it is also clear that the trace of the Polyakov loop $\text{tr} \Omega(\vec{x})$ is invariant under periodic $SU(2)$ gauge transformations obeying $g(\vec{x},  L) =  g(\vec{x}, 0)$. 

There are many properties concerning the Polyakov loop (\ref{polyakovloop1}) that are relevant to our future discussion and that we shall unravel as we go on. We first go back to enumerating the possible gauge invariant terms that can be added to our $L_{3d}$ of (\ref{3dlagrangian}). The above discussion implies that we can add the following, local in $R^3$ but not local on $S^1$, gauge invariant terms to (\ref{3dlagrangian}):\footnote{\label{omegafootnote}Note that the gauge invariants $\text{tr}(\Omega^k)$ are not independent and can be expressed through $(\text{tr} \Omega)^k$ via the characteristic equation of the matrix $\Omega$:  \begin{equation}
\label{chareqn}\Omega^2 - \Omega \;\text{tr}\Omega + \text{det} \Omega =0,\end{equation}  which is easily verified in our $2\times 2$ case.}
\begin{equation}
\label{nonlocalterms}
\ldots \supset \sum\limits_{k,p \in Z} {c_{kp} \over L^3}\; (\text{tr} \;\Omega)^k \; (\text{tr} \;\Omega^\dagger)^p ~,
\end{equation} 
where we wrote the $L^{-3}$ factor based on dimensional analysis and  $c_{kp}$ are unknown dimensionless coefficients (clearly, they should obey relations ensuring the reality of the effective action terms (\ref{nonlocalterms})). The origin and physical consequences of adding terms like (\ref{nonlocalterms}) to (\ref{3dlagrangian}) will be discussed later (the sum  over integer powers $k$ and $p$ will be seen to arise due to integration out of KK modes).

  Now we reveal one property that distinguishes the terms with $k+p$-even in (\ref{nonlocalterms}). The $k+p$-even terms are invariant under the following $Z_2$ transformation
\begin{equation}
\label{center1}
Z_2^{(1)}: \; \text{tr}\; \Omega \rightarrow - \text{tr}\; \Omega~
\end{equation}
while terms with odd $k+p$ violate (\ref{center1}). We shall use the notation $Z_2^{(1)}$ for (\ref{center1}), in some departure from the literature where it is sometimes called ``0-form center symmetry.'' This is because from the point of view of the 3d long-distance theory valid at scales $\mu \ll 1/L$,  $\Omega$ appears as local operator and the symmetry (\ref{center1}) acts as a ``normal'' 0-form symmetry. We shall continue using the notation $Z_2^{(1)}$ for (\ref{center1}) and hope that this will not cause undue confusion.

We claim that the transformation of (\ref{center1}) is our familiar 1-form center symmetry, $Z_2$ for gauge group $SU(2)$,  discussed at the end of the previous section.\footnote{Unfortunately, tying up all the lose ends and explaining the relation of tr$\Omega$ to confinement/deconfinement will have to wait for Section \ref{sec:adjointgpy}.} Here, it is denoted by $Z_2^{(1)}$ to emphasize the fact that this is a 1-form symmetry, i.e. it acts only on gauge invariant operators associated with lines ($\Omega$ is associated with a line,  the noncontractible loop along the $S^1$).
 There are many ways to define this $Z_2^{(1)}$ symmetry that we can not possibly go into. We shall only note the ``old-fashioned'' way of doing so. Comparing (\ref{omegatransform}) and (\ref{center1}) we note that the latter is equivalent to the transformation of tr $\Omega(\vec{x})$ under   an ``improper gauge transformation,''  on $R^3 \times S^1$, one obeying
 \begin{equation}
  \label{improper}
  g(\vec{x}, x_4+L) = - g(\vec{x}, x_4) ~~, ~~ {\text{for example}}, ~~ g_A (x_4) = e^{\;i {2 \pi x_4 \over L}{\sigma^3 \over 2}}.
  \end{equation}
Thus,  ``improper'' gauge transformations are ones periodic only up to elements of the $Z_2$ center of $SU(2)$. The transformations (\ref{improper}) represent global 1-form symmetries of the 
theory without fundamental fields.\footnote{\label{footnotefundam}The coupling of fundamental fields to gauge fields is inconsistent with such improper gauge transformations on $R^3 \times S^1$, as the fundamental fields $\Psi$ and their improper gauge  transformations  $\Psi \rightarrow g \Psi$, with $g$ obeying    (\ref{improper}), would obey different boundary conditions on $S^1$.
 On the other hand, adjoints transform as $\Phi \rightarrow g \Phi g^\dagger$, and  are insensitive to the  (mod $Z_2$) $g$ periodicity (\ref{improper}). We  refer   to \cite{Greensite:2011zz} and \cite{Gaiotto:2014kfa} for different descriptions of this 1-form symmetry. We also note that when the theory is considered on a four-torus, $T^4$, then there are four different center symmetries, hence the name ``1-form,'' each associated with one of the four types of noncontractible loops on $T^4$. Here, we shall mostly be concerned with the center symmetry associated with our $S^1$ and shall continue to label it $Z_2^{(1)}$.} Eq.~(\ref{improper}) is one way  to introduce the all-important $Z_2^{(1)}$ global center symmetry\footnote{The fact that (\ref{improper}) represent global symmetries was  recognized already in \cite{tHooft:1981sps,Luscher:1982ma}, called there ``central conjugations.''} responsible for the stability of the confining string, as per the discussion at the end of Section \ref{sec:polyakov_pass4}. It should be clear that local gauge invariant operators, which can only depend on $F_{MN}$ are automatically invariant under (\ref{improper}) and only line operators winding around ``the world'' are sensitive to it. 

Now to the point concerning the allowed terms in $L_{3d}$.  We conclude that since the theory with only adjoint fields is invariant under (\ref{improper}), so should the effective action (\ref{3dlagrangian}). Thus,  only terms  preserving the $Z_2^{(1)}$ symmetry (\ref{center1}) are allowed. Hence, in the $SU(2)$ theory, the  effective lagrangian (\ref{3dlagrangian}) should include terms of the form (\ref{nonlocalterms}) with $k+p$-even.\footnote{Running ahead, see eqn.~(\ref{gpygauge2}) for the  one-loop GPY potential, a function of tr$\Omega$ which can be massaged (using (\ref{chareqn})) into the form (\ref{nonlocalterms}).} 

Next, in a weak coupling set-up, it is convenient to work with the connection $A_4$ instead of $\text{tr} \Omega$ of  (\ref{polyakovloop1}). It is usual in gauge theory to study the temporal $A_0 =0$ gauge. Here, it is convenient to choose a similar perspective,   working in the ``$A_4 =0$'' gauge. The  quotation marks stand to remind us of the only difference between our case and the usual $A_0=0$ gauge: our $A_4$ is the component of the $SU(2)$ connection along a compact spatial direction. As such, it is characterized by the gauge invariant $\text{tr} \Omega$ (as per Footnote \ref{omegafootnote}, this is the only independent gauge invariant), which can not be removed by a choice of gauge. This means that one can choose a gauge eliminating ``most,'' but not all, of $A_4^a$. 

\begin{figure}[h]
\centerline{
\includegraphics[width=9.5 cm]{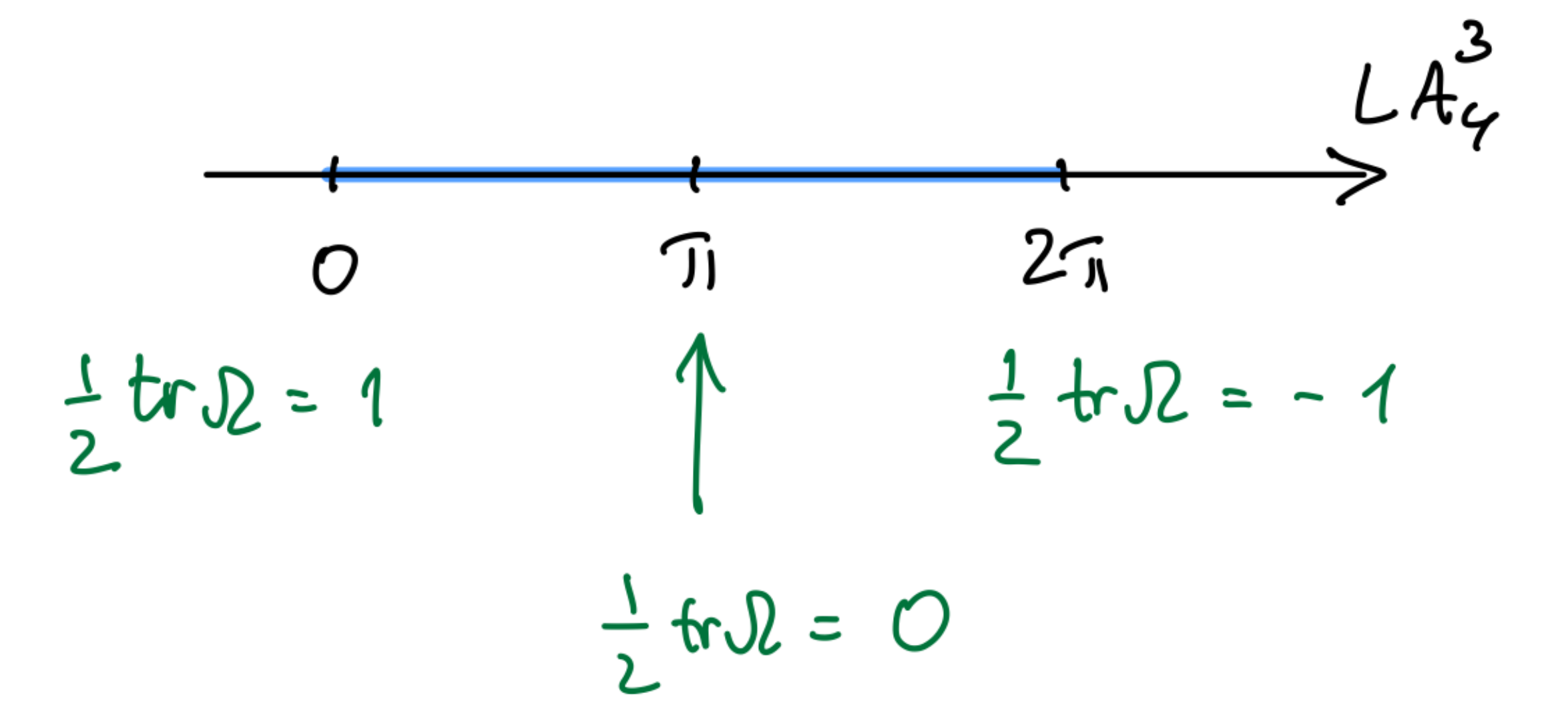}}
\caption{ The segment $\langle L A_4^3 \rangle \in [0, 2 \pi]$ known as the ``moduli space'' or ``Weyl chamber'' of the $S^1$-holonomy of the $SU(2)$ theory. All points inside the Weyl chamber are physically distinct. That there are no further gauge identifications of points inside the Weyl chamber follows by the fact that they are distinguished by the different values of the gauge invariant operator $\langle \text{tr} \Omega/2 \rangle \in [1,-1]$. As explained in the text, on the two edges of the Weyl chamber, the $Z_2^{(1)}$ center symmetry is maximally broken, see also Exercise 10. Center symmetry acts on the Weyl chamber as a reflection w.r.t. the middle point. This point corresponds to the center-symmetric value of the holonomy, $\langle \text{tr} \Omega\rangle=0$, a major player in our study of the dynamics.\label{fig:weylchamber}}
\end{figure}  

 The question about the possible values  $\text{tr} \Omega$ can take, the convenient gauge choice, as well as some other properties are the subject of the following important
\begin{quote}
{\flushleft{\bf{Exercise 9:}}} Prove as many of the statements below as you feel like.
\begin{enumerate}
\item A periodic gauge transformation (\ref{omegatransform}) can be used to bring $\Omega(\vec{x})$  to a diagonal form, i.e.
\begin{equation}\label{omega1}
\Omega' (\vec{x}) = e^{\; i \int\limits_{0}^L dx^4 A_4^{3 '}(\vec{x}, x^4) T^3}.
\end{equation} 
\item Using further periodic gauge transformations in the Cartan subgroup of $SU(2)$,  the $x^4$ dependence of $A_4^{3'}$ in $\Omega'$ can be eliminated to find
\begin{equation}
\label{omega21}
\Omega_{diag.}(\vec{x}) =e^{\; i L A_4^3(\vec{x}) {\sigma^3 \over 2}} = \text{diag}\;(e^{\; i {L A_4^3(\vec{x}) \over 2}}, e^{\; -i {L A_4^3(\vec{x}) \over 2}})~.
\end{equation} 
The gauge transformations done in
this and the previous part of this problem amount to what is known as ``unitary gauge'' for the case where the ``Higgs'' field's role is played by the holonomy. 
 
The form of $\Omega_{diag.}$ of (\ref{omega21}) will be very useful for us in what follows.
First, it shows that the gauge invariant ${1 \over 2} \text{tr} \Omega =  \cos{L A_4^3 \over 2}$ takes values between $1$ and $-1$.
 It also suggests that $L A_4^3$ is a compact variable of period $4 \pi$ (it is subject to further gauge identifications,  see item 4. below). 
\item Under the $Z_2^{(1)}$ global symmetry transformations with $g_A(x^4)$ from (\ref{improper}), $\Omega_{diag}.$ transforms as $\Omega_{diag}. \rightarrow - \Omega_{diag.}$, consistent with (\ref{center1}). Taken to act on $LA_4^3$, these center-symmetry transformations correspond to $2\pi$ shifts.
\item The periodic gauge transformation $g(x^4) = (g_A(x^4))^2$ leaves $\Omega_{diag.}$ invariant and gives rise to  $4 \pi$ shifts of $L A_4^3$. This, combined with the transformation of $\Omega_{diag.}$ under the so-called ``Weyl reflection,'' a gauge transformation with a constant $g= i \sigma_2 \in SU(2)$,  implies that $L A_4^3 \rightarrow - L A_4^3 \;(\text{mod} \;4 \pi)$ is a gauge identification.
\end{enumerate}
\end{quote}

The importance of the results of the above exercise can not be overstated. When studying the dynamics, we shall be very interested in the possible vevs of $LA_4^3$ (after all, we already noted that it acts as a kind of a scalar field in our 3d theory). In particular, we want to know which vevs of $L A_4^3$ are physically distinct, i.e. are not  gauge equivalent. From the information above, we see that there are gauge transformations that act as  $4 \pi$ shifts of $L A_4^3$. Combined with the Weyl  reflections $L A_4^3 \rightarrow - LA_4^3$, we find that the set of independent values, or the ``moduli space'' of $L A_4^3$ is the segment $L A_4^3 \in [0, 2 \pi]$. The two end points are distinct and are related by the global $Z_2^{(1)}$ symmetry. The action of $Z_2^{(1)}$ on the Weyl chamber can be pictured as a reflection of the interval with respect to the middle point.

These findings are depicted and summarized in the caption of Figure \ref{fig:weylchamber}, describing again what is also known as the ``Weyl chamber.'' 

\hfill\begin{minipage}{0.85\linewidth}

\textcolor{red}{
{\flushleft{\bf Summary of \ref{sec:holonomyandcenter}:}} Here, we introduced some  background relevant for our study of the circle-compactified theory. We began  by introducing the notion of an effective 3d  lagrangian (\ref{3dlagrangian}), valid at energy scales below the lowest KK mode mass, $\mu \ll 1/L$. We also introduced the gauge holonomy, or Polyakov loop (\ref{polyakovloop1}). We described how the $Z_2^{(1)}$ global center symmetry acts (\ref{center1}),  showing that 
the trace of the Polyakov loop is the gauge invariant order parameter for the $Z_2^{(1)}$   1-form symmetry. We also noted that the symmetries of the problem allow   holonomy-dependent terms like (\ref{nonlocalterms}) in the effective 3d lagrangian. Finally, we described the Weyl chamber, the space of physically distinct values of the holonomy, depicted on Figure \ref{fig:weylchamber}. We noted that there is one special point on it, the center-symmetric point $LA_4^3 = \pi$, or tr $\Omega = 0$. This point will play an important role in our further studies. }

\end{minipage}

\bigskip
 
\subsection{Spectrum near  center-symmetry. When does small-${L}$  imply weak coupling?}
\label{sec:perturbative3x1}

At the classical level, all values of $A_4^3$ on the Weyl chamber of Figure \ref{fig:weylchamber} (except for its edges, see Exercise 10 below) are possible starting points to studying the perturbative expansion. The all-important---because the dynamics depends on the value of $\langle A_4^3 \rangle$---question as to whether the chosen point on the Weyl chamber is stable in the quantum theory is postponed to Section \ref{sec:adjointgpy}.

In this  Section, we shall study the perturbative expansion on the Weyl chamber, focusing on the neighbourhood of the center symmetric point (i.e. sufficiently far away from the edge of the Weyl chamber, see the last part of Exercise 10)
\begin{equation}
\label{a4vev}
\langle A_4^3 \rangle = v \sim {\pi \over L},
\end{equation}
where we used $v$ to denote the scale of the near center-symmetric vev, as indicated by the $\sim$ sign.
In what follows,  we often drop the expectation values signs  $\langle ... \rangle$  when discussing the vev, hoping that this does not cause undue confusion and is clear from the context.

 The main point we want to make here is that the vev (\ref{a4vev}) plays the role of the vev of the Higgs field: it breaks $SU(2)$ to $U(1)$ at a scale $v$, leading to an abelianization of our 3d effective theory (\ref{3dlagrangian}). Notice that $v$ is of the same order as the lowest KK mass scale already mentioned when writing $L_{3d}$.
The novelty here, with respect to the Polyakov model, is that the Higgs field is part of the gauge connection, so you may not be familiar with its use to break the gauge symmetry (nonzero holonomies around extra spacetime dimensions have long been used to break the gauge symmetry, e.g. in  string theory). Thus, we shall flesh out some of the details. 

To see that the holonomy vev $v$ breaks $SU(2)$ to its Cartan subgroup, it suffices to consider the kinetic terms in (\ref{3dlagrangian}). Now, the $\mu 4$ components of the nonabelian field strength are  $F_{\mu 4} = \partial_\mu A_4 - \partial_4 A_\mu + i \left[ A_\mu, A_4 \right]$.
Next, recall that we are working in the gauge where $A_4^3$ is the only nonzero component of $A_4$, and to boot, it only depends on $\vec x \in R^3$. We can now separate out the various isospin components of $F_{\mu 4}$. In the process, we set $\partial_4 =0$, since we know that every mode with nonzero $x^4$ dependence has KK mass quantized in units of  $1/L$ and that such modes are already integrated out of (\ref{3dlagrangian}). We also expand $A_4^3 = v + a_4^3$ into its vev and $R^3$-dependent fluctuation, to find for the zero-KK modes of the field strength, $F_{\mu 4}^3 = \partial_\mu A_4^3$ for the Cartan component, while  the non-Cartan components are
\begin{eqnarray}
\label{fmu4}
F_{\mu 4}^1 {\sigma^1\over 2} = i A_\mu^2 (v + a_4^3) [{\sigma^2 \over 2}, {\sigma^3 \over 2}] \simeq - A_\mu^2 v {\sigma^1\over 2},~\text{and} ~~~ F_{\mu 4}^2 {\sigma^2\over 2} = i A_\mu^1 (v + a_4^3) [{\sigma^1 \over 2}, {\sigma^3 \over 2}]\simeq A_\mu^1 v {\sigma^2\over 2},
\end{eqnarray}
where we neglected the  terms  linear in the $a_4^3$ fluctuation in the last equality. It is clear now that the terms proportional to $v$   above give mass of order $v$ to the $W$-bosons $A_\mu^1$ and $A_\mu^2$. Those still skeptical, please consider the following
\begin{quote}
{\flushleft{\bf{Exercise 10:}}} Convince yourself of the validity of    (\ref{fmu4}). Show that, when plugged into $L_{3d}$ of (\ref{3dlagrangian}),  at the center-symmetric point they give mass  $m_W = {\pi \over L}$ to the non-Cartan subalgebra fields $A_\mu^{1,2}$, the $W$-bosons. At the same time, the fluctuation $a_4^3$ and the Cartan component $A_\mu^3$ remain massless (if no potential for $a_4^3$ is present). Argue that the $\mu \ll 1/L$ 3d lagrangian is now the abelian restriction of (\ref{3dlagrangian}) (omitting the isospin index as we did in the Polyakov model):
\begin{equation} \label{3dlagrangianabelian}
L_{3d} = {L \over 4 g_4^2} F_{\mu\nu} F^{\mu\nu } + {L \over 2 g_4^2} (\partial_\mu a_4) (\partial^\mu a_4) + \ldots.
\end{equation}
The dots denote terms that we shall investigate in the following Sections.
 
A bonus question is to answer: What happens to the perturbative spectrum\footnote{Hint: when $\langle A_4^3\rangle =0$, it  should be obvious that the theory remains nonabelian (as an aside, a 3d nonabelian theory  also flows to strong  coupling and is difficult to analyze). When $\langle A_4^3\rangle = 2 \pi/L$, it might seem that a mass is generated, yet again abelianizing the theory in the IR. However, since the theories at the  two edges are related by the $Z_2^{(1)}$ symmetry,  the spectra should be the same and one should be able to see this explicitly.} if the vev (\ref{a4vev}) is taken  towards the edges of the Weyl chamber? \end{quote}

In the remainder of this Section, we shall discuss the reason for the validity of the above weak-coupling analysis of the spectrum at the center-symmetric point---as well as near it, i.e. sufficiently far from the edges of the Weyl chamber.  We stress that to make the story self-consistent, we have to make sure that the (near) center-symmetric vev (\ref{a4vev}) is stable, i.e. it is the value of $\langle A_4^3 \rangle$  preferred by quantum corrections. We shall discuss, in Section \ref{sec:adjointgpy}, under what conditions (\ref{a4vev}) is, indeed, a minimum of the quantum effective potential. For now, we shall simply assume that this is so and argue that for sufficiently small $L$, the weak-coupling expansion is valid.\footnote{We also note that in  
 the case of SYM, there are no perturbative quantum corrections, so the Weyl chamber is, indeed, a ``moduli space''-- a manifold of exactly degenerate ground states, to all orders of perturbation theory.}
This follows from the observations made above: as $v$ plays the role of the $SU(2) \rightarrow U(1)$ breaking scale, it determines the mass of the $W$-bosons. In a theory with only adjoint fields, all components which are not in the Cartan subalgebra of $SU(2)$ obtain mass of order $v$ (notice that  non-Cartan components of adjoint-representation fields are the only ones charged under the unbroken $U(1)$). Thus the long-distance theory is a free $U(1)$ theory with no light charged fields. 
\begin{figure}[h]
\centerline{
\includegraphics[width=8.5 cm]{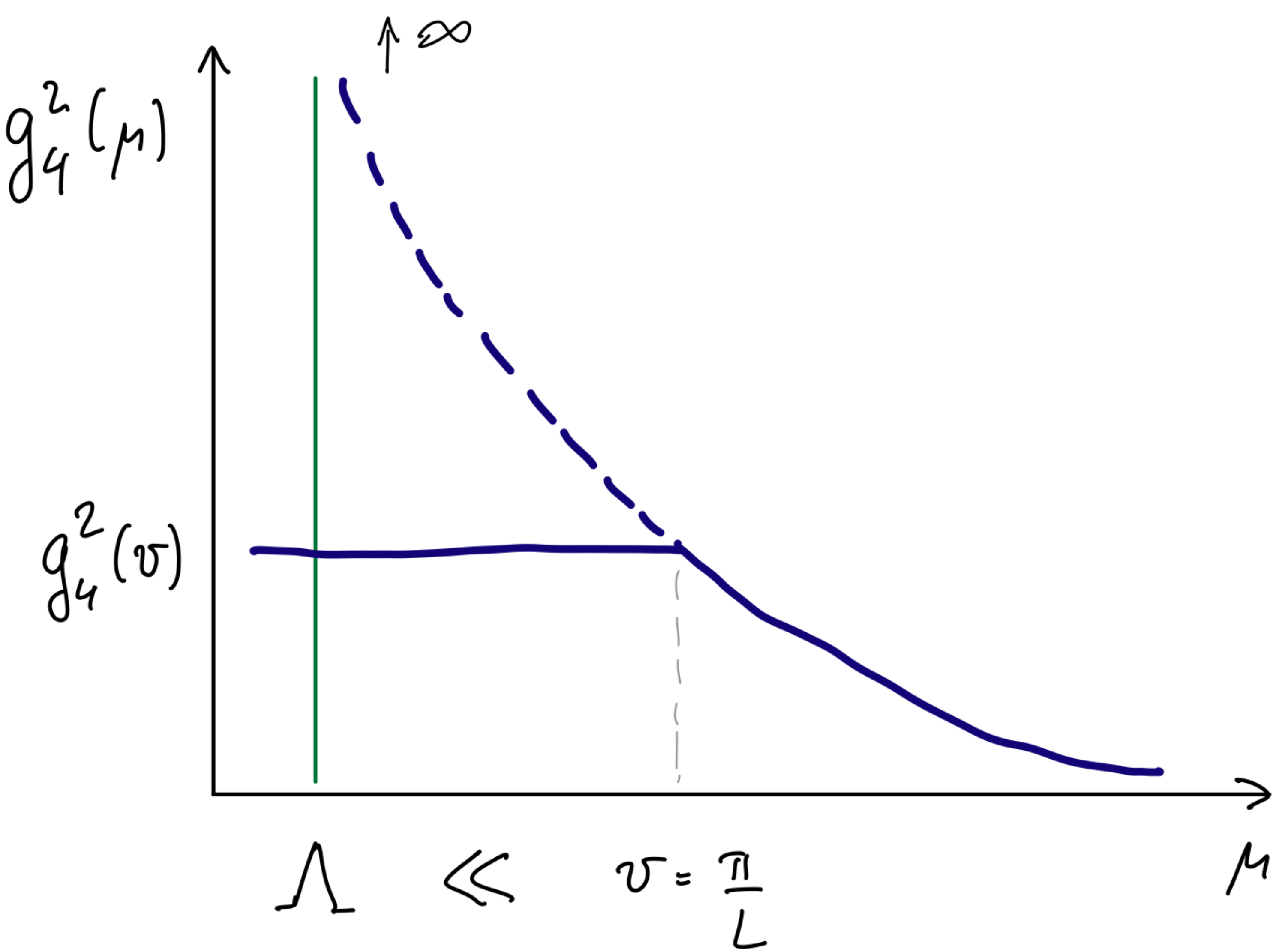}}
\caption{ The asymptotically-free running  coupling $g_4^2(\mu)$ of the 4d theory approaches strong coupling at IR scales of order $\Lambda$, as shown by the dashed line. At (or near) the center-symmetric point on $R^3 \times S^1$, $SU(2) \rightarrow U(1)$ Higgsing takes place at a  scale $v \sim { \pi\over L}$. The long-distance 3d $U(1)$ theory is free, with 3d gauge  coupling determined by matching to the 4d theory and given by $g_{3}^2 = g_4^2(\mu\sim v)/L$. The weak-coupling analysis is justified if the scale of the breaking is larger than $\Lambda$, i.e. when  $\Lambda L \ll \pi$. \label{fig:runningcoupling}}
\end{figure}

Let us now recall the logarithmic running of the coupling $g_4^2(\mu)$ in a 4d nonabelian asymptotically-free gauge theory, a cartoon of which is shown on Figure \ref{fig:runningcoupling}. At scales $\mu \gg v \sim \pi/L$, the $R^3 \times S^1$ theory is essentially 4d and the coupling runs logarithmically to zero at high energy. If no Higgsing (or compactification) would be present, the  4d coupling would continue to run according to the dashed line, hitting a ``Landau pole'' at an energy scale of order $\Lambda$, the dynamical  strong scale of the theory. But in our setup, at  the scale $v$, related to the size of the $S^1$, the gauge group breaks to $U(1)$. Further, as shown above, all states charged under the $U(1)$ obtain mass of order $v$, thus there is nothing to make the (now 3d) $U(1)$ coupling run at lower energy scales. 
The coupling is  frozen to its value at $\mu \sim v$ and does not evolve as the energy is  lowered. 

Thus, in order for our weak-coupling analysis to be consistent, we must ensure that the coupling $g_4^2(v)$, taken at the scale of the $SU(2)$-breaking $v$, is weak. From Figure \ref{fig:runningcoupling}, we see that this means that   $v \sim \pi/L \gg \Lambda$ must hold, as already stated in the Introduction. Notice that this condition is very similar to the one in the electroweak sector of the standard model: the scale of the Higgs vev must be much larger than the strong-coupling scale of the $SU(2)_L$ weak interaction theory.

The above argument is  a qualitative one, but it can be backed up by a calculation of the loop corrections and a determination of the scale of the coupling $g_4^2$ that enters in (\ref{3dlagrangianabelian}). The procedure of finding the coupling of the EFT (\ref{3dlagrangianabelian}) is called ``matching'' and we shall not discuss the details. (While the calculation is straightforward in principle, the details can be somewhat arduous. These have, so far, been done for theories with massless adjoints, with \cite{Poppitz:2012sw,Anber:2014lba} or without  \cite{Anber:2014sda}  help from supersymmetry. The calculations with massive adjoints have been done only recently \cite{Lai:2021}.) For our purposes, all that matters is that the answer, as suggested earlier, is that the coupling in $L_{3d}$ is the 4d running coupling of the $SU(2)$ theory, $g_4^2(\mu)$, with $\mu$ a scale of order $v \gg \Lambda$.

Later on, we shall sometimes express the parameters defining our long distance EFT on $R^3 \times S^1$ in terms of $\Lambda$ and $L$, with small but nonzero $\Lambda L \ll \pi$. There, we shall  give an equation for $\Lambda$. 
For now we note that the input parameters $\Lambda$ and $L$ can be expressed via a fixed small $g_4^2(1/L) \ll 1$ and $L$.\footnote{\label{footnotelimit}We stress that taking $L$ small and finite, so that $g_4^2(1/L)$ is small is a limit distinct from the dimensional reduction limit, where one takes $L \rightarrow 0$ while the three-dimensional coupling $g_3^2 = g_4^2(1/L)/L$ is fixed, which implies taking $g_4^2(1/L)$ to zero.}

Finally, we can rewrite the $\mu \ll 1/L$ lagrangian  (\ref{3dlagrangianabelian}) using the photon/dual-photon ($\sigma$) duality from Section \ref{sec:polyakov_pass2}. The only replacement we need to make is $g_3^2 \rightarrow g_4^2/L$. Thus, we obtain the perturbative dual effective lagrangian of the $R^3\times S^1$ theory near the center-symmetric point
\begin{equation} \label{3dlagrangiandual}
L_{3d} =  {1 \over 2} { g_4^2 \over L (4 \pi)^2}   (\partial_\lambda \sigma)^2  + {L \over 2 g_4^2} (\partial_\mu a_4) (\partial^\mu a_4) + \ldots.
\end{equation}
As before, $\sigma$ is a compact scalar of period $2\pi$ as argued in Section \ref{sec:polyakov_pass2}.

The next questions we need to address is the nature of the ``$\ldots$'' terms. Some of these terms, as already discussed, depend on the Polyakov loop (\ref{nonlocalterms}) and will be arranged to ensure stability of the center-symmetric vev. This is studied in Section \ref{sec:adjointgpy}. Other terms contributing to ``$\ldots$'' are not local in the original electric variables (\ref{3dlagrangian}) but are local in the $\sigma$-description. These arise from   corrections due to various  nonperturbative instanton fluctuations, as in the Polyakov model. The monopole-instantons  in the $R^3 \times S^1$ theory are studied in Section \ref{sec:bpsandkkmonopoles}. 

\hfill\begin{minipage}{0.85\linewidth}

\textcolor{red}{
{\flushleft{\bf Summary of \ref{sec:perturbative3x1}:}} Here, we found the  perturbative spectrum of the $SU(2)$ theory expanded close to the middle of the Weyl chamber. We showed that $SU(2)$ abelianizes, i.e. breaks to $U(1)$ at the scale $v \sim \pi/L$, due to the center symmetric expectation value of the holonomy (\ref{a4vev}). The long distance theory is the rather boring one of (\ref{3dlagrangianabelian}): the 3d abelian free Maxwell theory along with the massless neutral scalar $a_4$. We  rewrote it using the dual-photon description in (\ref{3dlagrangiandual}). We also argued that the weak-coupling analysis is valid provided $L \Lambda \ll \pi$ holds, i.e. the circle size is small compared to the inverse strong-coupling scale of the theory. (For $SU(N)$ theories, the  condition is now $\Lambda L N \ll 2 \pi$, since $SU(N) \rightarrow U(1)^{N-1}$ at a scale $2 \pi/(LN)$.) }

\end{minipage}

\bigskip

\subsection{$M$ and $KK$ monopole-instantons.}
\label{sec:bpsandkkmonopoles}

In this Section, we continue to work at (or near) the center-symmetric point (\ref{a4vev}). As in the Polyakov model, we found that the IR theory is the rather boring (\ref{3dlagrangianabelian}), in complete analogy with Section \ref{sec:pert_polyakov}. Also similar to what we did in Section \ref{sec:instantons_polyakov}, we proceed to study the finite action Euclidean solutions in our $R^3 \times S^1$ theory near the center-symmetric point. 

The solutions that we shall discuss and their properties were discovered in the late 1990's in a remarkable set  of papers \cite{Lee:1997vp,Kraan:1998pm}. What they found was that the 4d BPST instanton in an $SU(N)$ theory dissociates into $N$ constituents in the bulk of the Weyl chamber (its generalization to $SU(N)$). Here, we shall construct these solutions explicitly for $SU(2)$. By studying their properties, we shall show that they have the correct ``quantum numbers'' (remember they are not particles!) to be interpreted as instanton constituents. In the past, people have had reasons to suggest  the existence of instanton constituents (``instanton quarks''), but the present semiclassical incarnation is  a very concrete, well-defined, and useful way to uncover them.  

The paper of K. Lee and P. Yi \cite{Lee:1997vp}, in particular, studied the maximally supersymmetric ${\cal{N}}=4$ Yang-Mills theory  on $R^3 \times S^1$ and its finite action monopole-instantons occurring in the Weyl chamber in these theories. They used the $D$-brane realization  of ${\cal{N}}=4$ SYM in string theory. It turns out that it helps visualize these novel solutions and figure out many of their properties in a remarkably simple way. The reason these monopole-instantons, found in a highly supersymmetric setting,  also occur in nonsupersymmetric theories is that the nonzero bosonic backgrounds only involve a subset of the fields of ${\cal{N}}=4$ SYM, namely those already occurring in the nonsupersymmetric gauge kinetic term (\ref{4dlagrangian}).
We shall not use the stringy language as it entails introducing more background than we can possibly do here.  (We mention it for historical reasons and because it can be quite useful.) Instead, we shall 
use the QFT way of looking for these solutions.

\begin{figure}[h]
\centerline{
\includegraphics[width=10cm]{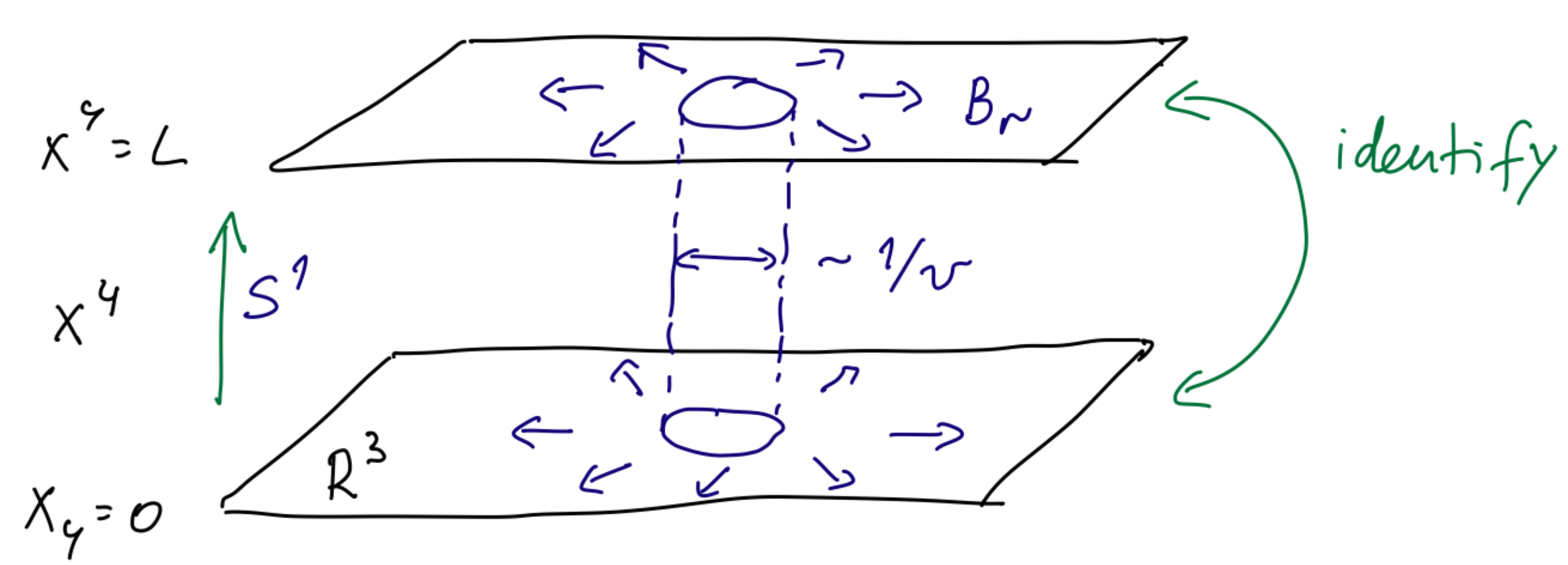}}
\caption{ A cartoon of the spacetime structure of the $x^4$-independent $M$ monopole-instanton solution on $R^3 \times S^1$.  This is essentially the solution localized in $R^3$ shown on Figure \ref{fig:monopole1}, now  trivially embedded in $R^3 \times S^1$, by allowing it to propagate in $x^4$ without change. The $KK$ monopole-instanton, on the other hand,  is twisted in the $S^1$ direction by  the improper gauge transform $g_A(x^4)$ of (\ref{improper}). \label{fig:mmonopole}}
\end{figure}

 Let us begin by recalling the self-dual BPS monopole-instanton solutions (\ref{bpsmonopole1}) of the Polyakov model. We shall call them $M$,  motivated by (\ref{thooftvertices}). Recall that these solve the equations of motion of the 3d $SU(2)$ gauge theory with an adjoint Higgs field $A_4^a$.  But
the BPS-limit 3d lagrangian (\ref{bps3dlagrangian}) equals precisely our tree-level $R^3 \times S^1$ lagrangian without $x^4$-dependence (\ref{3dlagrangian}). Thus, the BPS solutions of (\ref{bps3dlagrangian}) will also be solutions of the $x^4$-independent equations of motion following from (\ref{4dlagrangian}). We only need to replace the vev appearing in (\ref{bpsmonopole1}) by that appearing in (\ref{a4vev}). But since we judiciously used the same letter, there is nothing left to do. The only difference occurs in the overall normalization of the action. The BPS monopole instantons had action $S_0 = 4 \pi v/g_3^2$. Now we only need replace $v \rightarrow v \sim \pi/L$ and $g_3^2 \rightarrow g_4^2/L$, to obtain the action of the BPS monopole-instanton solutions $M$ on $R^3 \times S^1$ at a point $v$ on the Weyl chamber:
\begin{equation}
\label{Maction}
S_M(v) = {4 \pi v L\over g_4^2}, ~~ \text{at center symmetry:} ~S_M({\pi\over L}) = {1 \over 2}  {8 \pi^2 \over g_4^2} = {1 \over 2} S_{BPST}.
\end{equation}
We noted above the remarkable fact that the action of the $M$ monopole-instantons, evaluated at the center-symmetric point, is equal to one-half the famous BPST instanton action in 4d (we shall come to this point below). Also, from the construction above, these finite-action Euclidean solutions of (\ref{4dlagrangian}) on $R^3\times S^1$ are independent on the $x^4$ coordinate. Viewed from the perspective of $R^3\times S^1$, their core has the shape of a ``tube,'' localized on $R^3$, of   thickness $1/v$, and winding around the $S^1$ (i.e. they are extended objects on $R^3 \times S^1$, which is how they appear in string theory). The core of the $M$ monopole-instanton and its asymptotic field on $R^3\times S^1$ are sketched on Figure \ref{fig:mmonopole}.

\begin{figure}[h]
\centerline{
\includegraphics[width=12.5 cm]{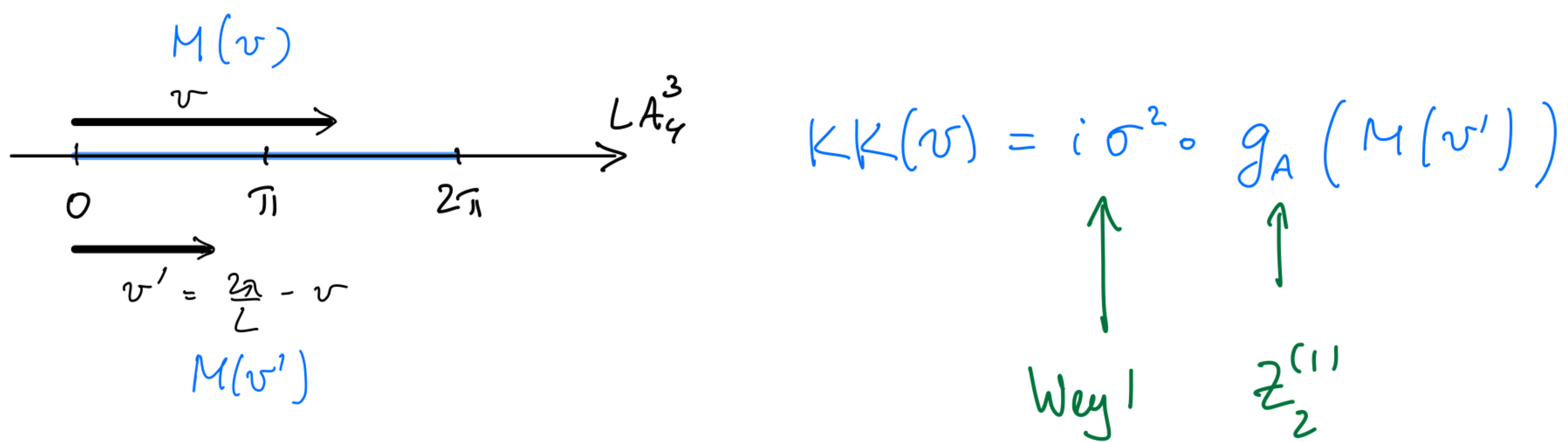}}
\caption{ Illustrating the construction of the $KK$ monopole-instanton. The arrow of length $v$ denoted by $M(v)$ shows the variation of the $A_4$ field  of the $M(v)$ solution: from $0$ at its center to its asymptotic value $v$ at $R^3$ infinity. To construct the $KK$ monopole, start with an $M(v')$ solution (denoted by $M'$ in the text), where $A_4$ varies from $0$  at the center to $v'$ asymptotically, as shown by the lower arrow. Then apply a global $Z_2^{(1)}$  (represented by the improper gauge transformation $g_A$ of (\ref{improper})) and a Weyl reflection on the $M(v')$ solution, a procedure schematically denoted by $ i \sigma^2 \circ g_A$ on the Figure. As symmetries map solutions into solutions, the result is  the monopole-instanton $KK(v)$, in the same vacuum as $M(v)$. \label{fig:kkmonopole}}
\end{figure}

The most interesting part of the story is only beginning. As advertised, it turns out that there is another self-dual solution, whose quantum numbers complement those of the $M$ solution found above, so that the two together have the properties of a 4d BPST instanton. We shall now show this using QFT tools. The procedure we are about to describe between eqns.~(\ref{vprime}) and (\ref{kk1}) is illustrated on Figure \ref{fig:kkmonopole}.

To construct this so-called $KK$ monopole-instanton solution, 
recall that $v$ (taken to denote the $A_4^3$ vev) is a compact variable parameterizing the Weyl chamber, recall Figure \ref{fig:weylchamber}. Consider now another vev, a reflection of $v$ across the midpoint of the Weyl chamber:
\begin{equation}\label{vprime}
v' = {2 \pi \over L} - v~. 
\end{equation}Clearly,  this is also a point on the Weyl chamber ($v'=v$ at the center symmetric point). Consider now the self-dual $M$ solutions constructed above, but with $v$ replaced by $v'$, so we call them $M'$. The asymptotics of $A_4$ and $B_r$ (understood to point in the 3rd isospin direction, recall (\ref{stringgauge})) far away from their respective cores, and the actions of the $M$ and $M'$ self-dual solutions are as follows:\footnote{We have chosen, for brevity, to not display the $1/r$ tail of $A_4$ asymptotics, as in (\ref{stringgauge}).}
\begin{eqnarray}\label{mprime}
M:&& A_4 \simeq v, ~ B_r \simeq {1\over r^2}, ~S_M = {4 \pi v L \over g_4^2}~,\\
M':&& A_4' \simeq v', ~ B_r' \simeq {1 \over r^2},~ S_{M'} = {4\pi v' L \over g_4^2} = {8 \pi^2 \over g_4^2} - {4 \pi v \over g_4^2}~. \nonumber
\end{eqnarray}
We stress that the two solutions live in different theories (or more precisely, different superselection sectors, defined around different vacuum states): one with $A_4^3$ vev  $v$, and the other $v'$. Of course, if   $v=v'$ these solutions are identical. 

But now we notice an interesting fact: we can use $M'$ to construct a solution in the same vacuum, i.e. with $A_4$ vev $v$. We shall achieve this in two steps. First, we  apply the already familiar, recall (\ref{improper}), 
 improper $SU(2)$ gauge transformation $g_A(x^4)= e^{ \; i{2 \pi x_4 \over L} {\sigma^3 \over 2}}$. Under such transformations $SU(2)$ connections transform as  $A_M' \rightarrow  A_M'' = U(A_M' - i \partial_M)U^{-1}$. Recall that $g_A$ is not a gauge transformation globally, but is one locally. Thus, it will map solutions to solutions, but will change the boundary conditions. On the other hand, local gauge invariants, such as the action, will remain the same. So, by transforming $M'$ via $g_A$, we   construct another solution, which we call $M''$. The asymptotic values (away from the core) of $M''$ are now \begin{eqnarray}\label{mtwoprime}
M'':&& A_4''\simeq v' - {2 \pi \over L} = - v, ~ B_r'' \simeq {1 \over r^2},~ S_{M''} = S_{M'} = {8 \pi^2 \over g_4^2} - {4 \pi v \over g_4^2} =  {8 \pi^2 \over g_4^2} - S_M~.
\end{eqnarray}
This asymptotic behaviour of $M''$ is due to the fact that\footnote{Acting with $g_A$ on the asymptotic forms of $A_4', B_r'$, one has to imagine them multiplied by $\sigma^3/2$.  Notice also that the full   nonabelian core of the solutions is severely affected, acquiring $x^4$ dependence (hence the ``tube'' stretching along $S^1$ (shown on Figure \ref{fig:mmonopole}) is twisted as it winds around the $x^4$). Finally, all of the construction here is in the string gauge. Luckily, it should be clear by now that we won't make use of  the core structures; needless to say, they can be worked out.}   $g_A$ commutes with the asymptotic abelian form of $A_4', B_r'$ in (\ref{mprime}). Only the asymptotic value of $A_4''$ is shifted due to the inhomogeneous term in its  gauge transform. 

We are not done, yet, because the $A_4''$ asymptotics in (\ref{mtwoprime}) is still $-v$, rather than $v$. The second step in constructing the so-called KK (for Kaluza-Klein) monopole-instanton solutions is to use the already familiar Weyl reflection to change the sign of $-v$ (this is the constant gauge transformation $g = i \sigma^2$). This has the effect of mapping solutions to solutions as well as of reversing both $A_4''$ and $B_r''$ above (as they are both along $\sigma^3$), without changing the action. After this step, we finally obtain the asymptotic form of the self-dual $KK$ monopole-instanton solutions. There are now two sets of distinct solutions  at the same point on the Weyl chamber:
\begin{eqnarray}\label{kk1}
KK:&& A_4 \simeq    v, ~ B_r  \simeq -{1 \over r^2},~ S_{KK} = {8 \pi^2 \over g_4^2} - S_M~,\\
M:&& A_4 \simeq v, ~ B_r \simeq {1\over r^2}, ~S_M = {4 \pi v L \over g_4^2}~.\nonumber
\end{eqnarray}
We now notice a curious fact: the sum of the actions $S_{KK} + S_M$ equals the 4d BPST instanton action. So, it appears that in the bulk of the  Weyl chamber, the 4d instanton has dissociated into solutions of smaller action ($1/2$ each, at $v=\pi/L$), the $M$ and $KK$ monopole-instantons. This is quite remarkable and this interpretation will be strengthened below. 
A further argument for the consistency of this interpretation is to notice that the self-dual $M$ and $KK$ solutions have opposite magnetic charge (\ref{magneticcharge}), as follows from the opposite sign of the asymptotics of their $B_r$ fields. Thus, a collection  of an $M$ and $KK$ has no long-range magnetic field, as appropriate for 4d instantons which lack long-range tails.

Before we conclude this Section, we have to come to grips with yet another property of the $M$ and $KK$ monopole-instantons: their topological charge. In order to be viewed as constituents of BPST instantons, their topological charges have to sum to unity. We shall shortly show this.
This is a feature absent in the 3d Polyakov model and is  the reason why these $R^3 \times S^1$ theories are distinct from it,  ``remembering'' their 4d origin.\footnote{Recalling Footnote \ref{footnotelimit}, we stress that in the dimensional reduction limit with $g_4^2(1/L) \rightarrow 0$ and $v$ an arbitary scale, the KK monopole action goes to infinity, i.e. the theory loses the information about its 4d origin. This is not our small-$L$, fixed-$\Lambda$ limit.}

Recall that the topological charge is defined as
\begin{equation}
\label{topologicalcharge}
Q_T ={1 \over 32 \pi^2} \int\limits_{R^3 \times S^1} d^4 x F_{MN}^a \tilde F_{MN}^a~,~~ \tilde F_{MN} = {1 \over 2} \epsilon_{MNPQ} F_{PQ}~,
\end{equation}
where  all repeated indices are summed over. 
The topological charge comes together with a new parameter, the $\theta$ angle. Depending on the matter content, it may or may not be observable, but can always be included as a book-keeping device. In the Euclidean path integral, Euclidean field configurations of nonzero $Q_T$ contribute with an extra phase factor $e^{i \theta Q_T}$ multiplying their Boltzmann probability.

 On $R^3 \times S^1$, the topological charge can be expressed in terms of the asymptotics of the $B$-field at the $R^3$ infinity (i.e. the magnetic charge), the value of the holonomy (the $A_4$ vev $v$), and the integer Pontryagin index characterizing $\pi_3(G)$, the third homotopy class of the gauge group.  The topological classification of  finite action Euclidean solutions on $R^3 \times S^1$ was  given in the GPY paper  \cite{Gross:1980br} (our calculation below, yielding (\ref{topologicalcharges}), conforms with this and will suffice here). 
The calculation of the topological charge for the $M$ and $KK$ solutions will be done in 
\begin{quote}
{\flushleft{\bf Exercise 11:}} Use the  fact that the topological charge density $q$ is a total divergence, i.e. $q = {1 \over 32 \pi^2}   F_{MN}^a \tilde F_{MN}^a = {1 \over 16 \pi^2} \epsilon_{MNLK} \partial_M  \left( A_N^a \partial_L A_K^a - {1 \over 3} \epsilon^{abc} A_M^a A_L^b A_K^c\right)$.  Integrate by parts and, using the asymptotics  the $M$ and $KK$ monopole-instantons, show that 
\begin{eqnarray}\label{topologicalcharges}
Q_T(M) &=& {1 \over 8 \pi^2} \int\limits_0^L dx^4 \oint\limits_{S^2_\infty} d^2 s^\mu \; A_4^3 \; B_\mu^3 = {v L \over 2 \pi}  \rightarrow {1 \over 2} \; \text{at center symmetry} \nonumber \\
Q_T(KK) &=& Q_T(M') = {v' L \over 2 \pi} = 1 - {v L \over 2 \pi} \rightarrow {1 \over 2} \; \text{at center symmetry}.
\end{eqnarray}
Thus, $Q_T(M) + Q_T(KK) = 1$. 
This result can also be obtained from the fact that both solutions are self-dual in 4d sense, i.e. their action is proportional to the topological charge (compare (\ref{topologicalcharges}) with (\ref{kk1})).
\end{quote}
There is an infinite tower over the $M$ and $KK$ solutions, of increasingly larger action,\footnote{Obtained by ``adding BPST instantons'' to them---this can be achieved by starting with the simple BPS solutions with  vevs $v + {2 \pi k \over L}$ and then bringing the vev into the $[0, 2\pi]$ domain by applying  multiple $g_A$ transformations.} but  we shall not describe it, as we are only interested in the lowest action solutions. Near center symmetry, these are the  $M$ and $KK$.

To end this Section, we can now combine the information about our   Euclidean solutions of finite action into 't Hooft vertices similar to (\ref{thooftvertices}). 
We can use the insight from the Polyakov model,  because our theory is abelian near the center-symmetric point and the long-distance physics on $R^3 \times S^1$ at $\mu \ll 1/L$ is essentially 3d. 
Thus,  we can use a dual description of the unbroken $U(1)$ in terms of a dual photon $\sigma$ already given in (\ref{3dlagrangiandual}).

 Furthermore, since the cores of the $M$ and $KK$ monopole instantons are of size   $\sim v^{-1} \sim L$, the only property that matters is their action, topological charge, and long-distance interactions  via long-range magnetic fields.\footnote{We shall see in the next Section, that, as in the Polyakov model, the $A_4^3$-field is gapped due to the yet-unknown terms (\ref{nonlocalterms}), except in SYM.}
Thus, we can now simply list the four 't Hooft vertices (corresponding to the two solutions we found and their ``anti-particles'')   in the $SU(2)$ theory on $R^3 \times S^1$. Here, we take the actions and topological charges at the center symmetric point (as promised and postponed many times, we will justify this assumption in the next Section):
\begin{eqnarray}
\label{thooftdym}
M &:& L^{-3} \;e^{- S_0+ i {\theta \over 2}} \; e^{+ i \sigma},  ~\text{where, from now on} ~~ S_0 \equiv {4 \pi^2 \over g_4^2} = {S_{BPST} \over 2}~\nonumber \\
M^*&:& L^{-3} \; e^{- S_0 - i {\theta \over 2}} \; e^{- i \sigma}, \nonumber\\
KK &:& L^{-3} \; e^{- S_0+ i {\theta \over 2}} \; e^{- i \sigma },  \\
KK^* &:& L^{-3} \; e^{- S_0- i {\theta \over 2}}\;  e^{+ i \sigma },.\nonumber
\end{eqnarray}
The above mnemonic is as the one in the Polyakov model (we took the liberty to  omit the explicit $x \in R^3$ dependence). The replacement of $v$ with $1/L$ in the prefactor is due to the holonomy vev (\ref{a4vev}), working with exponential-only accuracy, as before. 

The most important new features of (\ref{thooftdym}), compared to (\ref{thooftvertices}), are as follows. First, the $\theta$-angle dependence, taking into account the fractional topological charge of the solutions: $+1/2$ for $M$ and $KK$ and $-1/2$ for their ``antiparticles.'' The second major difference with respect to the 3d Polyakov model is the appearance of a new set of monopole-instantons, the $KK$ solutions described earlier. As the $M$ and $KK$ have opposite magnetic charges, they come with opposite-sign $e^{\pm i \sigma}$ factors. 

A further property of the $R^3 \times S^1$ EFT distinct from the Polyakov model is the presence of the $Z_2^{(1)}$ center symmetry associated with the $S^1$. Now we recall that $KK$ monopole instantons were obtained from $M'$ using a global center-symmetry transformation and a gauge transformation (Weyl reflection). Thus, at the center symmetric vev, the two are center-symmetry images of each other, as $M=M'$. Thus, the $Z_2^{(1)}$ action on the 't Hooft vertices should be to exchange the $M$ and $KK$ vertices, $Z_2^{(1)}: M \rightarrow KK$,  or $Z_2^{(1)}: e^{- S_0 + i {\theta \over 2}} e^{i \sigma} \rightarrow e^{- S_0 + i {\theta \over 2}} e^{-i \sigma}$. We conclude that the dual photon transforms under the $Z_2^{(1)}$ center symmetry\footnote{To avoid confusion, we remind the reader (see also Appendix \ref{appx:notation}) that, as we elaborated after (\ref{center1}), in these notes we use $Z_2^{(1)}$ to denote the center symmetry along the $S^1$ direction. Hence it acts on local fields in the 3d theory valid at distances $\gg L$; in many papers it is sometimes called ``0-form center symmetry.''} as \begin{equation}
\label{centersigma}
 Z_2^{(1)}: \sigma \rightarrow 2 \pi - \sigma~,
\end{equation}
where the $2\pi$ shift  keeps the image in the fundamental domain. Earlier arguments for (\ref{centersigma}) are in \cite{Anber:2015wha,Aitken:2017ayq}. We shall find many uses of (\ref{centersigma}) in what follows.

Our plan is, in the next Section \ref{sec:adjointgpy}, to discuss the reasons for the stability of center symmetry in different classes of theories, closing our last loophole. After this we shall finally come to our main topic:  studying the calculable dynamics of confinement and chiral symmetry breaking.

\hfill\begin{minipage}{0.85\linewidth}

\textcolor{red}{
{\flushleft{\bf{Summary of \ref{sec:bpsandkkmonopoles}:}}} The main result obtained here is the classification of instantons of lowest action near the center-symmetric point on $R^3 \times S^1$. The appearance of the $KK$ monopole-instantons, in addition to the $M$ monopole-instantons (which, in our gauge, resemble those in the 3d Polyakov model) is a new feature related to the 4d nature of the theory and its topological structure.
These solutions carry fractional, compared to the 4d BPST instanton, actions and topological charges. 
 We introduced 't Hooft vertices (\ref{thooftdym}) for the  $M$, $M^*$, $KK$ and $KK^*$ monopole-instantons that we shall use to study the dynamics. Similar to (\ref{thooftvertices}), these encode the action and topological charge of the various instantons as well as their long-distance magnetic interactions. Further, we argued that the $Z_2^{(1)}$ symmetry interchanges the $M$ and $KK$ 't Hooft vertices.}
 
 \end{minipage}
 
 \bigskip

\subsection{Adjoint fermions, ``GPY'' potential, and center stability.}
\label{sec:adjointgpy}

Before continuing to analyze the dynamics of various models, we have to address the reason for our choice of center-symmetric vev (\ref{a4vev}). This is important, because as we explained in Section \ref{sec:perturbative3x1}, it is the (near) center stability which is responsible for abelianization and calculability of the IR physics on $R^3 \times S^1$.  We finally shall address this question here, by computing the one-loop effective potential on the Weyl chamber, i.e. the $v$-dependent contribution to the vacuum energy. It is often called ``the Gross-Pisarski-Yaffe, or GPY, potential'' in the pure-YM case \cite{Gross:1980br}. We shall see that in the pure-YM theory, center symmetry is maximally broken, while the addition of adjoint fermions naturally stabilizes it.

To this end, we right away modify our theory (\ref{4dlagrangian}) by adding fermions in the adjoint representation of the gauge group. We write the fermion action\footnote{See Appendix \ref{appx:notation} for an explanation of our notation, including warnings about possible confusions. } in Minkowski space with $(+,-,-,-)$ metric, as in (\ref{dual2}), where now $M, N= 0,1,2,3$ and the $S^1$ direction is $x^3 \equiv x^3 + L$:
\begin{equation}
\label{lagrangianadjoint}
L_{4d, adj.} =
- {1 \over 4 g_4^2} F_{MN}^a F^{MN \; a} + {i \over g_4^2} \bar\lambda^a_{\dot\alpha} \; \bar\sigma^{M \dot\alpha \alpha} (D_M \lambda_\alpha)^a + {1 \over g_4^2} ({m \over 2} \lambda^a_\alpha \lambda^a_\beta \epsilon^{\beta\alpha} + {m^*\over 2} \bar\lambda^a_{\dot\alpha} \epsilon^{\dot\alpha \dot\beta} \bar\lambda^a_{\dot\beta})~.
\end{equation}

We omitted the topological term from (\ref{4dlagrangian}) as it will play no role until later.
   The Weyl fermions are two-component $SL(2,C)$ spinor fields $\lambda^a_\alpha$, anticommuting variables representing a single flavour of adjoint Weyl fermion. Most importantly, the fermions satisfy periodic boundary conditions on the $S^1$, just like the gauge field, $\lambda_\alpha^a (x^3+L) = \lambda_\alpha^a(x^3)$. $m$ is the complex Majorana mass parameter of the adjoint fermion. 
For notational simplicity (essentially to avoid introducing a flavour index) we wrote (\ref{lagrangianadjoint}) for a single flavour, but the generalization to  $n_f$ flavours $\lambda_\alpha^{a, I}$, where $I=1,...,n_f$ is the flavour index, is trivial.  The one assumption we  make (for convenience) for $n_f > 1$ Weyl fermions is that the masses of the different fermion flavours are all equal. We shall study the global symmetries of the adjoint theory and their realization in the vacuum in detail in Section \ref{sec:adjsym} below.

\subsubsection{The adjoint spectrum on the Weyl chamber.} 
\label{sec:adjointspectrum}
But first, to the matter at hand, the center stability. Consider the theory (\ref{lagrangianadjoint}) in the  bulk of the Weyl chamber, i.e. for $A_4^3 \rightarrow A_3^3 \equiv v \ne \{0, {2 \pi  \over L} Z \}$, recalling (\ref{a4vev}) and the $4 \rightarrow 3$ replacement due to our index choice above. We showed in Section \ref{sec:perturbative3x1} that this abelianizes the gauge theory.
As far as the adjoint fermions are concerned, it is easy to see that $\lambda_\alpha^1$ and $\lambda_\alpha^2$ obtain a contribution to their mass due to $A_3^3 = v$ (since the vev is in the $T^3$ direction), while $\lambda^3_\alpha$ does not. In particular, for $m=0$, the Cartan component of the adjoint fermion remains massless, while the non-Cartan ones obtain mass of order $v$. 

To see this explicitly,  and for important use further  below, consider the following
\begin{quote}
{\flushleft{\bf Exercise 12:}} Expand the periodic fermions in a Fourier integral/series ($x^\mu = (x^0, x^1, x^2) \in R^3$)
\begin{equation}
\label{lambdaseries}
\lambda^a_\alpha(x^\mu, x^3) = {1 \over L} \sum\limits_{p \in Z} \int {d^3 k \over (2 \pi)^3} e^{i {2 \pi p \over L} x^3 + i k_\mu x^\mu} \lambda_\alpha^a (k^\mu, p)~,
\end{equation}
and likewise for the c.c. field $\bar\lambda$. Here $k^\mu$ denotes the $R^3$-momentum vector and $p$ is the KK number. Further, introduce $\sigma^\pm = (\sigma^1 \pm i \sigma^2)/2$ and define the corresponding components of the adjoint fermion as 
\begin{equation}
\label{lambdagroup}
\lambda_\alpha = T^3 \lambda^3_\alpha + {\sigma^+ \over 2} \lambda^+_\alpha + {\sigma^- \over 2} \lambda^-_\alpha~.
\end{equation}
Next, keep only the $A_3^3 = v$ gauge-field background, setting all other components to zero.
Convince yourself that in Fourier space (i.e. acting on above $\lambda$), the kinetic operator becomes
$\bar\sigma^M D_M = (\sigma^\mu k_\mu, \sigma^3({2 \pi p\over L} + v \left[ T^3, \circ\right])$ 
(the  $\left[ T^3, \circ\right]$ notation means that it acts as a commutator on the Lie algebra). Then, show that the fermion kinetic term in (\ref{lagrangianadjoint}) becomes
\begin{eqnarray}
\label{lambdakineticweyl}
S_{kin., \lambda} &=& {1 \over  g_4^2}{1 \over L} \sum\limits_{p \in Z} \int {d^3 k \over (2 \pi)^3}\left[ \bar\lambda^3(\vec{k},p) \left(\sigma^{\mu} k_\mu + \sigma^3 {2 \pi p \over L}\right) \lambda^3(\vec{k,p}) \right. \nonumber 
\\
&&  \left. \qquad \qquad \qquad+ {1 \over 2} \bar\lambda^+(\vec{k,p})\left(\sigma^{\mu} k_\mu + \sigma^3({2 \pi p \over L} - v)\right) \lambda^-(\vec{k},p) \right. \\
&& \left. \qquad \qquad \qquad+ {1 \over 2} \bar\lambda^-(\vec{k,p})\left(\sigma^{\mu} k_\mu + \sigma^3({2 \pi p \over L} + v)\right) \lambda^+(\vec{k},p)\right] \nonumber.
\end{eqnarray}
\end{quote}
The interpretation of the above is clear: the $p=0$ KK mode of $\lambda^3$ remains massless, while the $\lambda^\pm$ non-Cartan components, the ones with definite charges under the unbroken Cartan $U(1)$ all obtain mass due to the expectation value of $A_3^3$ (except when $v = 2 \pi Z/L$, revisit  Exercise 10). The mass obtained due to the coupling to $v$ is sometimes called ``real'' mass, a terminology coming from supersymmetry, to be contrasted with the complex mass parameter $m$ in (\ref{lagrangianadjoint}). It is important to note that this mass does not break the chiral symmetry (this is only possible in a setting when a direction of $R^4$ is compactified).

We shall see below that there are many uses  of the result (\ref{lambdakineticweyl}) of this simple Exercise. 

\subsubsection{The one-loop GPY potential via supersymmetry.}
\label{sec:oneloopgpy}

First, we note that the single-flavour theory with $m=0$ is actually ${\cal{N}}=1$ SYM. Our $S^1$ boundary conditions on the gauge field and $\lambda$ are both periodic, so the $R^3 \times S^1$ compactification preserves supersymmetry. Furthermore, the constant vev $A_3^3 = v$ also preserves supersymmetry. We shall now use  one important property of SYM (that you must have heard about, so we simply quote it), namely that in supersymmetric backgrounds the $vL$-dependent  loop-corrections to the vacuum energy cancel. At the one-loop level, this cancellation implies that the gauge boson loop contribution to the $v$-dependent vacuum energy is equal to the negative of the fermion-loop contribution. 

This statement is of great use already in the study of the pure-YM theory, i.e. with the fermions absent, as it will allow us to simply reproduce the classic GPY result about the one-loop ``GPY'' potential in the bosonic YM theory on $S^1$, which, by the above reasoning is equal to the negative of the $m=0$ Weyl fermion contribution.
The calculation of the adjoint Weyl fermion contribution to the vacuum energy is simpler than the original GPY calculation.\footnote{The interested reader can compare with the original calculation of \cite{Gross:1980br}; here, we won't have to think about ghosts, gauge fixing, etc. Also, for   a gauge invariant definition of the potential for tr$\Omega$, the constrained effective potential, see e.g. \cite{KorthalsAltes:1993ca}.}

\begin{figure}[h]
\centerline{
\includegraphics[width=9.5 cm]{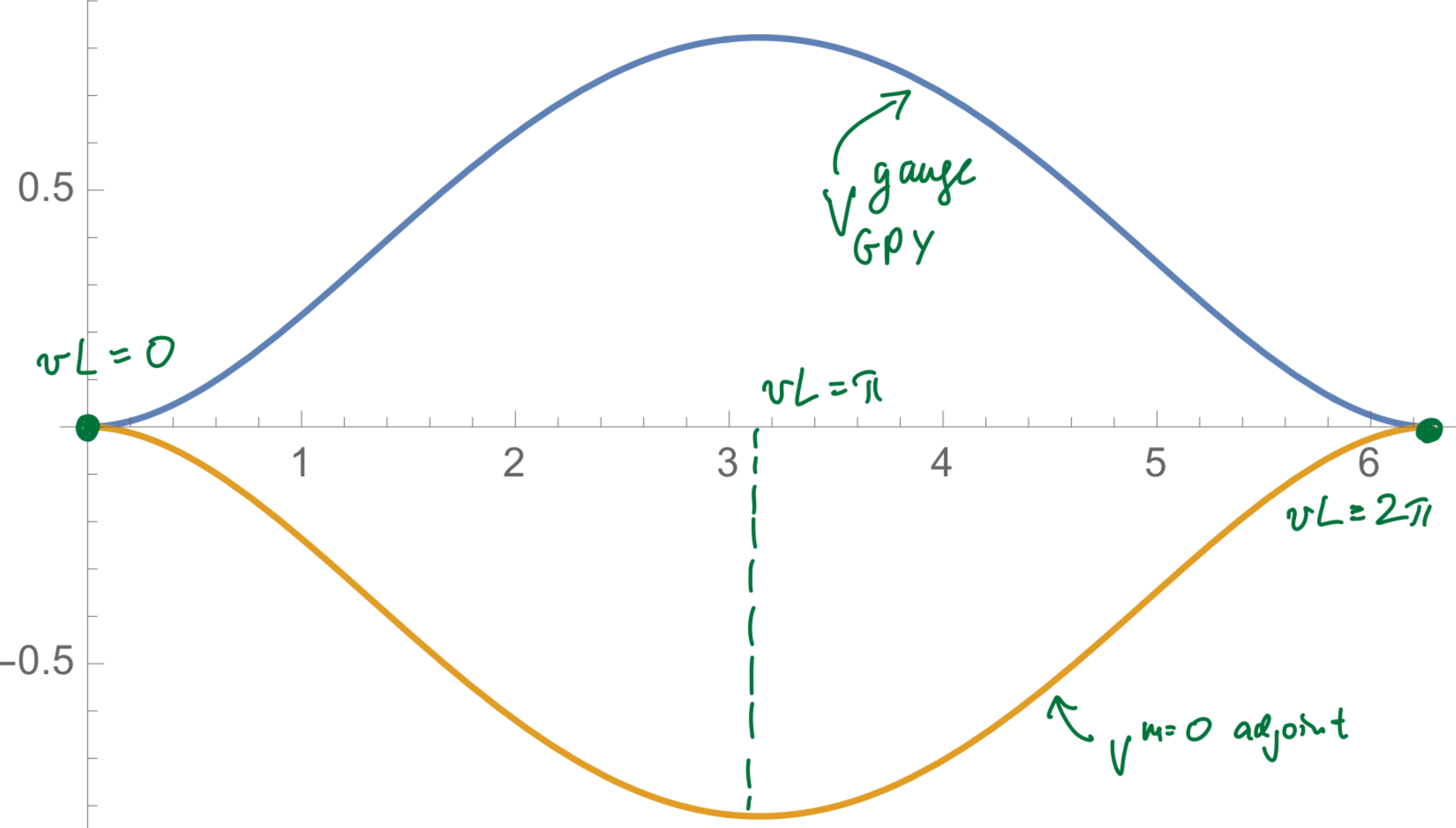}}
\caption{ The one-loop effective potential on the Weyl chamber in pure YM theory, computed using supersymmetry (eqn.~(\ref{gpygauge}) is plotted with $L=1$). In the pure-YM theory, the one-loop potential is given by the top curve (\ref{gpygauge}) and the energy is minimized at $v =0$ and $v={2\pi \over L}$, i.e. at ${1\over 2}\langle\tr \Omega\rangle = \pm 1$. The $Z_2^{(1)}$ symmetry is thus maximally broken. The physical interpretation of this phase is a deconfined phase of YM theory, see Section \ref{sec:digression}. \newline In SYM the potential is the sum of the two curves and thus vanishes  (in fact, it vanishes to all orders of perturbation theory). Thus,  one can study the quantum theory for any $v$. As we shall see, nonperturbative corrections are responsible for stabilizing center symmetry in SYM.\label{fig:gpy1}}
\end{figure}

Now recall that computing one-loop effective potentials amounts to computing  determinants in the corresponding background. Thus, in order to find the one-loop contribution to the $v$-dependent vacuum energy of the $m=0$ fermions, we see from (\ref{lambdakineticweyl}) that we have to compute the Gaussian path integral over $\lambda^+$ and $\lambda^-$ (the $\lambda^3$ components do not couple to $v$ and give no contribution to the one-loop effective potential). Computing the Grassmann path integral over the Fourier modes of $\lambda^{\pm}$ produces a determinant. To compute it, we integrate over the Minkowski space $\vec{k},p$ Fourier modes of $\lambda^{\pm}$ and then continue to Euclidean momenta to obtain (\ref{det1}) below. With $V$ denoting the $R^3$ spacetime volume, we have the usual definition of the effective potential
\begin{equation}\nonumber
e^{- S_{eff}}= e^{- V_{eff}(v) V}\bigg\vert_{\lambda^\pm \; contribution}  = \det (\text{operator of quadratic fluctuations in eqn.~(\ref{lambdakineticweyl})})~.
\end{equation}
Noting that we are after the $v$-dependent piece only, we can discard inessential overall constants. Computing the determinants, we find, switching to Euclidean $3$-momenta, the following formal expression
\begin{equation}\label{det1}\det (\text{operator of quadratic fluctuations}) = \prod\limits_{\vec{k}} \prod\limits_{p \in Z}\left( \vec{k}^2 + ({2 \pi p\over L}- v)^2\right) \left(\vec{k}^2 + ({2 \pi p\over L}+ v)^2\right) ~,
\end{equation}
and, thus, the final expression for the one-loop effective potential\footnote{The factor or 2 is because the two terms in the determinant give identical contributions after summing over $p$.}
\begin{eqnarray}
\label{gpyadjoint}
V_{m=0 \; \;adj.}(vL) &=& - 2 \sum\limits_{p\in Z} \int {d^3 k\over (2 \pi)^3} \ln \left(\vec{k}^2 + ({2 \pi p \over L} - v)^2 \right) = \ldots \text{see Appendix \ref{appx:gpy}}  = ~\nonumber \\
&=& -{1 \over L^3} \;{1 \over 12 \pi^2}\; [vL]^2 (2 \pi - [vL])^2 + const.~,~\text{where}~~ [vL] = vL  \; (\text{mod} 2\pi)~.
\end{eqnarray}
To obtain the second line, one needs to do a calculation. It should, however,  be clear that the overall scaling with $L$ and the periodicity w.r.t. to $vL$ follow from the expression in the first line. The calculation can be done, for example, via $\zeta$ function regularization,\footnote{Introduced in the framework of QFT path integrals by Hawking \cite{Hawking:1976ja}, in a setting more general than what is needed here.}  a standard procedure useful for computing such Casimir-like vacuum energies. The additive regulator-dependent $v$-independent constant can be dropped.
\begin{quote}
{\flushleft{\bf Exercise 13:}} Verify as much of eqns. (\ref{det1}) and (\ref{gpyadjoint}) as you feel like. Feel free to consult  Appendix \ref{appx:gpy} (this is recommended especially if you have not done $\zeta$-function calculations). \end{quote}
The single massless adjoint contribution (\ref{gpyadjoint})  to the effective potential on the Weyl chamber is shown on the bottom curve of Figure \ref{fig:gpy1}. Clearly, it alone favours the center-symmetric value $v=\pi/L$ (we shall come back to this point later).

As discussed above,  appealing to the supersymmetry of the $n_f=1$ massless theory, we   obtained the bosonic one-loop contribution for free, as it   simply equals  the negative of (\ref{gpyadjoint}). The result is known as the GPY potential
\begin{eqnarray}
\label{gpygauge}
V_{gauge}(vL) &=& {1 \over L^3} \;{1 \over 12 \pi^2}\; [vL]^2 (2 \pi - [vL])^2~.
\end{eqnarray}
As is clear from the top line of Figure \ref{fig:gpy1}, this equation implies that the vacuum energy is minimized at $vL=0$ or $2 \pi$, i.e. at  $\langle \text{tr} \Omega\rangle = \pm 1$,  where the $Z_2^{(1)}$ symmetry is maximally broken. 

For completeness, we note that the above GPY potential (\ref{gpygauge}) can be expressed in terms of a $Z_2^{(1)}$-invariant function of tr$\Omega$ of the general form (\ref{nonlocalterms}). Begin with the identity 
${1 \over 24 \pi^2} [vL]^2 (2 \pi - [vL])^2 = - {1 \over \pi^2} \sum\limits_{m \ne 0} {e^{i vL m} \over m^4} + {\pi^2\over 45}$ (which is nothing but a Fourier series) and  recall that $ \Omega^k = \text{diag}(e^{i k vL/2} , e^{-i k vL/2})$. Then, dropping $v$-independent terms, we obtain (\ref{gpygauge}) in the equivalent form:
\begin{eqnarray}
\label{gpygauge2}
V_{gauge}(vL) &=& -  {1 \over L^3} {2 \over \pi^2}  \sum\limits_{m > 0} {e^{i vL m} +  e^{- i vL m} \over m^4 } =  -  {1 \over L^3} {2 \over \pi^2} \sum\limits_{m > 0} { | \text{tr}( \Omega^m)|^2 \over m^4}.
\end{eqnarray}
This can be then further massaged into the form (\ref{nonlocalterms}) with $k+p$ even, using the characteristic equation (\ref{chareqn}). Most importantly, however, the last form in  equation  (\ref{gpygauge2}) is remarkable in that it also holds in the $SU(N)$ theory, see  \cite{Unsal:2008ch} (for $SU(N)$ it is  a function of the $N-1$ independent eigenvalues of $\Omega$ rather than just on $vL$). 

\subsubsection{Digression: $Z_2^{(1)}$-breaking and deconfinement in high-$T$ pure-YM theory.}
\label{sec:digression}

There is some important physics associated with the breaking of the $Z_2^{(1)}$ center symmetry and the thermal deconfinement transition that we shall discuss next. This constitutes a necessary and important digression from our main line of thought.

In the theory with only bosons, the $R^3 \times S^1$ theory can be given a thermal interpretation with the $S^1$ size   being identified with the inverse temperature $L = 1/T$. The Euclidean path integral on $R^3 \times S^1$ computes the thermal equilibrium partition function.\footnote{For future use, we also note that a thermal interpretation of the theory with fermions requires them to be antiperiodic on the thermal  $S^1$, i.e. $\lambda(L={1 \over T}) = - \lambda(0)$. This can be traced back to the difference between Bose-Einstein and Fermi-Dirac distribution functions.}
Thus, we can actually rewrite (\ref{gpygauge}) as
$V_{GPY} ={T^3 \over 12 \pi^2} ({v \over T})^2 (2 \pi - {v \over T})^2$. This expression was derived by GPY using high-temperature perturbation theory, valid at  $T \gg \Lambda$, where the one-loop contribution to the $v$-dependence of the free energy can be trusted, since, due to asymptotic freedom $g^2(T) \ll 1$ for $T \gg \Lambda$.  
The main conclusion of this calculation is that at $T \gg \Lambda$, the $Z_2^{(1)}$ center symmetry is spontaneously broken, as ${1\over 2}\langle \text{tr} \Omega\rangle = \pm 1$. 

Our goal here is to tie some remaining loose ends connecting the $Z_2^{(1)}$ symmetry to confinement and deconfinement. 
Recall that our mention of center symmetry  began in Section \ref{sec:polyakov_pass4}, where we qualitatively argued that in a theory with no center symmetry confining strings cannot be stable and an area law can not hold for arbitrarily large Wilson loops. We then continued with the more precise definition of $Z_2^{(1)}$ in Section \ref{sec:holonomyandcenter}, where we showed that  tr$\Omega$ is an order parameter for the center symmetry. 
From the GPY calculation, we then found that at $T \gg \Lambda$, center symmetry is broken.

The main point we want to make here is that the broken-$Z_2^{(1)}$ phase of high-$T$ YM theory is interpreted as a deconfined phase. 
One quick way to argue this is to note that the normalized expectation value of the fundamental representation Polyakov loop in the thermal ensemble can be interpreted as the change of the free energy of the  system due to the insertion of a fundamental static charges in the thermal bath,  $\langle {\text{tr} \Omega} \rangle \sim e^{- {F_{quark}(T)\over T}}$ (this is because of the physical interpretation of $\Omega$ as  inserting a static charge; see \cite{Greensite:2011zz} for a more formal derivation). We associate confinement with the notion that fundamental quarks are not free and expect that $F_{quark} = \infty$, implying that 
$\langle {\text{tr} \Omega} \rangle = 0$, i.e. confinement is associated with an unbroken center symmetry. 
Conversely, a finite free energy, associated with a deconfined phase, implies a nonzero $\langle {\text{tr} \Omega} \rangle$ and thus broken center symmetry.

To gain more intuition,  consider the correlator of two Polyakov loops a distance $R \in R^3$ apart. The physical interpretation is that a fundamental quark-antiquark pair separated by $R$ is inserted in the thermal bath.  There are two important properties of this correlator. The first is that the large-$R$ behaviour of the  correlation function
\begin{equation}
\label{polyakovcorr1}
\langle \text{tr} \Omega(x) \text{tr} \Omega^\dagger (x+R) \rangle \sim e^{- {V(R,T) \over T}}~, \end{equation}
determines the free-energy, or  $T$-dependent interaction potential $V(R,T)$, between a quark and an antiquark inserted a distance $R$ apart. The second is the general property of cluster decomposition\footnote{See S. Weinberg's book \cite{Weinberg:1995mt} (vol. 1, Ch. 4) for discussion in QFT language and  Zinn-Justin's \cite{Zinn-Justin:2002ecy} (Ch. 23.3) for statistical mechanics terms.}  of correlation functions at large spacelike separations. Clustering implies that  for $R\rightarrow \infty$, the correlator factorizes, i.e.  
\begin{equation}\label{polyakovcorr2}
\lim\limits_{R \rightarrow \infty} \langle \text{tr} \Omega(x) \text{tr} \Omega^\dagger (x+R) \rangle= |\langle \text{tr} \Omega(x)\rangle |^2~, 
\end{equation}
where the expectation values on both sides are taken in the same vacuum.
\begin{figure}[h]
\centerline{
\includegraphics[width=10.5 cm]{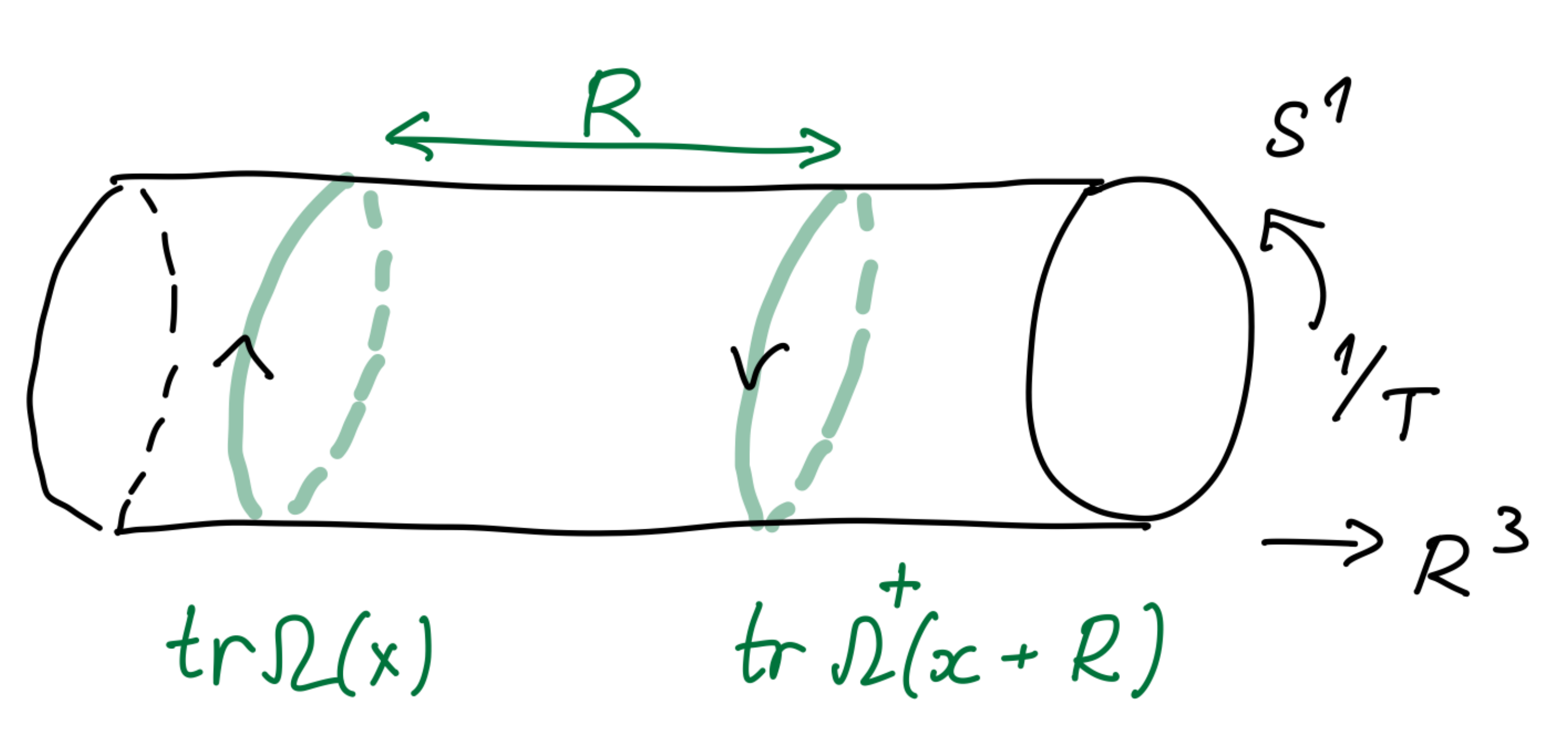}}
\caption{ A fundamental quark-antiquark pair inserted in the thermal $R^3 \times S^1$  theory (where $S^1$ size is $1/T$) is represented by two Polyakov loops  a distance $R \in R^3$ apart. It measures the $T$-dependent interaction potential $V(R,T)$ between the two sources. In the low-$T$ confined phase,  $V \sim \Sigma R$, giving an area-like behaviour of the correlation function and implying, by (\ref{polyakovcorr2}), $\langle \text{tr} \Omega(x)\rangle = 0$, an unbroken $Z_2^{(1)}$. In a deconfined phase, instead, the correlator is nonzero at large $R$ implying that $\langle \text{tr} \Omega(x)\rangle \ne0$, a broken center symmetry. \label{fig:pcorrelator}}
\end{figure}
In a confined phase, where $V(R,T) = \Sigma(T) R$, where $\Sigma(T) \ne 0$ is a $T$-dependent string tension, we conclude by comparing the infinite-$R$ limit of  (\ref{polyakovcorr1}) with (\ref{polyakovcorr2}),  that $\langle \text{tr} \Omega(x)\rangle  =0$, i.e. the $Z_2^{(1)}$ symmetry is unbroken (notice that the confining behaviour of $V \sim \Sigma R$ can be also associated with an area law, as $R/T$ is the area between the two Polyakov loop insertions, as is clear from  Figure \ref{fig:pcorrelator}).

Conversely in a phase when the quark potential is screened by the thermal plasma, we expect that   $V(R,T) \sim {e^{- R/\ell(T)}\over R}$
has a  Yukawa form with some screening length $\ell(T)$. We conclude that the correlator does not vanish as $R \rightarrow \infty$, hence $\langle \text{tr} \Omega(x)\rangle \ne 0$, i.e. center symmetry is broken. This phase is associated with a deconfined plasma of gluons (in pure YM theory). As gluons carry colour, this charged plasma screens the chromoelectric static potential over distances larger than $\ell(T)$.

Thus, we find that in a thermal $R^3 \times S^1$ theory with center symmetry, the behaviour of the trace of the  Polyakov loop as a function of $T=1/L$ plays the role of an order parameter of the confinement-deconfinement transition.  At low $T$ (large $L$) a confining phase respecting the center symmetry is expected, while at small $L$ (large $T$) there is a deconfinement phase transition to a phase with broken center symmetry.

 The GPY perturbative calculation is valid at $T \gg \Lambda$, i.e. only deep in the deconfined phase, where it leads to the conclusion of the breakdown of center symmetry. We stress that perturbative calculations in hot YM theory can not tell us about the value of $T=T_c$ where the transition to the phase with unbroken center symmetry takes place. This so-called confinement-deconfinement transition in thermal YM theory occurs at strong coupling and is unaccessible to controlled analytical means, as already stated in \cite{Gross:1980br}.
 
 A further remark is that even at $T = 1/L \gg \Lambda$, the $R^3$ theory is nonabelian, as per the discussion in Section 
 \ref{sec:perturbative3x1}, since ${1\over 2} \langle \text{tr} \Omega \rangle = \pm 1$. As mentioned many times (already in Section \ref{sec:pert_polyakov}), nonabelian theories on $R^3$ also become strongly coupled in the IR, making the long-distance physics in the center-broken phase inaccessible to analytical studies. Also note that a phase transition to a center-broken phase upon compactification is also expected (and is known to occur) when all directions are taken small: as stated in the Introduction the femto-universe theory on a  small spatial $T^3$ is separated from the $R^3$ theory by a deconfinement phase transition associated with breaking the center. 
 
 Luckily \cite{Unsal:2007jx,Unsal:2008ch}, as we discuss next, the addition of massive or massless adjoint fermions periodic on the $S^1$ allows to avoid the center-breaking deconfinement phase transition upon decrease of $L$. These periodic adjoint fermions stabilize center symmetry and make the analytical study of the IR   abelian theory possible.
 \begin{figure}[h]
\centerline{
\includegraphics[width=9.5 cm]{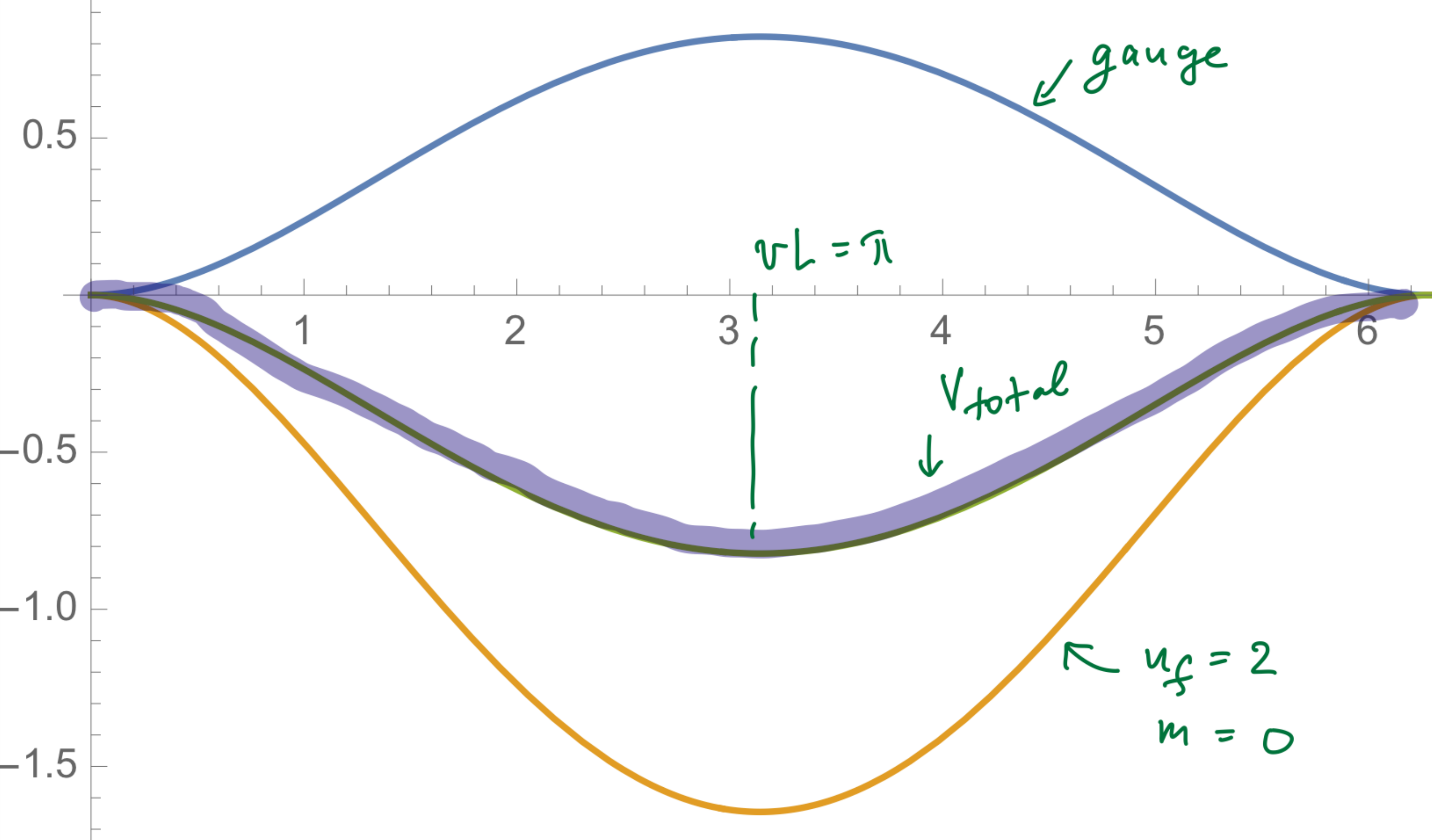}}
\caption{ The one-loop effective potential on the Weyl chamber for $n_f=2$ massless adjoint fermions, shown by the thick curve (the same  $L=1$ scale as on Fig. \ref{fig:gpy1} is used). For $n_f >1$, the massless fermions dominate  over the gauge contribution in  the total potential. The potential on the Weyl chamber is minimized at the center symmetric point $\langle{\text tr} \Omega \rangle= 0$. Thus the IR theory is abelian and weakly-coupled at $L\Lambda \ll \pi$, as discussed in Sections \ref{sec:perturbative3x1} and \ref{sec:adjointspectrum}.\label{fig:gpycenter}}
\end{figure}

\subsubsection{Stabilizing center symmetry via massless or massive adjoints.}
\label{sec:adjstability}

The calculation of Section \ref{sec:oneloopgpy} implies that   with $1 < n_f \le 5$ massless adjoint fermions periodic on the $S^1$, i.e. in QCD(adj), center symmetry is stable. (The theory loses asymptotic freedom at $n_f=6$.)
The potential on the Weyl chamber for $n_f=2$ massless adjoints is shown on  Figure \ref{fig:gpycenter}, showing the center stabilization effect.

Now, what if we give the adjoint fermions mass $m$, as in (\ref{lagrangianadjoint})?
We claim that eqn.~(\ref{gpyadjoint}) gets replaced by the following expression:
\begin{eqnarray}
\label{gpyadjointmass}
V_{m\ne0 \; \;adj.}(vL) &=& - 2 \sum\limits_{p\in Z} \int {d^3 k\over (2 \pi)^3} \ln \left(\vec{k}^2 + ({2 \pi p \over L} - v)^2 + m^2 \right) = \ldots \text{see Appendix \ref{appx:gpymass}}  = ~\nonumber \\
&=& {2   \over \pi^2 L^3} \sum\limits_{p=1}^\infty (m L p)^2 K_2 (m L p) {\cos p v L \over p^4}.
\end{eqnarray}

\begin{figure}[h]
\centerline{
\includegraphics[width=9.5 cm]{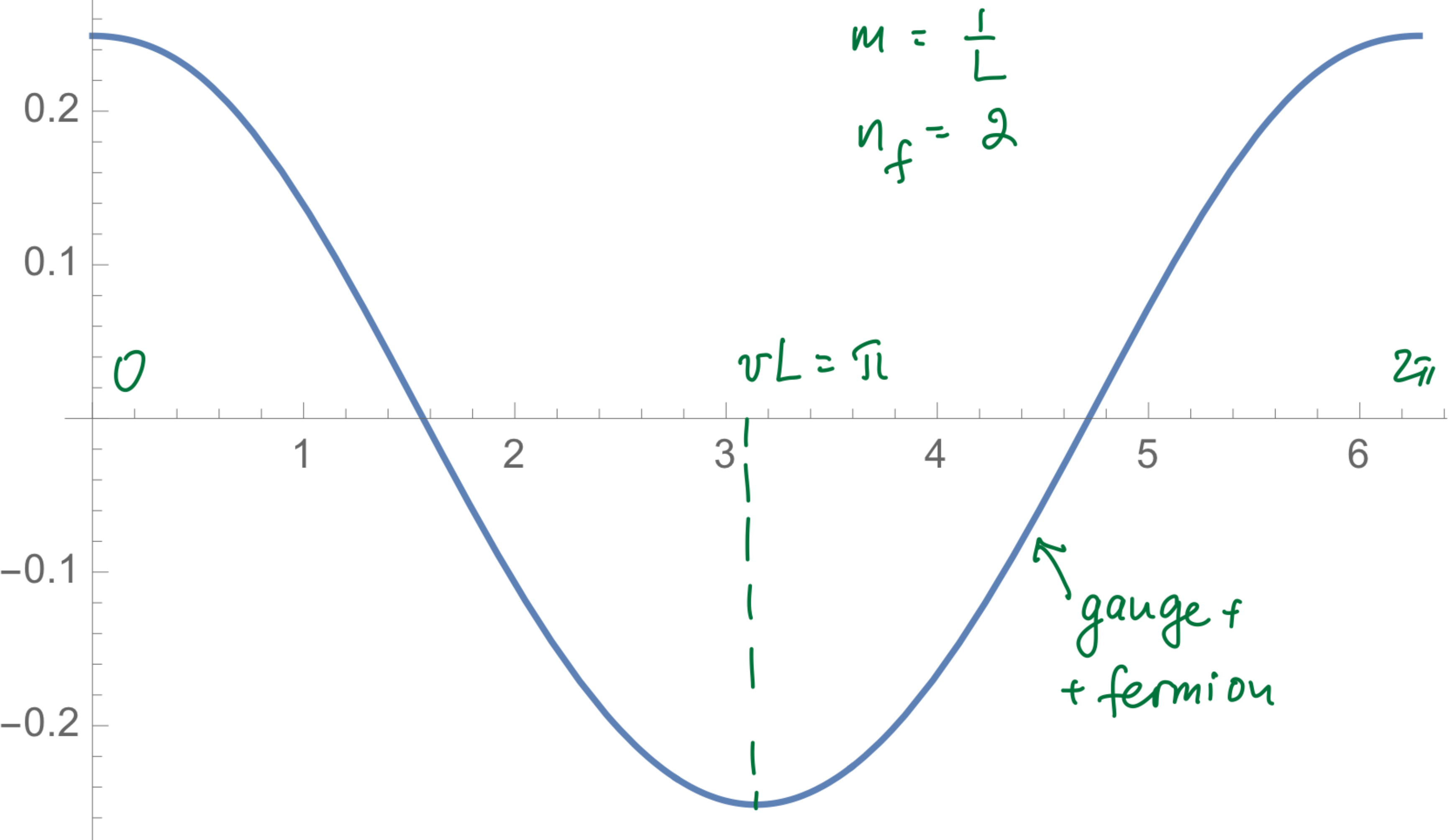}}
\caption{ The one-loop effective potential (\ref{gpyadjointstable}) on the Weyl chamber for the $SU(2)$ gauge theory with $n_f=2$ massive adjoint fermions, periodic on the $S^1$ with $m=1/L$.  The potential on the Weyl chamber is minimized at the center symmetric point ${\text tr} \Omega = 0$, and the abelianized IR theory is valid. The same scale, with $L=1$, as on Figs. \ref{fig:gpy1} and \ref{fig:gpycenter} is used. \label{fig:massivegpy}}
\end{figure}

\begin{quote}
{\flushleft{\bf Exercise 14:}} Starting from the massive-adjoint lagrangian (\ref{lagrangianadjoint}), derive the  generalization of (\ref{lambdakineticweyl}) by also including the mass term contribution to the action bilinear in $\lambda^\pm$. Use it to obtain the massive generalization of (\ref{det1}), and hence the  equation on the first line in (\ref{gpyadjointmass}). Then,  consulting  Appendix \ref{appx:gpymass}, verify the final result in (\ref{gpyadjointmass}). \end{quote}
Thus, with $n_f$ massive adjoints of the same mass $m$,  the total one-loop potential due to the gauge bosons  and fermions, using the first equality of    (\ref{gpygauge2})  for the gauge contribution, is
\begin{equation}
\label{gpyadjointstable}
V(vL)_{gauge + m\ne 0\; adj} =   {2   \over \pi^2 L^3} \sum\limits_{p=1}^\infty \left[ n_f (m L p)^2 K_2 (m L p) - 2\right] {\cos p v L \over p^4}.
\end{equation}
The reader can use (\ref{gpyadjointstable}) to do their own study of the center-stability as  a  function of $n_f$ and $mL$. 
An explicit example to keep in mind is to 
  consider $n_f=2$ and $mL =1$. We  plot the potential (\ref{gpyadjointstable}) on Figure \ref{fig:massivegpy}; notice that the infinite sum of mass-dependent terms is rapidly converging, so plotting a few terms suffices, while   the gauge field contribution is standard and sums to (\ref{gpygauge}). As the plot makes it clear, massive adjoints stabilize center symmetry  for the chosen $m = 1/L$.

The most important conclusion is the following qualitative one: that having $2 \le n_f \le 5$ fermions of mass $mL \le {\cal{O}}(1)$ is necessary to achieve center stability. 
For $mL \gg 1$, the fermions decouple and center symmetry breaks as in the pure-YM theory (the precise value of $m$ where center symmetry breaks depends on $n_f$). On the other hand, for $mL \sim 1$, the fermions ensure center stability and hence the abelian description of the IR physics. At the same time, the fermions of mass $m \sim 1/L$ decouple from the $\mu \ll 1/L$ dynamics. 

We shall make extensive use of this center stability in our study of nonperturbative dynamics on $R^3 \times S^1$, our main topic that we finally turn to next.

\hfill\begin{minipage}{0.85\linewidth}

 \textcolor{red}{
{\flushleft{\bf Summary of \ref{sec:adjointgpy}:}} This main goal of this Section was to show
that adding $n_f$ adjoint fermions, massive of massless, with $m L \le {\cal{O}}(1)$ and $n_f > 1$,  stabilizes the center-symmetric point on the Weyl chamber. The (near) center stability is crucial for the ability to perform a controlled semiclassical study of the IR dynamics on $R^3 \times S^1$: near the center symmetric point, the theory abelianizes and is in a weak-coupling regime. The perturbative physics is rather boring, but much like the Polyakov model of Section \ref{sec:polyakov}, we shall see that calculable nonperturbative effects completely change the IR physics on $R^3 \times S^1$ in new and interesting ways. Center stability at small $L$ also implies the absence of an associated phase transition (plaguing earlier femto-universe ideas) as the $R^4$ limit is approached.
}

\end{minipage}

\bigskip

 \section{Back to the 2010's: the properties of dYM.}  \label{sec:dym}
 
 We begin our study of nonperturbative dynamics on $R^3 \times S^1$ with dYM theory. 
 This theory  is often said to be in the ``universality class'' of 4d pure YM theory. What one means by this is that dYM has the same non-spacetime global symmetries\footnote{These are charge conjugation, for $N>2$ (for $SU(2)$, charge conjugation is part of the gauge group) and the $1$-form  $Z_N^{(1)}$ center symmetry. The addition of generic massive adjoint fermions to stabilize the center does not add any new symmetries (apart from $Z_2$ fermion number, irrelevant for the IR).  We also caution against thinking that the Higgsed $SU(2) \rightarrow U(1)$ phase in dYM is distinct from the unbroken-$SU(2)$ phase.  There is no gauge invariant distinction between the Higgsed phase and the unbroken phase---a gauge invariant description of the $U(1)$ phase can be given \cite{Cherman:2016jtu,Aitken:2018mbb}, but the gauge-fixed  $U(1)$ language is much more appropriate at weak coupling.} as 4d YM (the spacetime Lorentz symmetry is an exception, as it   is broken by the $S^1$-compactification). Furthermore, these symmetries are realized in the same mode in the two theories and thus the dYM and 4d YM theories are expected to be continuously connected. This continuous connection (without phase transition) between dYM and 4d YM is often called ``adiabatic continuity.''

In the subsections that follow, we shall  study the calculable properties of dYM and compare with some of the available lattice data for 4d YM. We shall observe qualitative (and in some cases even quantitative, notably for quantities concerning the $\theta$-angle dependence and other topological properties) agreement between the two theories.

dYM was already defined in the Introduction, see Section \ref{sec:whatabout}, as the YM theory with $n_f=2$ (or more) massive adjoint fermions ensuring center stability, as described in some detail in the previous Section. dYM was introduced in \cite{Unsal:2008ch} (see also \cite{Myers:2007vc,Myers:2009df}), where a potential for the Polyakov loop more general than our (\ref{gpyadjointstable}) was considered, the so-called ``double-trace deformation'' potential. It has the form, given here for $SU(N)$:
 \begin{equation}\label{doubletrace}
 V_{double \; trace}[\Omega] = \sum_{n=1}^{\lfloor {N \over 2} \rfloor} {c_n \over L^3} |\text{tr} \Omega^n|^2,
 \end{equation} with coefficients $c_n$ chosen such that center stability is ensured.\footnote{Adding the nonlocal double-trace deformation (\ref{doubletrace}) results in a nonrenormalizable theory.  On the lattice, this can be viewed as a finite-lattice spacing theory. The fixed lattice spacing formulation, apart from allowing numerical studies (as in e.g.~\cite{Bonati:2018rfg,Bonati:2019kmf}), permits one to write loop equations and prove various exact properties regarding the large-$N$ limit. We note that ``large-$N$ volume independence'' considerations played an important role in the introduction of $V_{double \; trace}$. We shall not pursue this direction here,  see  \cite{Unsal:2008ch} for references.} It is clear, remembering (\ref{gpygauge2}) and the characteristic equation for $\Omega$ of footnote \ref{omegafootnote}, that our center stabilizing potential due to adjoint fermions, eqn.~(\ref{gpyadjointstable}), can be cast into the above form. 
 
 For the purposes of studying the $\mu \ll 1/L$ physics, the precise choice of $c_n$ does not matter, as long as they are large enough to ensure center stability. We shall, however, stick with our massive adjoint fermion interpretation as the source of $V_{double \; trace}$, since it comes from a renormalizable asymptotically-free theory with a  well-defined continuum limit. Note that the adjoint fermion mass needed to stabilize center symmetry (recall Fig.~\ref{fig:massivegpy}) obeys 
 \begin{equation}\label{mstable}
 \Lambda \ll m \le {{\cal{O}}(1)\over L}~.
 \end{equation}
 To connect to the $R^4$ pure YM theory, the large-$L$, fixed-$\Lambda$ limit has to be taken while keeping $m$ fixed with $m \gg \Lambda$, as in the leftmost inequality in (\ref{mstable}).

 Thus, the perturbative IR lagrangian (Euclidean) valid at $\mu \ll 1/L$ is now the sum of (\ref{3dlagrangiandual}) and (\ref{gpyadjointstable})\begin{equation} \label{3dlagrangiandualdYM1}
L_{3d, dYM} =  {1 \over 2} { g_4^2 \over L (4 \pi)^2}   (\partial_\lambda \sigma)^2  + {L \over 2 g_4^2} (\partial_\mu a_4) (\partial^\mu a_4) +  V(\pi + a_4 L)_{gauge + m\ne 0\; adj}~,\end{equation}
where we replaced $vL $ by $\pi + a_4 L$, recalling that $v$ denotes the constant mode of $A_4^3$ while $a_4$ is the slowly varying fluctuation around the vev (\ref{a4vev}), with $vL=\pi$ the minimum of (\ref{gpyadjointstable}).
The potential on the Weyl chamber gives mass to $a_4$, which can be found by expanding the potential in (\ref{3dlagrangiandualdYM1}); parametrically, it is of order $g/L$.  
That this is so is clear from the fact that the second derivative of (\ref{gpyadjointstable}) is of order unity in $L=1$ units, see Figure \ref{fig:massivegpy}. We can thus integrate out the $a_4$ field to arrive at the IR lagrangian valid at $\mu \ll g/L$:\begin{equation} \label{3dlagrangiandualdYM2}
L_{3d, kin., dYM} =  {1 \over 2} { g_4^2 \over L (4 \pi)^2}   (\partial_\lambda \sigma)^2~,
\end{equation}
i.e. the same boring free dual-photon lagrangian (\ref{dualaction}) that we encountered in the Polyakov model. 
 In the next Section \ref{sec:dymvacua}, we shall study the nonperturbative corrections to (\ref{3dlagrangiandualdYM2}).

\subsection{Mass gap, string tension,  their $\theta$-dependence, and ``adiabatic continuity.''}
\label{sec:dymvacua}

The fun part begins when we remember eqn.~(\ref{thooftdym}), which gives the 
't Hooft vertices associated with the $M$ and $KK$ monopole-instantons and their ``anti-particles.'' Just like in the Polyakov model, these proliferate in the vacuum. For $S_0 = 4 \pi^2/g_4^2 \gg 1$, the gas is dilute (only it now has four constituents instead of two), and the insertion of  't Hooft vertices exponentiate into an action incorporating the leading order (in $e^{- S_0}$) nonperturbative effects in dYM:
\begin{equation} \label{3dlagrangiandualdYM3}
L_{3d, dYM} =  {1 \over 2} { g_4^2 \over L (4 \pi)^2}   (\partial_\lambda \sigma)^2 - {e^{- S_0} \over L^3}\left(e^{ i {\theta \over 2}} \; e^{+ i \sigma} + e^{ - i {\theta \over 2}} \; e^{- i \sigma}+e^{+ i {\theta \over 2}} \; e^{- i \sigma } + e^{- i {\theta \over 2}}\;  e^{+ i \sigma }\right) + \ldots.
\end{equation} The four terms are the contributions, in the order shown, of $M$, $M^*$, $KK$ and $KK^*$ monopole-instantons.\footnote{
We remind the reader that we work with exponential-only accuracy (ignoring a numerical factor and the $1/g_4^4$ prefactor due to integration over monopole-instanton zero modes, see \cite{Unsal:2008ch}).
The ``$\ldots$'' denote terms that are higher order in the $\sim e^{-S_0}$ semiclassical expansion as well as higher order terms involving powers of insertions of $\partial_\mu \sigma$ and suppressed by additional factors of the heavy mass scales $1/L$, $g_4/L$.} 
We have retained their $\theta$-angle dependence due to their fractional topological charges (\ref{topologicalcharges}). 

Next, we recall that if one  expands the exponentials corresponding to the potential terms in (\ref{3dlagrangiandualdYM3}) in the Euclidean path integral (e.g., expanding $e^{ \;e^{-S_0} L^{-3}  \int d^3 x e^{i \sigma + i \theta/2}}$ in Taylor series, which is equivalent to going from the second to the first line in (\ref{sumnmcontribution}); likewise for the other three terms), after integration over $\sigma$ one obtains the representation of the partition function in the form of a  dilute gas of $M, M^*, KK, KK^*$, exactly as we did in the Polyakov model in Section \ref{sec:polyakov_pass3}. The most important difference is that there are now two types of charged particles ($M, KK$) and their antiparticles ($M^*, KK^*$), which contribute with complex fugacities due to the $\theta$ angle. This is how the theory remembers its 4d origin and is qualitatively different from the 3d Polyakov model.
 
 Back to our EFT (\ref{3dlagrangiandualdYM3}), we note that, shifting the vacuum energy, we can rewrite it as
\begin{eqnarray} \label{dYM3}
L_{3d, dYM} &=&  {1 \over 2} { g_4^2 \over L (4 \pi)^2}   (\partial_\lambda \sigma)^2 + {2 e^{- S_0} \over L^3}\left(2- \cos(\sigma + {\theta \over 2}) - \cos(\sigma-{\theta \over 2})\right) + \ldots\nonumber \\
&=& {1 \over 2} { g_4^2 \over L (4 \pi)^2}   (\partial_\lambda \sigma)^2 + {4 e^{- S_0} \over L^3}(1 -  \cos {\theta \over 2} \cos\sigma)  + \ldots~.
\end{eqnarray}
All forms of the potential above---the two lines in (\ref{dYM3}) and the form shown in (\ref{3dlagrangiandualdYM3}))---have their use. The top line in (\ref{dYM3}) is similar to what one finds in the $SU(N)$ theory: the two cosines are due to the $M+M^*$ and $KK+KK^*$ contributions.\footnote{For $N>2$ one obtains $N$ terms in the analogue of the first line of (\ref{dYM3}) instead, due to the $N$ constituents of the BPST instanton there. For $N>2$,  one can not write such a simple formula as  in the second line. Studies of $\theta$-dependence in dYM for $N>2$ can be found in  \cite{Bhoonah:2014gpa,Anber:2017rch,Aitken:2018mbb,Aitken:2018kky}.}
 A look at the bottom line in (\ref{dYM3}) shows that the order $e^{-S_0}$   contribution to the dual-photon potential vanishes at $\theta =\pi$ \cite{Unsal:2012zj}. The    potential  in the form (\ref{3dlagrangiandualdYM2}) shows that the $M$ contributions cancel with the $KK^*$ ones (and the $M^*$ with the $KK$ ones). This $\theta=\pi$ cancellation of the leading semiclassical effect, to which we shall come back later, arises due to the complex fugacities of the various monopole-instantons. It was called  ``topological interference'' in 
\cite{Unsal:2012zj}.

\begin{figure}[h]
\centerline{
\includegraphics[width=9.5 cm]{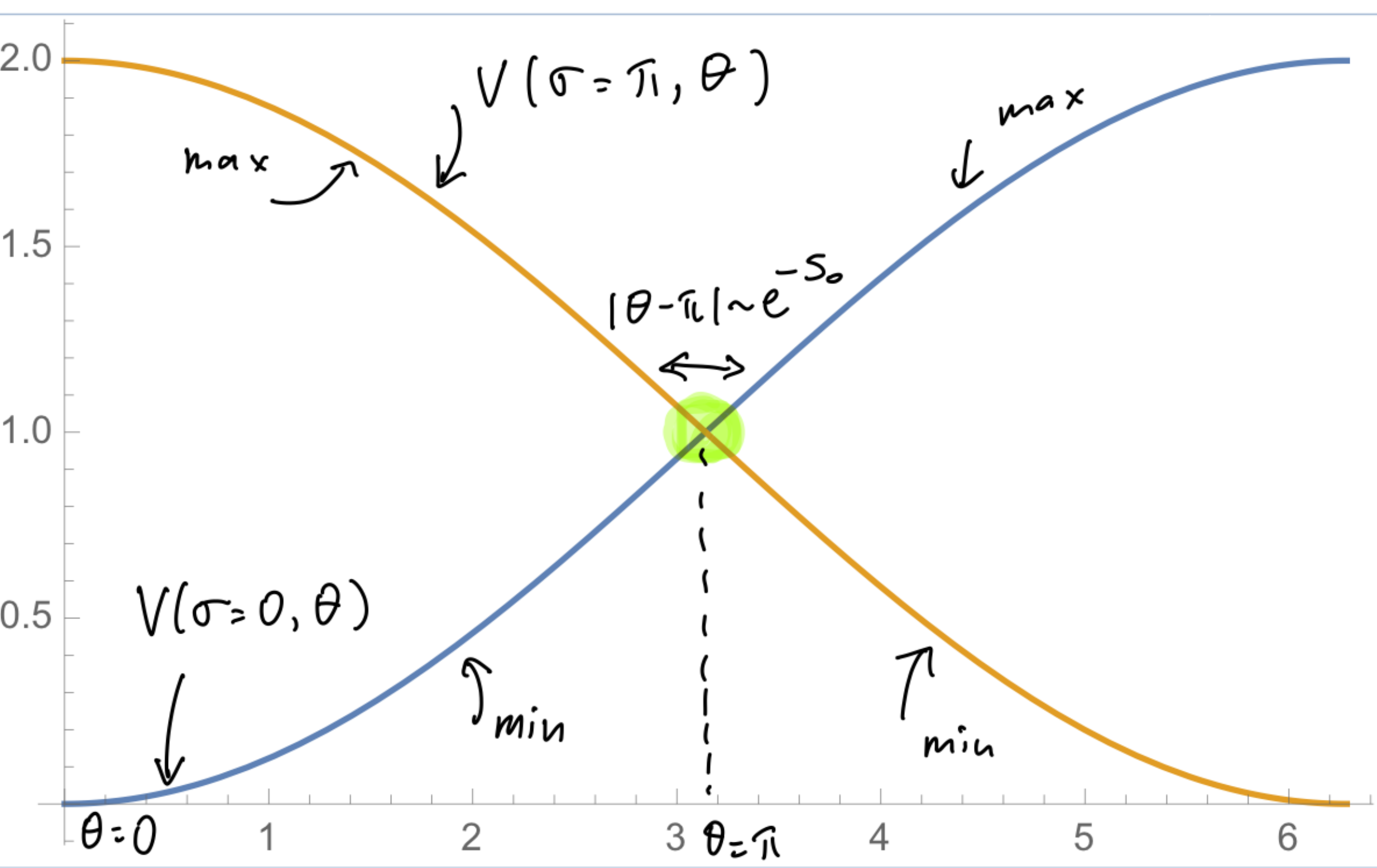}}
\caption{ The energies of the two extrema of the leading order semiclassical potential (\ref{dYM3}),   $\sigma = 0$ and $\sigma = \pi$, plotted as a function of $\theta$.  At $\theta=0$, the $\sigma =0$ extremum is a minimum, while the one at $\sigma = \pi$ is a maximum. Conversely, at $\theta = 2 \pi$, the minimum is at $\sigma = \pi$, while $\sigma=0$ is now a maximum of the potential. The level crossing occurs near $\theta = \pi$. To study the physics in a finite region $|\theta-\pi| = {\cal{O}}(e^{-S_0})$, one has to account of the next order in the semiclassical expansion, due to the fact that the leading-order potential in (\ref{dYM3}) vanishes at $\theta=\pi$. As discussed in Section \ref{sec:paritycenteranomaly},  the next-order contribution implies that there are two ground states at $\theta = \pi$ with broken parity. This reflects the generalized 't Hooft anomaly involving parity and the center symmetry. \label{fig:dymtheta}}
\end{figure}

For now, let us assume that $\theta \ne \pi$ and study the ground state and spectrum of dYM. The extrema of the $\cos\sigma$ potential are at $\sigma = 0$ and $\sigma = \pi$. A plot of the energies of the two extrema of the potential, as a function of $\theta$, is shown on Figure \ref{fig:dymtheta}. Concentrating on $0\le \theta < \pi$, the minimum is at $\sigma = 0$. We conclude (excluding the region near $\theta = \pi$ indicated on Figure \ref{fig:dymtheta} and working with exponential-only accuracy, i.e.  for brevity ignoring non-exponential $g_4^2$-dependence) that the dual photon mass scales as:
\begin{equation}
\label{msigmadYM}
m_\sigma^2 \sim {e^{- S_0} \over L^2} \cos {\theta \over 2},~~ ~ 0 \le \theta \le \pi - {\cal{O}}(e^{-S_0})~.
\end{equation}
The mass gap for the dual photon is thus of nonperturbative origin, as in the Polyakov model (\ref{msigma}). In the dYM EFT of (\ref{3dlagrangiandualdYM3}), we can repeat verbatim the derivation of the fundamental string tension $\Sigma_1$ of Section \ref{sec:polyakov_pass4}  and replace (\ref{wilsonsigma2}) with
\begin{equation}
\label{stringdYM}
\Sigma_1 \sim {g_4^2 \over L}\; m_\sigma \sim {e^{- {S_0\over 2}} \over L^2} \left(\cos {\theta \over 2}\right)^{1\over 2}~, ~ 0 \le \theta \le \pi - {\cal{O}}(e^{-S_0})~.
\end{equation}
The confining string is represented by the same type of semiclassical domain wall (domain line) configuration as the one shown on Figure \ref{fig:conf2} for the Polyakov model.

A new effect due to the four-dimensional nature of the theory is the fact that the fundamental string tension (\ref{stringdYM}) is a decreasing function of $\theta$: it reaches maximum at $\theta=0$ and decreases upon increasing $\theta$ above zero as
\begin{equation}\label{sigma1theta}
\Sigma_1(\theta) \simeq \Sigma_1(0)(1- {\theta^2 \over 16}+\ldots).
\end{equation} This behaviour of the confining string tension has been observed in lattice simulations  of 4d pure YM theory \cite{DelDebbio:2006yuf}, which  studied small variations of $\theta$ away from zero (it is a nontrivial matter to study $\theta$-dependence on the lattice, because of the notorious ``sign problem''). The  $\theta$-dependence found there  agrees with that motivated by large-$N$ arguments.\footnote{The large-$N$ expansion for YM theory has a long history, starting with \cite{tHooft:1973alw}. The $\theta$-angle dependence at large-$N$ was first discussed in \cite{Witten:1979vv,Witten:1980sp}. Unfortunately, we have no space to introduce large-$N$ ideas here and simply refer to the literature. However, we stress that our---decidedly small-$N$---discussion here shows that the implications of the large-$N$ arguments appear to extend down to $N=2$. This is not the first time we shall see an indication of this.}
For our $N=2$ case, the coefficient for the leading $\theta^2$ dependence in (\ref{sigma1theta}) is close to the lattice result for $N=3$, the smallest value of $N$ they study: ref. \cite{DelDebbio:2006yuf} obtained $0.08(1)$  for the coefficient of the $\theta^2$ term in $\Sigma_1$ for $N=3$, vs. our $0.0625$ for $N=2$.
This lends credence to the expectation that dYM captures qualitatively---and in some cases also quantitatively---the features of the 4d pure YM theory.

The agreement of (\ref{sigma1theta}) with the lattice results is quite remarkable (more recent lattice studies \cite{Bonati:2018rfg,Bonati:2019kmf}  of dYM
with a fixed deformation term like (\ref{doubletrace})
also reveal further  quantitative agreements of observables related to the $\theta$-dependence between dYM and pure 4d YM theories).  The arguments that led us  to (\ref{dYM3}) made no reference to the existence of a  large-$N$ limit, but relied solely on calculable semiclassics. Yet, they show good agreement with large-$N$ and, most importantly, with ``experiment''---the lattice data for the 4d YM theory.

Before we continue to study further properties of dYM, let us briefly comment on the adiabatic matching between the spectra of dYM and 4d YM.  Here we shall just mention the matching of states described by our EFT (\ref{dYM3}) (for a  detailed discussion, which also studies  states of masses $\sim 1/L, g_4/L$ that we integrated out on the way to (\ref{dYM3}),  see \cite{Aitken:2017ayq} and the recent lattice studies not related to $\theta$-dependence \cite{Athenodorou:2020clr}). We shall only stress some interesting and intriguing features seen in the dYM analysis, referring for more detail to the original papers.

The lightest perturbative state in dYM is the dual photon, of mass $m_\sigma$ (\ref{msigmadYM}). What state does it evolve into in the adiabatically connected 4d YM theory? As usual, symmetries offer a guide. Recall our discussion around eqn.~(\ref{centersigma}), where we argued 
that the dual photon is charged under the $Z_2^{(1)}$ center symmetry along the $S^1$. As we  discussed in  Section \ref{sec:holonomyandcenter}, no local operators in the 4d YM theory are charged under the $Z_2^{(1)}$ symmetry. The dual photon state then should evolve into a state created by a fundamental Wilson loop winding around the (now large) $S^1$, which transforms under $Z_2^{(1)}$ in the same manner as (\ref{centersigma}). Such states are called ``electric flux'' states, with energies  proportional to $\Sigma_1 L$ at large $L$, where $\Sigma_1$ is the  string tension in the $R^4$ theory.  We will not go into more details here, see \cite{tHooft:1979rtg,tHooft:1981sps}. The intuitive picture is that  the fundamental Wilson  loop winding around a spatial circle (\ref{wilsonloop2}), taken as an operator acting on the Hilbert space at fixed time, creates a thin chromoelectric flux along the loop. In a confining phase, as we saw in Section \ref{sec:polyakov_pass4}, this flux does not spread in the directions transverse to the loop, hence its energy grows linearly with the size of the circle. The conclusion is that the dual photon itself does not evolve into a localized excitation on $R^4$.

Now, one can ask whether there is a stable state in the small-$L$  EFT of dYM that evolves into a state created by local operators on $R^4$? It turns out that the answer is ``yes:'' these are bound states of two dual photons which have zero $Z_2^{(1)}$ charge and hence can be created by local operators. The physics of these bound states relies on the fact that ``$\phi^4$'' scalar interactions in $2+1$ dimensions create nonrelativistic bound states that are bound exponentially weakly. Such interactions appear in the expansion of the cosine potential in (\ref{dYM3}) and correspond in the nonrelativistic limit to attractive delta-function potentials. The nonrelativistic bound state of two dual photons in dYM  was studied in
\cite{Aitken:2017ayq} and its mass was found to be
\begin{equation}
\label{lightestglueball}
m_{glueball} = 2 m_\sigma\left( 1 - c \; e^{ - b \; e^{d/g_4^2} }\right),~\text{with calculable positive}~ {c, b, d \sim {\cal{O}}(1)}.
\end{equation}
The bound state is stable since its mass is smaller than that of the two constituents. 
 We denoted this state ``glueball'' as it is the lightest state in dYM which is a center-symmetry singlet and is a Lorentz scalar. Similarly, the lightest state created by localized gauge invariant operators in YM theory on $R^4$ is the scalar glueball and it is natural to conjecture that the two are adiabatically connected. 
 
Finally, we stress one  remarkable feature of (\ref{lightestglueball}): the doubly-exponential nonperturbative dependence on $g_4^2$. So far, we have seen that nonperturbative (with respect to a small coupling $\lambda$) effects  have a characteristic $e^{\; - {c/\lambda}}$ nonanalytic dependence on the small coupling $\lambda$. The binding energy found above has, instead, a doubly-exponential $e^{\; - c\; e^{ d /\lambda}}$ nonanalyticity. This double exponent can be thought as a nonperturbative effect in terms of a nonperturbative, $\sim e^{- 1/\lambda}$, coupling (e.g. the coupling $\sim e^{- S_0}$ in our dYM EFT). One can imagine, in principle, this tower of exponentiations continuing.
  The dependence found in (\ref{lightestglueball}) suggests that the analytic structure in the coupling constant of physical quantities in QFT is much more complicated that previously  thought and that much remains to be revealed.
  
  \hfill\begin{minipage}{0.85\linewidth}

 \textcolor{red}{ 
{\flushleft{\bf Summary of \ref{sec:dymvacua}:}} Here, we studied the  $\mu \ll g_4/L$ EFT of dYM, eqn.~(\ref{dYM3}). At $\theta \ne \pi$, the theory has a unique gapped vacuum. As our EFT is equivalent to the EFT of the 3d Polyakov model (\ref{iraction2}),  we  simply borrowed the calculation of the string tension.  The ``only'' new feature due to the 4d nature of the UV theory is the $\theta$-angle dependence. We showed that the fundamental string tension, eqn.~(\ref{sigma1theta}), decreases upon increasing $\theta$ away from zero, in remarkable agreement with available lattice data in 4d YM. We also discussed qualitatively the ``adiabatic continuity'' between the spectra of the 4d YM and dYM, argued that the lightest glueball is a bound state of two dual photons, and discovered an intriguing double-exponential nonperturbative dependence on the coupling (\ref{lightestglueball}).
 }
 
\end{minipage}
 
 \bigskip
 
  \subsection{$\theta =\pi$.}
\label{sec:dymthetapi}

We now go back to our dYM EFT (\ref{dYM3}) and focus on the region near $\theta = \pi$, where the potential is either exactly zero or very small. This is the region indicated by the yellow circle on Figure \ref{fig:dymtheta}.
At $\theta=\pi$, to leading order in the semiclassical expansion, there is no potential for $\sigma$ and thus it might appear that the dual photon does not obtain mass and the theory remains gapless. 
The question that naturally arises, then, is about the nature of the  ``$\ldots$'' terms in that equation and, in particular, whether these terms contribute a potential term for $\sigma$. 

That the ``$\ldots$'' terms are there follows from the usual EFT philosophy permitting all terms allowed by  symmetries to appear, with relative importance usually controlled by a power-counting rule. Our EFT is valid at scales $\mu \ll g_4/L$ and is derived using a weak coupling perturbative 
 expansion in  $g_4^2$ and a  semiclassical expansion in the exponentially small $e^{- S_0} = e^{- 4 \pi^2/g_4^2}$. Note that each term in the semiclassical expansion contains its own perturbative expansion around the relevant saddle points. This is the expansion in small fluctuations around monopole-instanton backgrounds that we have been neglecting (even to leading order, where it is well studied and contributes to the prefactor in the 't Hooft vertices like (\ref{thooftvertices}) and (\ref{thooftdym})).
Based on well-understood examples from semiclassical expansions of differential equations and quantum mechanics, the  two expansions (perturbative and semiclassical) are expected to combine in a nontrivial way  \cite{Dunne:2016nmc} into a so-called ``resurgent transseries'' that, yet again, we have no space and time to go into. 

Nonetheless, we will have to address the issue of higher order terms that can appear in the potential in (\ref{dYM3}) at $\theta = \pi$, to which we turn next.

\subsubsection{$\theta=\pi$: spontaneous breaking of parity.}
\label{sec:paritycenteranomaly}

Here we focus on the neighbourhood of $\theta = \pi$ indicated with the yellow circle on Figure \ref{fig:dymtheta}. The leading-order semiclassical potential (\ref{dYM3}) vanishes at $\theta=\pi$ due to the exact cancellation of $M$ and $KK^*$ amplitudes. The question that we shall address now is whether there are any other terms in the nonperturbative potential of dYM that do not vanish at $\theta=\pi$?

Symmetries offer a handle, as usual. As already mentioned, the only non-spacetime symmetry of $SU(2)$ dYM is center symmetry, which, by construction, is preserved in dYM. Center symmetry demands that  the dual-photon potential is an even periodic function of $\sigma$, i.e. contains only terms $\sim \cos (n \sigma)$, $n\in Z$,
invariant under $Z_2^{(1)}$ of (\ref{centersigma}). The continuous subgroup of the Lorentz group preserved by the $R^3 \times S^1$ compactification also imposes no further constraints.
Thus, we have to consider the spacetime discrete symmetries. Consider spatial parity $P$, a reflection of $x^{1,2,3}$, where $x^3$ is the compactified spatial direction, on which parity acts as $x^3 \rightarrow L- x^3$.\footnote{A more complete account of discrete symmetries, albeit using a slightly different basis, can be found in \cite{Aitken:2017ayq,Aitken:2018mbb,Aitken:2018kky}. }  Spatial parity  reverses the sign of the electric field, $P: \vec{\cal{E}} \rightarrow - \vec{\cal{E}}$ (${\cal{E}}_i = F_{0i}$) and preserves the magnetic field, $P: \vec{\cal{B}} \rightarrow \vec{\cal{B}}$ (${\cal{B}}_i = {1 \over 2} \epsilon_{ijk} F^{jk}$), where $i, j = 1,2,3$ are spatial indices. We use $\cal{E}$ and $\cal{B}$ to denote the electric and magnetic fields, in order to stress that these are the physical electric and magnetic fields on $R^{1,2} \times S^1$, rather than the Euclidean $R^3$ ones introduced in our study of monopole-instantons (\ref{eandb}) (for brevity, we also do not show the Lie-algebra index).
Next, we recall the duality relation (\ref{dualityrelation}). Skipping unimportant overall factors, this relation implies that $\partial_0 \sigma \sim {\cal{B}}_3^3$, $\partial_1 \sigma \sim {\cal{E}}_2^3$, $\partial_2 \sigma \sim {\cal{E}}_1^3$. Thus, we find that the dual photon is parity-even,  $P: \sigma \rightarrow \sigma$. 

At first sight, it would then appear that $P$ is a symmetry of our EFT (\ref{dYM3}) for any value of $\theta$. However,  we recall that $\theta$ couples to the topological charge (\ref{topologicalcharge}) as $e^{i \theta Q_T}$. The topological charge can be written as $Q_T \sim \int\limits_{R^3 \times S^1} \vec{\cal{B}}^a \cdot \vec{\cal{E}}^a$ showing that $Q_T$ changes sign upon a parity transformation. Since the topological charge is quantized,  invariance of the $e^{i \theta Q_T}$ factor in the path integral implies that $P$ is a symmetry of the $SU(2)$ YM theory only for $\theta=0$ and $\theta = \pi$; in the  latter case, however, parity has to be supplemented by a $2\pi$ shift of the $\theta$ angle. 

\begin{quote}
{\flushleft{\bf Exercise 15:}} A careful reader might object to the above appeal to the $2\pi$-periodicity of $\theta$, since the topological charge of monopole-instantons on $R^3 \times S^1$ is fractional, as per (\ref{topologicalcharges}). To restore justice,  starting from (\ref{3dlagrangiandualdYM2}) and recalling 
footnote \ref{footnoteneutral}, show that only integer $Q_T$ configurations contribute to the partition function.
\end{quote}
Thus,  $P$ acts differently at $\theta=0$ and $\theta=\pi$, incorporating a $2 \pi$ shift of $\theta$ for $\theta= \pi$:  
\begin{eqnarray}
\label{paritydYM}
P_{\theta=0}&:& \sigma(x^0,x^1,x^2) \rightarrow \sigma^P(x^0,x^1,x^2) = \sigma(x^0,-x^1,-x^2) \nonumber \\
P_{\theta=\pi}&:& \sigma(x^0,x^1,x^2) \rightarrow \sigma^P(x^0,x^1,x^2) = \sigma(x^0,-x^1,-x^2) + \pi \; (\text{mod} \; 2 \pi) ~.
\end{eqnarray}
In the second line, the $\pm\pi$ shift of $\sigma$ in $P_{\theta=\pi}$ is needed to compensate the $2\pi$ shift of $\theta$ in the 't Hooft vertices  $e^{i \sigma \pm i \theta/2}$ of the various monopole-instantons.\footnote{If this argument seems too quick, the interested reader can use the discussion of parity at $\theta=\pi$ of \cite{Cox:2021vsa} and, starting from the UV theory, arrive at an operator derivation of (\ref{paritydYM}). The logic is as follows: the canonical operator shifting the $\theta$-angle by $2\pi$  (which is part of the parity transform 
$\theta=\pi$) is the exponential of the integral of the spatial $SU(2)$ Chern-Simons 3-form. In the center-preserving holonomy background (\ref{a4vev}), the Chern-Simons operator simplifies when acting on low energy states. Correspondingly, the
operator implementing $2\pi$ shifts of $\theta$ becomes $e^{- i \pi \left( {1 \over 4 \pi} \int_{R^2} F_{12}^3 \right) }$. When rewritten in dual photon variables, from eqn.~(\ref{qoperator})   with
$g_3^2 \rightarrow g_4^2/L$, this is seen to generate the $\pi$-shift of the dual photon.}

It is now clear that a $\cos \sigma$ term is not invariant under the $P_{\theta=\pi}$ transformation (\ref{paritydYM}). dYM theory, however, is parity invariant at $\theta = \pi$ and  our leading-order potential (\ref{dYM3}) insures $P_{\theta=\pi}$ invariance simply by vanishing at $\theta=\pi$. 
It is clear from (\ref{paritydYM}) that $P_{\theta=\pi}$ restricts the potential by only allowing terms of the form 
\begin{equation}\label{dYMpi}
V^{dYM}_{\theta=\pi}(\sigma) = \sum\limits_{n=1}^\infty {a_n  \over L^3} \; \cos (2n \sigma).
\end{equation}
The semiclassical interpretation of these terms, implied by the $\sigma$-dependence, would be that they arise from a dilute gas of magnetic charge-$2n$ monopole-instanton-like objects, each such object contributing a factor of $e^{\pm i 2 n \sigma}$ to the path integral, as per Section (\ref{sec:polyakov_pass2}). The nature of these objects is very interesting and is not  understood in complete detail. For now, we note that ref.~\cite{Unsal:2012zj} gave arguments  that $a_n \sim e^{- 2 n S_0}$ and  argued that the leading order coefficient $a_1$ is nonzero (see Section \ref{sec:firstbions} below).

Accepting this and   only keeping the leading $n=1$ term, we arrive at the $\theta=\pi$ potential
\begin{equation}\label{dYMpi1}
V^{dYM}_{\theta=\pi}(\sigma) = {c  \over L^3}\; e^{-2 S_0} \; \cos (2 \sigma),
\end{equation}
and immediately
conclude that at $\theta = \pi$, dYM has two vacua, one at $\sigma=0$ and another at $\sigma = \pi$ (taking $c<0$). These vacua transform into each other under the $P_{\theta=\pi}$ transformation, thus parity is spontaneously broken. (Clearly, already the generic form of the potential (\ref{dYMpi})
implies that the potential has at least two minima related by parity.) Notice that the order parameter is $\langle e^{i \sigma} \rangle = \pm 1$, the expectation value of a monopole operator.

Independent arguments in favour of the breaking of parity at $\theta=\pi$ have come from recent lattice simulations of the 4d $SU(2)$ YM theory \cite{Kitano:2021jho}. A recent theoretical argument, based on the  operator algebra of the parity and center symmetries, was given in \cite{Cox:2021vsa}. It was shown there  that at $\theta = \pi$, this algebra implies an exact double degeneracy of all states related by $P_{\theta=\pi}$ in the Hilbert space of the  4d $SU(2k)$ YM theory formulated on a three torus of arbitrary finite lengths, with appropriately twisted boundary conditions. In the infinite volume limit (where the boundary conditions are irrelevant and which can be taken with one direction kept small, as in dYM) this double-degeneracy suggests\footnote{A possible caveat is the emergence of a gapless theory as $V \rightarrow \infty$, including the vanishing of all condensates in this limit.} the spontaneous parity breaking.

It is important to point out that the spontaneous parity breaking at $\theta=\pi$ and the operator-algebra argument mentioned above reflect the mixed 't Hooft anomaly between parity and the $Z_2^{(1)}$ center symmetry (we shall not review this subject here). This is an example of ``generalized'' 't Hooft anomalies,   discovered recently in \cite{Gaiotto:2014kfa,Gaiotto:2017yup,Gaiotto:2017tne}.
\begin{figure}[h]
\centerline{
\includegraphics[width=8.5 cm]{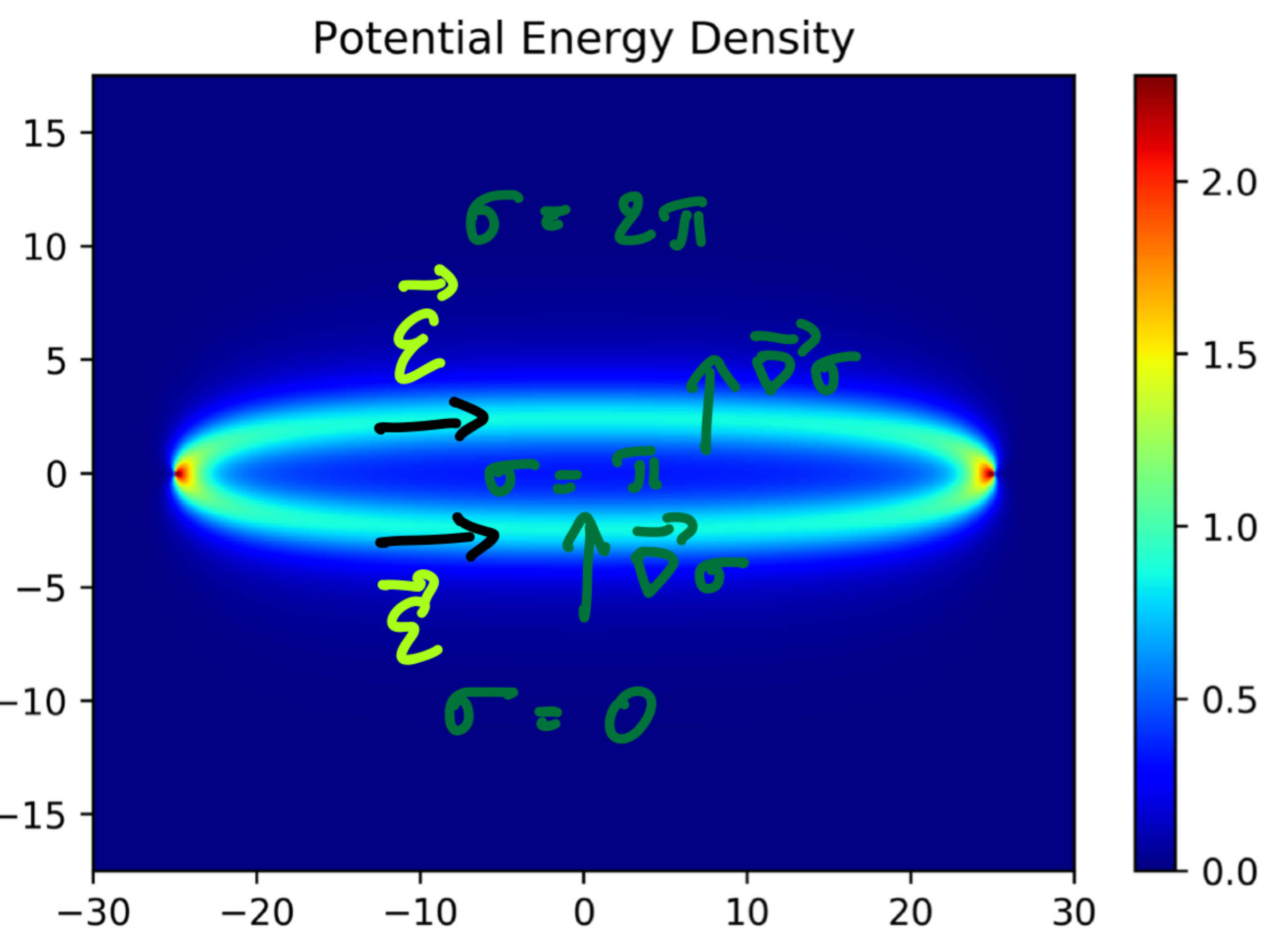}}
\caption{ The double-string configuration confining fundamental quarks in the $\theta=\pi$ dYM theory, embedded in the $\sigma = 0$ vacuum. The chromoelectric flux of the fundamental quarks is equally split between the two domain walls. Inside the double-string, the $\sigma$ field is in the other vacuum. The transverse size of the double string grows like $\ln R$ with the source separation $R$. (A similar configuration, but with $\sigma=2\pi$ inside the double-string and $\sigma = 4\pi$ at the top, is responsible for the confinement of $W^\pm$ bosons in the Polyakov model and in dYM at $\theta \ne \pi$. As discussed in Section \ref{sec:polyakov_pass4}, pair production of $W$-boson will lead to a breaking of the adjoint string.)\label{fig:doublestring1}}
\end{figure}

\subsubsection{$\theta=\pi$: ``double-string'' confinement and deconfinement on domain walls.}
\label{sec:doublestringdYM}

The breaking of parity at $\theta=\pi$, seen by studying the next-to-leading order potential  (\ref{dYMpi1}) has profound implications for the physics of confinement, which, after all, is our main topic.

For $\theta \ne \pi$, the confining strings in dYM (of tension (\ref{stringdYM})) are ``domain-wall like'' configurations of the $\sigma$ field, interpolating between the $\sigma = 0$ and $\sigma = 2\pi$ (gauge equivalent due the periodicity of the dual photon) minima of the potential in (\ref{dYM3}), recall Figure \ref{fig:conf2}. The $2\pi$ monodromy across the string ensures that these configurations carry the flux of a fundamental quark source, exactly as in the Polyakov model.

\begin{figure}[h]
\centerline{
\includegraphics[width=10.5 cm]{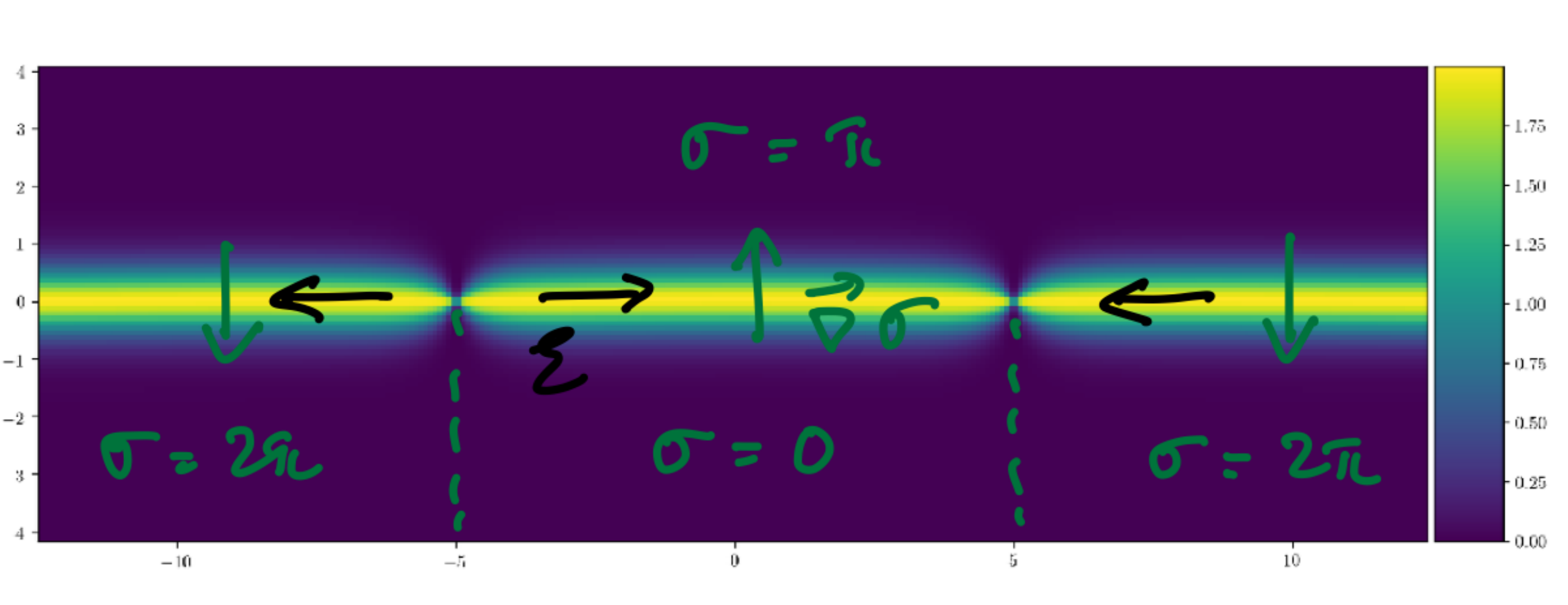}}
\caption{ Deconfinement of quarks on the domain wall between the two vacua of $\theta=\pi$ dYM. The  location of the quark sources is indicated by the two dots; the dotted lines are unphysical $2\pi$ discontinuities of $\sigma$. The gradient of the $\sigma$ field and the $\vec{\cal{E}}$ field are shown by arrows. The quarks exhibit no force while on the wall (provided their separation is larger than the screening length) since the tensions of the domain walls to the left and right of each quark are equal. \label{fig:decwall1}}
\end{figure}

At $\theta= \pi$, however, the potential (\ref{dYMpi1}) has two degenerate minima at $\sigma=0$ and $\sigma=\pi$. These are physically distinct and are related by the spontaneously broken parity. Thus, there exist genuine domain walls connecting the $\sigma=0$ to the $\sigma=\pi$ vacuum. These domain walls thus have a monodromy $\pi$, i.e. they carry electric flux of one-half a fundamental quark source (naturally, there are no such sources possible in the $SU(2)$ theory---adding them as very heavy dynamical fields would allow the domain walls to end, leading to contradiction). What configuration, then,  confines fundamental quarks? Since one requires a $2 \pi$ monodromy, it seems that the answer would be that the string is composed of two domain walls, each carrying half the flux of the fundamental quark \cite{Anber:2015kea}.  That this is so is shown by using   a numerical simulation, with results shown on Figure \ref{fig:doublestring1}.
The $\sigma$ field varies (far away from the sources) from the $\sigma=0$ vacuum to the $\sigma =\pi$ vacuum inside the string and then back to the $\sigma = 2\pi$ vacuum. By the duality relation, each domain wall carries half the quark chromoelectric flux. The separation between the two domain walls comprising the double string grows like $\ln R$ with the quark-antiquark separation $R$. (This is due to the exponential repulsion of the two domain walls at large distances, which wants the two walls to separate, but is  countered by the energy cost of stretching them. For a simple model relevant for the $SU(2)$ case, see \cite{Anber:2015kea}, while numerical checks are in \cite{Bub:2020mff}.)

The ``double-string'' mechanism of confinement in $\theta=\pi$ dYM also shows another interesting phenomenon: the deconfinement of quark on domain walls. To see this, simply cut the upper domain wall in  the double string on Figure \ref{fig:doublestring1} and take the cut ends to infinity to obtain the picture shown on Figure \ref{fig:decwall1}. The top of the figure is in the $\sigma=\pi$ vacuum and the bottom is in the $\theta=0$ vacuum. The quark and antiquark suspended on the domain wall experience no mutual attraction since the wall tensions of the walls on the two sides of each quark are equal.

\begin{figure}[h]
\centerline{
\includegraphics[width=9.5 cm]{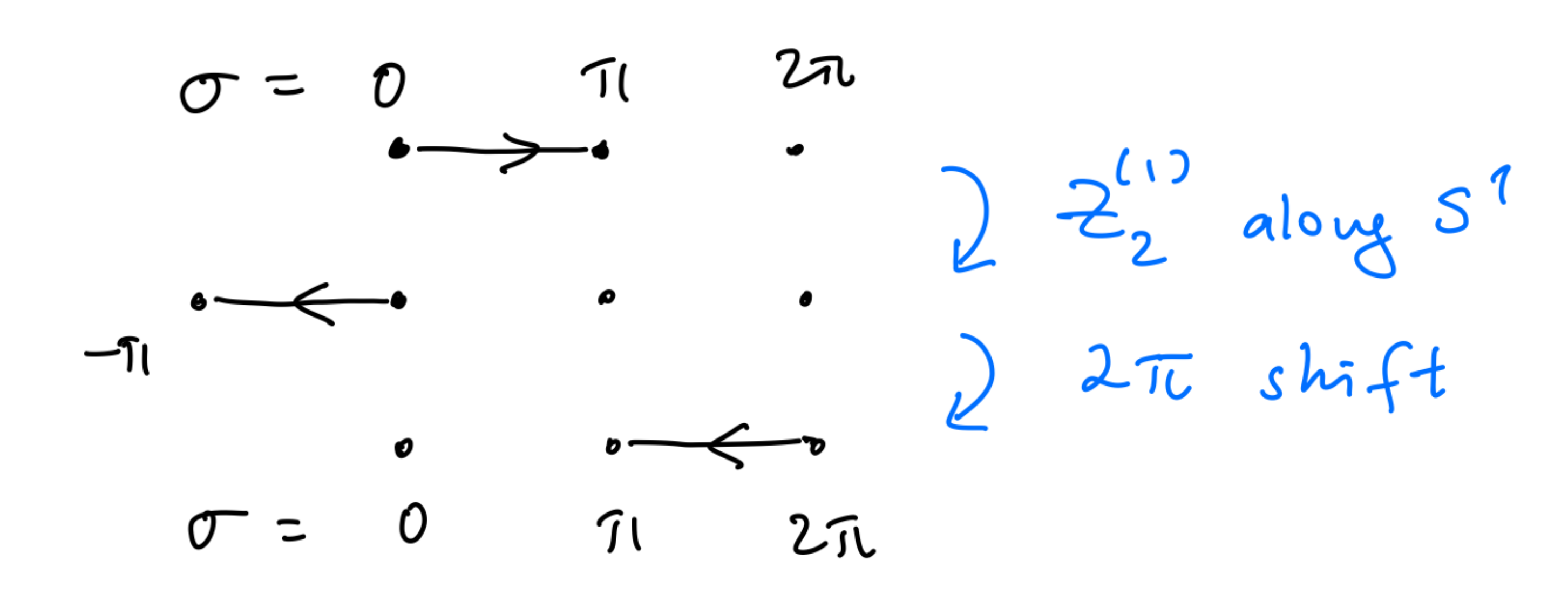}}
\caption{ The action of  a $Z_2^{(1)}$ center symmetry transformation and a subsequent $2\pi$ shift of $\sigma$ on the domain wall solution connecting $\sigma=0$ to $\sigma=\pi$ (the arrow shows the values $\sigma$ takes as the wall is  traversed). This  solution is thus mapped to  the domain wall interpolating from $\sigma=2\pi$ to $\sigma=\pi$. Since the walls are related by a symmetry, they  have the same tension, ensuring deconfinement of quarks due to tension balance, as shown on Figure \ref{fig:decwall1}. The existence of two inequivalent domain walls between the same vacua can be formalized by saying that  ``a topological QFT lives on the domain wall worldvolume'' (for the case at hand, this is a QFT with only two states in its Hilbert space). \label{fig:dwcenter}}
\end{figure}

It is important to note that there are two distinct domain walls in $\theta=\pi$ $SU(2)$ dYM, related by the $Z_2^{(1)}$ center symmetry (\ref{centersigma}). This can be seen by noting that in the middle of  the domain walls connecting the two vacua  on Figure \ref{fig:decwall1}, the $\sigma$ field takes values $\pi/2$ (the middle wall on the Figure) or $3 \pi/2$ (in the walls on the left and the right). The values of $\sigma$ in the centers of these domain walls are thus related by a $Z_2^{(1)}$ transformation, $\sigma \rightarrow - \sigma$ (mod $2\pi$). This means that the domain wall solutions are not invariant under $Z_2^{(1)}$ and are mapped into each other under the center symmetry; see Figure \ref{fig:dwcenter} for an illustration.\footnote{Recall that we use $Z_2^{(1)}$ to denote the center symmetry along $S^1$; often this is called ``zero-form'' center symmetry.}  

The equality of the middle- and left- (and right-) wall tensions is due to the fact that the two walls are $Z_2^{(1)}$ images of each other and is the reason behind the semiclassical explanation of deconfinement of quarks on the wall (the fact that quarks are deconfined on the wall  implies that the rest of the $1$-form center symmetry, the one acting on Wilson loops winding around the directions of the wall worldvolume is also broken on the wall). There are formal arguments relating this deconfinement on the wall to a discrete version of the ``anomaly inflow'' mechanism familiar from continuous anomalies \cite{Callan:1984sa}, generalized to the case of the mixed anomaly between discrete parity and  $1$-form symmetries. We can not discuss it any further and refer the reader to the literature (see \cite{Gaiotto:2017tne,Komargodski:2017smk}, and, in a context related to our discussion, \cite{Cox:2019aji,Bub:2020mff}; more references to the rather voluminous recent literature can be found there).\footnote{Let us mention also that there are phenomena related to the $\theta=\pi$ domain walls and the deconfinement of probe quarks on their worldvolume whose unraveling demands going beyond the $\mu \ll 1/L$ EFT, for example the braiding of fundamental Wilson loops  on domain walls at $\theta = \pi$ discussed in e.g. \cite{Hsin:2018vcg} (this remark is also relevant for deconfinement on the  QCD(adj) and SYM strings). Accounting for this is an outstanding open problem.}

A final historical remark is that the deconfinement on domain walls in related theories was first observed in realizations of (supersymmetric) Yang-Mills theory as low-energy limits of stringy $M$-theory constructions in \cite{Witten:1997ep} (and unpublished work by S.-J. Rey cited there \cite{SJRey:1998}). It was also argued for using large-$N$ arguments, see e.g. \cite{Armoni:2003ji}.
The semiclassical explanation of  \cite{Anber:2015kea}   using the $R^3 \times S^1$  tools described here  is the first one purely in a QFT framework, before any connection to anomalies was understood.

\subsubsection{$\theta=\pi$: a first encounter with magnetic bions.}
\label{sec:firstbions}

Here, we shall discuss the semiclassical configurations that are thought to be responsible for the generation of the higher-order terms in the potential (\ref{dYMpi}), put forward in \cite{Unsal:2012zj}.  The considerations there were strongly motivated by the behaviour of a quantum-mechanical model and we adapt them to our situation. 

We shall only focus on the charge-$2$ contributions generating the leading term $\sim \cos 2 \sigma$ in (\ref{dYMpi}). Here, we will argue (rather hand-wavingly; a better treatment awaits in Section  \ref{sec:adjsym}) that these terms are due to  $M$-$M$ and $M$-$KK^*$ composite objects  of charge-$2$ , called ``magnetic bions.'' 

To begin, we have the four ``fundamental'' monopole-instantons, $M \sim e^{- S_0} e^{i \sigma + i \theta/2}$, $KK \sim e^{- S_0} e^{- i \sigma + i \theta/2}$ and their antiparticles. As argued above, these are the lowest action Euclidean solution of dYM. Clearly, to find higher orders in $e^{-S_0}$, we need to study higher-action saddle points of the Euclidean path integral. 
As we are after   $\sigma$-dependent terms, we are only interested in saddle points with nonzero magnetic charge. 

One  can construct  charge-$2$ monopole-instantons out of the basic ``constituents'' as follows. Begin with a configuration that we denote $M$-$M$. Let us think of this object as a composite of an $M$ with another  $M$. Further, let us assume that the composite is localized in $R^3$ so that it can be represented by a local monopole operator. A dilute gas of such objects would then contribute a factor like $e^{- 2 S_0} e^{2 i \sigma + i \theta}$ in the action: 
the coefficient of $\sigma$ in the exponent is the magnetic charge, as we know from Section \ref{sec:polyakov_pass2}, and the coefficient of $\theta$ is the topological charge, with the two correlated as follows from (\ref{topologicalcharges}). From $M$-$M$, we can obtain a $KK$-$KK$ following the twisting procedure of Section \ref{sec:bpsandkkmonopoles}, which should thus contribute a factor $e^{- 2 S_0} e^{-2 i \sigma + i \theta}$, due to the opposite magnetic charge but equal topological charge. Naturally, we could also construct $M^*$-$M^*$ and $KK^*$-$KK^*$, whose 't Hooft vertices would be the complex conjugates. If we combine these four contributions, we would obtain the following potential
\begin{equation}
\label{MMpot}
V_{M-M, KK-KK} = {f(g_4^2) \over L^3} e^{- 2 S_0}(\cos(2 \sigma + \theta) + \cos(2 \sigma - \theta)) = {2 f(g_4^2) \over L^3} e^{- 2 S_0} \cos \theta \cos(2 \sigma)~.
\end{equation}
We notice that there is a $g_4^2$-dependent prefactor $f$ that we have not computed. Further, the potential $V_{M-M,KK-KK}$ has the same form as the potential in (\ref{dYM3}), except that both $\sigma$ and $\theta$ are multiplied by a factor of $2$, due to the double charge.
Next, a look at the $M$ and $KK$ 't Hooft vertices (\ref{thooftdym}) shows that we can also imagine a charge-$2$  localized configuration with the quantum numbers of $M$-$KK^*$ (as well as $M^*$-$KK$). These would contribute a factor $e^{-2S_0} e^{2 i \sigma}$ and $e^{-2S_0} e^{-2 i \sigma}$, with no $\theta$-dependence, as these saddle points would carry no topological charge. The corresponding contribution to the potential of a dilute gas of these objects would be simply
\begin{equation}
\label{MKKpot}
V_{M-KK^*} = {h(g_4^2) \over L^3} e^{- 2 S_0} \cos(2 \sigma) ~,
\end{equation}
with yet another unknown prefactor $h(g_4^2)$.
Thus, combining the two potentials, we find that a $\cos 2 \sigma$ potential for the dual photon at the next order in the semiclassical expansion is both allowed by the symmetries (recall Section \ref{sec:paritycenteranomaly}) and can be constructed out of the available leading-order semiclassical objects.

Taking the ``composite nature'' of the objects generating the $\cos 2 \sigma$ potential seriously, one may ask in what sense are the $M$-$M$ or $M$-$KK^*$ saddle points of the path integral? 
Motivated by quantum mechanical examples, ref.~\cite{Unsal:2012zj} argued that an analytic continuation of the path integral is required to define these higher-charge saddle  points.
Certainly, these $M$-$M$ or $M$-$KK^*$ configurations approach exact saddle points (classical solutions) ``at infinity,'' i.e. when their centers are infinitely far away from each other, as their interactions vanish at infinite separation. However, their magnetic charges are equal, hence these $M$-$M$ (or $M$-$KK^*$) constituents repel. How can one consider them as objects   localized well enough  to be represented by insertions of local terms $e^{i 2 \sigma}$ in the path integral? The answer  lies in the analytic continuation of the integral over the separation quasi-zeromode (the name stems from the fact that this becomes an exact zeromode at infinite separation). The fact that the solutions repel means that there is a saddle point of this integral at infinity, but analytic continuation into the complex plane leads to another saddle point at a finite complex distance. We refer to  \cite{Unsal:2012zj} and the review \cite{Dunne:2016nmc} for more discussion of these fascinating issues. We shall come back to this in our ``neutral bion'' example in Section  \ref{sec:neutralbions}.

The magnetic bion topological excitations were first found in \cite{Unsal:2007jx},  in  QCD(adj), where $M$-$KK^*$ composites arise without the need for  analytic continuation of the quasi-zero mode integral. We shall exhibit them in more detail in Section \ref{sec:adjsym}.

\hfill\begin{minipage}{0.85\linewidth}

\textcolor{red}{
{\flushleft{\bf Summary of \ref{sec:dymthetapi}:}} The main point of this Section was to study the behaviour of dYM at $\theta=\pi$, where the leading order semiclassical potential (\ref{dYM3}) vanishes. We used symmetries to argue that at $\theta=\pi$ terms of the form $\cos 2 n \sigma$, with $n\in Z$ are allowed as ``$\ldots$'' terms in (\ref{dYM3}). These terms lead one to conclude that parity is spontaneously broken at $\theta = \pi$. We studied the implications of the broken parity for confinement, explained the double-string confinement mechanism and the deconfinement on the domain walls in $\theta=\pi$ dYM. We noted that all of these phenomena reflect certain generalized 't Hooft anomalies. Finally, we reviewed the nature of magnetic bions, charge-$2$ saddle points of the path integral, which can be thought of as charge-$2$ composites of the $M$ and $KK$ monopole-instantons and should be responsible for the generation of the $\cos 2 \sigma$ potential. A study of these objects will resume in Section \ref{sec:adjsym}.}

\end{minipage}
\bigskip

  \subsection{Qualitative picture of the small-$L$, finite-$T$ deconfinement transition on $R^2 \times S^1 \times S^1_\beta$.}
\label{sec:dymtemperature}

In this Section, we shall discuss the  finite-$T$ equilibrium physics of the small-$L$ dYM theory. For simplicity, we concentrate on $\theta=0$ and will only mention results concerning the $\theta$-dependence in the end. To study thermodynamics, we replace $R^3 \times S^1$ with $ R^2 \times S^1_\beta \times S^1$. The last $S^1$ factor is our familiar spatial circle, while the $S^1_\beta$ is the thermal circle whose circumference is $\beta = 1/T$, the inverse temperature (this is  familiar from equilibrium thermodynamics and was already used in Section \ref{sec:digression}). 

We shall find  that upon increasing $T$ (making $S^1_\beta$ smaller), the confining property of 
dYM is lost at some critical $T_c$, still within the region of validity of the abelian small-$L$ description.
The physics is quite remarkable and unraveling the details requires more discussion than we can go into, as it relies on understanding the behaviour of 2d gases of both electric  and magnetic charges, whose relative dominance determines whether the theory confines or not. 
Thus, 
our discussion will be even more qualitative than usual, relying only on familiarity with the notion of thermal compactification and the monopole-instanton gas picture of the vacuum.\footnote{\label{f21}
The literature on this small-$L$, finite-$T$ transition originates in studies of a similar transition in the Polyakov model \cite{Dunne:2000vp,Kovchegov:2002vi}. The studies on $R^3 \times S^1 \rightarrow R^2 \times S^1 \times S^1_\beta$ in  dYM  were initiated in \cite{Simic:2010sv} and continued in \cite{Unsal:2012zj,Anber:2015wha}, while QCD(adj) and SYM were the subject of  \cite{Anber:2011gn,Anber:2012ig,Anber:2013doa}. Much more detail can be found in these references.}

\begin{figure}[h]
\centerline{
\includegraphics[width=16.5 cm]{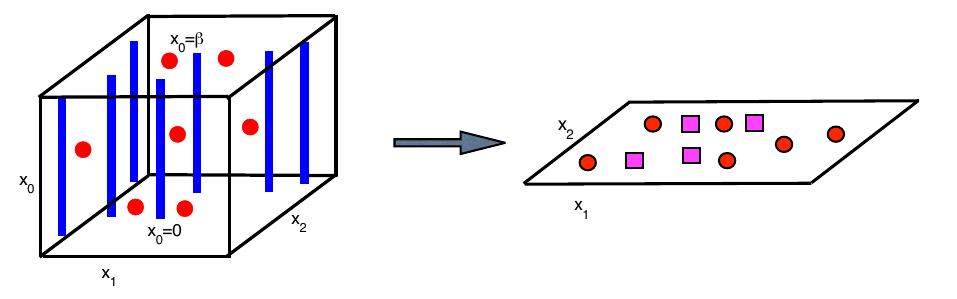}}
\caption{ A picture of the $R^2 \times S^1_\beta$ vacuum. The $M$ and $KK$ monopole-instantons, represented by dots, are much smaller than the size of the thermal circle at the temperatures of interest.  In addition, static electrically-charged $W$-bosons, whose worldlines are showed by the blue lines, are excited with nonzero probability. In the temperature range (\ref{regime1}), the typical distance between either kind of objects is much larger than the size of the thermal circle (this is not depicted to scale on the drawing and has to be imagined), allowing the gas of  to be considered as approximately 2d. The resulting 2d gas of electrically (squares) and magnetically (circles) charged particles is shown on the right. These interact by their respective (magnetic and electric) Coulomb interactions, and also by a mutual Aharonov-Bohm interaction. (Briefly, see the references from footnote \ref{f21}, the latter is a phase in the Boltzmann partition function that depends on the relative 2d angular positions of the electric and magnetic charges, but not on their distance.)  The physics can be studied by various means that we do not have time to discuss, resulting in a deconfinement phase transition at the critical temperature (\ref{tc}). \label{fig:twodimensional}}
\end{figure}

We begin  by noting that upon increasing $T$ from zero (i.e. taking $S^1_\beta$ finite), the monopole-instanton picture of the vacuum remains intact in the $T \ll m_W = {\pi \over L}$ low-temperature limit. The monopole-instanton core size is of order $L$, hence $M$ and $KK$ simply fit in the finite $S^1_\beta$-direction (since $\beta \gg L$), appearing with the already familiar Boltzmann probability from (\ref{thooftdym}), $\sim e^{-{4 \pi^2 \over g_4^2}}$. The heavy $W$-bosons that were integrated out at zero-$T$ can, due to thermal fluctuations, appear with Boltzmann-suppressed probability $\sim e^{- {m_W \over T}} =  e^{- { \pi \over LT}}$. Thus, we have a picture of a typical  vacuum configuration like the one shown on the left panel of Figure \ref{fig:twodimensional}. We can already learn some qualitative lessons if we simply compare these two factors. It is clear that for low enough $T$, the Boltzmann suppression of the $W$-bosons is larger than that of the monopole-instantons, so they should be irrelevant at low $T$ and not significantly affect confinement. 
Equating the two exponentials, we can find an estimate for the  ``critical'' temperature $T_c$, i.e. the temperature above which the Boltzmann suppression of the $W$-bosons is less than that of monopole instantons, so they dominate the vacuum:
\begin{equation}
\label{tc}
e^{- {4 \pi^2 \over g_4^2}} = e^{- { \pi \over LT_c}}, ~\text{at}~~ T_c = {g_4^2 \over 4 \pi L}~.
\end{equation}

Remarkably, this naive estimate can be substantiated in a much more quantitative way as we now briefly describe. 
Notice that at weak coupling $g_4^2 \ll 1$, the above $T_c \ll { \pi \over L}$, so the validity of the abelianized description is not suspect. Furthermore, temperatures near and below
$T_c$ clearly fall in the following regime:
\begin{equation}
\label{regime1}
m_\sigma^{-1} \sim \ell_D \gg \bar\ell \gg \beta = {1 \over T} \gg L~,~~\text{or} ~{1\over \bar\ell} \ll T \ll {1 \over L}
\end{equation}
The first inequality is already familiar, it says that the Debye screening length $\ell_D$ (or inverse dual photon mass) is much larger than the typical distance between monopole instantons $\bar\ell$, as familiar from the Polyakov model, recall Figure \ref{fig:MMbargas}.  The second inequality states that the extent of the time direction $\beta$ is much smaller than the typical separation between monopole-instantons $\bar\ell$. Then, the monopole-instanton  gas looks approximately two-dimensional, as is illustrated on Figure \ref{fig:twodimensional}. This is because the interaction energy of two monopole-instantons on $R^2 \times S^1_\beta$ placed  apart at a distance $r \in R^2$ is logarithmic for $r \gg \beta$,  rather than $\sim 1/r$.\footnote{The precise relation can be found either using the method of images or Gauss' law, showing that ${4 \pi L \over g_4^2}{q_1 q_2 \over r}$ is replaced by $- {8 \pi L T \over g_4^2} q_1 q_2 \log r$ for $r \gg T^{-1}$, similar to how static electric charges interact in $R^2$, recall Exercise 3.2 and eqn.~(\ref{chargesinteraction}).}

The $W$-boson gas is also dilute and nonrelativistic: the $W$ boson density is  given by $\sim m_W T e^{- {m_W \over T}}$, as follows from the nonrelativistic statistical mechanics of a 2d gas. Thus, the typical distance between $W$ bosons is also $\gg \beta$. This means that the entire system is approximately 2d and can be treated by studying the statistical mechanics of a classical 2d gas of electrically and magnetically charged particles, as the cartoon on the r.h.s. of Figure \ref{fig:twodimensional} shows.

These studies are quite nontrivial, as the models are not exactly solvable and not weakly coupled near $T_c$ (2d gases of only electric or only magnetic particles are well understood, but the phase structure of systems that have both is much less under control).
In many cases (see \cite{Anber:2011gn} for an exception), finding the order of the transition and determining various correllators in many cases has to be done numerically. For the case of $SU(2)$ refs.~\cite{Dunne:2000vp,Kovchegov:2002vi}  used fermionization to argue that the transition is in the Ising universality class, in agreement with universality arguments and lattice data for thermal pure-YM theory  \cite{Svetitsky:1982gs,Yaffe:1982qf}. Thus, at least for $SU(2)$ dYM, it appears that the small-$L$ deconfinement transition is smoothly connected to the pure-YM one.
To the best of our knowledge, the order of the transition in the 2d electric-magnetic Coulomb gas describing dYM for $SU(N)$ is not known at present.

Here, we shall only briefly discuss the center-symmetry realization. The center symmetry in question is the one in the $S^1_\beta$ direction, similar to the discussion in Section \ref{sec:digression} (the center-symmetry in the small-$L$ direction remains unbroken). One studies the correlator of the traces of two fundamental Polyakov loops winding around $S^1_\beta$  (we denote each such trace by  $\Omega_\beta$, to distinguish from the loops winding around the spatial $S^1$). These are  charged under the $Z_2^{(1)}$-center in the thermal direction, and their correlator obeys:
\begin{equation}
\label{poldym1}
\langle \Omega_\beta(r) \Omega_\beta(0)\rangle\vert_{r \rightarrow \infty} \rightarrow \left\{ \begin{array}{cc} e^{- {\Sigma r \over T}} \rightarrow 0,& T < T_c, \cr 1, & T  > T_c.\end{array} \right.
\end{equation}
In the high-$T$ deconfined phase, it is the dominance of the electric $W$-bosons that leads to screening of the fundamental charges, leading to the non zero correlator and center-symmetry breaking.
Conversely,  the $S^1_\beta$-center-symmetry is unbroken in the low-$T$ confined phase, due to the monopole-instanton dominance, leading to the confinement of the fundamental probes (and an area law, as in the top line of (\ref{poldym1})), as per our general discussion of deconfinement. 

There are qualitative features of this small-$L$ thermal transition that are borne out by thermal transitions in the $R^4$ theory: the order of the $SU(2)$-theory transition already mentioned and the $\theta$-dependence of the critical temperature (which decreases as $\theta$ increases away from zero, as follows from the fact that the M and KK fugacities decrease, due to the ``topological interference'' effect \cite{Unsal:2012zj}). 
In addition, this small-$L$ transition saturates an inequality between the deconfinement and chiral restoration transition temperatures in QCD(adj)/SYM \cite{Anber:2011gn,Anber:2012ig,Anber:2013doa} required by the matching of generalized 't Hooft anomalies \cite{Shimizu:2017asf,Komargodski:2017smk}. Finally, there are interesting connections to 2d condensed-matter systems---for example the system describing a similar transition in $SU(3)$ QCD(adj) (not discussed here, see \cite{Anber:2011gn}) is related to the theory of melting of 2d crystals \cite{Nelson:melt}.

\hfill\begin{minipage}{0.85\linewidth}

\textcolor{red}{
{\flushleft{\bf Summary of \ref{sec:dymtemperature}:}} In this Section, we discussed the thermal physics of  the small-$L$ theory, i.e. considered dYM on $R^2 \times S^1_\beta \times S_1$. We argued that there is a thermal deconfinement transition associated with the breaking of the $S^1_\beta$ center-symmetry above $T_c = g_4^2/(4 \pi L)$. The physics near $T_c$ can be described as that of a 2d gas of electrically and magnetically charged particles, with Coulomb and Aharonov-Bohm interactions. The electric charges dominate at high $T$, causing screening of fundamental charges, while the magnetic ones dominate at low $T$ and are responsible for confinement. This is an attractive picture and arises naturally in this calculable setup. Some qualitative aspects of the small-$L$ transition agree with what is known for large-$L$ theories and with constraints from generalized 't Hooft anomalies.}

\end{minipage}

\bigskip

 \section{QCD(adj). }
 \label{sec:adjsym}
 
 We begin our study of QCD(adj) with gauge group $SU(2)$ and $1<n_f \le 5$ massless flavours of Weyl fermions. 
  In the past, these theories have been studied with with varying motivation in mind. The $n_f=1$ massless theory is ${\cal{N}}=1$ SYM, studied for its tractability and relation to pure YM. The theory with $n_f=2$ is related to Seiberg-Witten theory (${\cal{N}}=2$ SYM) by decoupling the complex adjoint scalar field, while the $n_f=4$ theory is ${\cal{N}}=4$ SYM with the scalars decoupled. 
 The theories with various $n_f$ have been  the subject of lattice studies, motivated by ``walking technicolour'' extensions of the electroweak-breaking sector of the standard model: for early lattice work, see \cite{Catterall:2008qk,Hietanen:2008mr,DelDebbio:2009fd}, while a recent one with more references is \cite{Athenodorou:2021wom}. The $n_f=1$ theory (SYM) has also been studied on the lattice (see \cite{Bergner:2018unx} and references therein).

The QCD(adj) lagrangian was already given in (\ref{lagrangianadjoint}). The theory (for any $N$) is asymptotically free for $n_f < 6$. The quantum theory has a dynamical strong scale $\Lambda$. The running coupling $g_4^2(\mu)$ can be expressed, at one-loop order, via $\Lambda$ and the one-loop coefficient of the beta function, $\beta_0$ , as
 \begin{eqnarray}\label{lambdascale}
 {g^2_4(\mu) \over 4 \pi} &=& {4 \pi \over \beta_0}\left[ {1 \over \log(\mu^2/\Lambda^2)} + \ldots \right]~,~\text{where}~  \beta_0 = {22 - 4 n_f \over 3}~, ~ \text{or}\nonumber
 \\
 \Lambda^{\beta_0} &=& \mu^{\beta_0}\; e^{- {8 \pi^2 \over g_4^2(\mu)}}~,
 \end{eqnarray}
 where the dots denote terms depending on higher-loop coefficients of the beta function. The second line above is often also quite useful.
 
 In this Section, we shall study the $n_f > 1$ case of non-supersymmetric QCD(adj) on $R^3 \times S^1$. 
The perturbative dynamics of QCD(adj) was already considered in Section \ref{sec:adjstability}. We showed that the center symmetry along the $S^1$, $Z_2^{(1)}$, is preserved with $n_f \ge 2$ massless adjoints. Further, in Section \ref{sec:adjointspectrum}, we showed that at the center-symmetric point on the Weyl chamber, the non-Cartan components of the adjoints gain mass of order $1/L$ (they are the ones responsible for center stability generating the potential (\ref{gpyadjoint})), while the Cartan components remain massless. The $A_4^3$ field also obtains mass of order $g_4/L$. 

Thus, the $\mu \ll g_4/L$ spectrum of QCD(adj) consists of the Cartan gauge field of the unbroken $U(1)$ and the $n_f$ Cartan components of the adjoints, $\lambda_\alpha^{3 \; I}$, $I = 1,...,n_f$. The Cartan fermions are not charged under the Cartan $U(1)$ photon, so we can use the dual-photon description to describe the latter.
The kinetic term of the long distance theory is then given by the kinetic term of the dual photon plus the kinetic terms of the Cartan components of the adjoints:
\begin{equation}\label{adjoint1}
L_{kin}^{QCD(adj)}=  {1 \over 2} { g_4^2 \over L (4 \pi)^2}   (\partial_\lambda \sigma)^2+ i{ L \over g_4^2} \; \bar\lambda^{3 \; I}_{\dot\alpha} \; \bar\sigma^{\mu \dot\alpha \alpha} (\partial_\mu \lambda_\alpha)^{3 \; I} ~,
\end{equation}
giving rise, once again, to a rather boring long-distance EFT (the fields $\lambda^{3 \; I}$ of dimension $3/2$ are the $S^1$ zero modes of the 4d adjoints).

Our next question is: what  nonperturbative terms can be added to (\ref{adjoint1}) and what are the semiclassical objects generating them? 

 \subsection{Discrete chiral symmetry and its action in the EFT.}
\label{sec:discretechiral}

 We begin with a discussion of the symmetries, in an attempt to narrow down the possible terms that can be added to (\ref{adjoint1}). The $1$-form $Z_2$ center symmetry, which only acts on the dual photon in (\ref{adjoint1}) as (\ref{centersigma}), $Z_2^{(1)}: \sigma \rightarrow - \sigma$, is unchanged from dYM.  
 It does not act on $\lambda^{3\; I}$.
 
 The new element in massless QCD(adj)
 on $R^3 \times S^1$ is the classical $U(n_f)$ ``$0$-form'' global chiral symmetry, acting by a rotation of the $n_f$ flavours,
\begin{equation}\label{unfsymmetry}
\lambda^I \rightarrow U^I_{\;J} \; \lambda^{J}~,  ~~ \text{where} ~ U \in U(n_f) = \frac{U(1) \times SU(n_f)}{Z_{n_f}}~,
\end{equation} 
where we omit the spinor and Lie-algebra indices and the notation for $U(n_f)$ means simply that $U(1)$ and $SU(n_f)$ share a common   $Z_{n_f}$ subgroup of center elements. The $U(1)$ transformations act on the undotted Weyl spinors as $\lambda^I \rightarrow e^{i \alpha} \lambda^I$. The $SU(n_f)$ symmetry is anomaly free and thus remains a symmetry in the quantum theory, but the $U(1)$ is broken by the anomaly to $Z_{4 n_f}$. More precisely, the anomaly-free symmetry in the quantum theory is $(Z_{4n_f} \times SU(n_f))/{Z_{n_f}}$, where, as above, this means that 
 transformations in the $Z_{n_f}$ center of $SU(n_f)$ act in the same way as a $Z_{n_f}$ subgroup of $Z_{4 n_f}$.
There are various ways, which are related to each other,  to see this symmetry reduction by anomalies and we discuss them below, as they are important for our considerations.\footnote{The discussion that follows is very similar to the $U(1)_A$ symmetry breaking by the anomaly in massless QCD with fundamental fermions. There, one can also find a discrete subgroup  of $U(1)_A$ unbroken by the anomaly. The reason it is never discussed in textbooks is that  it is not an independent symmetry and can embedded in the other global symmetries of fundamental QCD, which is easy  to show.}

That 
the $U(1)$ factor suffers from an anomaly, similar to the $U(1)_A$ symmetry in QCD, is well known. The classical $U(1) \in U(n_f)$ chiral current of the massless theory (\ref{lagrangianadjoint}), ${j}^M_f = {1\over g^2} {\bar\lambda}^{a \;I } \bar\sigma^M \lambda^{a \; I}$, with a sum over $a$ and flavour $I$ understood, obeys the (Heisenberg picture) operator equation, the anomaly equation:\footnote{To the reader not familiar with anomalies we recommend   \cite{Shifman:2012zz} (especially the sections on anomalies and on the anomaly-free chiral symmetry in SYM), whose conventions we follow below. Recall that, as in (\ref{lagrangianadjoint}), $M,N$ are 4d spacetime indices.  }
\begin{equation}\label{anomaly1}
\partial_M j^M_f =  4 n_f  \partial_M K^M~,
 \end{equation}
 where $
K^M = {1 \over 16 \pi^2} \epsilon^{MNPQ} \left( A_N^a \partial_P A_Q^a - {1 \over 3} \epsilon^{abc} A_M^a A_P^b A_Q^c\right)$ was already mentioned in arriving at (\ref{topologicalcharges}). Notice that the r.h.s. of (\ref{anomaly1}) is nothing but $4n_f $ times the topological charge density $q$ appearing in (\ref{4dlagrangian}).

The first argument---somewhat quicker and helpful with writing EFT terms---for the existence of an anomaly-free $Z_{4n_f} \subset U(1)$ is as follows. In 4d, the effect of the anomaly can be seen by inspecting the contribution of fluctuations with $Q_T=1$ (i.e. the ones corresponding to a BPST instanton)   and examining the corresponding 't Hooft vertex. In the theory with massless fermions and an anomalous $U(1)$, fermion fields must appear in the corresponding 't Hooft vertex \cite{tHooft:1976snw,tHooft:1976rip}. This can be understood  by integrating the anomaly equation (\ref{anomaly1}) over spacetime between an initial and a final time slice. The integral of the r.h.s. is $4 n_f Q_T$, while the integral of the l.h.s. gives the change of $U(1)$-charge,  $Q_5 \equiv \int d^3 x j^0_f(\vec{x},t)$, between the initial and final times. Thus, integrating (\ref{anomaly1}) shows that the $Q_5$ charge changes  in units of $4 n_f Q_T$   in a nonzero $Q_T$ background:
\begin{equation}
\label{anomaly2}
\Delta Q_5 = \int d^3 x \; j^0_f(\vec{x},t)\bigg\vert_{t=-\infty}^{t=+\infty} = 4 n_f Q_T ~.
\end{equation}
Hence, for integer $Q_T$, the $U(1)$ charge $Q_5$ is violated in multiples of $4 n_f$,  i.e. $Q_5$ is conserved (mod $4n_f$).

From (\ref{anomaly2}) above, we can conclude the following: for $SU(2)$ QCD(adj) with $n_f$ massless $\lambda^{I}$, the local 't Hooft vertex\footnote{In the electroweak sector of the standard model where $\rho$ is of order the electroweak breaking scale, the $B+L$ violating 't Hooft vertex has above schematic form with  $\lambda^{4n_f}$ replaced by $qqql$, three quark and one lepton fields  (for a single generation) giving $\Delta B = \Delta L = 1$ \cite{tHooft:1976rip}.} due to a BPST instanton fluctuation centered at $x\in R^4$  must have the schematic form
\begin{equation}
\label{bpstr4}
  e^{- {8 \pi^2 \over g^2(1/\rho)}} \; e^{ i \theta}\;  \lambda^{4 n_f}(x)~ = (\Lambda \rho)^{\beta_0}\; e^{ i \theta} \;  \lambda^{4 n_f}(x)~
\end{equation}
where $\lambda^{4 n_f}(x)$ denotes an $SU(2)$ gauge-, $SU(n_f)$ flavour-, and $SL(2,C)$ Lorentz-invariant contraction of all indices $ \lambda(x)$ is endowed with.\footnote{A local invariant $\sim \lambda^{4n_f}$ that does not vanish due to Fermi statistics exists, but its form is not crucial to us here, it can be taken to be the square of (\ref{Mchiral}) below.  There is also a complex conjugate term for anti-BPST instantons involving $\bar\lambda$. } The fermion-field insertions are the minimal number necessary to ensure that $Q_5$ is violated by the required amount (\ref{anomaly2}). The fermion insertions in the BPST-instanton 't Hooft vertex can also be understood as arising due to an index theorem. We shall not further discuss this and only note that integrating the anomaly equation is a physicists' way to understand this. 

 The local term (\ref{bpstr4}) represents the contribution of an instanton   of  size $\rho$ located at $x \in R^4$ and should be understood to be appropriate in an EFT valid at $\mu \ll 1/\rho$, as it does not resolve the instanton size. In (\ref{bpstr4}), we did not include any factors having to do with the integration over the size of the instantons, which should make (\ref{bpstr4}) have  dimension $4$.\footnote{This is because in our $R^3\times S^1$ setup, monopole-instantons have fixed size and no integral over $\rho$ will appear. In the electroweak sector of the standard model the instanton size integral is cut off at the electroweak-breaking scale, while in an unbroken gauge theory such as $SU(3)$-colour of the standard model the integral over $\rho$ diverges as $\rho \rightarrow \infty$, related to the ``Landau pole'' IR problem and making quantitative predictions difficult \cite{Shifman:2012zz,Schafer:1996wv}. 
} 

Let us now further dwell on the various factors in (\ref{bpstr4}). 
 The exponential factor is the familiar BPST instanton action. We also used (\ref{lambdascale}) to write the 't Hooft vertex in terms of the strong coupling scale $\Lambda$. The $\theta$-angle dependent factor $e^{i \theta}$ is included for book-keeping purposes (even though $\theta$ is not observable in the massless-fermion theory), indicating that  (\ref{bpstr4}) is due to field configurations with topological charge unity. The main point of our writing eqn.~(\ref{bpstr4}) is to argue that  the 't Hooft vertex violates $U(1)$, i.e. leads to processes obeying (\ref{anomaly2}), $\Delta Q_5 = 4 n_f$, but clearly preserves a $Z_{4n_f} \subset U(1)$ symmetry.  The form (\ref{bpstr4}), including all prefactors, was first computed in  \cite{tHooft:1976snw} for $SU(N)$ gauge theories with fundamentals and the interested reader  (bound to notice that the anomaly-equation-based argument just given offers a significant shortcut---but no control over the pre-factors) is invited to study the rather long-winded calculation. 

To connect (\ref{bpstr4})  to our second description of the $U(1) \rightarrow Z_{4n_f}$ anomalous breaking, we notice that the 't Hooft vertex is invariant under the $U(1)$ transformation $\lambda \rightarrow e^{ i \alpha} \lambda$ provided we also allow the parameter $\theta$ to transform as  $\theta \rightarrow \theta - 4 n_f \alpha$.
In fact, 
 it is convenient\footnote{This line of reasoning played an important role in the ``power of holomorphy'' arguments  \cite{Seiberg:1994bp}. However, thinking of parameters as vevs of nondynamical fields is also useful (but less powerful) beyond supersymmetry.}  to think of the parameter $\theta$ as the vacuum expectation value of a nondynamical  ``axion''  field, whose shift compensates for the anomalous-$U(1)$ transformation of $\lambda$.  The fact that $\theta$ is a parameter that can not be transformed means that the 't Hooft vertex is only invariant for $\alpha = 2 \pi k/(4 n_f)$, $k \in Z$, i.e. under $Z_{4 n_f}$. Notice that this  corresponds to integer-$2\pi$ shifts of $\theta$, as seen from the above. This discussion will now naturally connect  to the one following below and leading to eqn.~(\ref{z4nf}). 

The second argument that we shall give for the existence of a discrete chiral symmetry $Z_{4 n_f}$ will be to note that the anomaly equation (\ref{anomaly1}) allows one  to define a conserved (but gauge variant, see below) current which we label $ J^\mu_5$ for historical reasons:
$
{{J}}^M_5 = j^M_f - 4 n_f  K^M~,
$ $\partial_M J^M_5 = 0$.
The corresponding $U(1)$ charge operator,  
\begin{equation}\label{u1charge}\tilde Q_5 = \int d^3 x {{J}}^0_5 = \int d^3 x {j}_f^0 - 4 n_f  \int d^3 x K^0~,
\end{equation} is conserved  (as opposed to the non-conserved $Q_5$ entering (\ref{anomaly2})) but is not gauge invariant. The gauge noninvariance of the charge $\tilde Q_5$ is due to the fact that under large $SU(2)$ gauge transformations, the ones with a nontrivial winding\footnote{Here, $S^3$ is the  ``one-point compactification'' of $R^3$, due to the boundary condition on gauge transformations approaching unity at spatial infinity. See the standard construction of $\theta$-vacua in e.g. \cite{Shifman:2012zz}.} number $S^3 \rightarrow SU(2)$, the integral $\int d^3 x K^0$ shifts by an integer, as you will be reminded in the following exercise, adapted to our spatial-$R^{2} \times S^1$ setup.
\begin{quote}
{\flushleft{\bf Exercise 16:}} Show that under a gauge transformation $C$,  $A_k^C = C (A_k - i \partial_k) C^{-1}$, where $k=1,2,3$ is a spatial index ($x^3 = x^3 + L$),  with gauge fields periodic on the spatial $S^1$, and $C=1$  outside a compact region (say a disc $D^2$) in $R^2$
\begin{eqnarray}\label{winding}
\int\limits_{R^2 \times S^1} d^3 x (K^0\left(A^C \right) -  K^0\left(A\right)) 
= {1 \over 24 \pi^2} \int\limits_{R^2 \times S^1}d^3 x\; \epsilon^{ijk}\;  \tr (C \partial_i C^{-1}) (C \partial_j C^{-1}) (C \partial_k C^{-1}).\end{eqnarray}
Further, taking  $C$ to also approach unity at the ``boundary'' of $S^1$ (e.g. $x^3 =0, L$), the combined $R^2 \times S^1$ boundary conditions above make $C$ into a map from $D^2 \times I$, where $I$ is the one-dimensional interval on $S^1$ where $C \ne 1$. As $C$ is unity on the boundary of $D^2 \times I$, it is effectively a map   $S^3 \rightarrow SU(2)$. The r.h.s. above is the integer winding number of this map, familiar from  Skyrmion physics and the construction of $\theta$ vacua in YM theory, e.g. \cite{Vainshtein:1981wh,Coleman:1985rnk,Shifman:2012zz}.
\end{quote}
The point of the above exercise was to remind you of the fact that the conserved charge (\ref{u1charge}) corresponding to the anomalous $U(1)$ is not gauge invariant. Under a large gauge transformation $C$ with unit winding number (\ref{winding}), the charge changes as $C: \tilde Q_5 \rightarrow \tilde Q_5 -4 n_f $, by $4n_f$ units.\footnote{To dispel any doubts that this is the minimum amount of change of $\tilde Q_5$ under any gauge transformation on $R^2 \times S^1$ requires some extra work. One way to proceed is by studying the case where  all three direction of space are taken compact, i.e. $T^3$. When two of the $T^3$ directions are taken infinite, we obtain our $R^2 \times S^1$ of interest. The construction of the $\theta$ vacua and large gauge transformations can be generalized to the $T^3$ case. For an $SU(N)$ gauge theory on a spatial $T^3$ of arbitrary sides (``an $SU(N)$ bundle''), the result is that the minimum violation of $Q_5$ is controlled by maps of integer winding (\ref{winding}) \cite{vanBaal:1982ag,Luscher:1982ma}. A recent discussion (and more references) looking also at the more general $SU(N)/Z_N$ bundles on $T^3$ is in \cite{Cox:2021vsa}.}

As usual in QFT, having a conserved charge, one defines a corresponding unitary symmetry operator. For the $U(1)$ case of interest to us, this unitary symmetry operator would be 
$X_\alpha = e^{i \alpha Q}$. For a conserved and gauge invariant $Q$ (this would apply, for example to $Q$ of (\ref{qoperator})), this would generate an $e^{i \alpha} \in U(1)$ transformation when acting on the field operators of unit charge. But we just showed that our conserved $\tilde Q_5$ changes by $4 n_f $ under a large gauge transformation. Thus, it is clear that while $e^{i \alpha \tilde Q_5}$ is not gauge invariant, the operator 
\begin{equation}
\label{discrete1}
X_{4 n_f } = e^{i {2 \pi \over 4 n_f } \tilde Q_5} = e^{i {2 \pi \over 4 n_f} \int d^3 x j_f^0} \; e^{\;- i 2 \pi \int d^3 x K^0}
\end{equation}
 is gauge invariant.  This is because the second factor in (\ref{discrete1}), the $e^{- i 2 \pi \int d^3 x K^0}$ operator, is gauge invariant---it is  multiplied by unity after integer-winding gauge transformations. The operator (\ref{discrete1}) generates a $Z_{4 n_f} \in U(1)$ discrete chiral symmetry whose action on the fermions is shown in (\ref{z4nf}).
  
 The operator $e^{-i 2 \pi \int d^3 x K^0}$ generates $2\pi$ shifts of the $\theta$ angle (in the canonical $A_0=0$ gauge  quantization of the gauge theory\footnote{Very quickly, this is because $e^{ i\theta}$ is the eigenvalue of a physical state $|\psi\rangle$ under unit-winding large gauge transformation $C_1: | \psi \rangle \rightarrow e^{ i \theta} |\psi \rangle$. Thus, acting with $e^{-i 2 \pi \int d^3 x K^0(A)}$ on a physical state shifts $\theta$ by   $-2 \pi$, owing to (\ref{winding}).}) and the above reasoning shows that such $2\pi$ shifts of $\theta$ are now part of the anomaly-free discrete chiral symmetry, just like they were part of parity at $\theta=\pi$ in the pure gauge theory.
 In fact, the reason we gave this somewhat long-winded operator point of view on the discrete chiral symmetry was precisely to note that $X_{4 n_f }$ involves a shift of the $\theta$-angle of the $SU(2)$ theory by $2\pi$.  Now we recall our 
  discussion of parity in dYM, near eqn.~(\ref{paritydYM}), where we saw that the dual photon shifts by $\pi$ under such $2\pi$ shifts of the $\theta$-angle. We conclude that in the abelianized phase of QCD(adj) on $R^3 \times S^1$, such a $\pi$ shift of the dual photon will accompany the discrete chiral $Z_{4 n_f}$ transformation: 
 \begin{eqnarray}
 \label{z4nf}
Z_{4 n_f}:  \lambda^{3 \; I}_\alpha &\rightarrow& e^{i {2 \pi \over 4 n_f}} \; \lambda^{3 \; I}_\alpha~, \nonumber\\
 \sigma &\rightarrow & \sigma + \pi ~.
 \end{eqnarray}
The top line is simply the restriction of (\ref{unfsymmetry}) to $Z_{4n_f}$.

We shall see in the next Section that invariance under the $Z_{4n_f}$ transformations (\ref{z4nf}) along with the $Z_2^{(1)}$ center symmetry $Z_2^{(1)}: \sigma \rightarrow - \sigma$ will severely constrain the  nonderivative terms that can be added to (\ref{adjoint1}). In fact, combined with semiclassical power-counting rules, similar to the ones we discussed in dYM, the effective lagrangian will be essentially determined.

  \hfill\begin{minipage}{0.85\linewidth}

\textcolor{red}{
{\flushleft{\bf Summary of \ref{sec:discretechiral}:}} Here, we argued that the $Z_{4n_f}$ subgroup of the $U(1)$ chiral symmetry remains anomaly free. This follows from the anomaly equation (\ref{anomaly1}) by constructing the unitary and gauge invariant symmetry operator (\ref{discrete1}), or, equivalently, be examining the 't Hooft vertex of a BPST instanton (\ref{bpstr4}). Most importantly, the anomaly-free discrete chiral symmetry acts as (\ref{z4nf}) on the IR degrees of freedom,  shifting the dual photon by  $\pi$, similar to parity  in dYM at $\theta = \pi$. It is important to note that the results of this Section regarding the $Z_{4n_f}$ transforms also hold for $n_f=1$, i.e. SYM.
}

\end{minipage}

\bigskip

\subsection{Symmetry constraints on the QCD(adj) EFT.}
\label{sec:symmadj}

We shall now use the symmetry transformations  under $Z_{4n_f}$ and $Z_2^{(1)}$ to constrain the form of the non derivative terms that can be added to (\ref{adjoint1}). We recall that $\sigma$ is a $2\pi$ periodic field, therefore whatever we write should be a periodic function of $\sigma$. We shall organize our study of the possible nonperturbatively-generated terms by their $\sigma$-periodicity, i.e. the magnetic charge $Q_M$ of the semiclassical objects that generate them.

Begin with the simplest possibility of  terms $\sim e^{i \sigma}$, due to monopole instantons with unit  magnetic charge. Under $Z_{4 n_f}$ this term changes sign, $Z_{4 n_f}: e^{i \sigma} \rightarrow - e^{i\sigma}$. The only other fields that transform are the fermions $\lambda^{3 \; I}$. Under $Z_{4 n_f}$ we have that $(\lambda^3)^{2 n_f}$ transforms by a factor of $e^{ i {2 \pi \over 4 n_f} 2 n_f} = -1$, i.e. has the correct transformation property to make an $Z_{4n_f}$ invariant. In fact, we can now write the $Z_{4n_f}$-invariant proportional to $e^{i \sigma}$ as  \begin{equation}
\label{Mchiral}
M \sim e^{i \sigma} \; {\text det}_{\small IJ} \;   \lambda^{3 \;I}_{ \alpha}  \lambda^{3 \;J}_{ \beta} \epsilon^{\beta\alpha}  \equiv e^{i \sigma} (  \lambda \cdot   \lambda)^{n_f}~,
\end{equation}
where in the first term, we showed explicitly how all $SL(2,C)$ and flavour indices are contracted (as indicated, the determinant is taken over the flavour indices $IJ$), while in the second equality we introduced a short-hand notation for the determinant. 
We called (\ref{Mchiral})   an $M$ term, alluding to the fact that $M$ vertices in dYM came with a factor of $e^{i \sigma}$. In fact, just like the BPST instanton on $R^4$ has a 't Hooft vertex that comes with fermions attached (\ref{bpstr4}), an $M$ monopole-instanton on $R^3 \times S^1$ has to have fermions attached to its 't Hooft vertex.\footnote{In addition to our $Z_{4n_f}$ symmetry-based argument, 
this can be argued based on  a generalization of the index theorem to $R^3 \times S^1$ with nontrivial holonomy, discussed by mathematicians in \cite{Nye:2000eg} and in physicist-friendly terms in \cite{Poppitz:2008hr}. We shall not need to discuss it here, as our symmetry arguments suffice. The $R^3 \times S^1$ index theorem is a generalization of the Callias index theorem on $R^3$ \cite{Callias:1977kg}, see also \cite{Weinberg:2012pjx}. For a flavour of the index theorem, see Appendix \ref{appx:index}.} In fact the calculation of (\ref{Mchiral}), similar to 't Hooft \cite{tHooft:1976rip}, but on  $R^3 \times S^1$ was first done for $n_f=1$ in \cite{Davies:1999uw} (\cite{Poppitz:2012sw} added some imporant details), and for $n_f>1$ in \cite{Unsal:2007jx,Anber:2011de}; as with (\ref{bpstr4}) the details are not important for a qualitative understanding.

Further, we learned in our study of dYM that ``$M+KK = BPST$.'' Thus, the fermions of the BPST instanton vertex have to somehow split between the $M$ and $KK$. Furthermore, since $M$ and $KK$ are related by the $Z_2^{(1)}$ center symmetry (\ref{centersigma}), which does not act on fermions, the number of fermions appearing in each 't Hooft vertex should be the same, i.e. an $M$ 't Hooft vertex should have half the fermions appearing in the BPST-instanton 't Hooft vertex (\ref{bpstr4})---which is precisely what we see in (\ref{Mchiral})---with the other half residing at the $KK$ vertex.

Thus, what about $e^{- i \sigma}$ terms? Clearly we can have, in a way consistent with the $Z_{4n_f}$ symmetry, 
\begin{equation}
\label{KKchiral}
KK \sim e^{- i \sigma}   ( \lambda \cdot   \lambda)^{n_f}.
\end{equation}
We shouldn't forget the $Z_2^{(1)}$ center symmetry, $\sigma \rightarrow - \sigma$. It interchanges $M$ with $KK$  (and $M^*$ with $KK^*$), exactly as in dYM; thus, these terms should appear as a sum in the EFT. Furthermore, we observe that the product of (\ref{Mchiral}) and (\ref{KKchiral}) is precisely the BPST 't Hooft vertex (\ref{bpstr4}): the factors of $e^{\pm i \sigma}$ cancel out and fermion charge is violated by the required $4 n_f$ units. Thus, the splitting of a BPST instanton into $M$ and $KK$ components familiar from dYM persists in the QCD(adj) theory.

The c.c. of (\ref{KKchiral}) would be precisely the $KK^*$ contribution:
\begin{equation}
\label{KKchiralbar}
KK^* \sim e^{ i \sigma} \; {\text det}_{IJ} (\bar\lambda^{3 \;I}_{\dot\alpha} \bar\lambda^{3 \;J}_{\dot\beta} \epsilon^{\dot\alpha\dot\beta}) \equiv e^{ i \sigma} (\bar\lambda \cdot  \bar\lambda)^{n_f}~,
\end{equation}
while the c.c. of (\ref{Mchiral}) would be the $M^*$ contribution
\begin{equation}
\label{Mchiralbar}
M^* \sim  e^{-i \sigma} ( \bar\lambda \cdot  \bar\lambda)^{n_f}~.
\end{equation}
 What we conclude, then, combining the four terms above, is that the terms of lowest dimension in the EFT,  proportional to $e^{ \pm i \sigma}$ are
\begin{equation}
\label{eftunitcharge}
L_{|Q_M| = 1}  = e^{- S_0} {1 \over L^3} \left[(e^{i \sigma}+ e^{- i \sigma}) \; L^{3 n_f}  ( \lambda \cdot\lambda)^{n_f} + (e^{i \sigma}+ e^{- i \sigma}) L^{3n_f} ( \bar\lambda \cdot  \bar\lambda)^{n_f}\; \right]~.
\end{equation}
As usual, we work with exponential-only accuracy neglecting overall powerlaw $g_4^2$ dependence (they can be found in the already mentioned \cite{Davies:1999uw,Poppitz:2012sw,Unsal:2007jx,Anber:2011de}). We kept the $e^{\pm i\sigma}$ factors to indicate the origin of the four terms above.
The terms in (\ref{eftunitcharge}) are not particularly interesting: for $n_f>1$, they appear to be quite unimportant higher-dimensional operators, irrelevant in our weakly-coupled 3d IR EFT. 

In particular, no potential for the dual photon is possible at the $Q_M = \pm 1$ level. Does the theory confine, then?
As from our discussion of magnetic bions in dYM, we suspect that we have to study higher magnetic-charge contributions to find out. It is clear that, as in dYM at $\theta = \pi$, the only potential terms for the dual photon   invariant under the discrete chiral symmetry $\sigma \rightarrow \sigma + \pi$ are those of even magnetic charge, $e^{ \pm i 2n \sigma}$. We expect that they appear at order $e^{- 2 n S_0}$ in the semiclassical expansion. Thus, the simplest possibility \cite{Unsal:2007jx} is a charge-$2$ magnetic-bion term similar to (\ref{MKKpot}):
\begin{equation}
\label{eftunitcharge2}
L_{|Q_M| = 2}  = - e^{- 2 S_0} {1 \over L^3} (e^{i 2 \sigma} + e^{- i 2 \sigma})~.
\end{equation}
We shall study the dynamical origin and the dynamical implications of this term in Section \ref{sec:bions}. To summarize, our effective Lagrangian to order $e^{-2 S_0}$ is now the sum of (\ref{adjoint1}), (\ref{eftunitcharge}) and (\ref{eftunitcharge2})
\begin{eqnarray}\label{adjoint2}
L_{kin}^{QCD(adj)} &=&  {1 \over 2} { g_4^2 \over L (4 \pi)^2}   (\partial_\lambda \sigma)^2+ i{ L \over g_4^2} \;  \bar\lambda^{\; I}_{\dot\alpha} \; \bar\sigma^{\mu \dot\alpha \alpha} (\partial_\mu \lambda_\alpha)^{ \; I} ~, \nonumber \\
&&  + \;{2 e^{-S_0}   L^{3 n_f} \over L^3}\; \cos \sigma\; \left[(\lambda \cdot \lambda)^{n_f} + (\bar\lambda \cdot \bar\lambda)^{n_f} \right] +{2 e^{-2 S_0} \over L^{3}}\;(1-  \cos 2 \sigma)~,
\end{eqnarray}
where we omitted the Cartan isospin index of the fermions.
In the next Section, we shall study the microscopic origin of the $e^{-2S_0}$ term above as well as the symmetry realization and implications for the IR physics of QCD(adj).

\hfill\begin{minipage}{0.85\linewidth}

\textcolor{red}{
{\flushleft{\bf Summary of \ref{sec:symmadj}:}} Here, we studied the lowest dimensional terms allowed in the QCD(adj) EFT by the $Z_{4n_f}$ chiral and the $Z_2^{(1)}$ center symmetries. We found that the symmetries are very restrictive: combined  with dimensional analysis, they imply that to leading order in the semiclassical expansion only multi-fermion-like terms (\ref{eftunitcharge}) are allowed (due to unit magnetic charge monopole instantons). A potential term (\ref{eftunitcharge2}) can appear only at the next-to-leading order of the semiclassical expansion and is due to charge-$2$ magnetic-bion-like configurations. Much of the insight found here applies to $n_f=1$, but the absence of the $A_4^3$-field (which remains massless for $n_f=1$) means that some our considerations, notably the final result for the IR EFT (\ref{adjoint2}), will require substantial modification for the case of SYM. }

\end{minipage}

\bigskip

 \subsection{The infrared dynamics of QCD(adj) on the circle.}
 \label{sec:bions}

\subsubsection{The composite nature of magnetic bions and a picture of the QCD(adj) vacuum.}
\label{sec:bionstructure}

In this subsection, we shall 
 elucidate the nature of the objects generating the magnetic bion terms in (\ref{adjoint2}). This can be done in more than one way. We shall follow the original EFT approach of \cite{Unsal:2007jx}, since it suffices for a qualitative understanding. Alternatively, one can also proceed \cite{Anber:2011de} using the instanton calculus developed in QCD \cite{Schafer:1996wv}, but this is unnecessarily technical and long winded (this is useful when one wants to compute further corrections, including determinants, the effect of the $A_4$-mass, etc.). 

To begin, recall that as we already discussed, the $M$ and $KK$ terms, apart from the $e^{\pm i \sigma}$ factors taking into account the Coulomb-like interactions of the monopole instantons, are morally similar to 't Hooft's original calculation of the $B+L$ violating vertex in the standard model. The length scale relevant to the generation of the $M$ and $KK$  terms is of order $L/g_4$, the size of the $M$ and $KK$ classical solutions (this is the size due to the $A_4$-cloud around the monopole instantons' core of size $L$). As it turns out, and as we shall see shortly, the magnetic bions are generated at larger length scales. Thus, we can use the $|Q_M|=1$ terms in (\ref{adjoint2}) to construct the $|Q_M|=2$ magnetic bion terms (generated at the larger length scales $L/g_4^2$), a posteriori justifying our scale-by-scale approach.

To begin, as we did when we discussed the Polyakov model, we can expand the contribution of the $M$ terms (as well as $M^*$, $KK$, and $KK^*$)  to the Euclidean path integral, e.g. $e^{- \int d^3 x e^{-S_0} L^{3n_f-3} e^{i \sigma} (\lambda\lambda)^{n_f}}$, into a Taylor series. This gives rise to a grand-canonical gas of $M$, $KK$, $M^*$ and $KK^*$ Euclidean objects, each coming with its own 't Hooft vertex which now contains fermionic insertions. The insertions of $ (\lambda\lambda)^{n_f}$ and $e^{i \sigma}$ in the 't Hooft vertices stand to indicate that these local objects in $R^3$ interact by exchange of dual photons---as in the Polyakov model---and by exchanging fermions. The latter interaction is a new feature in QCD(adj) and is ultimately responsible for the appearance of magnetic bions. 

Consider the effect of an $M$ fluctuation at $x$ (with 't Hooft vertex (\ref{Mchiral})), and $KK^*$ one at $y$ (with 't Hooft vertex (\ref{KKchiralbar})), represented by the following two-point function contribution\footnote{We shall neglect the spinor, flavour, etc., index contractions of the spinors in the 't Hooft vertices, hence the schematic notation in (\ref{mkk}).  It will become clear shortly that these details  (which can be found \cite{Anber:2011de}) only affect the pre-factor of the $M$-$KK^*$ amplitude. } to the Euclidean path integral
\begin{eqnarray} \label{mkk}
\langle M\text{-}KK^* \rangle =\int d^3 x \int d^3 y  e^{-2 S_0} L^{6n_f-6} \big\langle e^{i \sigma(x)} \;  \lambda(x)^{2 n_f} \;
e^{i \sigma(y)}  \; 
\bar\lambda(y)^{2n_f} \big\rangle~.
\end{eqnarray}
The correlator is computed using the free dual photon $+$ fermion action (\ref{adjoint1}). 
According to our interpretation above, this correlator informs us about the long-distance interaction between $M$ and $KK^*$.  In eqn.~(\ref{interaction1}) we already showed that  
\begin{equation}\label{sigmacorrelator}
\big\langle e^{i \sigma(x)}  
 e^{i \sigma(y)}\rangle = e^{- {4 \pi \over g_4^2}{L \over |x-y|}},
 \end{equation} giving rise to a repulsive ``force''---meaning, as the r.h.s. indicates, that the probability for $|x-y| \sim L$ is much smaller than the one for $|x-y| \rightarrow \infty$.

 What remains to be seen is the effect of the $\lambda(x)\bar\lambda(y)$ correlator. The free fermion action in (\ref{adjoint1}) implies that 
\begin{equation}\label{lambdacorrelator}
\big\langle \lambda(x)  
\bar\lambda(y) \rangle = {g_4^2 \over L}\;{f(\hat{n}) \over |x - y|^2} = {g_4^2 \over L^3}\; f(\hat{n})\; e^{ -2 \log {|x-y|\over L}}~.
\end{equation}
Notice that this is just the inverse of the operator appearing in the fermion kinetic term.\footnote{The spinor fields have dimension $3/2$ and $R^3$-propagator in momentum space  $\sim {\sigma^\mu k_\mu \over k^2}$. The $x$-space form shown in (\ref{lambdacorrelator}) is just the Fourier transform, with spinor indices hidden in the dimensionless $f(\hat{n})$.} We used $\hat{n}$ to denote the angular components of a unit vector from $x$ to $y$ and we hid all index dependence in $f(\hat{n})$.

 The r.h.s. above reveals the most important role of the fermions: as the $M$ and $KK^*$ separation $|x-y|$ approaches $L$ (its lowest limit), the logarithm and the exponent are of order unity, while as the separation approaches infinity, the exponent tends to zero. Thus, the most important conclusion from (\ref{lambdacorrelator}) is that the fermions attached to the $M$ and $KK^*$ generate a long-range attractive ``force'' between the objects that experience magnetic repulsion. Thus the possibility of  ``bound states,'' or more appropriately, correlated tunneling events, localized in $R^3$, leading to a change of magnetic charge by $2$ units.

\begin{figure}[h]
\centerline{
\includegraphics[width=9.5 cm]{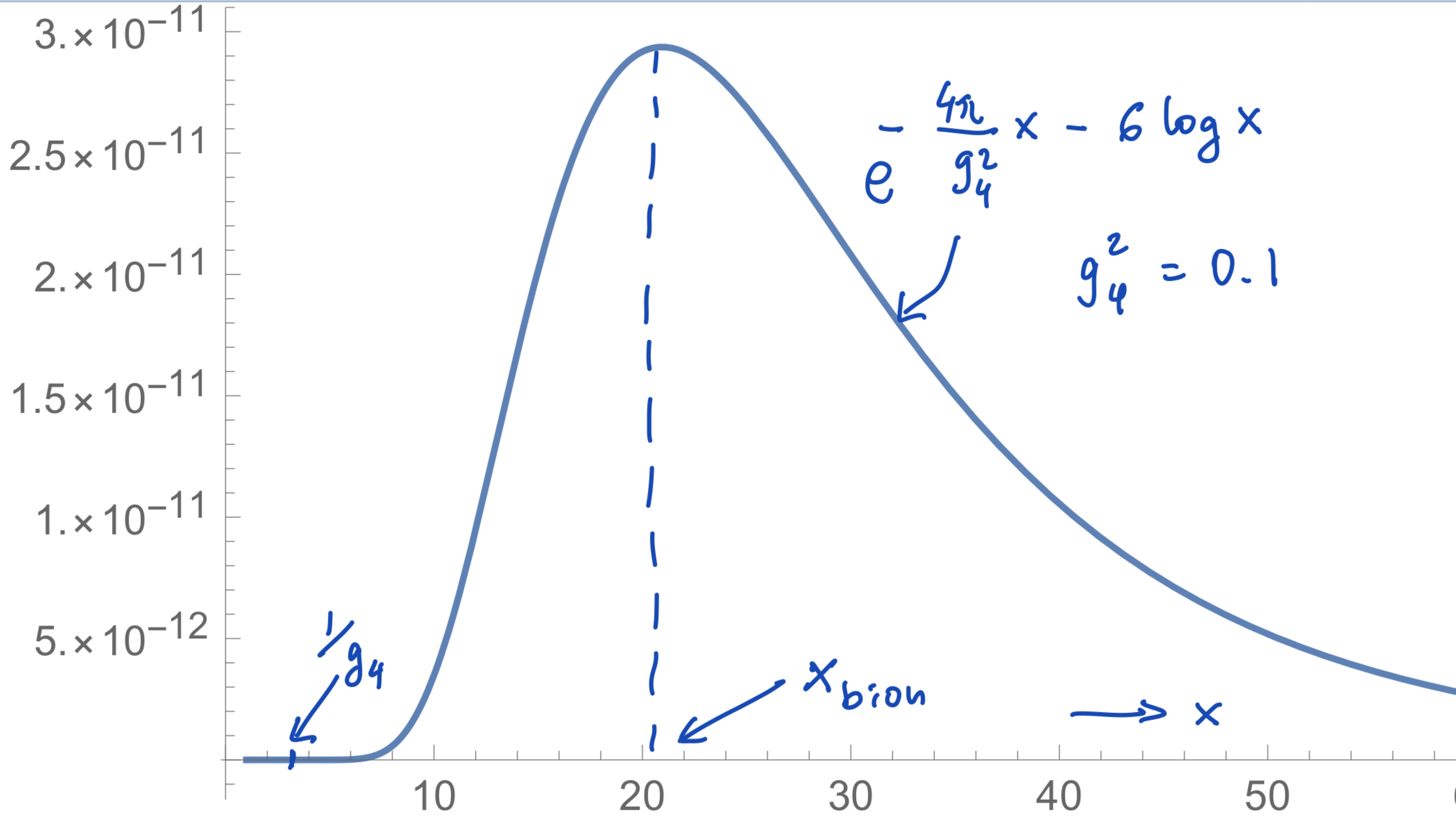}}
\caption{ The separation quasizero mode integrand in (\ref{mkk2}), plotted for $g_4^2 = 0.1$ and $n_f = 2$ QCD(adj) in units of $L=1$. The bion size, $x_{bion}$, is  $\sim 1/g_4$ times larger than the size of the constituent $M$ and $KK^*$, indicated by $1/g_4$ on the figure. \label{fig:bionsize}}
\end{figure}

To see this in more detail, 
let us now use (\ref{sigmacorrelator}), (\ref{lambdacorrelator}) to rewrite (\ref{mkk}). On the way, we introduce the ``center-of-mass'' coordinate $z = {x+y \over 2}$ (the integrand is independent thereof) and the relative separation coordinate, often called ``separation quasizero mode,'' whose length we denote $r = |x-y|$. We do not show the result of integration over the angular part of the relative separation coordinate of the function $f(\hat{n})^{2n_f}$, as it only contributes to the prefactor (the overall sign is positive) and obtain  
\begin{eqnarray} \label{mkk2}
\langle M\text{-}KK^* \rangle &=&\int d^3 x \int d^3 y\;  e^{-2 S_0} L^{6n_f-6} \left({g_4^2 \over L^3}\right)^{2 n_f} f(\hat{n})^{2 n_f}\; e^{\; - {4 \pi \over g_4^2}{L \over |x-y|} - 4 n_f \log {|x-y|\over L}}~\nonumber \\
&\sim&  {e^{-2 S_0} \over L^{3}} \int d^3 z \int\limits_{0}^{\infty} {dr \over L} e^{\;- {4 \pi \over g_4^2}{L \over r} - (4 n_f - 2) \log {r \over L}}= {e^{-2 S_0} \over L^{3}} \;  I_{m.b.}(n_f, g_4^2) \int d^3 z~,
\end{eqnarray}
where we implicitly defined the finite integral $I_{m.b.}(n_f,g_4^2)$.\footnote{This integral is finite, positive, and can be easily computed (we return to a more interesting version thereof  in the ``neutral bion'' Section \ref{sec:neutralbions}). We extended the lower limit of integration over $r$ to $r=0$, since the integrand is essentially zero there, see Figure \ref{fig:bionsize}.}
Clearly the Coulomb repulsion and the fermion-induced attraction balance each other, leading to a stable value,
\begin{equation}\label{rbion}
r_{bion} = { \pi \over g_4^2} {L \over  n_f -{1/2}} \gg {L \over g_4} \gg L~,
\end{equation}
called the ``bion size''. Notice that, as promised, at small $g_4^2$, $r_{bion} \gg L/g_4$, the Compton wavelength of the $A_4^3$ field. The integral over $r$ can thus be taken, computing $I(L,n_f,g_4^2)$. It is  saturated near $r_{bion}$, in the region of validity of the EFT (see Figure \ref{fig:bionsize}). 

\begin{figure}[h]
\centerline{
\includegraphics[width=8.5 cm]{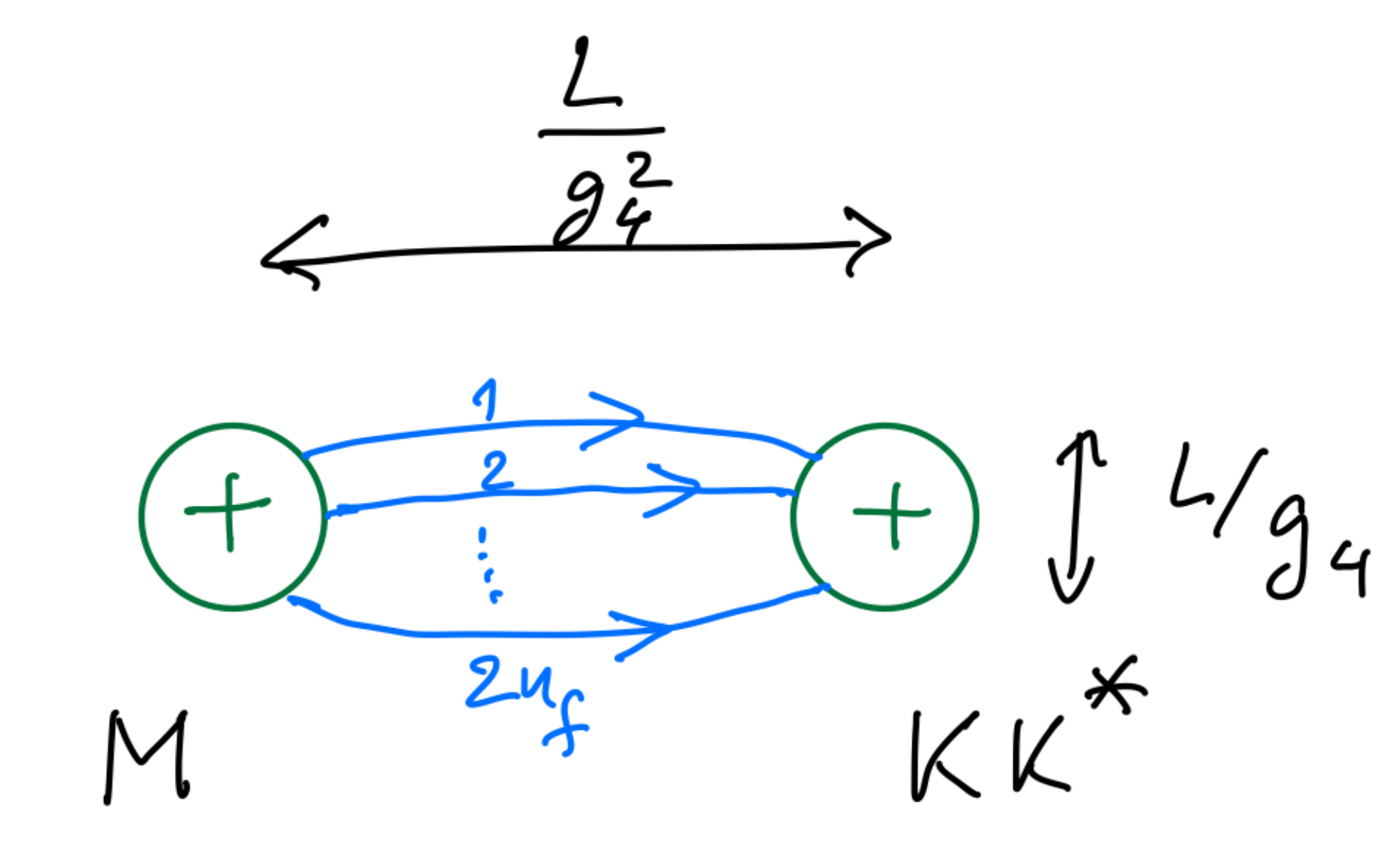}}
\caption{ The correlated tunnelling events known as ``magnetic bions'' $B$: the $M$ and $KK^*$ constituents have equal magnetic charge and  thus repel at long distances. However, the hopping of the  $2n_f$ fermions associated to each of the constituents induce attraction, leading to a stable ``molecule'' of size $r_{bion} \sim L/g_4^2$. The $B$ (and $B^*$) have magnetic charge $2$ but have no topological charge and hence no fermion zero modes. Their proliferation in the vacuum is responsible for the generation of mass gap, see Figure \ref{fig:magneticbiongas}. \label{fig:magnbionsize}}
\end{figure}

We have thus found that a new kind of  ``topological molecule,'' a magnetic bion of charge-2 exists, owing to the compensation between Coulomb repulsion and fermion-hopping-induced attraction. Its contribution to the Euclidean partition function is represented by a 't Hooft vertex,
obtained from (\ref{mkk2}) by dropping the $d^3 z$ integral, ignoring the non-exponential $g_4^2$ factors arising after the integration over $r$, and attaching an $e^{i 2 \sigma(z)}$ factor to signify that the object has charge-$2$ (to see this factor, consider the correlations between two magnetic bions at distances  $\gg r_{bion}$):
\begin{equation}
\label{bion}
B = M\text{-}KK^*\; {\text{at}} \; z:\; {e^{-2S_0} \over L^3} \; e^{i 2 \sigma(z)}~.
\end{equation}
As usual, the magnetic bion 't Hooft vertex above is given up to a positive $g_4^2$-dependent constant. Notice that this 't Hooft vertex has the form of the charge-$2$ term that we argued for using symmetries. What we gained is an understanding of its origin as due to correlated $M$-$KK^*$ tunnelling events: a new topological molecule of size $r_{bion}$, action $\sim 2S_0$, magnetic charge-$2$, and topological charge zero---hence without any fermion zero modes. The absence of zero modes (related to $Q_T=0$) means that bions can generate a potential for the bosonic dual photon.
Clearly, there are also $B^*$ (= $M^*$-$KK$) molecules with 't Hooft vertices given by the complex conjugate of (\ref{bion}). We also note that, as in the discussion of Section \ref{sec:polyakov_pass3} the fugacity of the magnetic bions is positive (apart from the $e^{i 2 \sigma}$ factor responsible for incorporating interactions), thus these are ``real saddles'' according to Footnote \ref{footnotenegative}.

 Exponentiating the $B$ and $B^*$ contributions (\ref{bion}), as in the Polyakov model (recalling Section \ref{sec:polyakov_pass3}), we obtain the bion contribution to the Euclidean path integral 
\begin{equation}\label{bionpotential1}
e^{- S_{bion}} \equiv e^{- \int d^3 x V_{bion}} = e^{ - \int d^3 x \left( - {e^{-2S_0} \over L^3} \; 2 \cos 2 \sigma(x)\right)}~,
\end{equation}
showing that the charge-$2$ bions generate a $- \cos 2 \sigma$ potential.

\begin{figure}[h]
\centerline{
\includegraphics[width=9.5 cm]{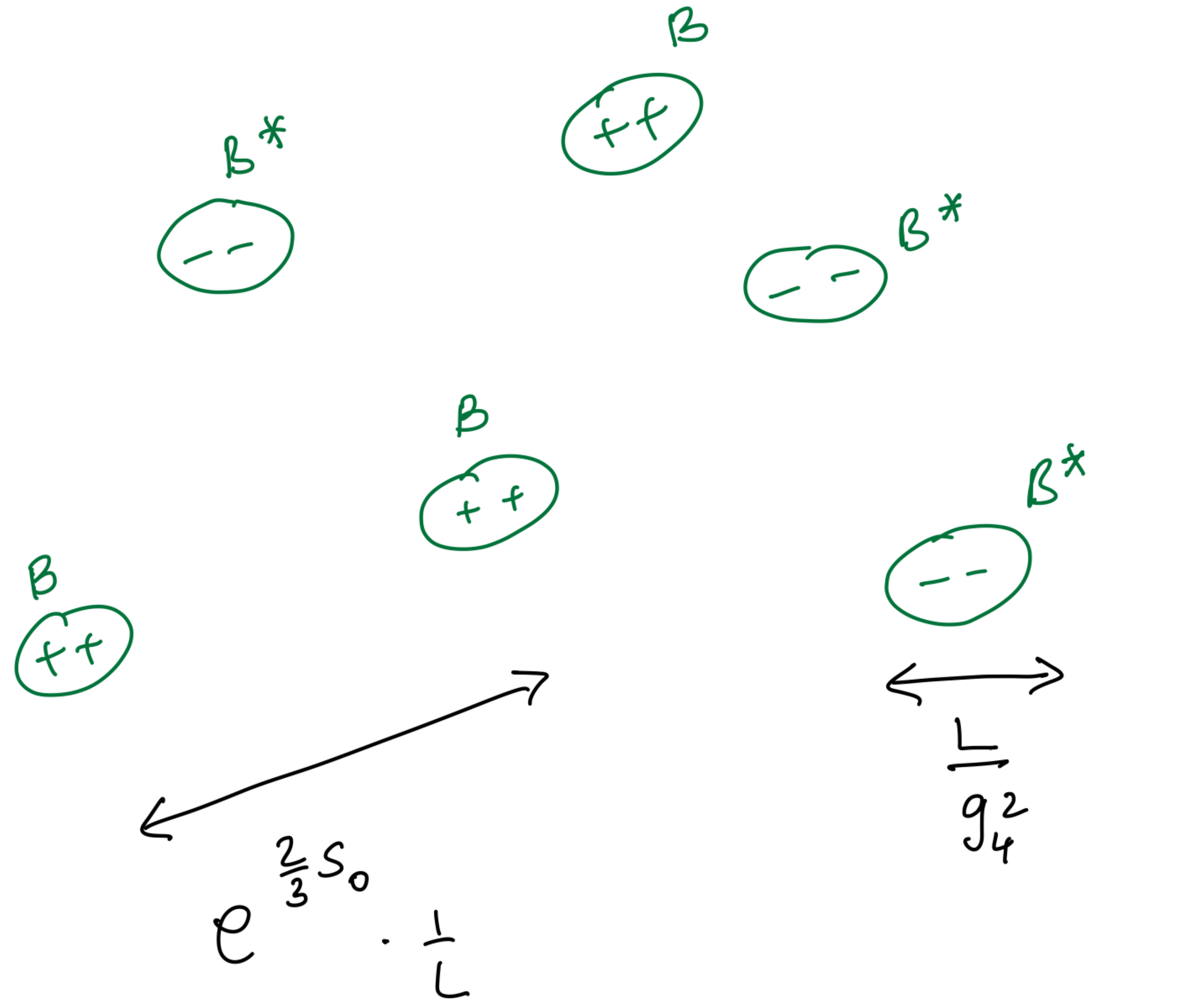}}
\caption{ A typical configuration contributing to the vacuum to vacuum amplitude in  QCD(adj): a dilute grand-canonical gas of bions $B$ and anti-bions $B^*$ and the hierarchy of characteristic scales involved. Compare with Fig.~\ref{fig:MMbargas}  for the Polyakov model. \label{fig:magneticbiongas}}
\end{figure}

Let us also address the question of other possible ``di-atomic'' molecules that one can contemplate. An $M$-$M$ object    would experience magnetic repulsion between its constituents, and, most importantly, no fermion-induced long-distance interaction at leading order, as the propagator $\langle \lambda \lambda\rangle$ vanishes (due to the chiral symmetry of the massless theory). Further, this object would carry twice the topological charge of $M$ and therefore have twice as many fermion zero modes (from the index theorem, or, equivalently, the anomaly equation). Thus, it could not be responsible to generate a bosonic potential. The same comment applies to a $KK$-$KK$ made object. Another possibility is an $M$-$KK$ (and c.c.) molecule---but this is our familiar BPST instanton with $4n_f$ zero modes, no magnetic charge---and no $\sigma$ dependence. Finally, we could also consider the possibility of $M$-$M^*$ and $KK$-$KK^*$ composites, but for the purposes of generating a potential for $\sigma$, these can not be relevant, as they have no magnetic charge (these so-called ``neutral bions'' will, however, play important role in SYM).
We  conclude that, to the order of our calculation, the $M$-$KK^*$ (and c.c.) object, leading to the magnetic bion $B$ and $B^*$ molecules, is the only one responsible for a dual photon potential.

The picture of the vacuum of QCD(adj) is thus similar to the one in the Polyakov model, albeit the dilute gas is one of charge-$2$ objects, the magnetic-bion monopole-instantons (the charge-$2$ nature of the confining objects is not an ``innocent'' small change, as it drastically affects the nature of confining strings, as we already saw in $\theta=\pi$ dYM).  A cartoon is shown on Figure \ref{fig:magneticbiongas}, where we also indicate the exponential scale hierarchy between the bion size and the typical distance between bions.
 
  \subsubsection{Symmetry realization, domain walls, and confinement.}
  \label{sec:chiralqcdadj}
  
  Let us now  study the physics of QCD(adj) with $n_f >1$ on $R^3 \times S^1$. Our summary of the IR phase will be rather quick, as we already did all the preparatory work, having explained the origin of all terms in (\ref{adjoint2}). We also studied the confining string in a theory with a $\cos 2\sigma$ potential, recall Section \ref{sec:dymthetapi} (and the $\theta=\pi$ dYM potential (\ref{dYMpi1})), and we shall use this insight here. Thus, we shall now simply enumerate the symmetry realization of QCD(adj) at small $L$:
  \begin{enumerate}
\item The bion-induced potential for the dual photon  in (\ref{adjoint2}) has two minima, at $\langle\sigma\rangle = 0$ and $\langle\sigma\rangle = \pi$. The $Z_{4n_f}$ discrete chiral symmetry (\ref{z4nf}),
 $\sigma \rightarrow \sigma + \pi$, is thus broken to $Z_{2 n_f}$ by the expectation value of the dual photon. The $Z_2^{(1)}$ center symmetry is preserved in both vacua (at $\langle\sigma\rangle = \pi$, due to the $2\pi$ periodicity of $\sigma$).
 Notice that the order parameter for the broken chiral symmetry is  the expectation value of a monopole operator, $\langle e^{i \sigma} \rangle=\pm 1$
 
  There are are domain walls (lines) between the  $\langle e^{i \sigma} \rangle= 1$ and $\langle e^{i \sigma} \rangle = -1$ vacua. There are two different domain walls between the $e^{i\sigma} = \pm 1$ vacua, as in $\theta=\pi$ dYM, related by the action of $Z_2^{(1)}$, as in Figure \ref{fig:dwcenter}.
Likewise, each domain wall carries electric flux appropriate to confining ``half'' a fundamental quark.

 \item The dual photon acquires mass (below, we use (\ref{lambdascale}) to rewrite $e^{-S_0} = e^{ - {4 \pi^2 \over g_4^2(1/L)}}$ in terms of $\Lambda$)
 \begin{equation}\label{massgapadj}
 m_\sigma \sim {e^{- S_0} \over L} = {1 \over L} \left({\Lambda L} \right)^{11- 2 n_f \over 3} = \Lambda \; (\Lambda L)^{8 - 2 n_f \over 3},
 \end{equation} but the Cartan subalgebra fermions $\lambda^I$   remain massless ($I=1,...,n_f>1$).\footnote{
  For those familiar with 't Hooft anomaly matching of traditional or $0$-form symmetries, one can ask how are the various 't Hooft anomalies involving the unbroken $SU(n_f)$ chiral symmetry matched by our small-$L$ theory?  The answer is that they are matched by a combination of contributions from the massless $\lambda^I$ fermions and  global-symmetry background-field-dependent terms, local in $R^3$, induced upon integrating the non-Cartan and Kaluza-Klein components of all fermions, in a  slight generalization of the discussion in \cite{Poppitz:2020tto}. We shall not discuss this in detail here, nor shall we dwell on the matching of the generalized 't Hooft anomalies involving the center symmetry and  the discrete chiral symmetry.}
 
 At this point, notice the interesting fact \cite{Poppitz:2009uq,Anber:2011de} that in the approximation we are using, $m_\sigma$ is an increasing function of $L$ (at fixed $\Lambda$) for $n_f < 4$ and a decreasing function of $L$ for $n_f =5$. For more discussion of the large-$L$, fixed-$\Lambda$ limit, see the end of this Section.
 
 \item The continuous $SU(n_f)$ chiral symmetry is unbroken and the massless $\lambda^I$ transform in the fundamental representation.

\item  Confinement of fundamental quarks proceeds as in dYM at $\theta=\pi$, via the formation of a double string composed of two domain walls. Thus, center symmetry in the $R^3$ directions is also unbroken. 
Furthermore, as in dYM at $\theta=\pi$, quarks are deconfined on the domain walls (and center symmetry is broken there).\footnote{Just as in dYM this has to do with matching the mixed parity/center generalized 't Hooft anomaly \cite{Gaiotto:2017yup}, here it has to do with the mixed discrete chiral/center anomaly, e.g. \cite{Komargodski:2017smk,Cox:2021vsa,Cox:2019aji}.} The qualitative pictures of confining strings shown on Figures \ref{fig:doublestring1} and \ref{fig:decwall1} apply verbatim.
  \item 
  Expanding the multifermion terms in (\ref{adjoint2})  around the vev for $\sigma$, we find that irrelevant (at weak coupling, as in our small-$L$ theory) interactions between the fermions (and the fluctuations of $\sigma$) are induced by the $M$ and $KK$ terms. \end{enumerate}
  Overall,  the main difference of the QCD(adj) small-$L$ phase compared to dYM at $\theta=\pi$ is the presence of the massless $\lambda^I$ fermions transforming as a fundamental under the unbroken $SU(n_f)$ chiral symmetry. These fermions affect the interactions between the domain walls in the double-string, see \cite{Anber:2015kea}.
 
   \subsubsection{Possible (new)  large-$L$ phases.}
  \label{sec:chiralqcdadjphasesj}
 
 Finally, what about the relation between the small-$L$ calculable regime of QCD(adj) and the large-$L$ $R^4$ regime? Our discussion of this topic  will be qualitative and even more hand-waving than usual. 
 At the outset, let us state that not much is known with certainty about the phases of the $n_f = 2,3,4,5$ theories on $R^4$.  
 
 We begin with $n_f = 5$ (and perhaps $n_f=4$). It is thought that these theories  flow to an interacting fixed point in the IR. An indication for this can be seen by studying the two-loop beta function (not shown in (\ref{lambdascale}), but easily found in the literature,  e.g. the appendix of \cite{Poppitz:2009uq}). All global symmetries, including the discrete chiral $Z_{4n_f}$ remain unbroken and their 't Hooft anomalies are assumed to be matched by the IR fixed-point theory (a CFT). This scenario is not necessarily in conflict with our small-$L$ analysis, especially for $n_f=5$, as discussed in \cite{Poppitz:2009uq} (see also the fine print disclaimer there). The idea is that the fixed-point coupling $g_4^2$ for $n_f=5$ is small (in fact, to two-loop order $g_4^2/(4 \pi) \simeq 0.13 \ll 1$ at the fixed point) thus we shall assume that the physics of this theory can be analyzed using weak coupling means at any scale, as the coupling never becomes large in the IR. Then, beginning with $R^4$, let us compactify on $R^3 \times S^1$ with now arbitrary $L$. Since the theory never reaches strong coupling, our weak coupling analysis does not require small $L$. Abelianization occurs  with  the $W$-boson mass $\sim 1/L$ at any $L$, and the conclusion that there are two vacua with broken $Z_{4n_f} \rightarrow Z_{2 n_f}$ holds for any $L$, as does the fact that $m_\sigma \sim e^{- S_0}/L$ ($S_0$ is now evaluated at the small fixed-point coupling). Thus,  in the limit $L\rightarrow \infty$, the mass gap vanishes, the two vacua merge, and the theory becomes gapless consistent with an IR CFT scenario.
 
 For lower numbers of flavours,  the most conventional ``vanilla'' possibility for the $n_f=2,3$ theories (and maybe $n_f=4$)  is that they break the continuous chiral symmetry, $SU(n_f) \rightarrow SO(n_f)$. This is the QCD(adj) ``cousin'' of what happens in QCD with fundamental massless quarks, QCD(F) (of course the symmetry breaking pattern is different in QCD(F)). In this scenario, the $SU(n_f)$ chiral symmetry is broken by the formation of a fermion-bilinear condensate  \begin{equation}\label{bilinearcondensate}
 \langle \lambda^{I}_\alpha \lambda^{J}_\beta \epsilon^{\beta \alpha} \rangle \sim \delta^{IJ} \Lambda^3.
 \end{equation} 
 This is in marked contrast with our small-$L$ analysis, which   showed that for any $n_f$, the continuous chiral symmetry $SU(n_f)$ remains unbroken. The fermion bilinear (\ref{bilinearcondensate}) also breaks the discrete chiral symmetry to $Z_2$ fermion number, as has been studied in detail in \cite{Cordova:2018acb}. If that is the correct $R^4$ phase, then there would have to be a phase transition associated with the spontaneous breaking of $SU(n_f)$ upon increase of $L$. Since, upon increase of $L$ the $R^3 \times S^1$ theory enters a strongly-coupled regime (as the condition $\Lambda L \ll \pi$ is violated, recall Figure \ref{fig:runningcoupling}), a possible scenario is that the multi-fermion terms due to $M$ and $KK$ in (\ref{adjoint2}) become relevant  and, as in the Nambu-Jona-Lasinio model (see \cite{Nambu:1961tp,Nambu:1961fr} for the original papers and \cite{Hill:2002ap} for a modern review), lead to spontaneous breaking of $SU(n_f)$. We also note that the calculable small-$L$ limit offers a dynamical reason for the appearance of the multifermion terms, connecting to the ``instanton-liquid'' models of the vacuum in QCD(F) \cite{Schafer:1996wv}.

However, it was recently realized \cite{Anber:2018iof,Poppitz:2019fnp} that there are possible  ``unconventional'' phases for the $n_f=2,3$ theories on $R^4$.  We saw that at small $L$ the $Z_{4n_f}$ discrete chiral symmetry was broken by the expectation value of the $\sigma$ field, or equivalently, by $\langle e^{ i\sigma}\rangle = \pm 1$. As we increase $L$, the dual photon field ceases to make sense, as there is no meaningful scale separation between Cartan and non-Cartan gauge bosons and thus no abelianization. However, we argued above that $e^{i \sigma}$ has the same transformation properties as the multifermion $SU(n_f)$-invariant, but $Z_{4n_f}$-charged, multi-fermion operator ${\text det}_{IJ} (\lambda^{a \;I}_{\alpha} \lambda^{a \;J}_{\beta} \epsilon^{\beta\alpha})$. Thus, while on $R^4$ an $e^{i \sigma}$ condensate does not make sense, the multi-fermion object is sensible and it is possible that there is a phase with a nonzero multifermion condensate but vanishing bilinear condensate
 \begin{equation}\label{multicondensate}
\langle {\text det}_{IJ} (\lambda^{a \;I}_{\alpha} \lambda^{a \;J}_\beta \epsilon^{\beta\alpha})\rangle \sim \Lambda^{3 n_f} ~,~ \text{while} ~~\langle \lambda^{I}_\alpha \lambda^{J}_\beta \epsilon^{\beta \alpha} \rangle = 0~.
 \end{equation}
 Thus, in such a phase, $SU(n_f)$ would be unbroken and the symmetry realizations at small $L$ and large-$L$ would be continuously connected, much like it was argued to be the case in dYM. 
 
It should be stressed that, at the moment, we do not have a rigorous argument favouring (\ref{bilinearcondensate}) or (\ref{multicondensate}). For recent discussions showing the consistency of this scenario with various anomalies, see \cite{Cordova:2018acb,Cordova:2019bsd,Cordova:2019jqi,Anber:2019nze,Anber:2020gig}, while 
 \cite{Athenodorou:2021wom} has the most recent lattice results, unfortunately not yet decisive.
 Our main point is that the suggested phase with condensates (\ref{multicondensate}) is one example where a new possibility for a $R^4$ phase was suggested and motivated by an analytically solvable $R^3 \times S^1$ example. 
 
 \hfill\begin{minipage}{0.85\linewidth}

 \textcolor{red}{
 {\flushleft{\bf Summary of \ref{sec:bions}:}} Here, we studied the dynamical aspects of QCD(adj) on $R^3 \times S^1$. We began by studying the composite nature of the magnetic bions, the topological excitations composed of $M$ and $KK^*$ monopole-instantons.  These novel topological molecules have magnetic charge $2$ and carry no topological charge and hence no fermion zero modes.  Their proliferation in the vacuum leads to a potential for the dual photon, generates the mass gap for gauge excitations, and is the cause for confinement in this theory---in marked contrast with the charge-$1$ $M$ monopole-instantons of the Polyakov model. The proliferation of the magnetic bions also leads to a spontaneous breakdown of the discrete chiral symmetry $Z_{4n_f} \rightarrow Z_{2n_f}$, but leaves $SU(n_f)$ unbroken.  Similar to   dYM at $\theta=\pi$  (here, instead, reflecting the mixed $Z_{4n_f}$/$1$-form center 't Hooft anomaly), confinement of fundamental charges proceeds via the double string mechanism of Fig.~\ref{fig:doublestring1} and quarks are deconfined on domain walls, as on Fig.~\ref{fig:decwall1}. We also discussed two possible scenarios for the behaviour of the $n_f = 2,3,...$ theory in the $R^4$ limit, one with a $SU(n_f)$-breaking phase transition upon increase of $L$, with a nonvanishing bilinear condensate (\ref{bilinearcondensate}), and the other obeying ``adiabatic continuity,'' with the multifermion condensate of (\ref{multicondensate}) and an unbroken $SU(n_f)$ on $R^4$.}
 
 \end{minipage}
 
 \bigskip

  \section{SYM.}
 \label{sec:softthermal}
 
 We now go on to study the $SU(2)$ QCD(adj) with $n_f=1$, whose 4d Lagrangian is  (\ref{lagrangianadjoint}) with $m=0$. This is also known as SYM (super-Yang-Mills). In fact,  ${\cal{N}}=1$ 4d supersymmetry is an accidental symmetry of this theory (``accidental'' means that supersymmetry automatically emerges once the action is made chirally symmetric, i.e. once the $Z_4 \rightarrow Z_2$ chiral-symmetry-breaking Majorana fermion mass $m$  in (\ref{lagrangianadjoint}) is set to zero; $m$  is the only relevant coupling that breaks supersymmetry). 
 
 The reason SYM is of great interest is that, owing to the ``power of holomorphy'' constraints of supersymmetry, many (but not all!) aspects  of its dynamics can be understood also at strong coupling, as already mentioned in the Introduction. The discussion of the supersymmetric formalism will take us far afield here and we shall not use it (see Shifman's textbook \cite{Shifman:2012zz} for a pedagogical introduction). For us, of utmost importance is the fact that, when compactified on $R^3 \times S^1$ with periodic boundary conditions for $\lambda$, the small-$L$ SYM also becomes weakly-coupled and semiclassically calculable (provided the $S^1$ holonomy is near the middle of the Weyl chamber of Fig.~\ref{fig:weylchamber}, as per our discussion in Section \ref{sec:perturbative3x1}).

On one hand, weak coupling at small $L$ will allow us to study properties not accessible via the holomorphy tools, such as the nature of the confining string. On the other hand, 
we can confront the exact results obtained using holomorphy (which remain valid  at small $L$, as our $S^1$ boundary conditions do not break supersymmetry) with the non-supersymmetric semiclassical path integral tools  that we have been using throughout. In this process many interesting and sometimes puzzling observations will be seen to arise. 
The details will be discussed below, but the overarching conclusion worth stating at the outset is that they point to the need of deeper understanding of path integrals and to a related interesting interplay of perturbative and nonperturbative properties of QFT. In particular, they indicate that an analytic continuation of the path integral is necessary to properly account for nonperturbative effects. Needless to say, the travel along this road is not completed yet!

\subsection{Fields, symmetries,  't Hooft vertices, and the allowed terms in the EFT.}
\label{sec:symfields}
    
On $R^3 \times S^1$ with periodic boundary conditions for $\lambda$, the major difference from $n_f>1$ is the vanishing of the perturbative GPY potential, as illustrated on Figure \ref{fig:gpy1}. This vanishing holds to all orders of perturbation theory and is due to supersymmetry nonrenormalization theorems. It implies that the $a_4$-field in (\ref{3dlagrangiandual}), the fluctuation of $A_4^3$ around the center-symmetric vev (\ref{a4vev}) remains perturbatively massless and should be included in the $\mu \ll 1/L$ EFT.
We shall now define a dimensionless field $\phi$ as follows
\begin{equation}
\label{phi}
\phi = {4 \pi \over g_4^2} \; L a_4 = {4 \pi \over g_4^2}  \;  L (A_4^3 - {\pi \over L})~,~~ {\text{such that}} ~~ \langle \phi \rangle = 0 \leftrightarrow \langle \text{tr} \; \Omega \rangle = 0,
\end{equation}
i.e., $\phi$ measures the deviation of the holonomy vev from the center symmetric point (the reason for the  overall normalization factor will become clear below). The kinetic term of the $\mu \ll 1/L$ EFT is given by (\ref{adjoint1}), with the kinetic term for the $a_4$ from (\ref{3dlagrangiandualdYM1}) added and rewritten via $\phi$:
 \begin{eqnarray}
  \label{3dlagrangiandualdSYM1}
L_{3d, SYM} =  {1 \over 2} { g_4^2 \over L (4 \pi)^2}   (\partial_\lambda \sigma)^2  + {1 \over 2} { g_4^2 \over L (4 \pi)^2}   (\partial_\lambda \phi)^2 + i{ L \over g_4^2} \; \bar\lambda_{\dot\alpha} \; \bar\sigma^{\mu \dot\alpha \alpha} (\partial_\mu \lambda_\alpha)~.
 \end{eqnarray}
As there is only one adjoint Weyl fermion we dropped the flavour index $I$ (as well as the Cartan subalgebra index). 
The global $Z_4$ chiral (as per Section \ref{sec:discretechiral}) and $Z_2^{(1)}$ symmetries (recall from Section \ref{sec:bpsandkkmonopoles} that center symmetry is a reflection of the Weyl chamber w.r.t. the middle point) act as follows
\begin{eqnarray}\label{symsymmetries}
Z_4:&& e^{i \sigma} \rightarrow - e^{i  \sigma},\nonumber \\
&& \lambda \rightarrow e^{i {\pi \over 2}} \lambda, \nonumber \\
&& \phi \rightarrow \phi,  \\
Z_2^{(1)}:&& e^{i \sigma} \rightarrow e^{ - i \sigma}, \nonumber\\
&& \lambda \rightarrow \lambda,~ \nonumber \\
&&\phi\rightarrow - \phi~.
\end{eqnarray}
Now, we could refer to supersymmetry and just declare that the EFT should be described in terms of  the complex combinations $-\phi + i \sigma$ (or $-\phi - i \sigma$), which are the lowest components of chiral (or antichiral) superfields\footnote{For a   superfield  derivation of the kinetic term (\ref{3dlagrangiandualdSYM1}), using linear-chiral superfield duality, for a general gauge group, and including the nontrivial K\" ahler metric, see \cite{Anber:2014lba}.} which also contain the fermions.  
We shall not need all this technology and shall take a slight shortcut to give an alternative view on the need to consider the combinations $-\phi \pm i\sigma$. 

It goes at follows.
Recall that the $M$ monopole instanton contributes to the Euclidean partition function with a Boltzmann factor $e^{- S_0} = e^{- {4 \pi^2 \over g_4^2}}$ at the center symmetric point. 
We shall reserve the symbol $e^{-S_0}$ to mean precisely the center-symmetric point action. However, in SYM we can study the semiclassical expansion anywhere on the Weyl chamber, which is a ``vacuum moduli space'' to all orders of perturbation theory. Here, an  $M$ fluctuation comes  with  a Boltzmann factor given in (\ref{kk1}). Recalling (\ref{a4vev}), this factor can be rewritten in terms of  (\ref{phi}) as follows:
\begin{equation}\label{Mboltzmann}
M: e^{ - {4 \pi \over g_4^2} A_4^3 L} =e^{ - (\phi + {4 \pi^2 \over g_4^2})}= e^{ - S_0} \;e^{- \phi}~.
\end{equation}
Likewise, a $KK$ fluctuation has a Boltzmann factor, also taken from (\ref{kk1}),
\begin{equation}\label{KKboltzmann}
KK: e^{- {8 \pi^2 \over g_4^2} + {4 \pi \over g_4^2} A_4^3 L} =e^{ - 2S_0} e^{ \phi + {4 \pi^2 \over g_4^2}}= e^{ - S_0} \;e^{+ \phi}~.
\end{equation}
Next we recall that $M$ ($KK$)  come with a factor of $e^{i \sigma}$ ($e^{- i \sigma}$) due to their magnetic charge. In addition, invariance under the $Z_4$ discrete chiral symmetry forces us to include the fermions, as we discussed in Section \ref{sec:symmadj}. Putting everything together, we  conclude that the 't Hooft vertices of $M$ (\ref{Mchiral}) and $KK$ (\ref{KKchiral}), invariant under the chiral $Z_4$ should now read
\begin{eqnarray}\label{susythooftchiral}
M:&&   e^{- S_0} \; e^{i \sigma - \phi}\;   \lambda\cdot \lambda~, \nonumber \\
KK: &&    e^{-S_0} \; e^{- i \sigma + \phi} \;  \lambda\cdot  \lambda~.
\end{eqnarray}
Notice that, as before, $Z_2^{(1)}$ of (\ref{symsymmetries}) exchanges $M$ with $KK$.
In addition, the c.c. vertices are:
\begin{eqnarray}\label{susythooftantichiral}
M^*:&&  e^{- S_0} \; e^{-i \sigma - \phi}\; \bar\lambda\cdot \bar\lambda~, \nonumber \\
KK^*: &&    e^{-S_0} \; e^{ i \sigma + \phi} \; \bar\lambda\cdot \bar\lambda~.
\end{eqnarray}
The only modification compared to the QCD(adj) 
't Hooft vertices with $n_f>1$ of Section \ref{sec:symmadj} is that the ones in SYM include extra $e^{\pm \phi}$ factors. These are important because:
\begin{enumerate}
\item These factors indicate that away from the center symmetric point, the $M$ and $KK$ actions differ from $e^{-S_0}$. These vertices also show that both the $M$ and $KK$ monopole-instantons insertions depend on $-\phi + i\sigma$, but not on the c.c. $-\phi - i\sigma$. Thus, these vertices are holomorphic functions of the chiral superfield mentioned earlier, reflecting the ``power of holomorphy.'' In fact, (\ref{susythooftchiral}) arise from contributions to the superpotential, which is known to be a holomorphic function of the chiral superfields (while the c.c. vertices contribute to the conjugate superpotential).\footnote{The calculation of (\ref{susythooftchiral}) in $SU(2)$ SYM on  $R^3 \times S^1$ was first done in \cite{Davies:1999uw}, with the noncancelling one-loop determinants and the correction to the K\" ahler metric calculated in \cite{Poppitz:2012sw}. }
\item
In addition to reflecting holomorphy, the $e^{\pm \phi}$ factors signify that both $M$ and $KK$ carry ``scalar charge,''  or, more prosaically, that these monopole-instantons interact via long-range exchange of  massless $\phi$ quanta, in addition to the $\sigma$-mediated magnetic interaction. We shall make great use of this in what follows, so let us elaborate. Already in the Polyakov model, and more recently in (\ref{sigmacorrelator}), we used the fact that \begin{equation}
\label{sigmaexchange}
\big\langle e^{i \sigma(x)}  
 e^{\pm i \sigma(y)}\rangle = e^{\mp {4 \pi \over g_4^2}{L \over |x-y|}},\end{equation}
 where the signs on the two sides are correlated, showing that the like-magnetic charge monopole instantons repel, while those of opposite magnetic charge attract.  Similarly, we can now consider the two-point function of $e^{\pm \phi}$.
 \begin{quote}
{\flushleft{\bf Exercise 17:}} Show that the $e^{\pm \phi}$ two-point correlator,  computed using (\ref{3dlagrangiandualdSYM1}), yields
 \begin{equation}\label{phiexchange}
 \big\langle e^{ \phi(x)}  
 e^{\pm\phi(y)}\rangle = e^{\pm {4 \pi \over g_4^2}{L \over |x-y|}}~,
 \end{equation}
 where, again, the signs are correlated, showing that like scalar charges attract and opposite charges repel. 
Convince yourself that this leads to cancellation of ``forces'' due to $\sigma$ and $\phi$ exchanges between $M$ and $KK$ or $M$ monopole-instantons.
\end{quote}
At the technical level, it is the absence of factors of $i$ in the exponents on the l.h.s.  of  (\ref{phiexchange}) that makes for the crucial difference with (\ref{sigmaexchange}). As (\ref{phiexchange}) shows, two objects of the same scalar charge attract (as the probability to find them close to each other is larger than being far away) while those of opposite scalar charge repel, in exact opposite to (\ref{sigmaexchange}). This leads to the conclusion that there is no force between an $M$ and an $M$, a $KK$ and a $KK$, as well as an $M$ and $KK$ monopole-instantons, as the magnetic and scalar exchanges exactly cancel each other. The ultimate reason is that all these objects are all self-dual, or BPS.  However, there are forces between self-dual and anti-self-dual objects, i.e. between $M$ and $KK^*$ (as well as $M$ and $M^*$), which will become important below.\footnote{The  static force between (BPS or not) monopoles is discussed in the book by Manton and Sutcliffe \cite{Manton:2004tk}. For a discussion within our setup, see  Section 2.3 of \cite{Poppitz:2017ivi} and references therein.}
 \end{enumerate}
Coming back to our 't Hooft vertices, we conclude that symmetries imply that the leading $e^{-S_0}$ order EFT of SYM on $R^3\times S^1$ is given by the kinetic term (\ref{3dlagrangiandualdSYM1}) plus the sum of the four 't Hooft vertices in (\ref{susythooftchiral}) and (\ref{susythooftantichiral}), ensuring that the action is $Z_2^{(1)}$ invariant. These interactions   have the form of Yukawa interactions---or field-dependent mass terms---as they are bilinear with respect to the fermions:
\begin{equation}\label{susyyukawa}
L_{Yuk.} =  e^{- S_0} \left( (e^{i \sigma - \phi} +e^{- i \sigma + \phi}) \; \lambda\cdot \lambda +(e^{-i \sigma - \phi} +e^{ i \sigma + \phi}) \; \bar\lambda\cdot \bar\lambda\right)~.
\end{equation}

But what about potential terms, i.e. ones that do not involve fermions? Let us follow our experience with $n_f>1$ QCD(adj) and construct the possible terms by multiplying the Boltzmann factors of the various objects in (\ref{susythooftchiral}, \ref{susythooftantichiral}), ensuring that the resulting action is both $Z_4$ and $Z_2^{(1)}$ invariant. 

We  thus begin   as in QCD(adj), recall Sections \ref{sec:symmadj}, \ref{sec:bionstructure}. 
 We can  multiply the prefactors of $M$ to those of $KK^*$ to obtain $e^{-2 S_0} e^{ 2 i \sigma}
$, while the c.c. $M^*$ times $KK$ gives the term $e^{- 2 S_0} e^{-2 i \sigma}$. These bosonic terms are separately $Z_4$ chirally invariant, but need to be added to each other to respect $Z_2^{(1)}$. Thus, we obtain the  possible bosonic potential term
\begin{equation}
 \label{mb1}
 L_{m.b.} = - L^{-3} e^{-2 S_0} \; 2 \cos 2 \sigma~.
 \end{equation}
This kind of  term is already familiar to us; it is due to the magnetic bions we studied in Section \ref{sec:bionstructure} in the dilute gas approximation, eqn.~(\ref{bionpotential1}).\footnote{Its overall sign is negative, similar to the one found for charge-1 objects in the Polyakov model in Section \ref{sec:polyakov_pass3}. Recall that, as noted in Footnote \ref{footnotenegative} there, a dilute-gas summation of the contributions of objects of positive fugacity (``real saddles'') contributes a term with negative sign to the potential.} Moreover, the structure of the magnetic bions in SYM is similar to the one discussed for QCD(adj): the only difference is that the $M$ and $KK^*$ have mutual repulsion twice as strong as the one appearing in (\ref{mkk2}). This is  due to the extra repulsive contribution due to $\phi$ exchange (as per (\ref{phiexchange}) and because $M$ and $KK^*$ have opposite scalar charge). The fermion-induced attraction is as in (\ref{mkk2}) with $n_f=1$. Otherwise, the calculation of (\ref{mb1}) is identical to the one given in Section \ref{sec:bionstructure} and there is no need to repeat it (the bion size is two times bigger than $r_{bion}$ of (\ref{rbion}) with $n_f=1$).

In SYM, however,  other possible terms, different from (\ref{mb1}), exist. To construct them, we 
 multiply the bosonic prefactors of $M$ and $M^*$,  obtaining a term $e^{-2 S_0} e^{- 2\phi}$, while $KK$ times $KK^*$ yields $e^{-2 S_0} e^{+ 2 \phi}$. The sum of two such terms is invariant under $Z_2^{(1)}$, giving a possible bosonic potential term \begin{equation}
 \label{nb1}
 L_{n.b.} \sim L^{-3} e^{-2 S_0} \; 2 \cosh 2 \phi. 
 \end{equation}
 Here the subscript $n.b.$ stands for ``neutral bion,'' showing that whatever object generates $L_{n.b.}$ has no $\sigma$-dependence and so carries no magnetic charge. We leave the study   of the objects generating (\ref{nb1}) for Section \ref{sec:neutralbions}. 
 
 The proportionality sign in (\ref{nb1}) indicates our ignorance of the overall sign of this term. 
Thus,  we now ask the question: how should (\ref{nb1}) and (\ref{mb1}) be added? The first answer that we shall give is suggested by the unbroken supersymmetry of SYM:  as the vacuum energy in a supersymmetric theory must vanish, the two terms should simply be added as written, giving
\begin{equation}
\label{potentialsym}
L_{pot.} = L^{-3} e^{-2 S_0} 2 \left(\cosh 2 \phi - \cos 2 \sigma \right)~,
\end{equation}
so that, at the ($\phi=0$, $\sigma=0,\pi$) minimum, the potential vanishes.\footnote{Running ahead, the fact that at $\phi=0$ (the center symmetric point minimizing the potential) the first term in (\ref{potentialsym}) is positive should make one suspect that the $\cosh 2 \phi$ contribution---a positive semiclassical contribution to the ground state energy---is not due to a dilute gas summation over ``real saddles.'' See Section \ref{sec:neutralbions}.}
In fact the answer for the bosonic potential of the $\mu \ll 1/L$ EFT of SYM on $R^3\times S^1$, obtained by using the ``power of holomorphy,'' is simply  the above $L_{pot.}$.\footnote{Given here with our usual pre-exponential accuracy (we shall discuss the prefactor later). For those familiar with supersymmetry, $L_{pot.}$ can be obtained from the exact holomorphic superpotential $W \sim e^{X} + e^{-X}$, where $X$ is the chiral superfield with lowest component $-\phi + i \sigma$, see \cite{Anber:2014lba} for a discussion more general than we need.} 

In what follows, we shall  slightly reverse the order we proceeded in QCD(adj). First, in Section \ref{sec:symvacuum}, we shall take the EFT given by the sum of kinetic (\ref{3dlagrangiandualdSYM1}), Yukawa (\ref{susyyukawa}), and potential (\ref{potentialsym}) terms at face value and study the IR physics  implied. Only then, in Section \ref{sec:neutralbions}, we turn to the microscopic explanation of the neutral bion terms and the deeper issues they raise. In Section \ref{sec:thermalsym}, we shall discuss the role neutral bions play in a center-symmetry breaking/restoration phase transition believed to be continuously connected to the thermal deconfinement transition in pure YM theory.

\hfill\begin{minipage}{0.85\linewidth}

\textcolor{red}{
{\flushleft{\bf Summary of \ref{sec:symfields}:}} Here, we studied the symmetry transformation of the fluctuation of the holonomy around the center-symmetric value, the scalar field $\phi$ (\ref{phi}), and dual photon $\sigma$ under the $Z_4$ chiral and $Z_2^{(1)}$ center symmetries. We used the symmetries to constrain the possible terms that can appear in the SYM EFT on $R^3 \times S^1$. The upshot is that, up to order $e^{-2 S_0}$, Yukawa-like terms (\ref{susyyukawa}) and potential terms (\ref{potentialsym}) are allowed. The microscopic origin of the Yukawa and magnetic bion terms
are as the ones discussed for QCD(adj), but SYM presents us with a novel ingredient: the neutral bion terms (\ref{nb1}), whose microscopic origin needs to be understood. }

\end{minipage}

\bigskip

\subsection{Vacua, domain walls,  ``double-string'' confinement, and  liberation of quarks on domain walls.}
 \label{sec:symvacuum}
 
 Here, as in QCD(adj) of Section \ref{sec:chiralqcdadj}, we shall be equally quick in our discussion, as we have done all preparatory work. Begin by summarizing the EFT we obtained in the previous Section:
 \begin{eqnarray}
 \label{sym2}
 L_{SYM} &=&  {1 \over 2} { g_4^2 \over L (4 \pi)^2}   (\partial_\lambda \sigma)^2  + {1 \over 2} { g_4^2 \over L (4 \pi)^2}   (\partial_\lambda \phi)^2 + i{ L \over g_4^2} \; \bar\lambda_{\dot\alpha} \; \bar\sigma^{\mu \dot\alpha \alpha} (\partial_\mu \lambda_\alpha)~ \nonumber \\
 &&+ {\alpha }\; e^{- S_0} \left( (e^{i \sigma - \phi} +e^{- i \sigma + \phi}) \;  \lambda\cdot \lambda +(e^{-i \sigma - \phi} +e^{ i \sigma + \phi}) \; \bar\lambda\cdot \bar\lambda\right)~\\
 && + {\beta \over L^{3}}\; e^{-2 S_0} 2 \left(\cosh 2 \phi - \cos 2 \sigma \right)~.\nonumber
 \end{eqnarray}
 Here $\alpha$ and $\beta$ ($\beta>0$, as per (\ref{bionpotential1})) are coefficients with power-law $g_4$-dependence and can be found in \cite{Poppitz:2012sw,Anber:2014lba}. The expression shown is the leading one in a combined small-$g_4$ and $e^{-S_0}$ expansion.\footnote{The functional form the potential in the last term can only be altered by an overall $\phi$-dependent prefactor due to the nontrivial K\" ahler metric, but this correction is small at small $L$. } We can now, as for QCD(adj), summarize the main lessons we learn by studying the weakly-coupled IR EFT (\ref{sym2}):
 \begin{enumerate}
 \item SYM on $R^3 \times S^1$ has two vacuum states:
 \begin{eqnarray}\label{vacuasym}\text{vacuum}\; 1:~ \langle \phi \rangle = 0&,& \langle \sigma \rangle = 0,  \nonumber\\
 \text{vacuum}\; 2:~ \langle \phi \rangle = 0&,& \langle \sigma \rangle = \pi ~.
 \end{eqnarray}
Thus, since the vev of $\phi$ vanishes, $Z_2^{(1)}$ is preserved ($\langle\sigma\rangle = \pi$ preserves $Z_2^{(1)}$ owing to the compact nature of the dual photon). The chiral symmetry, on the other hand, is broken, $Z_4 \rightarrow Z_2$, as in QCD(adj), by the expectation value of the monopole operator $\langle e^{i \sigma} \rangle = \pm 1$.

We stress that the stability of center symmetry in SYM is not due to the perturbative GPY potential, but is due to nonperturbative effects---the mysterious ``neutral bions'' generating the $\cosh 2 \phi$ potential on the Weyl chamber.
\item The vacuum energy vanishes, as it should due to unbroken supersymmetry.\footnote{For those familiar with supersymmetry, the Witten index \cite{Witten:1982df} of $SU(2)$ SYM equals $2$, exactly the number of zero energy vacua we found. The index does not change upon supersymmetry-preserving compactification, guaranteeing that the chiral symmetry realization is the same at small $L$ and large $L$.} The masses of the $\sigma$ and $\phi$ excitations around the vev are  (with exponential-only accuracy) equal to \begin{equation}
m_\sigma = \Lambda \; (\Lambda L)^2,\label{msigmasym}
\end{equation} 
which is just (\ref{massgapadj}) with $n_f=1$.
\item The fermions also obtain mass $m_\sigma$
from the Yukawa terms (reversing the argument and demanding  that all masses be  equal can be used to fix the ratio between $\alpha$ and $\beta$). Thus, all excitations in SYM on $R^3 \times S^1$ are massive, as opposed to QCD(adj), where the $\lambda^{I}$ remain massless. The four states of mass (\ref{msigmasym}), two bosonic and two fermionic,  fill in a so-called chiral supermultiplet. 
\item There are domain walls between the two vacua (\ref{vacuasym}). As in dYM at $\theta=\pi$, there are two different domain walls (both BPS), as can be seen by repeating the arguments of Section \ref{sec:doublestringdYM}. The simple semiclassical domain-wall counting agrees with  elaborate general BPS-wall counting arguments (for an introduction, see   Ch. 18 of \cite{Hori:2003ic}). 
\item Because the dual photon potential is $\cos 2 \sigma$, due to the proliferation of charge-$2$ magnetic bions, the argument that  quarks are confined by the double-string mechanism is identical to the one in dYM at $\theta=\pi$. In addition, quarks become liberated on domain walls, exactly as shown on Figures \ref{fig:doublestring1} and \ref{fig:decwall1}. The deconfinement on domain walls reflects the mixed chiral/center anomaly. 

Extensive numerical studies of domain walls,  double-strings, and deconfinement of quarks on domain walls in SYM   at small $L$ were performed in \cite{Cox:2019aji,Bub:2020mff}\footnote{All our pictures are taken from  \cite{Cox:2019aji,Bub:2020mff}; there are also many references there regarding the 't Hooft anomalies.} 
These studies show that properties that are quite difficult to study at strong coupling, such as the BPS domain wall multiplicities, the nature of the confining string, and the deconfinement of quarks on domain walls, are captured in a very intuitive manner in the semiclassical small-$L$ regime. This simplicity extends beyond $SU(2)$ and applies to all $SU(N)$ groups \cite{Cox:2019aji}. 
\item The small-$L$ and large-$L$ regimes of SYM are not separated by a phase transition\footnote{The absence of a phase transition does not imply that all quantities evolve monotonically upon increasing $L$ towards $R^4$. In particular, ref.~\cite{Bub:2020mff} found that $k$-string tension ratios (only defined for $SU(N)$, $N>4$ SYM)  have non-monotonic dependence on $L$.}
as the symmetry realization on $R^4$, see \cite{Shifman:2012zz}, is identical to the one found at small $L$. This is in contrast  to $n_f >1$ QCD(adj) where an $SU(n_f)$-breaking phase transition is possible for some $n_f$, as discussed in Section \ref{sec:chiralqcdadj}.  For a recent lattice study devoted to the continuity in SYM, see \cite{Bergner:2018unx}.

\item In addition to the stable dual-photon, holonomy scalar, and fermion  states, forming a chiral supermultiplet of mass $m_\sigma \sim e^{- S_0}/L$, there are stable nonrelativistic bound states of two dual photons (plus their superpartners, one bosonic and two fermionic bound states). This is similar to dYM, recall (\ref{lightestglueball}). Their mass is $2 m_\sigma$ minus their doubly-nonperturbarive binding energy:
\begin{equation}\label{glueballsym}
m_{glueball(-ino)} = 2 m_\sigma (1 - a e^{- c \;e^{\;{4 \pi^2 \over g_4^2}}})~.
\end{equation}
The calculation of $m_{glueball(-ino)}$ is given in \cite{Anber:2017ezt}, where one can also find $a,c>0$. The four nonrelativistic bound states with mass (\ref{glueballsym}), comprising a chiral supermultiplet whose precise nature is described in the reference,  are $Z_2^{(1)}$ center-symmetry singlets and are expected to map to the glueball-glueballino supermultiplet in SYM on $R^4$.
 \end{enumerate}
 
 \hfill\begin{minipage}{0.85\linewidth}

 \textcolor{red}{
 {\flushleft{\bf Summary of \ref{sec:symvacuum}:}} The summary of this brief Section is in the itemized list above. The main conclusion is that in SYM at small $L$, the symmetry realization is the same as on $R^4$. Semiclassical calculability, however, allows us to study questions not addressable by the powerful  supersymmetric tools of \cite{Seiberg:1994bp}, such as the nature of the confining string. We now move on to study the most  unusual such finding and elucidate, in Section \ref{sec:neutralbions}, the microscopic picture behind the ``neutral bion''-induced $\cosh 2 \phi$ potential in  (\ref{sym2}). Then, we shall discuss their possible role in the pure-YM thermal  deconfinement transition, in Section \ref{sec:thermalsym}.}
 
 \end{minipage}
 
 \bigskip

\subsection{Neutral bions and the need for analytic continuation of path integrals.}
 \label{sec:neutralbions}

In this Section, we shall discuss the microscopic origin of the $2 \cosh 2\phi$ potential in (\ref{sym2}). The $\phi$- and $\sigma$-dependent scalar potential was first obtained by purely supersymmetric tools. Holomorphy and symmetries allow one to determine the form of a holomorphic function of $z = -\phi + i \sigma$, the superpotential $W(z) \sim e^{z} + e^{-z}$, as first done in \cite{Seiberg:1996nz}. Supersymmetry then implies that to find the scalar $2 \cosh 2\phi -2 \cos 2 \sigma$ potential, clearly not a holomorphic function, one should compute $|dW/dz|^2$ (it is easy to see that one indeed obtains the potential in (\ref{sym2})). The semiclassical calculations on $R^3\times S^1$ \cite{Davies:1999uw,Davies:2000nw} have been focused only on calculating $W$, or equivalently the fermion-bilinear term in (\ref{sym2}), whose form has then been used to  infer the potential. 

Our goal here is to understand the appearance of the potential (\ref{sym2}) directly, not relying on the power of supersymmetry. After all, for us, SYM is only a particular case of QCD(adj), a set of theories at weak coupling at small $L$.  It is just that for SYM we have some other, non-semiclassical, tools at our disposal. The hope is  that this comparison will yield some interesting insight about semiclassics in QFT and that this insight transcends supersymmetry.

As we suggested in the previous Section, the microscopic origin of the neutral-bion potential (\ref{nb1})
 are ``topological molecules'' of the $M$-$M^*$ and $KK$-$KK^*$ type. 
The physical importance of neutral bions is that they ensure center stability, i.e. vanishing expectation value for $\phi$.
 
 However, as opposed to our treatment of the magnetic bion $M$-$KK^*$ (and c.c.) molecules, where the magnetic Coulomb repulsion was compensated by fermion-hopping induced attraction, here we face a problem: since $M$ and $M^*$ have the same magnetic charge and opposite scalar charge, there is attraction in both these channels. In addition, the fermion-hopping also generates attraction. Taking all of this into account and proceeding blindly, we can write, similar to (\ref{mkk2}), but taking the doubled attraction into account and taking $n_f=1$, the following amplitude for an $M$-$M^*$ tunneling event
 \begin{eqnarray} \label{mmbar1}
\langle M\text{-}M^* \rangle &=&\int d^3 x \int d^3 y\;  e^{-2 S_0}  \left({g_4^2 \over L^3}\right)^{2} f(\hat{n})^{2}\; e^{\;  {4 \pi \over g_4^2}{ 2 L \over |x-y|} - 4\log {|x-y|\over L}}~\nonumber \\
&\sim&  {e^{-2 S_0} \over L^{3}} \int d^3 z \int\limits_{0}^{\infty} {dr \over L} \;e^{\; {4 \pi \over g_4^2}{2 L \over r} - 2 \log {r \over L}}= {e^{-2 S_0} \over L^{3}} \;  I_{n.b.}(g_4^2) \int d^3 z~.
\end{eqnarray}
We defined the integral over the relative separation quasizero mode:
\begin{equation}
\label{Inb}
I_{n.b.}(g_4^2)\; ``=" \int\limits_{0}^{\infty} {dz} \; e^{\; {4 \pi \over g_4^2}{2  \over z} - 2 \log {z}}~.
\end{equation}
Notice that, as written, the integrand in $I_{n.b}$ only makes sense at large $r \gg L$ ($z \gg 1$), where the approximation of a long-range ``force'' between the $M$ and $M^*$ constituents makes sense. Nonetheless, we formally extended the lower limit to ``$z=0$'', where the  $M$ and $M^*$ are on top of each other and the interactions written are not sensible. It is clear that the integrand can not be trusted  near the origin and the expression $I_{n.b.}(g_4^2)$ needs to be properly defined (this is what the $``="$ sign above is meant to remind us of). The $KK$-$KK^*$ amplitude leads to an expression identical to (\ref{mmbar1}).

Let us now consider the magnetic-bion $M$-$KK^*$  tunnelling event in SYM, with amplitude given essentially in (\ref{mkk2}), but with $n_f=1$ and an extra factor of $2$ due to the scalar repulsion:
   \begin{eqnarray} \label{mmbar2}
\langle M\text{-}KK^* \rangle &=&\int d^3 x \int d^3 y\;  e^{-2 S_0}  \left({g_4^2 \over L^3}\right)^{2 n_f} f(\hat{n})^{2 }\; e^{\;-  {4 \pi \over g_4^2}{ 2 L \over |x-y|} - 4  \log {|x-y|\over L}}~\nonumber \\
&\sim&  {e^{-2 S_0} \over L^{3}} \int d^3 z \int\limits_{0}^{\infty} {dr \over L}\; e^{\;- {4 \pi \over g_4^2}{2 L \over r} - 2 \log {r \over L}}= {e^{-2 S_0} \over L^{3}} \;  I_{m.b.}(g_4^2) \int d^3 z~.
\end{eqnarray}
We defined, similar to $I_{m.b.}(n_f, g_4^2)$ of (\ref{mkk2}), the integral $I_{m.b.}$
\begin{eqnarray}
\label{Imb}
I_{m.b.}(g_4^2) = \int\limits_{0}^{\infty} {dz} \;e^{\; -{4 \pi \over g_4^2}{2   \over z} - 2 \log {z}} = \int\limits_{0}^\infty {d t} \; e^{- {8 \pi \over g_4^2} t} = {g_4^2 \over 8 \pi^2} ~.
\end{eqnarray}
As promised earlier, the above demonstrates that $I_{m.b.}$ is easy to calculate, especially for $n_f=1$. 

Now, upon comparison of (\ref{Imb}) to (\ref{Inb}), we notice that the integrands of the two expressions are simply related by reversing the sign $g_4^2 \rightarrow - g_4^2$. Hence (formally), we have a relation between the integrals
$I_{m.b.}(g_4^2) = I_{n.b.}(- g_4^2)$. Now, we come to an idea of Bogomolnyi and Zinn-Justin (a.k.a. ``the BZJ prescription'') originating in similar instanton---antiinstanton problems in double-well quantum mechanics \cite{Bogomolny:1980ur,Zinn-Justin:1981qzi}, see also \cite{Schafer:1996wv} and Zinn-Justin's book \cite{Zinn-Justin:2002ecy}. The idea is to {\it define} the $M$-$M^*$ amplitude using the relation between the integrands in (\ref{Inb}) and (\ref{Imb}). BZJ defined $I_{n.b}(g_4^2)$ as equal to  $I_{m.b.}(g_4^2)$ analytically continued to negative, $g_4^2 \rightarrow - g_4^2$. We write this prescription defining the neutral bion amplitude (\ref{Inb}) as $I_{n.b., BZJ}(g_4^2) = I_{m.b}(- g_4^2)$. From (\ref{Imb}), we then immediately find that $I_{n.b., BZJ}(g_4^2) = - {g_4^2 \over 8\pi}= - I_{m.b.}(g_4^2)$. 

Next, we recall that $M$-$M^*$ comes with a factor $e^{- 2 \phi}$ due to its overall scalar charge and $M$-$KK^*$ comes with $e^{i 2 \sigma}$ due to the magnetic charge (the c.c. amplitudes come with $e^{2 \phi}$ and $e^{- 2i \sigma}$, respectively). Thus, the relative minus sign between the magnetic bion amplitude $I_{m.b.}$ and neutral bion $I_{n.b., BZJ}$ obtained via the BZJ prescription is precisely the negative sign that supersymmetry and the vanishing vacuum energy forced upon us (recall eqn.~(\ref{sym2})).\footnote{In addition to the relative minus sign between the two terms in (\ref{sym2}) obtained by the BZJ prescription, the overall power of $g_4^2$ (which we did not show) in both terms also comes out right \cite{Poppitz:2011wy,Poppitz:2012sw,Argyres:2012ka}, agreeing with the holomorphy predictions.}

As we discuss below, BZJ were not guided by supersymmetry at all, so this ``right sign'' answer is even more remarkable and we believe that it is hardly an accident! To elucidate, briefly, BZJ's motivation was to understand the relation between perturbation theory in the double-well quantum mechanics, which is a divergent non-Borel-summable series, and the exponentially small nonperturbative semiclassical effects.\footnote{An introduction to the large order behaviour of perturbation theory and the nature of the perturbative series can be found in Chapter 41 of \cite{Zinn-Justin:2002ecy} (various aspects of the BZJ ideas are also discussed there, see Chapter 43).} In the double-well problem, BZJ found that the analytic continuation (equivalent to our  $g_4^2 \rightarrow - g_4^2$) in the instanton-anti-instanton amplitude leads to an ambiguous imaginary part depending on the way  the analytic continuation is done.\footnote{This imaginary part is absent in our SYM example, but would be present in dYM, where neutral bion molecules contribute to the ``gluon condensate.'' Explaining this would take us far afield, however; see the recent discussion and more references in \cite{Unsal:2021cch}.} BZJ showed that this ambiguous imaginary part is exactly cancelled by an ambiguity associated with the resummation of the (non-Borel-summable) perturbative series. This cancellation of ambiguities is a welcome feature, as one expects that the physical quantities, represented by the resummed series, are free of ambiguities.

The cancellations of ambiguities associated with resumming the perturbative series (the perturbative expansion in $g_4^2$) and with nonperturbative effects (the semiclassical $e^{-S_0}$ expansion), to all orders in both expansions, are the subject of ``resurgence theory''. We can not discuss this fascinating subject in any more detail (among other reasons, we would have to write another even longer set of notes!). These resurgent cancellations  of ambiguities are best understood in the theory of  certain nonlinear differential equations and in quantum mechanics of a single real degree of freedom, see \cite{Dunne:2016nmc} for an extensive list of references.

Here, we shall only note that the BZJ prescription above was formulated as an analytic continuation in the coupling $g_4^2$. This can be  rephrased as a deformation of the contour of the quasizero mode integral (the integral over the relative separation $z$) in (\ref{Inb}) into the complex plane. The point we wish to make is that this is, indeed, an analytic continuation of the path integral, since the quasizero mode direction is one of the (infinitely many) directions of field space. To this end, let us consider the integrand of $I_{n.b.}$ of (\ref{Inb}) as a function in the complex $z$ plane. The integrand has the form $e^{f(z)}$, with $f(z) = {8 \pi \over g_4^2 z} - 2 \log z$. The critical points of $f(z)$ are the solutions of $f'(z)=0 = {8\pi \over g_4^2 z^2} + {2 \over z}$, showing that one critical point is at  infinity and the other at $z_* = -{4\pi \over g_4^2}$. The critical point at infinity corresponds to noninteracting $M$ and $M^*$ and its contribution is taken into account by including the separate contributions of $M$ and $M^*$ which give rise to the fermion bilinear terms in (\ref{sym2}).  The other critical point of $f(z)$, the one at ``negative radius'' at $z_*=- 4\pi/g_4^2$  corresponds to a complex separation between the $M$ and $M^*$. We next recall from complex analysis that the steepest descent method requires us to deform the contour of integration passing through a critical point $z_*$ along the steepest descent path (a baby version of what is more generally known as ``Lefshetz thimbles''). This is the path passing through $z_*$ along which   Im$f(z)=$ Im$f(z_*)$ and Re$f(z)$ decreases away from $z_*$ (i.e. the action increases away from the critical point, ensuring convergence). For our $z_*$, this is the negative real axis. It is now easy to see that if we calculate $I_{n.b.}$ in eqn.~(\ref{Inb}) by integrating from $0$ to $-\infty$ instead, we obtain the result required by supersymmetry. The integrand is essentially the one shown on Figure~\ref{fig:bionsize}, except that the $x$-axis now runs from zero to minus infinity.  A ``physical picture'' of the neutral bion molecule is thus one where the $M$ and $M^*$ constituents are separated by a complex distance. However,   the absolute value of the distance is the same as $r_{bion}$, still within the validity of small-$L$ semiclassics (see also Footnote \ref{qmfootnote}).

Before we conclude this Section, we note that, as of the day of this writing, there is no known systematic way to perform  calculations similar to the calculation of the neutral bion amplitude to further higher orders, in both the perturbative and semiclassical expansions. Formulating the BZJ prescription in the $R^3 \times S^1$ theories starting from first principles, e.g. from the QFT path integral, is also an open problem.\footnote{\label{qmfootnote}There has, however, been a substantial amount of progress in understanding similar calculations in quantum mechanics, with or without supersymmetry. It is a remarkable fact that in   quantum mechanics, the analogues of the neutral bions can be found as either exact solutions of the theory with the fermions integrated out (because the fermion determinant there is computable in an arbitrary bosonic background) or by using the Lefshetz thimble analytic continuation procedures we briefly outlined here. See refs.~\cite{Behtash:2015zha,Behtash:2015kva,Behtash:2015loa,Behtash:2018voa} for a comparison between these approaches and for a detailed discussion of the necessity of analytic continuation in different quantum mechanics examples. See also the recent proposals \cite{Unsal:2020yeh,Unsal:2021cch,Pazarbasi:2021ifb} relating the quantum mechanical and QFT discussions.
 }  Ref.~\cite{Argyres:2012ka} outlined such procedures  for multi-bion amplitudes and proposed a general scheme, the ``resurgence triangle,'' of how the cancellation of ambiguities between the various orders of the semiclassical expansion (including the perturbative expansion around the various saddles) would proceed to all orders. There has not, however, been any progress showing how these resurgent cancellations work in practice in the class of $R^3 \times S^1$ theories we discuss here, in part because such calculations are technically challenging, see \cite{Anber:2014sda} for an example.  There has, however, been progress for BPS observables in  supersymmetric theories (where the path integral localizes to a finite dimensional one), in  topological, or in integrable theories; for some recent studies see e.g. \cite{Fujimori:2021oqg,DiPietro:2021yxb}, while more references and discussion can be found in the review article \cite{Dunne:2016nmc}. 
To conclude, despite the difficulties mentioned above, we nonetheless find the agreement of the result of the BZJ procedure with SYM (as further stressed in \cite{Behtash:2015kna}) and with quantum mechanical examples impressive and offering an important consistency check. We   take this as an argument in favour of the validity of the  manipulations  outlined here and as an impetus for the further pursuit of the Lefshetz thimble decomposition ideas. 

\hfill\begin{minipage}{0.85\linewidth}

\textcolor{red}{
{\flushleft{\bf Summary of \ref{sec:neutralbions}:}} In this Section, we considered in some detail the structure of the new type of topological molecule, the ``center-stabilizing neutral bion.'' The neutral-bion example considered above is important since it shows the need for analytic continuation of path integrals (infinite-dimensional)  and their Lefshetz thimble (think: steepest-descent contours) decomposition. Again, this goes beyond our topic---and the author's competence.  Our main point, however, is that the $R^3 \times S^1$ studies gave fascinating hints of the relevance of Lefshetz thimbles to path integrals and, ultimately, to the calculation of physical properties. Further studies and insight are definitely welcome. }

\end{minipage}
\bigskip

 \subsection{Neutral bions and the pure-YM thermal deconfinement transition: the continuity conjecture.}
 \label{sec:thermalsym}
 
In the previous Section, we studied the nature of the nonperturbative objects leading to center stability in SYM, the neutral bions. 
These are novel nonperturbative objects that generate a potential for the holonomy $\phi$, equivalently, tr $\Omega$, ensuring $Z_2^{(1)}$ stability.  The question about their possible relevance to the thermal deconfinement transition naturally arises; here, we shall explore this within a calculable setup.

To motivate the relevance of neutral bions to the deconfinement transition, recall that, as per Section \ref{sec:digression}, the thermal deconfinement transition in pure YM theory is associated with the $Z_2^{(1)}$ order parameter, $\langle \text{tr} \Omega \rangle$ (the unbroken $Z_2^{(1)}$ implies a low-$T$ confined phase while the broken-$Z_2^{(1)}$ phase is the high-$T$ deconfined one). We also saw that at high-$T$ in pure YM theory, the perturbative loop contribution, which is  dominant at  $T \gg \Lambda$, generates a center-breaking potential for this order parameter, signifying a deconfined high-$T$ phase. 
As $T$ is lowered towards $\Lambda$, one expects that $Z_2^{(1)}$ will be restored at some critical  $T_c \sim \Lambda$. However, this transition occurs at strong coupling and is not accessible to weak-coupling analytic tools, as already remarked by GPY \cite{Gross:1980br}.  

In this Section, we shall find a weakly-coupled $Z_2^{(1)}$-breaking transition in ``softly-broken'' SYM on $R^3 \times S^1$ (softly-broken refers to the addition of mass $m$ for the adjoint fermion, the ``gaugino''). 
We shall argue that the competition between neutral bions and monopole-instantons---two kinds of nonperturbative effects, the former center-stabilizing and the latter center-breaking---determines the $Z_2^{(1)}$-symmetry realization, the order of the phase transition, and its $\theta$-angle dependence. 
We shall conjecture that this transition, which is a semiclassically calculable quantum phase transition in the small-$m$, $L$ regime,  is continuously connected to the strongly-coupled thermal deconfinement transition in pure YM theory, by decoupling the gaugino. We shall also discuss the available evidence from lattice simulations.\footnote{While our discussion here is restricted to softly-broken $SU(2)$ SYM, we stress that this conjecture can be made---and agrees with the available lattice evidence---for all gauge groups in a similar setup \cite{Poppitz:2012nz,Anber:2014lba}.} The reason this  ``continuity conjecture'' is  natural is that when we take $m \rightarrow \infty$, we obtain pure YM theory on $R^3 \times S^1$; but this, now bosonic, theory has a thermal interpretation, where  $L$ is the inverse temperature $1/T$.

We  illustrate the conjecture on Figure \ref{fig:continuity}, but before discussing the evidence for continuity, let us flesh out the details of the $Z_2^{(1)}$-breaking quantum phase transition.
We begin by exploring the behaviour of SYM on $R^3 \times S^1$ upon adding  mass for the adjoint fermion (the gaugino). Turning on the mass $m$ in the QCD(adj) Lagrangian (\ref{lagrangianadjoint}) with $n_f=1$ has many effects:
\begin{enumerate}
\item The gaugino mass explicitly breaks supersymmetry, i.e. leads to the appearance of a nonzero GPY potential on the Weyl chamber. This is easy to infer from (\ref{gpyadjointstable}), and results in the potential given in (\ref{gpygaugino}), as we discuss below.
\item The gaugino mass also breaks the anomalous $U(1)$ chiral symmetry (as well its anomaly-free $Z_4$ subgroup) down to fermion parity and so introduces $\theta$-angle dependence. 
We shall adopt a convention such that the mass $m$ is real, i.e. take $m=m^*>0$ in (\ref{lagrangianadjoint}), absorbing its phase into the $\theta$-angle by a chiral rotation of $\lambda$. This means that we have to include $e^{i \theta/2}$ factors\footnote{This $\theta$ is what is usually called ``effective'' or $\bar\theta$ in the standard model, as it incorporates the phase of the determinant of the quark mass matrix.} in the $M$ and $KK$ 't Hooft vertices, and c.c. for $M^*$, $KK^*$, as we do in (\ref{Mmass}) below.

\item The gaugino mass, as already alluded above, also induces new nonperturbative terms in the scalar potential for $\sigma$ and $\phi$, since the monopole instanton zero modes contributing to the fermion bilinear terms in (\ref{sym2}) are lifted by mass insertions. These $M$, $KK$, etc., terms are worked out below and shown in eqn.~(\ref{scalargaugino}).
\end{enumerate}

 As we  show below, the various effects of soft breaking combine in interesting ways. They will lead to the conclusion that for a range of $m$, there is a phase transition to a center-symmetry restored phase in the small-$L$ theory. This transition occurs in the calculable small-$L$, small-$m$ regime. 
We shall estimate the critical value of $mL$ (at given $\Lambda$) and  argue that for $SU(2)$ the transition is second order, i.e. is a continuous transition. We shall then put forward a conjecture that this small-$L$ calculable transition is continuously connected to the thermal deconfinement transition in pure YM theory on $R^4$. There are many checks that appear to validate this conjecture (it holds for SYM with all gauge groups) and we shall briefly summarize them at the end. 

Consider the effects described above one by one. First, the GPY potential of the theory with a single massive adjoint, eqn.~(\ref{gpyadjointstable}) with $n_f=1$, can be expanded for $mL \ll 1$ using $K_2(x) \sim {2\over x^2} - {1\over 2}$. The first term cancels the gauge contribution, while the second gives the leading small-$m$ term in the one-loop potential on the Weyl chamber\footnote{
Recalling our definition of $\phi$ (\ref{phi}), we have $vL =\pi +  {g_4^2 \over 4\pi}\; \phi$, so this is really a potential for $\phi$. We also used the Fourier transform relation $\sum\limits_{p=1}^\infty {\cos(2 \pi p x)\over p^2}=\pi^2 \left(\lfloor x \rfloor(\lfloor x \rfloor^2 -1) - 1/6\right)$. Also, the reader may notice that in our calculation of (\ref{gpyadjointstable}), the mass $m$ was assumed real. In fact, the GPY potential, being perturbative, only depends on $|m|$ (as is easy to see following its derivation).} \begin{equation}
\label{gpygaugino}
V_{GPY}^{m^2}(\phi) = - {(mL)^2 \over \pi^2 L^3} \sum\limits_{p=1}^\infty {\cos p vL \over p^2} = - {(mL)^2 \over  L^3}\iless{{vL \over 2 \pi}} \left( \iless{{vL\over 2 \pi}}^2-1 \right)+ const.
\end{equation}This potential  is minimized at $vL =0$ or $2\pi$, i.e. at the center-broken edges of the Weyl chamber. Notice, however, that $V_{GPY}^{m^2}(\phi)$ is of order $(mL)^2$, hence for small enough $m$ it can compete with other contributions to the holonomy potential, as we shall now show. 
It is not common that perturbative center-destabilizing effects, such as the above $V_{GPY}$, can be balanced by nonperturbative effects that tend to stabilize center symmetry, but the small $L$ and small $m$ setup is unique in allowing   for this in a controllable way. At small $m$, there are two other contributions to the potential for $\phi, \sigma$, coming from the nonperturbative terms in (\ref{sym2}). 

The first kind of terms are  the neutral- and magnetic-bion induced term. These terms will not be affected, to leading order in $m$, provided the gaugino mass $m$ is taken to be smaller than the inverse size of the bion. This is because it is the gaugino hopping that caused the binding, and the attractive potential induced by gaugino hopping will not be affected if $m \ll r_{bion}^{-1} \sim g_4^2/L$.

\begin{figure}[h]
\centerline{
\includegraphics[width=6.5 cm]{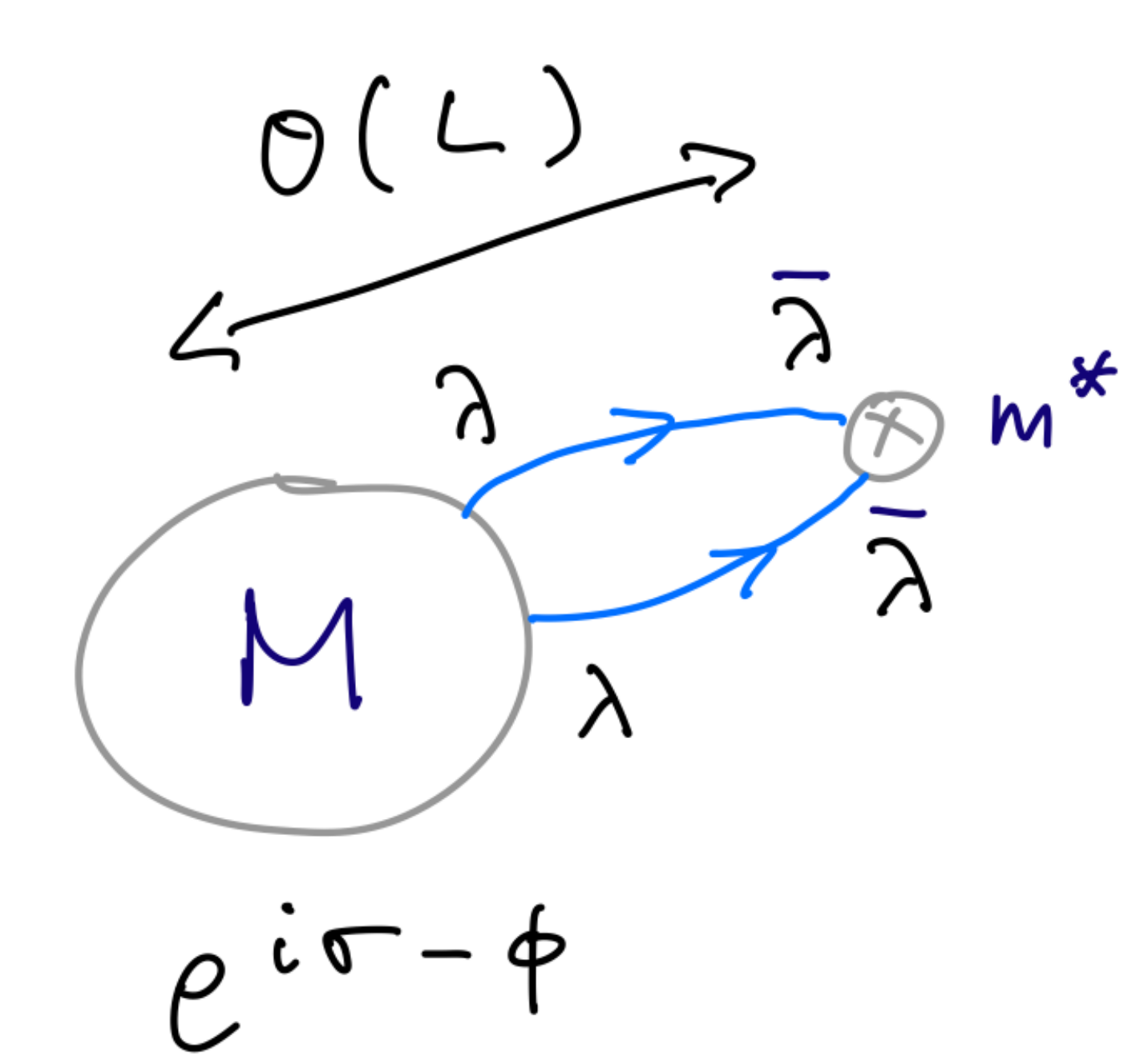}}
\caption{ The lifting of monopole-instanton fermion zero modes by a small fermion mass term. The fermion zero modes are localized over a distance of order $L$, the $M$ size. As $m \ll 1/L$, the propagators shown by the blue arrows (the same ones responsible for the bion binding of Figure \ref{fig:magnbionsize}) are massless propagators over a distance $L$, with the divergence cut off by the monopole-instanton core of size $L$. The calculation is sketched in eqn.~(\ref{Mmass}). \label{fig:liftzero}}
\end{figure}

The second kind of terms arises because in the presence of a small gaugino mass, the $M$ and $KK$ terms in (\ref{sym2})  have their zero modes lifted. The lifting of fermion zero modes by a mass or Yukawa term is a rather standard (but long) calculation in instanton calculus. We shall short-circuit it in several ways. First, we note that, without regard to pre-exponential $g_4$ dependence, one can argue based on symmetries alone, that we should simply replace $\lambda \lambda$ by $m^*/L^2$ and $\bar\lambda\bar\lambda$ by $m/L^2$. That $\lambda\lambda$ should be replaced by $m^*$ is justified by chiral symmetry, because of their identical transformation property, while the $L$-dependence follows on dimensional grounds as it sets all scales in the problem.\footnote{This follows, yet again, by thinking of $m$ as the vev of a field breaking the chiral symmetry. There could be further corrections that go like $m^* f(|m|L)$, for some unknown function $f$, but we are only interested in the leading small-$m$ term.}

 A further pictorial justification and the relevant scales are shown on Figure \ref{fig:liftzero}. We can estimate the answer by taking the $M$ 't Hooft vertex (\ref{susythooftchiral}) and computing the leading small-$m$ correction due the insertion of a gaugino mass term,  the $m^* L \bar\lambda\bar\lambda/g_4^2$ term in (\ref{lagrangianadjoint}). This results in an  $M$ 't Hooft vertex without fermion zero-mode insertions. Taking $m$ real, as discussed above, we obtain
\begin{eqnarray}
\label{Mmass}
M: &&{e^{- S_0} } e^{i \sigma(x) - \phi(x)} \; e^{i {\theta \over 2}}
\; \big\langle  \lambda(x) \lambda(x)  \; {mL\over g_4^2}\int d^3 y  \bar\lambda(y)\bar\lambda(y)\big\rangle  \sim {e^{- S_0}} {mL\over g_4^2} \left({g_4^2 \over L}\right)^2e^{i \sigma(x) - \phi(x)} \; e^{i {\theta \over 2}} \int d^3 y\; ``{1\over |x-y|^4}" ~\nonumber \\
&& \sim {e^{- S_0}} {mL}  \; {g_4^2 \over L^3} \; e^{i \sigma(x) - \phi(x)} \; e^{i {\theta \over 2}}\sim {m L \over L^3} e^{- S_0}  \; e^{i \sigma(x) - \phi(x)} \; e^{i {\theta \over 2}}. 
\end{eqnarray}
The derivation above is clearly schematic, but the final answer is correct, with our usual exponential-only accuracy. To obtain it, we used   the propagator (\ref{lambdacorrelator}) on the first line, simply dropping the angular part. Furthermore,  the integral as shown on the top line appears divergent at $y$ approaching $x$, the center of the  monopole. However, as the $M$ vertex is only local at distances  $\gg L$ from the monopole core, the divergence should be cut at a scale $L$,\footnote{The real calculation requires knowledge of the wavefunctions of the fermion zero modes in the $M$ background, see \cite{Poppitz:2012sw} and references therein. The same calculation could also be done using  recent ideas to include supersymmetry-breaking effects in exact results via anomaly mediation, see \cite{Murayama:2021xfj} (some work is required, however, to adapt this to the dual description on $R^3 \times S^1$).} thus replacing the integral over $y$ by $1/L$.

The upshot of the above discussion is that, combining all small-$m$ effects, we arrive at the leading-order scalar   potential  in the small gaugino mass theory:
\begin{eqnarray}
\label{scalargaugino}
V(\phi, \sigma) &=& {\beta \over L^{3}} e^{-2 S_0} \left(\cosh 2 \phi - \cos 2 \sigma \right)\nonumber \\
&+&  {\alpha (m L) \over L^{3}} e^{- S_0} \left( (e^{i \sigma - \phi} +e^{- i \sigma + \phi}) e^{i {\theta \over 2}} +(e^{-i \sigma - \phi} +e^{ i \sigma + \phi})e^{-i {\theta \over 2}}\right)~\\
  &-& {(mL)^2 \over  L^3}\iless{{vL \over 2 \pi}} \left( \iless{{vL\over 2 \pi}}^2-1 \right),~~ ({\text{with}}  \;vL \rightarrow \pi +  {g_4^2 \over 4\pi}\; \phi  ~\text{in the last term}).\nonumber 
\end{eqnarray} 
The coefficient $\alpha$ here is different from that in (\ref{sym2}) due to factors not shown in (\ref{Mmass}).
We have ordered the terms above in powers of $mL$: the first, center-stabilizing term is of order $(mL)^0$, the second term is of order $(mL)$ and the third term is of order $(mL)^2$. Of course, the third term has no $e^{-S_0}$ suppression, signifying its perturbative nature. Notice, however, that if $mL \sim e^{-S_0}$ all three kinds of terms can compete. 

Our next task is, using (\ref{scalargaugino}), to find the vacuum state---and its symmetries---as a function of $mL$, while dialing $mL$ from the supersymmetric and $Z_2^{(1)}$-preserving limit, $mL=0$, towards $mL \sim e^{-S_0}$. In the process, we shall ignore the  perturbative GPY potential, the last line in (\ref{scalargaugino}), assuring the reader that this is a consistent assumption.\footnote{This is one place where the ignored powerlaw dependence on $g_4^2$ in the $\alpha$ and $\beta$ terms is relevant. We advise the reader to either trust us or to consult \cite{Poppitz:2012sw}, where it is shown that there are further $1/g_4^2$ terms in the nonperturbative potential terms that make them more important than the GPY potential for $mL \le {\cal{O}}(e^{-S_0})$, showing that the perturbative term can be consistently ignored.}

Qualitatively, it is clear that if $mL=0$, the theory is supersymmetric, with two vacuum states (\ref{vacuasym}) breaking $Z_4 \rightarrow Z_2$. But what about nonzero values of $mL$? Qualitatively, we expect that at small $mL$, the vacuum degeneracy will be lifted (since $m$ explicitly breaks the $Z_4$ chiral symmetry) and one of these ground states will be the vacuum. To do the further analysis as simply as possible, we proceed as follows. We ignore the GPY term and rewrite the other two terms using (\ref{lambdascale}) to write $(\Lambda L)^3 = e^{- S_0}$.
Then we can have fun and write a dimensionless potential $\hat V$, introducing dimensionless $\hat m = mL$ and $\hat \Lambda = \Lambda L$. Also, for the purposes of illustrating the physics, we set $\alpha = \beta =1$. Thus proceeding, from (\ref{scalargaugino}), we obtain the following dimensionless potential:
\begin{equation}
\label{pot3}
{\hat V\over \hat\Lambda^6} = \cosh 2 \phi - \cos 2 \sigma +  {2\hat m \over \hat\Lambda^3} \left[e^{-\phi}  \cos(\sigma + {\theta \over 2}) + e^{\phi} \cos(\sigma - {\theta \over 2})\right], ~\text{where}~\hat m = mL, \; \hat \Lambda = \Lambda L.
\end{equation}
This form makes it  clear that the ratio $\hat m \over \hat\Lambda^3$ controls the relative importance of neutral and magnetic bions  and the monopole-instanton terms. At small values of this parameter, the center-symmetric neutral-bion contribution dominates, while, as we will see below, upon increasing  $\hat m \over \hat\Lambda^3$, the second, monopole-instanton induced term causes the spontaneous breakdown of $Z_2^{(1)}$.

To see this, 
for simplicity, let us also set $\theta = 0$, obtaining \begin{equation}
\label{pot4}
{\hat V\over \hat\Lambda^6}\big\vert_{\theta=0} = \cosh 2 \phi - \cos 2 \sigma +  {4 \hat m \over \hat\Lambda^3} \cos \sigma  \cosh \phi.~
\end{equation}
This form is as simple as it gets and is sufficient to getting the main idea across (see \cite{Poppitz:2012sw} for a more detailed study). 
It is now immediately clear, from the above potential, that at small $\hat m/\hat\Lambda^3$, the $\phi=0$ center-symmetric minimum favoured is the one at $\sigma = \pi$, as it has lower energy than the $\sigma=0$ one. However, upon increasing $\hat m/\hat\Lambda^3$ above, this $Z_2^{(1)}$-symmetric minimum is destabilized. This is easiest  to see by computing the second derivative of (\ref{pot4}) w.r.t. $\phi$, evaluated at $\sigma=\pi$, $\phi=0$. This yields $m_\phi^2\vert_{\phi=0,\sigma=\pi} \sim 4 -  4\hat m/\hat\Lambda^3$. Thus, the mass squared of  $\phi$ is positive for small $\hat m$, ensuring stability of the center symmetric vacuum. However, $m_\phi^2\vert_{\phi=0,\sigma=\pi}$ vanishes at a critical value  $\hat m_c = \hat\Lambda^3$,  indicative of a second order phase transition to a $Z_2^{(1)}$ broken phase. This second order transition occurs upon increasing $mL$ to a critical value   $(m L)_c \sim e^{- S_0}$, as promised. 
At fixed $\Lambda$, this transition occurs at the line $m = \Lambda L^2$, shown in the lower l.h. corner of  Figure \ref{fig:continuity}. As we just showed, the quantum $Z_2^{(1)}$-breaking transition occurs due to the competition between center-stabilizing neutral-bion and center-destabilizing monopole-instanton effects. 
 \begin{figure}[h]
\centerline{
\includegraphics[width=8.5 cm]{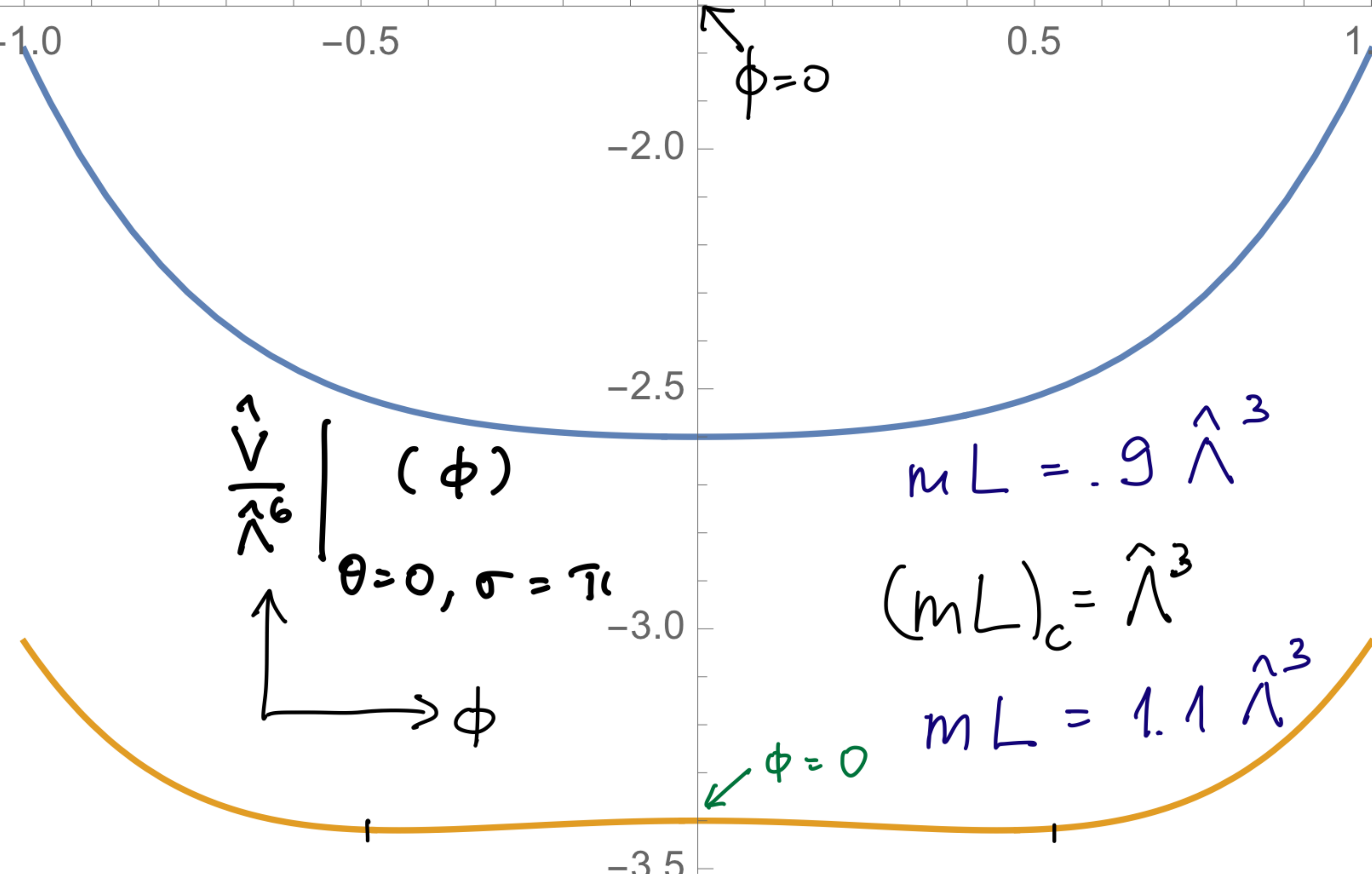}}
\caption{ The potential (\ref{pot4}) for $\sigma=\pi$ as a function of $\phi$ for two values of $mL$, above $(mL)_c=\hat\Lambda^3$ (lower curve) with broken $Z_2^{(1)}$, and below (top curve) with center-symmetric holonomy vev. The top and bottom curve illustrate the 2nd order of the center-symmetry breaking phase transition, corresponding to the unbroken- and broken-$Z_2^{(1)}$ phases, respectively. A study of the $\theta$-dependence reveals that $(mL)_c$ is a decreasing function of $\theta$. \label{fig:thermalphase}}
\end{figure}

The potential (\ref{pot4}) as a function of $\phi$ is shown on Figure \ref{fig:thermalphase}, for two values of $mL$, one  just above and one just below $(mL)_c$. 
One can also study the behaviour of $(mL)_c$ upon increasing $\theta$ away from $0$ and find that the critical value $(mL)_c$ decreases with $\theta$, see \cite{Poppitz:2012nz}.
We shall not further dwell on the study of this potential here. Instead, let us qualitatively discuss  its conjectured relation to the thermal deconfinement transition and the available evidence. 

On Figure \ref{fig:continuity} we plot the phase diagram of SYM with gaugino mass as a function of $m$ and $L$, the $S^1$ radius. The calculable regime of small $L$ and small $m$ is in the l.h. lower corner of plot. For $0< mL < \Lambda^3 L^3$, above the thick red line on the plot, the vacuum is one of the two SYM vacua with $Z_2^{(1)}$ center symmetry intact. The phase transition line is $m = \Lambda^3 L^2$. Thus, if one keeps $m$ fixed, one crosses the red line vertically down, from the center-preserved phase to the center broken one, by decreasing $L$ at fixed $m$. As smaller $L$ corresponds to   higher-$T$ when a thermal interpretation of the $S^1$ holds, it is natural to make the following continuity conjecture:
 it states that the phase diagram looks like the one shown on the figure: the quantum phase transition on the l.h. lower corner is continuously connected to the thermal deconfinement transition of pure YM theory, shown on the r.h.s. with a thick black line. For $m \rightarrow \infty$ ($m\gg \Lambda_{YM}$ should suffice), the IR physics is that of pure YM theory, which exhibits deconfinement at a critical $S^1$ radius $L_c \sim \Lambda_{YM}^{-1}$ (as is known from the lattice).  Here, $\Lambda_{YM}$ is the strong coupling scale of the pure YM theory. 
 
 Thus the center-symmetry preserving region of the SYM$+ m$ phase diagram is smoothly connected to the confined phase of pure YM and the center-broken region to the deconfined phase.
\begin{figure}[h]
\centerline{
\includegraphics[width=10.5 cm]{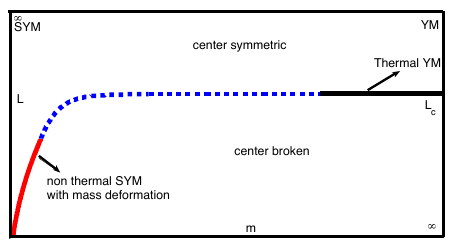}}
\caption{ The conjectured phase diagram of softly-broken SYM in the $m$-$L$ plane. The calculable center-symmetry
breaking quantum phase transition, occurring at small-$m$,$L$ and shown on Figure \ref{fig:thermalphase} is in the left-hand corner of the diagram,
shown by a thick red line. 
The center-breaking transition occurs, in the approximation of (\ref{pot4}), at the critical value $\hat m = \hat\Lambda^3$. More appropriately, this is written as $m = \Lambda^3 L^2$, which is the equation of the thick red curve shown on the plot. If one keeps $m$ fixed, it is clear that the center-broken phase is entered by decreasing $L$ (which, in the thermal theory means going to higher temperature). 
This weak-coupling quantum phase transition is conjectured to be continuously connected, upon decoupling the gaugino,
to the thermal deconfinement transition in pure YM theory, shown by the thick black line on the right.
\label{fig:continuity}}
\end{figure}
What evidence do we have for this phase diagram? It should be clear that we have no analytic proof, as we lose semiclassical calculability once we leave the small-$m$ (and small-$L$) regime, entering the strong-coupling regime of pure YM. Our knowledge of the properties of the deconfinement phase transition in pure YM come from lattice simulations and from some theoretical arguments concerning its order \cite{Svetitsky:1982gs,Yaffe:1982qf} (in particular, they argued that the $SU(2)$ $Z_2^{(1)}$-breaking transition is second order, using $Z_2$ Ising universality class arguments). Let us now summarize the evidence in favour of the continuity conjecture, including evidence for groups other than $SU(2)$ (studied in \cite{Poppitz:2012nz,Anber:2014lba}):
\begin{enumerate}
\item The order of the transition: for all gauge groups with nontrivial center ($SU(N)$, $Spin(N)$, $Sp(N)$, $E_6$, and $E_7$), the corresponding small-$L$ quantum phase transition in SYM with gaugino mass $m$ is a discontinuous 1st order phase transition. The calculable transition is continuous, 2nd order, only for $SU(2)$. This agrees with the lattice studies for all groups with center for which they exist \cite{Holland:2003kg}, see \cite{Anber:2014lba} for detailed comparison.
\item A discontinuous jump of tr $\Omega$ at a critical value of $mL$ (from a smaller value at small $mL$ to a larger value) is found for all gauge groups without a center $G_2, F_4, E_8$. This first order transition without order parameter has been seen in lattice studies of $G_2$ pure YM theory (no lattice studies of other center-less groups have been performed). Further (even more quantitative) comparison with $G_2$ lattice studies \cite{Pepe:2006er} can be found in \cite{Poppitz:2012nz,Anber:2014lba}.
\item The  critical value of $\hat{m}\over \hat\Lambda^3$ where the transition occurs (this is unity in our approximation) is a decreasing function of $\theta$ upon increasing $\theta$ away from $0$. At fixed $m, \Lambda$ this corresponds to $1/L_c$, i.e. $T_c$ decreasing with increasing $\theta$. On the other hand, the discontinuity of tr$\Omega$ across the first order quantum transition is, in all cases, an increasing function of $\theta$. Both these features are in agreement with the lattice \cite{DElia:2012pvq,DElia:2013uaf} (and with large-$N$ arguments regarding the $\theta$-dependence given there).
\end{enumerate}

Once again, we see that the small-$L$ studies offer a qualitative description of properties of the strongly-coupled 4d theory, in this case thermal pure-YM theory. 
The semiclassical physics behind the SYM$+m$ small-$L$ transition---the competition between center-stabilizing bions and center-breaking monopole-instantons---can not be, in a straightforward manner, argued to hold in the pure YM thermal case. However, there is direction of thermal-YM studies that uses interacting ensembles of monopole-instantons---akin to the instanton liquid at zero temperature, reviewed in \cite{Schafer:1996wv}---in pure YM theory in the $T_c \sim \Lambda$ strong coupling region to model the phase transition. For a description of these models and a discussion of their agreement with lattice data, see the early work \cite{Shuryak:2013tka} and, for work  including also fundamental matter and more references, the recent  \cite{DeMartini:2021xkg}.\footnote{Shuryak's notes on nonperturbative QCD \cite{Shuryak:2018fjr,Shuryak:2021vnj}   contain a discussion, often from a different perspective, of many of the topics  touched upon in these notes.}

Finally, the recent work \cite{Chen:2020syd} used the present small-$L$ SYM+$m$ framework and the continuity conjecture to study the realization of $CP$ symmetry and center-symmetry in pure-YM at $\theta=\pi$, conjecturing the existence of a novel  deconfined and $CP$-broken phase for $SU(2)$.

\hfill\begin{minipage}{0.85\linewidth}

\textcolor{red}{
{\flushleft{\bf Summary of \ref{sec:thermalsym}:}} In this Section, we explored the center-symmetry realization in SYM at small $L$ with added gaugino mass $m$. We found a $Z_2^{(1)}$-breaking quantum phase transition, driven by the competition  between center-stabilizing bion and center-destabilizing monopole-instanton effects. The agreement between the order of the transition and its $\theta$-dependence with available lattice studies suggest that this quantum phase transition is continuously connected, upon decoupling the gaugino, to the thermal deconfinement transition in pure YM theory, for all gauge groups, not just the $SU(2)$ case studied here. }

\end{minipage}

\bigskip

\section{QCD(F) and ``colour-flavour-center'' symmetry.}
\label{sec:fundamental}

As discussed in Section \ref{sec:holonomyandcenter}, the addition of fundamental quarks explicitly breaks the $Z_2^{(1)}$ center symmetry; we explained this in Footnote \ref{footnotefundam}, using our way of introducing the center symmetry via the improper gauge transformation (\ref{improper}).
Thus, if we  consider QCD(F) with fundamentals on $R^3 \times S^1$, we shall find that for either periodic and antiperiodic $S^1$-boundary conditions on the fundamental fermions, center symmetry is broken due to both the fermion and gauge contributions and the dynamics of the theory  is nonabelian also at small $L$. In addition, one expects a chiral restoration phase transition upon decreasing $L$, at least for the thermal case. The center-symmetry breaking effect of periodic or antiperiodic fundamental fermions is demonstrated on Figure \ref{fig:cfc1} in Section \ref{sec:cfc12} below.

 In this Section, we shall study an example of an interesting development in QCD with fundamental, rather than adjoint quarks, QCD(F), originated in \cite{Iritani:2015ara,Cherman:2016hcd,Cherman:2017tey}. We shall find a  set of $S^1$-boundary conditions, for theories where the number of Dirac flavours of fundamental quarks, $N_F$, equals the number of colours $N$ (or is an integer times $N$). These boundary conditions preserve a ``colour-flavour-center'' symmetry even in the presence of fundamentals, as explained in Section \ref{sec:cfc12}. We take $N=2$, as in the rest of these notes.\footnote{We couldn't find a discussion of the small-$L$/large-$L$ continuity for $SU(2)$ worked out in the literature (safe for some remarks in \cite{Aitken:2018mbb}),  so to stick with the $SU(2)$ spirit of these notes, we did it  ourselves. This Section, therefore, contains the only not previously published result of these notes. Compared to the  discussion for $N>2$ of \cite{Cherman:2016hcd}, the new elements  are the (well-known) chiral symmetry enhancement for $SU(2)$-fundamentals and the use of the loop-induced Chern-Simons couplings on $R^3 \times S^1$ (computed recently in \cite{Poppitz:2020tto} using fully gauge invariant regulators) in the argument.} We shall see that this suffices to illustrate many important features of the construction:
 \begin{enumerate}
 \item The most remarkable fact is that this construction offers 
a continuous connection between the low-lying (massless) states---the pions---on  $R^3 \times S^1$ at small $L$ and at large $L$. In particular, one can show that the chiral  symmetry realization is the same in both cases, offering evidence that there is no phase transition on the way from small to large $L$.
\item The Goldstone boson(s) is (are) the dual photon(s), i.e.~a sort of ``dual gluon(s).'' The dual gluon acquires charge under the continuous chiral symmetry, in contrast to usual descriptions of chiral symmetry breaking on $R^4$. This has been known from earlier studies on $R^3$ \cite{Affleck:1982as} as well as in other theories on $R^3 \times S^1$ \cite{Poppitz:2009tw}, but a connection to a theory so close to QCD(F) is new. The order parameter for continuous chiral symmetry breaking, as for the discrete symmetry cases of Sections \ref{sec:paritycenteranomaly}, \ref{sec:symvacuum}, is the expectation value of a monopole operator $\langle e^{i \sigma}\rangle$.
\item At the moment, it is not known how to generalize this to $N_F \ne Z N$ or to unequal quark masses. The construction that we shall describe is part of a general framework called ``distillation of Hilbert spaces'' by the authors of \cite{Dunne:2018hog,Kanazawa:2019tnf,Unsal:2021xay}. This is inspired, on one hand, by the supersymmetric Witten index \cite{Witten:1982df}, and on the other---by large-$N$ volume independence ideas, see the discussion in  \cite{Sulejmanpasic:2016llc}.  We won't have the time and space to describe this development. It is important to stress, however,   that it highlights the importance of the global-symmetry-twisted $S^1$-boundary conditions (the ``CFC'' ones, defined below) for the continuity between the small-$L$ and large-$L$  chiral-symmetry realization. 

Below, we explicitly argue for this continuity for our (decidedly small-$N$!) $SU(2)$ QCD(F) theory.
\end{enumerate}
 After summarizing the main points, we now explain the details.

\subsection{Colour-flavour-center symmetry.}
\label{sec:cfc12}

To describe the construction, consider the 4d fermion kinetic Lagrangian of the $I=1,...,N_F$ flavours of $SU(2)$-fundamental Dirac fermions (we suppress the explicit colour indices in $L_{fund}$, but show the gauge transformation of the fermions on the second line):
\begin{eqnarray}\label{fundlagr}
L_{fund} &=& 
 i \sum\limits_{I = 1}^{2 N_F}  \bar\psi^I_{\dot\alpha} \; \bar\sigma^{M \dot\alpha \alpha} (\partial_M   + i A_M^a T^a) \psi_{I \; \alpha}, \\
 \psi_I &\equiv & \left(\begin{array}{c} \psi_I^1 \cr \psi_I^2 \end{array} \right),  ~\psi_I \rightarrow g\; \psi_I, ~ g \in SU(2).~\nonumber\end{eqnarray}
 We use the same $SL(2,C)$-spinor notation as we used for the adjoint in (\ref{lagrangianadjoint}). As shown, all fermions   $\psi_I$ transform in the fundamental doublet representation of $SU(2)$ (recall that for $SU(2)$ the fundamental and antifundamental representations are equivalent). We have shown the $SU(2)$ gauge transforms of $\psi_I$  above, while $A_M^a T^a$ transforms as in Appendix \ref{appx:notation}. 
The number of Weyl fundamentals is $2 N_F$; it has to be even to avoid the Witten anomaly \cite{Witten:1982fp}.  A single fundamental Dirac fermion is composed of two of the undotted Weyl fermions $\psi_I$, say $\psi_1$ and $\psi_2$, with a Dirac mass term $\psi_{1 \alpha}^T i \sigma^2 \psi_{2 \; \beta} \epsilon^{\beta\alpha}$ (the colour indices, on which $\sigma^2$ and the transposition act, are not shown).  The form (\ref{fundlagr}) shows that the massless theory has  a classical $U(2N_F)$ nonabelian flavour symmetry under which $(\psi_1,...\psi_{2N_F})$ transforms as a fundamental.

We gave the detailed form (\ref{fundlagr}) in order to facilitate comparison with the Weyl adjoint  kinetic term (\ref{lagrangianadjoint}) and to aid the calculation of the fundamental fermions' contribution to the GPY potential (similar to the one done in Section \ref{sec:oneloopgpy} for adjoints).  
Now, we consider the  following important Exercise, where we introduce the notion of ``colour-flavour-center'' (or CFC) symmetry and make other important remarks:
 
 \begin{quote}
{\flushleft{\bf Exercise 18:}} Here, we study the contribution of massless fundamental Dirac fermions to the potential for the holonomy and use the result to introduce the CFC symmetry. As in Section \ref{sec:oneloopgpy}, consider only  the Polyakov line background of $A_3^3 = v$. \begin{enumerate}
\item
Proceed as in the calculation of the adjoint fermion determinant. Notice that now both colour-space components (shown on the second line in (\ref{fundlagr})) of the fundamental fermions contribute, since they both  couple to the holonomy $v$ (it is instructive to compare eqn.~(\ref{fundperiodic}) below  with the periodic adjoint contribution (\ref{gpyadjoint})).

Show that for a single massless fundamental Dirac fermion, say with Weyl components that we shall denote $\psi_1, \psi_2$, with periodic boundary conditions on $S^1$, the GPY potential is 
\begin{equation}\label{fundperiodic}
V_{P, \;F}(vL)= - 2 \sum\limits_{p\in Z} \int {d^3 k\over (2 \pi)^3} \ln\left[ \left(\vec{k}^2 + ({2 \pi p \over L} + {v \over 2})^2 \right)\left(\vec{k}^2 + ({2 \pi p \over L} - {v \over 2})^2 \right)\right].
\end{equation}
The two contributions inside the logarithm come from the two colour components of the e.g.~$\psi_1$ doublet. The other Weyl component of the Dirac fermion, $\psi_2$, leads to an identical contribution and the overall factor of $2$. 
\item
Likewise, show that for  fundamental Dirac fermion, with Weyl components that we denote  $\psi_3, \psi_4$, but now with antiperiodic boundary conditions, we instead obtain
\begin{equation}\label{fundantiperiodic}
V_{A, \;F}(vL)= - 2 \sum\limits_{p\in Z} \int {d^3 k\over (2 \pi)^3} \ln\left[ \left(\vec{k}^2 + ({2 \pi p \over L} +{\pi \over L}+  {v \over 2})^2 \right)\left(\vec{k}^2 + ({2 \pi p \over L} + {\pi \over L}- {v \over 2})^2 \right)\right].
\end{equation}
Notice that in both periodic and antiperiodic expressions above, no massless modes of any KK number  appear near  the center symmetric point $v = \pi/L$. Also note, 
comparing (\ref{fundperiodic}) to (\ref{fundantiperiodic}) that the different boundary conditions manifest themselves in the $\pi/L$ Wilson line which appears in the A (antiperiodic) fermion contribution to the GPY potential, but does not appear   in the P (periodic) fermion contribution. \item Treating the expressions above as if they converge (imagine them defined via zeta-function) show that
\begin{equation}\label{cfc1}
V_{P, \;F}(v)=V_{A, \;F}({2\pi \over L} - v).
\end{equation}
 This relation implies that the GPY potential for a system with two fundamental Dirac fermions (massless, or more generally, of the same mass), one periodic ($\psi_1, \psi_2$) and the other antiperiodic ($\psi_3,\psi_4$) on the $S^1$, is invariant under the $Z_2^{(1)}$ center symmetry (which reflects the Weyl chamber of Figure \ref{fig:weylchamber} around the middle). From now on, we shall call this theory the ``P$+$A'' theory.
 
  In fact, the relation (\ref{cfc1}) is due to the ``colour-flavour-center'' global symmetry, observed first in \cite{Poppitz:2013zqa}. 
To elucidate, recall that, as explained in Footnote \ref{footnotefundam}, introducing fundamental fermions breaks $Z_2^{(1)}$, because a center transformation (\ref{improper}) applied to a periodic fermion makes it antiperiodic and is thus not consistent with the boundary conditions. However, if there is both a periodic and an antiperiodic fundamental Dirac fermion, of the same mass, combining the center transformation with an exchange of the two leaves the action invariant and preserves the boundary conditions. The exchange of the two flavours is simply a discrete flavour transformation, hence the name ``colour-flavour-center'' symmetry.\footnote{\label{footnotecfc}The generalization to $SU(N)$, $N>2$, is discussed in \cite{Cherman:2016hcd}. The analysis of CFC stabilization for arbitrary $N$ is more complicated than for $SU(2)$; see the recent fun two-loop analysis of \cite{Unsal:2021xay}.}

\item
 The different boundary conditions in the P$+$A Dirac theory can be interpreted as due to the turning on of an $S^1$-Wilson line $\Omega$ for a (vectorlike) subgroup of the  global $U(4)$ chiral flavour symmetry. For example, with $I,J=1,...,4$, the P$+$A theory fields obey the $S^1$ boundary conditions $\psi_I(x^3=L) = \Omega_{I}^{\; J} \psi_J(x^3=0)$, where $\Omega = \text{diag}(1,1,e^{i\pi}, e^{-i \pi})$. An $x^3$-dependent field redefinition $\psi_1 = \psi_1'$, $\psi_2 = \psi_2'$, $\psi_3 = e^{i \pi {x^3 \over L}} \psi_3', \psi_4 = e^{-i \pi {x^3 \over L}} \psi_4' $, where all $\psi'$ are now periodic, removes $\Omega$ from the boundary condition, but brings in the $\pi/L$ factors (as in (\ref{fundantiperiodic})) into the $A$ Dirac fermion kinetic term. Convince yourself that these factors can be interpreted as background $S^1$-Wilson lines, in the Cartan subalgebra of an $SU(2)$ embedded in the lower right corner of $U(4)$. This point will be important below. 
\item Show that eqns.~(\ref{fundperiodic},\ref{fundantiperiodic}), using Appendix \ref{appx:gpy} and the identities leading to (\ref{gpygauge2}), imply that, for the $SU(2)$ P$+$A theory,
\begin{eqnarray}\label{cfc11}
V_{A, \;F}(vL)+V_{P, \;F}(vL)= {8 \over L^3 \pi^2} \sum\limits_{m=1}^\infty {1 \over m^4}  (1 + (-1)^m) \cos {mL v\over 2} ~.
\end{eqnarray}
Since this expression is manifestly invariant under $v \rightarrow {2 \pi\over L} - v$, the center symmetric point $v ={ \pi\over L}$ is bound to be an extremum. 

To find if it is a minimum, it is  easiest to plot the result  and examine if $v = \pi/L$ is a minimum. See Figure \ref{fig:cfc1} below, showing that CFC symmetry is preserved by the one-loop potential due to fundamental $A+P$ fermion flavours.
\end{enumerate}
\end{quote}
\begin{figure}[h]
\centerline{
\includegraphics[width=8.5 cm]{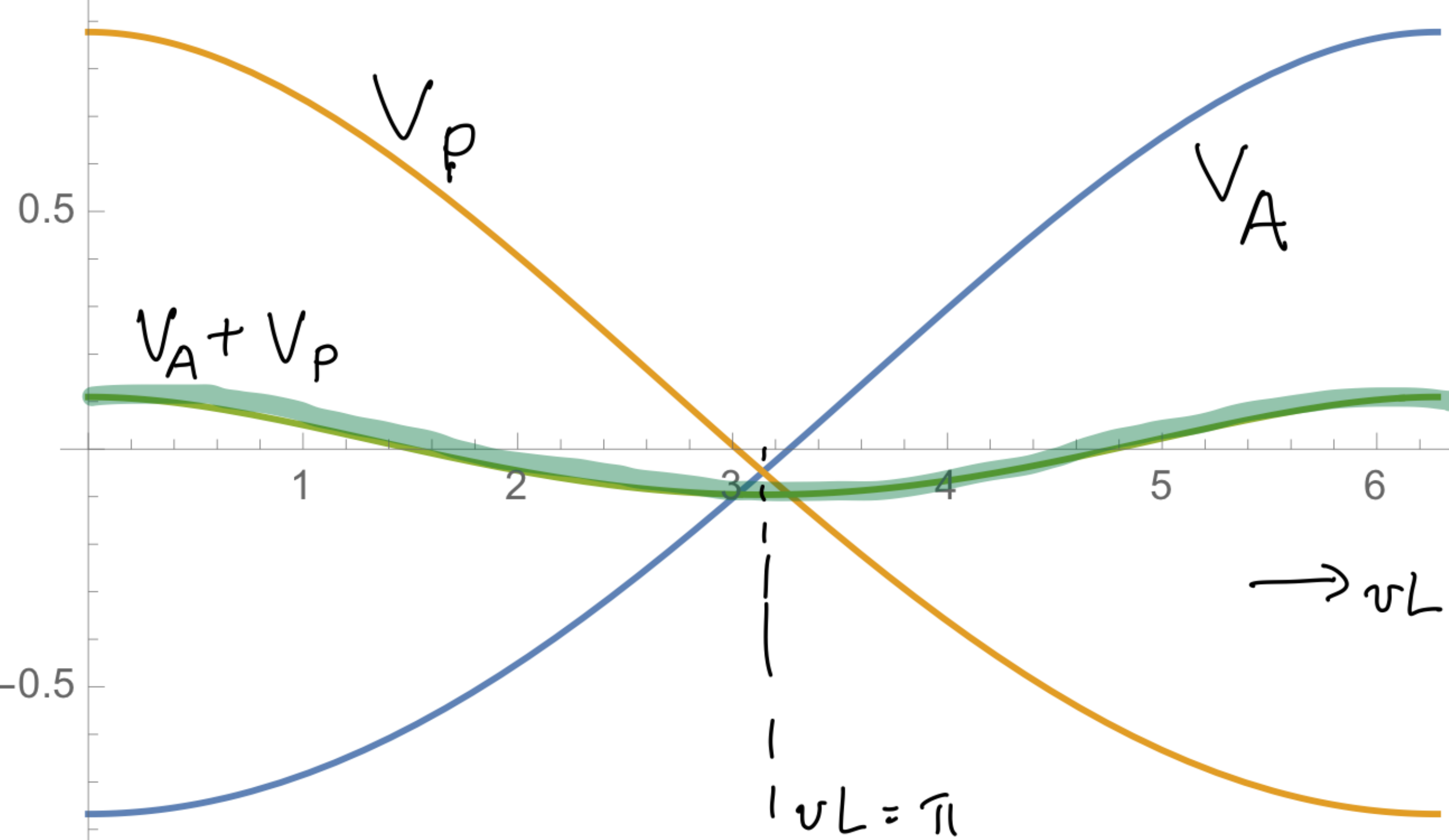}}
\caption{ The one-loop holonomy potential (\ref{cfc11}), the thick line,  in an $SU(2)$ gauge theory with a periodic and an antiperiodic massless  fundamental flavours. The same scale $L=1$ is used, to facilitate comparison with Figures \ref{fig:gpycenter} and \ref{fig:massivegpy}. The P and A contributions, favouring the broken $Z_2^{(1)}$, are also shown separately.
\label{fig:cfc1}}
\end{figure}
We need, of course, still worry about the gauge contribution (\ref{gpygauge}) which destabilizes the center symmetry. Comparing the absolute values of the potentials on Figure \ref{fig:gpycenter} and \ref{fig:cfc1}, drawn to the same scale, clearly shows that the gauge contribution is dominant. Here, we shall use our old trick and add massive adjoints, with $m \sim 1/L$ as on Figure \ref{fig:massivegpy}, to counter the gauge contribution, ensuring that the global minimum preserves CFC. 

Let us now concentrate on the $SU(2)$ P$+$A theory with $N_F=2$ Dirac fundamentals, with massive adjoints added to ``remove'' the center-breaking gauge contribution to the GPY potential.
The upshot of our analysis so far is this $R^3 \times S^1$ theory preserves center symmetry in its CFC disguise. The holonomy vev is $v=\pi/L$ and hence the theory abelianizes. What, then, are the light degrees of freedom? There is, of course, the dual (to $A_3^3$) photon $\sigma$. The adjoints have mass $\sim 1/L$ and decouple. 
The A and P fundamentals also have mass of at least ${\pi \over 2 L}$, as follows by inspecting the integrands in (\ref{fundperiodic}, \ref{fundantiperiodic}). Notice that this mass does not break the chiral symmetry, as it arises due to the coupling to the holonomy, i.e. from the 4d kinetic term. It would then appear tempting to  drop the A and P fundamentals from our IR considerations. Indeed, as we now discuss, this offers a sensible way to proceed at first (then we can revisit and ask if there is any other effect, or backreaction of the IR physics,  on the properties of the heavy A and P fermions).

\hfill\begin{minipage}{0.85\linewidth}

\textcolor{red}{
{\flushleft{\bf Summary of \ref{sec:cfc12}:} }Here, we explored a particular way to add fundamental quarks while still preserving a notion of center symmetry, abelianization, and semiclassical calculability on $R^3 \times S^1$. We introduced the notion of ``colour-flavour-center'' (CFC) symmetry and showed that center stability holds even in the theory with fundamentals, once the CFC boundary conditions are imposed. Thus, the theory abelianizes, and we showed that the dual photon is the only perturbatively massless degree of freedom. The nonperturbative physics is analyzed next.}

\end{minipage}

\bigskip

\subsection{Chiral symmetry breaking on $R^4$ and $R^3 \times S^1$.}
\label{sec:cfc13}

As usual, our goal is to find the symmetry realization in the vacuum  on $R^3 \times S^1$ and the relation to the $R^4$ theory.

To this end, let us go back to the chiral symmetries of  our  P$+$A theory. On $R^4$, where boundary conditions are irrelevant, the $SU(2)$ $N_F=2$ Dirac theory has a $U(2 N_F = 4)$ classical symmetry, with $\psi_I$ transforming as a fundamental. The overall $U(1)$ factor is anomalous and we shall not discuss it further.\footnote{There is a host of 't Hooft anomalies whose matching in the IR theory one can study. This is an interesting exercise that we shall not attempt here;  instead, we concentrate on the massless spectrum and the continuous symmetry realization.} The $SU(4)$ chiral symmetry is anomaly free and thus remains a symmetry of the quantum theory. On $R^4$, the expected behaviour of the theory is the spontaneous symmetry breaking $SU(4) \rightarrow SP(2)$ by the expectation value of the gauge-invariant fermion bilinear condensate. The fermion bilinear whose vev spontaneously breaks the chiral symmetry has the form of a general mass term (again we do not show the colour-space indices on which the transposition and $\sigma_2$ act):
\begin{equation}
\label{fermioncfc}
\langle\psi_{I \alpha}^T i \sigma_2 \psi_{J \beta} \epsilon^{\beta\alpha}\rangle \sim {\bf{J}}_{IJ} \Lambda^3: ~ SU(4) \rightarrow SP(2)\; (\text{or} \; USp(4)), ~\text{where}~~ {\bf{J}} \equiv  1_{2 \times 2} \otimes i \sigma_2 = \left(\begin{array}{cc}i \sigma_2 & 0\cr 0 & i \sigma_2 \end{array}\right).
\end{equation} 
As indicated above, the fermion bilinear  
 is antisymmetric in the  $SU(4)$  indices (owing to the anticommuting nature of the fermions and the antisymmetry of $\epsilon^{\beta\alpha}$ and  $i \sigma_2$). We used ${\bf{J}}_{IJ}$ to denote the elements of the $4\times 4$ antisymmetric matrix $
{\bf{J}}$, invariant under $SP(2)$ (this group is sometimes also called $USp(4)$, the set of $SU(4)$ matrices $U$ preserving $J$, i.e.~obeying $U^T J U = J$, and has dimension $10$).

Thus, the long-distance physics on $R^4$, assuming the symmetry breaking pattern (\ref{fermioncfc}), is described in terms the chiral Lagrangian of $15-10=5$ massless Goldstone bosons that parameterize the $SU(4)/USp(4)$ coset, including Wess-Zumino terms to match the various 't Hooft anomalies.
This coset space can also be written as $SO(6)/SO(5)$, owing to the Lie-algebra equivalences of $SU(4) \sim SO(6)$ and $USp(4)\sim SO(5)$. It is used as one of the ``little Higgs'' models of particle physics. For recent work and references, see \cite{Drach:2017btk}, which studies this theory on the lattice, and \cite{Davighi:2018xwn}, which works out the topological Wess-Zumino terms.

Now we go back to our CFC-preserving $R^3 \times S^1$ compactification. The different $S^1$ boundary conditions on the P and A fermions explicitly break the chiral $SU(4)$ to  $SU(2)_A \times SU(2)_P \times U(1)_X$.
 The P fermions   $(\psi_1, \psi_2) \sim (2,1)$ under $SU(2)_P \times U(1)_X$, while  the A $(\psi_3,\psi_4) \sim (2,-1)$ under $SU(2)_A \times U(1)_X$. From  the explicit form of $\bf{J}$ (\ref{fermioncfc}), it is clear that the  $SU(2)_P$ and $SU(2)_A$ parts of the global symmetries preserved by the CFC boundary conditions are in the unbroken $USp(4)$ (recall that for any $SU(2)$ matrix $V$, $V^T i \sigma_2 V =  i \sigma_2$).  $U(1)_X$, on the other hand,  is the only  global symmetry respected by the boundary conditions which is broken in the vacuum with the condensate (\ref{fermioncfc}).\footnote{
 Notice that Dirac masses for P ($\psi_1^T i\sigma_2 \psi_2$) and A ($\psi_3^T i \sigma_2 \psi_4$), not showing  the $SL(2,C)$ contraction, are invariant under the respective $SU(2)_{A,P}$ and have charge $\pm 2$ under $U(1)_X$, thus the $U(1)_X$ is part of the ``usual'' chiral symmetries as defined in QCD(F) with $N>2$.} This remark will be important for showing small-$L$/large-$L$ continuity further below.

 Next, we shall show  that the spontaneous breakdown of the $U(1)_X$ symmetry preserved by the compactification can be demonstrated in the small-$L$ CFC symmetric theory.  Furthermore, 
the dual photon $\sigma$---the only massless field left after the A,P-compactification of our theory---is   the Goldstone boson of the broken $U(1)_X$. 
There are several ways to show this and we shall allude to them all.

 First, we  proceed following the recent work \cite{Poppitz:2020tto}, as it offers the shortest way to our goal. As described above, in the CFC vacuum all fermions are massive, so they can  be integrated out. 
We shall do so and ask for the terms in the Wilsonian effective action generated upon integrating out the massive fields. These terms can depend on the light Cartan subalgebra gauge field $A_\mu^3$. In addition, to study the global-symmetry properties of the theory, we 
introduce a background field for the $U(1)_X$ symmetry, $X_M$, coupling to the A and P fermions charged under it. Under $U(1)_X$, we have $X_M\rightarrow X_M + \partial_M \omega$ and $\psi_1 \rightarrow e^{i \omega} \psi_1$, etc.
\begin{figure}[h]
\centerline{
\includegraphics[width=7.0 cm]{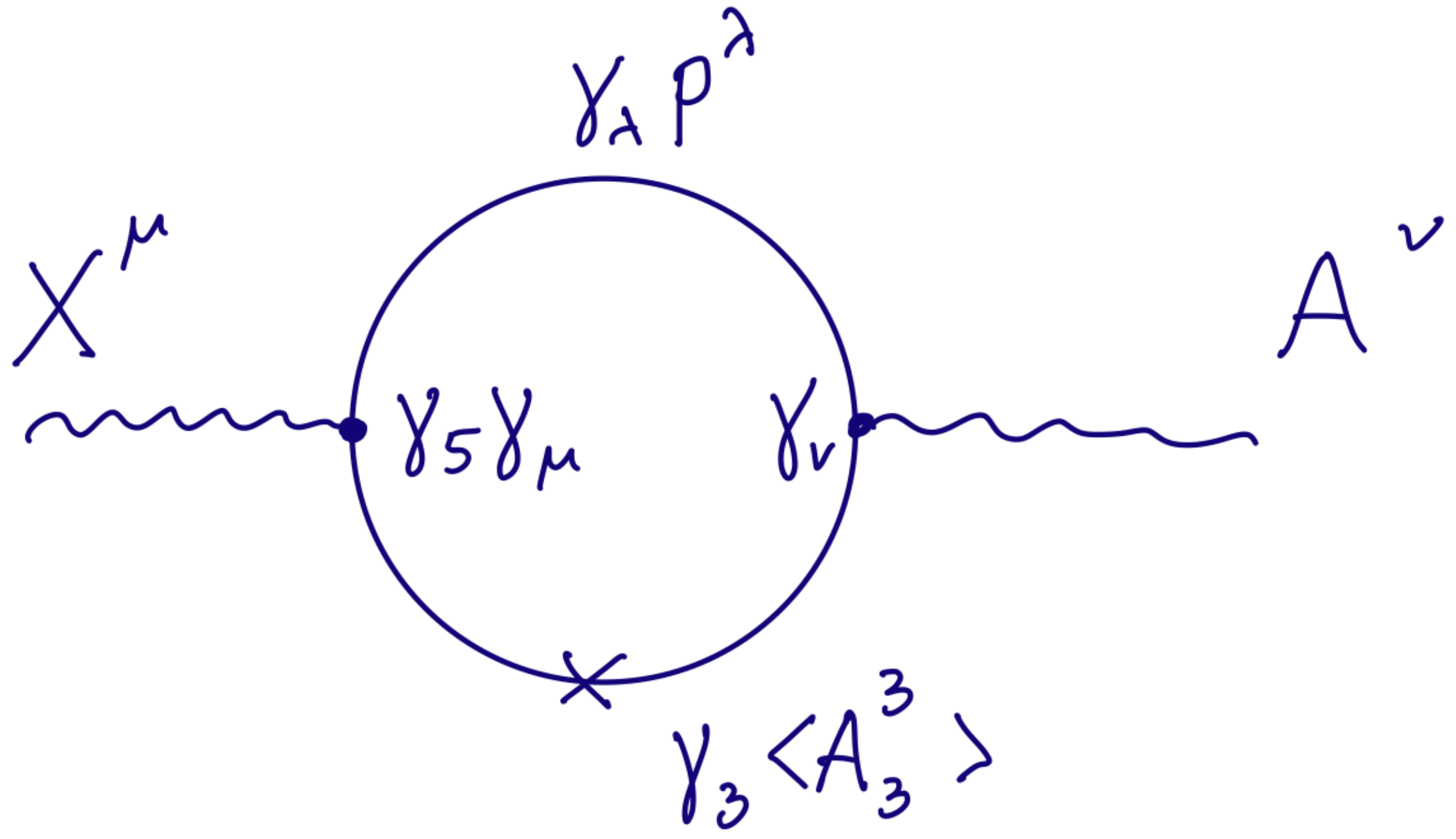}}
\caption{ The one-loop graph generating mixed Chern-Simons (CS) coupling of the dynamical Cartan gauge field $A_\nu$ and $X_\mu$, the $U(1)_X$ chiral-symmetry background field. To make the result (\ref{cscoupling}) plausible, using 4d $\gamma$-matrix notation, we have schematically indicated the relevant factors leading to the 3d CS coupling. The $X_\mu$ coupling to the A and P Dirac fermions is chiral, while an insertion of the holonomy vev in the numerator of one of the fermion propagators gives a $\gamma_3$ factor. Altogether, this leads to $\text{tr} \gamma_5 \gamma_3 \gamma_\mu \gamma_\nu \gamma_\lambda \sim \epsilon_{\mu\nu\lambda}$. To obtain eqn.~(\ref{cscoupling}), one has to carefully sum the loop contributions of the heavy A and P Dirac fermions, including their KK mode tower  \cite{Poppitz:2020tto}. }
\label{fig:csgraph}
\end{figure}

Then, we
 ask for the terms that depend on both the dynamical $A_\mu^3$ and the background $X_\mu$ in the effective long-distance theory.
We have two P Weyl fermions (a single Dirac) $(\psi_1,\psi_2)$ with charge $+1$ under $U(1)_X$, and two A Weyl fermions $(\psi_3,\psi_4)$ with charge $-1$, all of which  are $SU(2)$-gauge fundamentals. The $X_M$ couplings are precisely the couplings of a $U(1)$-chiral-symmetry background field to Dirac fermions on $R^3\times S^1$ studied in \cite{Poppitz:2020tto}.
Integrating these fermions out produces, via the one-loop diagrams shown on Figure~\ref{fig:csgraph}, Chern-Simons-like couplings between the $U(1)_X$ field and $A_\mu^3$ . 

To compute these couplings, one has to add the $A$ and $P$ Dirac fermion contributions to the $X$-$A$ Chern-Simons coupling; see the caption of Figure \ref{fig:csgraph} for a brief explanation of the result of \cite{Poppitz:2020tto}.\footnote{The loop contribution of a Dirac fermion is  
given in eqn.~(B.4) in \cite{Poppitz:2020tto}. One has to substitute   $q=\pm1$ for the $U(1)_X$ charges, use the $SU(2)$ fundamental-representation weights, and account for the different A-fermion boundary condition  by  inserting the $\mu = {\pi\over L}$ $U(1)_V$-Wilson line, similar to (\ref{fundantiperiodic}).} This results in the following $U(1)_X$-Cartan subalgebra $A_\mu^3$ (omitting the isospin index on $A_\mu, F_{\mu\nu}$ below) mixed CS term: \begin{equation}\label{cscoupling}
L_{CS} = -{1 \over 4 \pi} \epsilon^{\mu\nu\lambda} X_\mu F_{\nu\lambda}~.
\end{equation}
Now we perform the duality as in (\ref{dual2}), but with the above CS term included: start with
$
L_{Mink.}[F_{\mu\nu}, \sigma, X_\mu]$ $= - {L \over 4 g_4^2}   F_{\mu\nu} F^{\mu\nu}$ $- {1 \over 8 \pi} \partial_\mu \sigma  F_{\nu\lambda} \epsilon^{\mu\nu\lambda} - {1 \over 4\pi}\epsilon^{\mu\nu\lambda} X_\mu F_{\nu\lambda}~,
$
finding $F^{\mu\nu} = - {g_4^2 \over 4 \pi L} \epsilon^{\mu\nu\lambda} (\partial_\lambda \sigma + 2 X_\lambda)$, and ending up with the dual-photon Lagrangian coupled to $X_\lambda$:
\begin{equation}
\label{dual4}
L_{Mink.}[\sigma, X_\mu] = {1 \over 2} {g_4^2 \over (4 \pi)^2 L} (\partial_\lambda \sigma + 2 X_\lambda)(\partial^\lambda \sigma + 2 X^\lambda)~.
\end{equation}

Since the UV Lagrangian with $X_M$ coupled to the P and A fermions is $U(1)_X$ invariant, so must be the  IR effective lagrangian obtained after integrating them out, as their mass is chirally invariant.
Thus, $U(1)_X$ invariance of (\ref{dual4})   requires that the dual photon field shift under the $U(1)_X$ symmetry, namely:
\begin{equation}\label{dualphotonshift}
U(1)_X: ~ \sigma \rightarrow \sigma - 2 \omega~.
\end{equation}

Since $U(1)_X$ is an anomaly-free symmetry of the quantum theory, eqn.~(\ref{dualphotonshift}) implies that no potential for $\sigma$ can be generated. The masslessness of the dual photon is due to its being a Goldstone boson of the $U(1)_X$. The fact that the dual photon can be a Goldstone boson has been known since \cite{Affleck:1982as} in an $R^3$ framework and in some related $R^3 \times S^1$ theories  \cite{Poppitz:2009tw}, but the new realization here is its connection to $R^4$ via the CFC-symmetric compactification in QCD-like theories.\footnote{For a discussion in the  ``generalized global symmetry'' framework, see Section 5 of \cite{Gaiotto:2017yup}  and the earlier work   \cite{Kovner:1990nr,Kovner:1990pz,Kovner:1992pu}.} As in the breaking of parity in dYM at $\theta = \pi$ or of the discrete chiral symmetry in QCD(adj)/SYM, the order parameter for the breaking of the continuous chiral symmetry here is the expectation value of a monopole operator $\langle e^{i \sigma}\rangle$.

Second, a different point of view on the same phenomenon arises if one asks the question of the possible contributions of M and KK monopole instantons. The shift (\ref{dualphotonshift}) of the dual photon can be seen to arise from a consideration similar to our study of the QCD(adj) M and KK 't Hooft vertices (only there, the relevant symmetry was a discrete $Z_{4 n_f}$, recall (\ref{z4nf})). This is because the massive fermions A and P have zero modes in the M and KK background, as follows from the index theorem on $R^3 \times S^1$ already referred to, see \cite{Poppitz:2008hr}. As we shun to discuss details of the index theorem in these notes, we can argue slightly differently: accepting the shift symmetry (\ref{dualphotonshift}), we can ask what 't Hooft vertices of M and KK are allowed? 
 Clearly, an $e^{\pm i \sigma}$ 't Hooft vertex is not allowed by the shift symmetry.  Now we recall that CFC takes $\sigma \rightarrow - \sigma$, as it acts as $Z_2^{(1)}$,  and also interchanges $A \leftrightarrow P$. Thus, the following M and KK contributions 
\begin{equation} \label{chiralmass}
 e^{- S_0} \left( e^{i \sigma}\psi_1^T i \sigma_2 \psi_2  +  e^{- i \sigma}\psi_3^T i\sigma_2 \psi_4   + \text{c.c.}\right)~.
\end{equation}
are consistent with both $U(1)_X$ and with CFC.\footnote{As well as with the index theorem in monopole-instanton backgrounds, which implies that the P and A fermions reside on different monopole-instantons \cite{Poppitz:2008hr}; see also Appendix \ref{appx:index}.} Notice that there is one substantial difference from the corresponding bilinear fermion terms in SYM (\ref{susyyukawa}): here, the fermions are heavy, of chirally-symmetric mass $\pi/(2L)$ from the CFC compactification. The mass terms (\ref{chiralmass}) are nonperturbative chiral-symmetry breaking contributions to the fermion mass. The chiral-breaking mass terms (\ref{chiralmass}) represent, in our $\mu \ll 1/L$ EFT framework, a backreaction of the long-distance nonperturbative dynamics on the heavy states of mass $\sim 1/L$.

 In real-world QCD, the $L^{-1} e^{-S_0}$ chirality-breaking fermion mass from (\ref{chiralmass}) would represent the so-called ``constituent'' quark mass generation due to spontaneous chiral symmetry breaking. Also as in QCD, adding explicit chiral-symmetry-breaking small ``current''
 quark masses lifts the zero modes (as in Figure \ref{fig:liftzero}) producing an $M$ and $KK$ potential for $\sigma$ ($V \sim \cos \sigma$) and thus a nonzero ``pion'' mass. There is then, exactly as in  Section \ref{sec:polyakov_pass4}, a string confining fundamental quarks with tension calculable in the EFT. However, as also discussed in Section \ref{sec:polyakov_pass4}, a sufficiently long string is unstable to quark-antiquark pair creation. Of course, this is expected, as QCD(F) has no center symmetry in the $R^3$-directions.
 
 \hfill\begin{minipage}{0.85\linewidth}

 \textcolor{red}{
 {\flushleft{\bf Summary of \ref{sec:cfc13}:}} Here, we first outlined the QCD-like IR physics on $R^4$ and the nonabelian chiral-symmetry breaking pattern (\ref{fermioncfc}). Then, we studied the small-$L$ IR physics with ``CFC'' twisted boundary conditions. We showed that the dual photon shifts under the $U(1)_X$ chiral symmetry and is the Goldstone boson. The order parameter for the chiral-symmetry breaking is the expectation value of the monopole operator. The chiral-symmetry breaking leads to constituent mass of the A$+$P fundamental fermions, through the $M$ and $KK$ 't Hooft vertices.}
 
 \end{minipage}
 
 \bigskip

\subsection{Continuity between the large-$L$ and small-$L$ symmetry realization.}
\label{sec:cfc14}

Finally, we explain the remarkable feature of the CFC-symmetric compactification noted in the beginning of this Section: the small-$L$/large-$L$ continuity. Consider the low-energy limit of the $R^4$ theory, described by the $SU(4)/USp(4)$ chiral Lagrangian of the Goldstone bosons. Compactifying on an $S^1$ of size $L \gg \Lambda^{-1}$,  we have to remember that the different boundary conditions for the A and P fermions can be rephrased as the turning on background Wilson lines for the  vector global symmetry, or equivalently, imposing $S^1$-boundary conditions twisted by $\Omega$. In the low-energy chiral lagrangian, valid at $L \gg \Lambda^{-1}$,  these same $\Omega$-twisted boundary conditions on the $S^1$ have to be imposed on the fields in the chiral Lagrangian of the Goldstone bosons. 

Now, we recall our earlier symmetry considerations, showing that the CFC boundary conditions leave the $SU(2)_P \times SU(2)_A \times U(1)_X \in SU(4)$ part of the broken global symmetry intact, but that only  the $U(1)_X$ is part of the broken symmetry, i.e. one under which the goldstones transform nonlinearly. In other words, our CFC boundary conditions explicitly break all the spontaneously broken (by (\ref{fermioncfc})) symmetries but $U(1)_X$.  This implies that of the $5$ $SU(4)/USp(4)$ Goldstone bosons, only the $U(1)_X$-Goldstone will remain exactly massless at finite $L$, while the others should obtain mass of order $1/L$ from the CFC boundary conditions.\footnote{This can be worked out explicitly if needed, but it requires introducing the chiral Lagrangian for $SU(4)/USp(4)$. For a more familiar example in an $SU(3)$ QCD(F), using the familiar QCD $(SU(3)_L\times SU(3)_R)/SU(3)_V$ pion Lagrangian, see \cite{Cherman:2016hcd}.} This single massless Goldstone boson becomes, in the small-$L$ limit, the dual photon. 
The upshot is that   the massless spectrum of our P$+$A theory is
the same at $L \ll \Lambda^{-1}$ and $L \gg \Lambda^{-1}$. This provides evidence for the  
continuous connection between the small-$L$ and large-$L$ limit, i.e. for ``adiabatic continuity'' in QCD(F) with CFC boundary conditions. Notice, however, that the $L \sim \Lambda^{-1}$ regime is one where neither  the $R^4$ nor the small-$L$ descriptions used above applies, so we do not have proof of the continuity.

The reader may feel a certain amount of unhappiness after understanding the main points of this Section. This may arise from the realization that only the breaking of the $U(1)_X$ subgroup  of the nonabelian $SU(4)$ chiral symmetry can be seen in the calculable small-$L$ regime. Indeed, weak coupling on small $R^3\times S^1$ is  designed so that the physics abelianizes. The massless dual photons (which are Goldstone bosons also for the $N>2$ generalizations \cite{Cherman:2016hcd} of our CFC story) comprise only a subset of the $R^4$-pions which transform, linearly or nonlinearly, under nonabelian global symmetries. The dual photon(s), on the other hand, only transform under abelian global symmetries.  This may, indeed, be the best that is possible in a calculable semiclassical regime. 

At this point, we leave the judgment and possible applications of this construction for the future, but should not fail to notice its many interesting and thought-provoking features, notably the role monopole-instantons play in the constituent mass generation. We now rest our case and refer the interested reader to
recent work on this continuity for $N>2$, discussing many aspects not touched upon here \cite{Unsal:2021xay}. 

\hfill\begin{minipage}{0.85\linewidth}

\textcolor{red}{
{\flushleft{\bf Summary of \ref{sec:cfc14}:}} Here, we showed that considering ``CFC'' twisted boundary conditions allows one to exhibit a continuous connection between the IR spectra of QCD(F) at small and large $L$: the continuous chiral symmetry realization is the same at large $L \gg \Lambda^{-1}$ and at small $L \ll \Lambda^{-1}$.}

\end{minipage}

\bigskip

\section{A quick guide to the literature on  other theories.}

\label{sec:summary}

\subsection{dYM, QCD(adj)/SYM, QCD(F) with gauge group $SU(N)$ and $N>2$.}
\label{sec:biggerN}

Having familiarized ourselves with the detailed working of the various $R^3 \times S^1$-compactified $SU(2)$ theories, it is now natural to ask about larger numbers of colours. The only reason we stayed away from discussing $SU(N)$ was to not have to introduce more group theory notation. This would require venturing into Lie-algebraic territory: using roots, weights, etc., which are all required for the study of the circle compactified theories for general $N$ and for other gauge groups. A pedagogical approach would have required us to introduce these notions, which are not standard fare in QFT classes. This would have made the lectures  even longer. 

We believe, however, that  the $SU(2)$ case illustrated much of the small-$L$ physics quite nicely. Briefly, the general feature, abelianization on $R^3 \times S^1$, holds also in dYM, QCD(adj), and SYM for $N>2$ and appears as spontaneous breaking $SU(N) \rightarrow U(1)^{N-1}$ at a scale $m_W =  2 \pi/(NL)$. Thus, calculability requires $\Lambda N L \ll 2 \pi$. There are $N-1$ dual photons as well as $N$ different ``M and KK'' monopole instantons; these are now most appropriately labeled by the $N-1$ simple roots and the lowest, or affine, root of $SU(N)$. Their 't Hooft vertices (and molecules made thereof) generate the appropriate potentials for the dual photons. Magnetic bions and neutral bions appear, as in $SU(2)$, but have a more diverse nature. Overall, the small-$L$ physics of dYM, QCD(adj) and SYM is quite similar to the $SU(2)$ case. As alluded to in the last Section, QCD(F) differs because of the fact that it is only $SU(2)$ where the fundamental and antifundamental quarks are equivalent (for QCD(F) with $N>2$, see \cite{Unsal:2021xay}). 

Our $SU(2)$ considerations do not allow us to discuss the small-$L$, so-called ``abelian large-$N$'' limit. This consists in taking $N \rightarrow \infty$ and $L \rightarrow 0$ but keeping $NL\Lambda \ll 2 \pi$ fixed, in order to have weak coupling calculability \cite{Poppitz:2011wy}. For $SU(N)$ one can study the properties of confining strings of various $N$-alities \cite{Poppitz:2017ivi,Anber:2017tug,Bub:2020mff}. The  most unusual feature of $SU(N)$ theories in this limit is the appearance of an emergent dimension: even though one takes $L \rightarrow 0$, thus expecting the long-distance theory to be 3d, a latticized dimension of $N$ sites and lattice spacing $m_W^{-1} e^{S_0}$  (for SYM) emerges from the Cartan space of the gauge group. For more discussion of this phenomenon, somewhat similar to T-duality in string theory, see \cite{Cherman:2016jtu}.

For the reader who wishes to understand the details of the generalizations to $SU(N)$ (for other gauge groups, see further below) we  offer a  brief guide through the literature. The original papers on dYM \cite{Unsal:2008ch} and QCD(adj) \cite{Unsal:2007jx} all considered $SU(N)$ for arbitrary $N$, assuming familiarity with the Lie-algebraic notation. This can be learned by studying one of the many group-theory books, for example Ramond's \cite{Ramond:2010zz}. 
 There are also several papers that introduce aspects of this technology adapted to the $R^3\times S^1$ compactification. We especially recommend the voluminous appendix of \cite{Argyres:2012ka}, where the generalization of the Weyl chamber, shown on Figure \ref{fig:weylchamber} for $SU(2)$,   for all gauge groups is explicitly worked out; another paper whose appendix introduces some of the relevant technology is \cite{Anber:2014lba}.
 
An aspect we didn't discuss even for $SU(2)$ was the matching of various 't Hooft anomalies in the small-$L$ theory. These include both the traditional $0$-form symmetry anomalies, which are standard material, reviewed in \cite{Shifman:2012zz}, as well as the more recent so-called generalized 't Hooft anomalies involving higher-form global symmetries \cite{Gaiotto:2014kfa,Gaiotto:2017yup,Gaiotto:2017tne}. This is a big subject that we can not possibly do justice to here; unfortunately, an introductory set of notes does not appear to have been written yet. We already mentioned that the liberation of  quarks on domain walls (exhibited explicitly  in our semiclassical domain, see Sections \ref{sec:doublestringdYM}, \ref{sec:chiralqcdadj}, and 
\ref{sec:symvacuum}) is a manifestation of the generalized 't Hooft anomaly \cite{Komargodski:2017smk,Cox:2019aji,Anber:2018jdf,Anber:2018xek}. The matching on $R^3 \times S^1$ is discussed, in various ways, in \cite{Aitken:2018kky,Tanizaki:2019rbk,Poppitz:2020tto,Anber:2020xfk}, as well as in the recent \cite{Cox:2021vsa}. The appendix of the latter reference has a physicist-friendly discussion of the gauging of the $1$-form center symmetry for all gauge groups on the four-torus, relevant for  the   't Hooft anomaly between center symmetry and parity or discrete chiral symmetry. Further related recent work is in \cite{Unsal:2020yeh,Unsal:2021cch}.
 
 \subsection{Other matter representations with $SU(N)$ gauge group.}
\label{sec:other1}

The study of $SU(N)$ gauge theories with fermions in other representations  began soon after the work \cite{Unsal:2007jx,Unsal:2008ch} appeared, in the framework  which was called ``deformation theory.''  This consists of adding the center stabilizing potential (\ref{doubletrace}) by hand, to ensure abelianization, and then studying the ``M and KK'' vertices, taking into account the zero-modes of the various fermions via the index theorem \cite{Poppitz:2008hr}. The presence or absence of dual photon potentials can be then deduced from the relevant symmetries
\cite{Shifman:2008cx,Shifman:2008ja,Poppitz:2009uq,Poppitz:2009tw}. In this manner, the small-$L$ phases of theories with fermions in varying numbers of vectorlike fundamental, symmetric, antisymmetric, etc., representations were studied, as well as some chiral theories \cite{Poppitz:2009kz} (see \cite{ArabiArdehali:2019zac}
for interesting remarks on the supersymmetric version of the latter).

An uncomfortable aspect of this approach is the fact that the nonlocal term (\ref{doubletrace}) is added by hand, leading to a nonrenormalizable UV-incomplete theory.
It is desirable if 
the theory abelianizes as a result of the dynamics, as we saw happens naturally in QCD(adj). This was   seen to occur in some cases with mixed-representation fermions \cite{Poppitz:2009uq}. A general study was performed by Anber et al. in 
\cite{Anber:2017pak}. They studied general $SU(N)$ asymptotically-free and anomaly-free theories  with fermions in general representations. When taken on $R^3 \times S^1$, all fermions were taken periodic and their ``GPY'' potentials were studied. The conclusion was that theories with only a single-type of representation do not abelianize on $S^1$ and are thus generally strongly-coupled in the IR. However in  a large class of mixed-representation theories, abelianization occurs. A complete list of mixed-representation theories was given, see Table 8 of \cite{Anber:2017pak}. Some of these theories   were studied further in \cite{Anber:2019nfu,Anber:2021lzb} and were also used to suggest novel  possible phases on $R^4$.

\subsection{Other gauge groups.}
\label{sec:other2}

A natural question that arises is whether the abelianization---and the resulting calculability of the IR dynamics on $R^3 \times S^1$---in $SU(N)$ with $n_f$ adjoint fermions also occurs for other gauge groups with adjoints. Many of the simple Lie groups ($Sp(N), Spin(N), E_6, E_7$) have nontrivial centers. In  all cases, theories with only adjoint fermions respect the $1$-form center symmetry. 
A big difference compared to $SU(N)$, however, is that the order of the center of the gauge group is usually much smaller than the rank of the group. Let us compare $SU(N)$, where the order of $Z_N$ is $N$ and the rank is $N-1$ with, for example, $Sp(N)$, also known as $USp(2N)$, whose center is $Z_2$ while the rank is $N$. In each case the dimension of the Weyl chamber is equal to the rank, as was the case for $SU(2)$. The center symmetry maps  the Weyl chamber to itself and the center symmetric vevs are the fixed points of this map. If the theory is center-symmetric, this fixed-point set is bound to be an extremum. For $SU(N)$, as for $N=2$, there is a unique center-symmetric point on the Weyl chamber, where complete abelianization occurs. As we showed, this point is the minimum of the GPY potential for $n_f > 1$. 

In contrast, for $Sp(N)$\footnote{The situation for all other groups with centers (the maximal order of center is $4$) is similar \cite{Argyres:2012ka}.} there is an $\lfloor N/2 \rfloor$ dimensional fixed-point set which also passes through the edges of the Weyl chamber, where there is unbroken nonabelian gauge symmetry. A  priori, it is not obvious which point on this fixed-point set is the actual minimum of the perturbative ``GPY'' potential in  the theory with $n_f>1$ adjoints.
The question about the behaviour of the one-loop ``GPY'' potentials in such theories was studied in the already-mentioned 
ref.~\cite{Argyres:2012ka}. It turns out that only QCD(adj) with $n_f > 1$ and gauge groups  $SU(N)$ and $Sp(N)$ abelianizes at small $S^1$ (the small-$L$ IR physics of $Sp(N)$ QCD(adj)  was studied in \cite{Argyres:2012ka,Golkar:2009aq}). For all other gauge groups, there is an unbroken nonabelian gauge group, leading to a strongly-coupled IR physics even at small $S^1$.

The most remarkable feature, however, arises for the $n_f=1$ case of SYM with general gauge group $G$
\cite{Davies:2000nw}. There, as for our $SU(2)$ case, due to supersymmetry, the ``GPY'' potential vanishes to all orders of perturbation theory. The nonperturbative physics determines where the holonomy vev lies. It turns out that for SYM theories, abelianization  occurs for arbitrary gauge group: $G \rightarrow U(1)^r$, due to the ``M + KK'' nonperturbative potentials on the Weyl chamber (here $r$ is the rank of $G$). The minimum occurs away from the boundaries of the Weyl chamber at a particular point described in \cite{Davies:2000nw}. In each case, the small-$L$ IR theory is the magnetic dual of an $U(1)^r$ 3d gauge theory. 

In the case where $G$ has a center, this point on the Weyl chamber preserves center symmetry and the abelianization allows one, upon adding a gaugino mass, to investigate the  continuity conjecture to the thermal-YM theory, as in Section \ref{sec:thermalsym}, see also \cite{Poppitz:2012nz,Anber:2014lba,Chen:2020syd}. This allows also a study of other nonperturbative properties, not accessible to the usual holomorphy tools, such as the nature of the various domain walls and of the confining strings.

\hfill\begin{minipage}{0.85\linewidth}
\textcolor{red}{
{\flushleft{\bf Summary of \ref{sec:summary}:}} The small-$L$ dynamics of all dYM, QCD(adj)/SYM, QCD(F) theories studied here proceeds along   lines similar to $SU(2)$, for all $SU(N)$ gauge groups. However, in the nonsupersymmetric case of QCD(adj) only   $Sp(N)$ abelianizes at small $L$, while the other groups retain some unbroken nonabelian subgroup. In the case of SYM,  abelianization occurs for all gauge groups, with or without a center. Regarding other matter representations, abelianization holds in many cases of mixed-representation periodic fermions, as per the classification of  \cite{Anber:2017pak}.}

\end{minipage}

\bigskip

{\flushleft{\bf Acknowledgments:}} I am grateful to Andreas Wipf for the invitation to lecture at  the 2016 ``Saalburg'' Summer School and to Simon Catterall and Simon Hands for inviting me to contribute to the special issue  of the journal ``Symmetry'' on ``New Applications of Symmetry in Lattice Field Theory.''  Special thanks are due to Mohamed Anber, John Lai, and F. David Wandler for their comments and suggestions on the manuscript. This work is supported by an NSERC Discovery Grant.

\appendix
\section{Notations and conventions.}
\label{appx:notation}

Here, we summarize our conventions and warn about possible confusions caused by our notation. 

We use $\mu=1,2,3$ to label the $R^3$ Euclidean spacetime coordinates.
In $R^4$ we use instead $M=1,2,3,4$ to label the Euclidean coordinates. We use the same labeling on $R^3 \times S^1$ with $x^4 \equiv x^4 + L$ being the compact $S^1$. In each case, the Euclidean metric is positive definite $(+,+,...)$. On several occasions in the text, we fail to explicitly show lower/raised repeated indices that are summed over. The 4d Levi-Cevita tensor is $\epsilon^{1234}=+1$.

We use
 $a = 1,2,3$  to label $SU(2)$ Lie algebra indices.
The generators of the gauge group are $T^a$, obeying Tr($T^a T^b) = {1 \over 2} \delta^{ab}$ and can be taken to be one-half the Pauli matrices, $T^a = {\sigma^a/2}$. The gauge fields and adjoint  scalar transform as usual under $SU(2)$ gauge transformations $g$: $A_M = A_M^a T^a \rightarrow g(A_M - i \partial_M)g^{-1}$. An adjoint field transforms as  $\Phi = \Phi^a T^a \rightarrow g \Phi g^{-1}$. The field strength tensor is $F_{MN} = \partial_M A_N - \partial_N A_M + i \left[ A_M, A_N\right]$ and the adjoint covariant derivative is $D_M \Phi = \partial_M \Phi + i [A_M, \Phi]$. A fundamental field $\Psi$ transforms as $\Psi \rightarrow g \Psi$ and has covariant derivative $D_M \Psi = \partial_M \Psi + i A_M \Psi$.

On several occasions, we use $R^{1,3}$ or $R^{1,2} \times S^1$ Minkowski space. We use the same labels $M,N$ (and $\mu,\nu$) for the coordinates, but with $(+,-,-,..)$ metric. In addition, their ranges are now $M, N= 0,1,2,3$ (and $\mu, \nu = 0,1,2$), where $x^0$ is the time direction while the $S^1$ direction is $x^3 \equiv x^3 + L$.

For fermions, we use the ``God given''  two-component  $SL(2,C)$ spinor notation used in \cite{Shifman:2012zz}; for the most detailed introduction, see \cite{Dreiner:2008tw}.   Lagrangians involving fermions are written in Minkowski metric.
The minimal amount of information on the two component $SL(2,C)$ spinor notation we need is as follows. The Weyl spinor is an anticommuting complex object $\lambda_\alpha$ ($\alpha=1,2$), with $\bar\lambda_{\dot\alpha} = (\lambda_\alpha)^*$, $\epsilon^{12}= -\epsilon^{21} =\epsilon^{\dot{1}\dot{2}}=-\epsilon^{\dot{2}\dot{1}}=+1$. Instead of the usual gamma matrices, one uses the two-by-two matrices $\bar\sigma^0 = \sigma^0 = \bf{1}_2$, $\bar\sigma^{i} = - \sigma^i$, $i=1,2,3$, where $\sigma^i$ are the Pauli matrices (when using matrix notation there is no reason to explicitly show the indices $\bar\sigma^{\dot\alpha \alpha}$, as we  do in e.g.~(\ref{lambdakineticweyl})).
For a massless Weyl fermion, the kinetic term is ${i } \bar\lambda_{\dot\alpha} \; \bar\sigma^{M \dot\alpha \alpha} (\partial_M \lambda_\alpha) $ while a Majorana mass term is ${m \over 2} \lambda_\alpha \lambda_\beta \epsilon^{\beta\alpha} + {m^*\over 2} \bar\lambda_{\dot\alpha} \epsilon^{\dot\alpha \dot\beta} \bar\lambda_{\dot\beta}$. A Dirac mass term requires two Weyl fermions $\psi_{1,2}$ with the same kinetic terms and mass term $m \psi_{1 \; \alpha} \psi_{2 \; \beta} \epsilon^{\beta\alpha} + \text{c.c.}$. The Dirac mass term is symmetric w.r.t. interchanging $1\leftrightarrow 2$ due to anticommutativity of the spinors and the antisymmetry of $\epsilon$. In each case, gauge invariance is imposed as required by the fermions' representation.

{\flushleft{\bf Possible notational pitfalls:}} 
\begin{itemize}
\item The reader should be aware of the relabeling of the $S^1$ coordinate $x^4 \rightarrow x^3$ upon transition from Euclidean to Minkowski space. 
\item Throughout, we use $Z_2^{(1)}$ to denote the 1-form $Z_2$ (center) symmetry along the $S^1$ coordinate. Explicitly, this symmetry  appears first in eqn.~(\ref{center1}) as acting on the fundamental Wilson line wrapping the $S^1$. This  symmetry  is often called ``0-form center symmetry,'' a name justified from the point of view of the compactified theory.
 We avoid this terminology and, in the few occasions where we need to mention the 1-form symmetry in the noncompact directions, this is explicitly stated.
 \end{itemize}

\section{The  massless adjoint fermion contribution to ``GPY'' potential.}
\label{appx:gpy}
Here we compute the fermion determinant needed to obtain the finite $vL$-dependent part of (\ref{gpyadjoint}). Introducing dimensionless variables ($\vec q$ and $x = vL$), we define the quantity $F_\lambda[x]$, proportional to our formal expression for the determinant over the non-Cartan fermions $\lambda^\pm$:
\begin{eqnarray}
\label{gpy1}
 F_\lambda[x] &\equiv& \sum\limits_{p\in Z} \int {d^3 q\over (2 \pi)^3} \ln \left(\vec{q}^{\; 2} + ({2 \pi p } - x)^2\right) ~.
  \end{eqnarray}
We next define  the $\zeta$-function of the operator\footnote{In addition to \cite{Hawking:1976ja}, zeta-function regularization  is also discussed in the textbooks by Birrell and Davies \cite{Birrell:1982ix} and Schwartz \cite{Schwartz:2014sze} (in the latter, in the related context of flat-space Casimir-energy-like problems). The interested reader can also look up the original GPY calculation which  did not use zeta-function  \cite{Gross:1980br}.} whose eigenvalues appear inside the logarithm above: 
\begin{eqnarray}\label{zeta1}
  \zeta_\lambda[s, x]&\equiv& \sum\limits_{p \in Z}\int {d^3 q\over (2 \pi)^3} {1\over \left(\vec{q}^{\; 2} + ({2 \pi p } - x)^2 \right)^s}~,
\end{eqnarray}
and note that the r.h.s. converges for $Re[s]$ large enough. For other values of $s$, the function $\zeta_\lambda[s,x]$ is defined by analytic continuation. In terms of  $\zeta_\lambda[s]$ thus defined, we have 
\begin{equation}
\label{gpy11}
F_\lambda[x] = - \lim\limits_{s \rightarrow 0} {d \over d s} \zeta_\lambda[s,x]~.
\end{equation}
We compute $\zeta_\lambda[s,x]$ by first integrating over $d^3 q$ to obtain (below, $\lfloor a \rfloor$ denotes the largest integer smaller than $a$)
\begin{eqnarray}
\label{gpy2}
\zeta_\lambda[s,x] = { \Gamma[s-{3 \over 2}] \over 8 \pi^{3/2}\;\; \Gamma[s] \;(2 \pi)^{2s -3}}\; \sum\limits_{p\in Z} {1\over |p + \hat x|^{2s -3}}~, ~\text{where} ~~ \hat x \equiv {x \over 2 \pi}-   \iless{{x \over 2 \pi}} \in (0,1)~.
\end{eqnarray}
Then, we separate the sums over $p \ge 0$ and $p <0$, shifting variables in the latter sum, to obtain an expression for $\zeta_\lambda[s,x]$ in terms of the incomplete $\zeta$ function:
\begin{eqnarray}
\label{gpy3}
\zeta_\lambda[s,x] &=& {\Gamma[s-{3 \over 2}] \over 8 \pi^{3/2}\; \; \Gamma[s] \;(2 \pi)^{2s -3} } \left(\sum\limits_{p \ge 0} {1\over (p + \hat x)^{2s -3} } + \sum\limits_{p \ge 0} {1 \over (p+1- \hat{x})^{2s -3}}~, \right) \nonumber \\
&=&  { \Gamma[s-{3 \over 2}] \over 8 \pi^{3/2}\; \Gamma[s] \;(2 \pi)^{2s -3} }\left( \zeta(2s-3, \hat{x}) + \zeta(2s-3, 1 - \hat{x})\right)
\end{eqnarray}
Next, we realize that   to obtain a nonzero $F_\lambda[x]$, the derivative in (\ref{gpy11}) should act  on $1/\Gamma[s]$ (otherwise the contribution vanishes due to the pole of $\Gamma(s)$ at $s=0$). Then, using $\lim\limits_{s \rightarrow 0} {d \over d s}{1\over \Gamma[s]}= 1$ gives the only nonvanishing contribution to $F_\lambda[x]$, recalling that $[vL] \equiv vL \;(\text{mod} 2\pi)$,
\begin{eqnarray}
\label{gpy4}
 F_\lambda[x] &=&  -{4 \pi^2 \over 3 }\left( \zeta( -3, \hat{x}) + \zeta( -3, 1 - \hat{x})\right) = - {\pi^2 \over 45} + {1 \over 24 \pi^2} [vL]^2(2 \pi - [vL])^2~,
\end{eqnarray}
where to obtain the last equality we recalled that $x = vL$, the definition of $\hat x$ in (\ref{gpy2}), and used the relation between the incomplete zeta-function and the Bernoulli polynomials,  $\zeta(-3,a) = -  ({d \over d a} B_5(a))/20$, where $B_5(a) = a^5 - 5 a^4/2 + 5 a^3/3 - a/6$. 

\section{The  massive adjoint fermion contribution to ``GPY'' potential.}
\label{appx:gpymass}

Here, in order to compute (\ref{gpyadjointmass}), we consider the massive generalization of (\ref{gpy1}). We define, in terms of the dimensionless $x = vL$ and $\mu=mL$,
\begin{equation}\label{13}
F_\lambda[x,\mu] =
 \sum\limits_{p\in Z} \int {d^3 q\over (2 \pi)^3} \ln \left(\vec{q}^{\;2} + ({2 \pi p } - x)^2 + \mu^2 \right)~ = - \lim\limits_{s \rightarrow 0}{d \over d s}  \zeta_\lambda[s,x, \mu] ~,
 \end{equation}
where we proceeded in complete analogy with (\ref{zeta1}) to define a zeta-function
 \begin{eqnarray}
  \zeta_\lambda[s,x,\mu]&\equiv& \sum\limits_{p \in Z}\int {d^3 q\over (2 \pi)^3} {1\over \left(\vec{q}^{\; 2} + ({2 \pi p } - x)^2  + \mu^2\right)^s}~\nonumber \\
  &=& {\Gamma(s - {3 \over 2}) \over 8 \pi^{3/2} \Gamma(s) (2 \pi)^{2s -3}} \sum\limits_{p\in Z}{1 \over ((  p - \hat x)^2 + ({\mu \over 2 \pi})^2)^{s- 3/2}}~.
\end{eqnarray}
The sum over $p$ is now more complicated and is performed using the identity\footnote{The  derivation can be found in \cite{DiFrancesco:1997nk} (for easier access, note that it is also reproduced in the Appendix of \cite{Ponton:2001hq}).}
$$
f[t;a,c] =  \sum\limits_{n \in Z} {1 \over [(n+a)^2 + c^2]^t} = {\sqrt{\pi} |c|^{1 - 2 t} \over \Gamma(t)}\left[\Gamma(t-{1\over 2}) + 4 \sum\limits_{p=1}^\infty (\pi p |c|)^{t - {1 \over 2}} \cos(2 \pi p a) K_{t-{1\over 2}}(2 \pi p |c|)\right]~.
$$
The result is then substituted in (\ref{13}) to obtain, after careful manipulations similar to those in Appendix \ref{appx:gpy}, the desired form (\ref{gpyadjointmass}).

\section{A flavour of the $R^3 \times S^1$ index theorem: $SU(2)$ spin-$j$ representations.}
\label{appx:index}

Here, we provide the reader with a flavour of the index theorem, without considering the  details of its derivation (see \cite{Poppitz:2008hr} and the Appendix of \cite{Anber:2014lba} where Lie-algebraic notation valid for general gauge groups is used). In the text, we avoided making explicit use of the index theorem and instead relied on symmetry arguments to write terms involving fermions in the $\mu \ll 1/L$ EFT. Our main point here is simply to illustrate the fact that these arguments are backed up by the index theorem---unfortunately too technical a subject to present in detail.

The $R^3 \times S^1$ index theorem for monopole-instantons in a vacuum with a nonzero holonomy at infinity in $R^3$ (i.e. one which completely abelianizies the gauge group, as everywhere in these notes) is a generalization of the $R^3$ Callias index theorem \cite{Callias:1977kg} and has been derived in the mathematical literature \cite{Nye:2000eg}. Using terms accessible to physicists, the Callias index theorem was derived by E. Weinberg \cite{Weinberg:1979ma,Weinberg:2012pjx}. Ref. \cite{Poppitz:2008hr} generalized his derivation to $R^3 \times S^1$ to obtain the result of \cite{Nye:2000eg} in a physicist-friendly manner.
Below, we only outline the main ideas.
Begin
 by considering the quantity
\begin{equation}\label{indexdef}
I_{j} = \lim\limits_{\mu^2 \rightarrow 0} {\text{tr}} {\mu^2 \over D^\dagger D + \mu^2}- {\text{tr}} {\mu^2 \over D D^\dagger + \mu^2}~,
\end{equation}
where $D$ is the Weyl operator on $R^3 \times S^1$ in the appropriate ``spin-$j$'' representation of $SU(2)$. We explicitly showed this operator for the $j=1/2$  fundamental in (\ref{fundlagr}) and $j=1$  adjoint representations in (\ref{lagrangianadjoint}). The trace is taken over the space of two-component spinor functions  the operators $D^\dagger D$ and $D D^\dagger$ act on. Naturally, the Weyl operator is taken in  the monopole-instanton gauge background of interest. 

The most relevant  observation regarding (\ref{indexdef}) is  that nonzero discrete eigenvalues do not contribute to $I_j$. To see this, note that if $\Psi$ is a normalizable eigenfunction of $D^\dagger D$ with nonzero eigenvalue, $D^\dagger D \Psi = \lambda \Psi$, then $D\Psi$, also normalizable, obeys $D D^\dagger (D \Psi)= \lambda (D \Psi)$, i.e. is an eigenfunction with the same eigenvalue. Thus, the $\Psi$ and $D\Psi$ respective contributions to the first and second term in (\ref{indexdef}) cancel out already for finite $\mu^2$.  The detailed analysis of  \cite{Weinberg:1979ma}, which also holds on $R^3\times S^1$, showed that the continuous spectrum also does not contribute to (\ref{indexdef}), provided the vev of the holonomy is not at the edges of the Weyl chamber---hence the need for abelianization of the gauge group. Accepting this, the  conclusion is that only zero eigenvalues of $D^\dagger D$ and $D D^\dagger$ contribute to  (\ref{indexdef}). Thus, $I_j$ counts the number of zero modes of $D^\dagger D$ minus the number of zero modes of $D D^\dagger$. For the $M$ and $KK$ backgrounds, $I_j$ counts the number of  zero modes of the Weyl equation,   the normalizable solutions of  $D\Psi =0$.

The quantity $I_j$ can be seen to depend only on the topological properties of the gauge field background: on the topological charge, the magnetic charge, and the value of the holonomy at infinity (this follows from the explicit calculation, requiring some regularization---for which we refer the reader to the references).
For the particular  $M$ and $KK$  monopole-instanton backgrounds of interest to us, the index (\ref{indexdef}) is found to be, for fermions taken periodic on the $S^1$:
\begin{equation}\label{indexj}
I_j^M =  \sum\limits_{m=-j}^j 2 m \iless{ - {m v L \over 2 \pi}} ~,~~ I_j^{KK} = {2 j (j+1) (2j+1) \over 3} - I_j^M~.
\end{equation}
To unpack this formula, notice that for the fundamental and the adjoint, the two cases studied in this paper, we find $I_{1/2}^M$ 
= $- \lfloor{- {vL\over 4 \pi}}\rfloor + \lfloor{{vL \over 4 \pi}}\rfloor$ and   $I_{1}^M$ =$ 
- 2\lfloor{ - {vL \over 2 \pi}}\rfloor + 2 \lfloor{{vL \over 2 \pi}}\rfloor$. 
As an example of using (\ref{indexj}), let us now take $vL = \pi$, the center symmetric point. We immediately obtain from the above formulae $I_{1/2}^M = 1$ (hence $I_{1/2}^{KK}=0$) and $I_1^{M} = 2$ (hence $I_{1}^{KK} = 2$), exactly as found in the text by other means. Recall that we used center symmetry of the EFT as well as the shift of the dual photon under the chiral symmetry. 

One can also find expressions for fermions antiperiodic on the $S^1$, by turning on an appropriate $U(1)_V$ Wilson line. For the case of the fundamental  representation relevant for Section \ref{sec:fundamental},  one finds $I_{1/2}^M = 0$ (hence $I_{1/2}^{KK}=1$), i.e. upon changing the $U(1)_V$ background to dial the boundary condition from periodic to an antiperiodic, the fundamental fermion zero mode jumps from the $M$ to the $KK$ monopole instanton, as we already found in our discussion of QCD(F). A discussion of this jump and other jumps (seen to occur upon changing $vL$   in (\ref{indexj})) is given in \cite{Poppitz:2008hr}.

\bibliography{draftbiblio.bib}

\providecommand{\href}[2]{#2}\begingroup\raggedright\begin{thebibliography}{100}

\bibitem{Seiberg:1994bp}
N.~Seiberg, {\it {The Power of holomorphy: Exact results in 4-D SUSY field
  theories}},  in {\em {Particles, Strings, and Cosmology (PASCOS 94)}}, 5,
  1994.
\newblock \href{http://arxiv.org/abs/hep-th/9408013}{{\tt hep-th/9408013}}.

\bibitem{Greensite:2011zz}
J.~Greensite, {\it {An introduction to the confinement problem}},  {\em Lect.
  Notes Phys.} {\bf 821} (2011) 1--211.

\bibitem{Aharony:1999ti}
O.~Aharony, S.~S. Gubser, J.~M. Maldacena, H.~Ooguri, and Y.~Oz, {\it {Large N
  field theories, string theory and gravity}},  {\em Phys. Rept.} {\bf 323}
  (2000) 183--386, [\href{http://arxiv.org/abs/hep-th/9905111}{{\tt
  hep-th/9905111}}].

\bibitem{Seiberg:1994rs}
N.~Seiberg and E.~Witten, {\it {Electric - magnetic duality, monopole
  condensation, and confinement in N=2 supersymmetric Yang-Mills theory}},
  {\em Nucl. Phys. B} {\bf 426} (1994) 19--52,
  [\href{http://arxiv.org/abs/hep-th/9407087}{{\tt hep-th/9407087}}]. [Erratum:
  Nucl.Phys.B 430, 485--486 (1994)].

\bibitem{Gorsky:2007ip}
A.~Gorsky, M.~Shifman, and A.~Yung, {\it {N = 1 supersymmetric quantum
  chromodynamics: How confined non-Abelian monopoles emerge from quark
  condensation}},  {\em Phys. Rev. D} {\bf 75} (2007) 065032,
  [\href{http://arxiv.org/abs/hep-th/0701040}{{\tt hep-th/0701040}}].

\bibitem{Bjorken:1979hv}
J.~D. Bjorken, {\it {Elements of Quantum Chromodynamics}},  {\em Prog. Math.
  Phys.} {\bf 4} (12, 1979) 423--561.

\bibitem{Luscher:1982ma}
M.~Luscher, {\it {Some Analytic Results Concerning the Mass Spectrum of
  Yang-Mills Gauge Theories on a Torus}},  {\em Nucl. Phys. B} {\bf 219} (1983)
  233--261.

\bibitem{vanBaal:2000zc}
P.~van Baal, {\it {QCD in a finite volume}},
  \href{http://arxiv.org/abs/hep-ph/0008206}{{\tt hep-ph/0008206}}.

\bibitem{Lee:1997vp}
K.-M. Lee and P.~Yi, {\it {Monopoles and instantons on partially compactified
  D-branes}},  {\em Phys. Rev. D} {\bf 56} (1997) 3711--3717,
  [\href{http://arxiv.org/abs/hep-th/9702107}{{\tt hep-th/9702107}}].

\bibitem{Kraan:1998pm}
T.~C. Kraan and P.~van Baal, {\it {Periodic instantons with nontrivial
  holonomy}},  {\em Nucl. Phys. B} {\bf 533} (1998) 627--659,
  [\href{http://arxiv.org/abs/hep-th/9805168}{{\tt hep-th/9805168}}].

\bibitem{RTN:1993ilw}
{\bf RTN} Collaboration, M.~Garcia~Perez et~al., {\it {Instanton like
  contributions to the dynamics of Yang-Mills fields on the twisted torus}},
  {\em Phys. Lett. B} {\bf 305} (1993) 366--374,
  [\href{http://arxiv.org/abs/hep-lat/9302007}{{\tt hep-lat/9302007}}].

\bibitem{Gonzalez-Arroyo:1995ynx}
A.~Gonzalez-Arroyo and P.~Martinez, {\it {Investigating Yang-Mills theory and
  confinement as a function of the spatial volume}},  {\em Nucl. Phys. B} {\bf
  459} (1996) 337--354, [\href{http://arxiv.org/abs/hep-lat/9507001}{{\tt
  hep-lat/9507001}}].

\bibitem{Gonzalez-Arroyo:2023kqv}
A.~Gonzalez-Arroyo, {\it {On the fractional instanton liquid picture of the
  Yang-Mills vacuum and Confinement}},
  \href{http://arxiv.org/abs/2302.12356}{{\tt arXiv:2302.12356}}.

\bibitem{Unsal:2007jx}
M.~Unsal, {\it {Magnetic bion condensation: A New mechanism of confinement and
  mass gap in four dimensions}},  {\em Phys. Rev. D} {\bf 80} (2009) 065001,
  [\href{http://arxiv.org/abs/0709.3269}{{\tt arXiv:0709.3269}}].

\bibitem{Unsal:2007vu}
M.~Unsal, {\it {Abelian duality, confinement, and chiral symmetry breaking in
  QCD(adj)}},  {\em Phys. Rev. Lett.} {\bf 100} (2008) 032005,
  [\href{http://arxiv.org/abs/0708.1772}{{\tt arXiv:0708.1772}}].

\bibitem{Unsal:2008ch}
M.~Unsal and L.~G. Yaffe, {\it {Center-stabilized Yang-Mills theory:
  Confinement and large N volume independence}},  {\em Phys. Rev. D} {\bf 78}
  (2008) 065035, [\href{http://arxiv.org/abs/0803.0344}{{\tt
  arXiv:0803.0344}}].

\bibitem{Shifman:2008ja}
M.~Shifman and M.~Unsal, {\it {QCD-like Theories on R(3) x S(1): A Smooth
  Journey from Small to Large r(S(1)) with Double-Trace Deformations}},  {\em
  Phys. Rev. D} {\bf 78} (2008) 065004,
  [\href{http://arxiv.org/abs/0802.1232}{{\tt arXiv:0802.1232}}].

\bibitem{Gaiotto:2014kfa}
D.~Gaiotto, A.~Kapustin, N.~Seiberg, and B.~Willett, {\it {Generalized Global
  Symmetries}},  {\em JHEP} {\bf 02} (2015) 172,
  [\href{http://arxiv.org/abs/1412.5148}{{\tt arXiv:1412.5148}}].

\bibitem{Gaiotto:2017yup}
D.~Gaiotto, A.~Kapustin, Z.~Komargodski, and N.~Seiberg, {\it {Theta, Time
  Reversal, and Temperature}},  {\em JHEP} {\bf 05} (2017) 091,
  [\href{http://arxiv.org/abs/1703.00501}{{\tt arXiv:1703.00501}}].

\bibitem{Gaiotto:2017tne}
D.~Gaiotto, Z.~Komargodski, and N.~Seiberg, {\it {Time-reversal breaking in
  QCD$_{4}$, walls, and dualities in 2 + 1 dimensions}},  {\em JHEP} {\bf 01}
  (2018) 110, [\href{http://arxiv.org/abs/1708.06806}{{\tt arXiv:1708.06806}}].

\bibitem{Poppitz:2011wy}
E.~Poppitz and M.~Unsal, {\it {Seiberg-Witten and 'Polyakov-like' magnetic bion
  confinements are continuously connected}},  {\em JHEP} {\bf 07} (2011) 082,
  [\href{http://arxiv.org/abs/1105.3969}{{\tt arXiv:1105.3969}}].

\bibitem{Argyres:2012ka}
P.~C. Argyres and M.~Unsal, {\it {The semi-classical expansion and resurgence
  in gauge theories: new perturbative, instanton, bion, and renormalon
  effects}},  {\em JHEP} {\bf 08} (2012) 063,
  [\href{http://arxiv.org/abs/1206.1890}{{\tt arXiv:1206.1890}}].

\bibitem{Unsal:2012zj}
M.~Unsal, {\it {Theta dependence, sign problems and topological interference}},
   {\em Phys. Rev. D} {\bf 86} (2012) 105012,
  [\href{http://arxiv.org/abs/1201.6426}{{\tt arXiv:1201.6426}}].

\bibitem{Poppitz:2012sw}
E.~Poppitz, T.~Sch\"afer, and M.~Unsal, {\it {Continuity, Deconfinement, and
  (Super) Yang-Mills Theory}},  {\em JHEP} {\bf 10} (2012) 115,
  [\href{http://arxiv.org/abs/1205.0290}{{\tt arXiv:1205.0290}}].

\bibitem{Dunne:2016nmc}
G.~V. Dunne and M.~\"Unsal, {\it {New Nonperturbative Methods in Quantum Field
  Theory: From Large-N Orbifold Equivalence to Bions and Resurgence}},  {\em
  Ann. Rev. Nucl. Part. Sci.} {\bf 66} (2016) 245--272,
  [\href{http://arxiv.org/abs/1601.03414}{{\tt arXiv:1601.03414}}].

\bibitem{Cherman:2016jtu}
A.~Cherman and E.~Poppitz, {\it {Emergent dimensions and branes from large-$N$
  confinement}},  {\em Phys. Rev. D} {\bf 94} (2016), no.~12 125008,
  [\href{http://arxiv.org/abs/1606.01902}{{\tt arXiv:1606.01902}}].

\bibitem{Polyakov:1976fu}
A.~M. Polyakov, {\it {Quark Confinement and Topology of Gauge Groups}},  {\em
  Nucl. Phys. B} {\bf 120} (1977) 429--458.

\bibitem{Polyakov:1987ez}
A.~M. Polyakov, {\em {Gauge Fields and Strings}}, vol.~3.
\newblock 1987.

\bibitem{Deligne:1999qp}
P.~Deligne, P.~Etingof, D.~S. Freed, L.~C. Jeffrey, D.~Kazhdan, J.~W. Morgan,
  D.~R. Morrison, and E.~Witten, eds., {\em {Quantum fields and strings: A
  course for mathematicians. Vol. 1, 2}}.
\newblock 1999.

\bibitem{Shifman:2012zz}
M.~Shifman, {\em {Advanced topics in quantum field theory.}: {A lecture
  course}}.
\newblock Cambridge Univ. Press, Cambridge, UK, 2, 2012.

\bibitem{Banks:2014twn}
T.~Banks, {\em {Modern Quantum Field Theory}: {A Concise Introduction}}.
\newblock Cambridge University Press, 12, 2014.

\bibitem{Smilga:2004zr}
A.~V. Smilga and A.~Vainshtein, {\it {Background field calculations and
  nonrenormalization theorems in 4-D supersymmetric gauge theories and their
  low-dimensional descendants}},  {\em Nucl. Phys. B} {\bf 704} (2005)
  445--474, [\href{http://arxiv.org/abs/hep-th/0405142}{{\tt hep-th/0405142}}].

\bibitem{Anber:2014sda}
M.~M. Anber and T.~Sulejmanpasic, {\it {The renormalon diagram in gauge
  theories on $ {\mathrm{\mathbb{R}}}3\times {\mathbb{S}}1 $}},  {\em JHEP}
  {\bf 01} (2015) 139, [\href{http://arxiv.org/abs/1410.0121}{{\tt
  arXiv:1410.0121}}].

\bibitem{Harvey:1996ur}
J.~A. Harvey, {\it {Magnetic monopoles, duality and supersymmetry}},  in {\em
  {ICTP Summer School in High-energy Physics and Cosmology}}, 3, 1996.
\newblock \href{http://arxiv.org/abs/hep-th/9603086}{{\tt hep-th/9603086}}.

\bibitem{Weinberg:2012pjx}
E.~J. Weinberg, {\em {Classical solutions in quantum field theory}: {Solitons
  and Instantons in High Energy Physics}}.
\newblock Cambridge Monographs on Mathematical Physics. Cambridge University
  Press, 9, 2012.

\bibitem{Anber:2011de}
M.~M. Anber and E.~Poppitz, {\it {Microscopic Structure of Magnetic Bions}},
  {\em JHEP} {\bf 06} (2011) 136, [\href{http://arxiv.org/abs/1105.0940}{{\tt
  arXiv:1105.0940}}].

\bibitem{Vainshtein:1981wh}
A.~I. Vainshtein, V.~I. Zakharov, V.~A. Novikov, and M.~A. Shifman, {\it {ABC's
  of Instantons}},  {\em Sov. Phys. Usp.} {\bf 25} (1982) 195.

\bibitem{Coleman:1985rnk}
S.~Coleman, {\em {Aspects of Symmetry}: {Selected Erice Lectures}}.
\newblock Cambridge University Press, Cambridge, U.K., 1985.

\bibitem{Schafer:1996wv}
T.~Sch\"afer and E.~V. Shuryak, {\it {Instantons in QCD}},  {\em Rev. Mod.
  Phys.} {\bf 70} (1998) 323--426,
  [\href{http://arxiv.org/abs/hep-ph/9610451}{{\tt hep-ph/9610451}}].

\bibitem{Turner:2019wnh}
C.~Turner, {\it {Dualities in 2+1 Dimensions}},  {\em PoS} {\bf Modave2018}
  (2019) 001, [\href{http://arxiv.org/abs/1905.12656}{{\tt arXiv:1905.12656}}].

\bibitem{Polyakov:1975rs}
A.~M. Polyakov, {\it {Compact Gauge Fields and the Infrared Catastrophe}},
  {\em Phys. Lett. B} {\bf 59} (1975) 82--84.

\bibitem{Kogan:2002au}
I.~I. Kogan and A.~Kovner, {\it {Monopoles, vortices and strings: Confinement
  and deconfinement in (2+1)-dimensions at weak coupling}},
  \href{http://arxiv.org/abs/hep-th/0205026}{{\tt hep-th/0205026}}.

\bibitem{Poppitz:2017ivi}
E.~Poppitz and M.~E. Shalchian~T., {\it {String tensions in deformed Yang-Mills
  theory}},  {\em JHEP} {\bf 01} (2018) 029,
  [\href{http://arxiv.org/abs/1708.08821}{{\tt arXiv:1708.08821}}].

\bibitem{tHooft:1976snw}
G.~'t~Hooft, {\it {Computation of the Quantum Effects Due to a Four-Dimensional
  Pseudoparticle}},  {\em Phys. Rev. D} {\bf 14} (1976) 3432--3450. [Erratum:
  Phys.Rev.D 18, 2199 (1978)].

\bibitem{tHooft:1976rip}
G.~'t~Hooft, {\it {Symmetry Breaking Through Bell-Jackiw Anomalies}},  {\em
  Phys. Rev. Lett.} {\bf 37} (1976) 8--11.

\bibitem{Anber:2014lba}
M.~M. Anber, E.~Poppitz, and B.~Teeple, {\it {Deconfinement and continuity
  between thermal and (super) Yang-Mills theory for all gauge groups}},  {\em
  JHEP} {\bf 09} (2014) 040, [\href{http://arxiv.org/abs/1406.1199}{{\tt
  arXiv:1406.1199}}].

\bibitem{Pazarbasi:2021ifb}
C.~Pazarba\c{s}\i{} and M.~\"Unsal, {\it {Cluster expansion and resurgence in
  Polyakov model}},  \href{http://arxiv.org/abs/2110.05612}{{\tt
  arXiv:2110.05612}}.

\bibitem{Peskin:1995ev}
M.~E. Peskin and D.~V. Schroeder, {\em {An Introduction to quantum field
  theory}}.
\newblock Addison-Wesley, Reading, USA, 1995.

\bibitem{Bachas:1985xs}
C.~Bachas, {\it {Convexity of the Quarkonium Potential}},  {\em Phys. Rev. D}
  {\bf 33} (1986) 2723.

\bibitem{Fradkin:1978dv}
E.~H. Fradkin and S.~H. Shenker, {\it {Phase Diagrams of Lattice Gauge Theories
  with Higgs Fields}},  {\em Phys. Rev. D} {\bf 19} (1979) 3682--3697.

\bibitem{Cox:2019aji}
A.~A. Cox, E.~Poppitz, and S.~S.~Y. Wong, {\it {Domain walls and deconfinement:
  a semiclassical picture of discrete anomaly inflow}},  {\em JHEP} {\bf 12}
  (2019) 011, [\href{http://arxiv.org/abs/1909.10979}{{\tt arXiv:1909.10979}}].

\bibitem{Bub:2020mff}
M.~W. Bub, E.~Poppitz, and S.~S.~Y. Wong, {\it {Confinement on $\mathbb{R}^3
  \times \mathbb{S}^1$ and double-string collapse}},  {\em JHEP} {\bf 01}
  (2021) 044, [\href{http://arxiv.org/abs/2010.04330}{{\tt arXiv:2010.04330}}].

\bibitem{Anber:2013xfa}
M.~M. Anber, {\it {The abelian confinement mechanism revisited: new aspects of
  the Georgi-Glashow model}},  {\em Annals Phys.} {\bf 341} (2014) 21--55,
  [\href{http://arxiv.org/abs/1308.0027}{{\tt arXiv:1308.0027}}].

\bibitem{Cherman:2020hbe}
A.~Cherman, T.~Jacobson, S.~Sen, and L.~G. Yaffe, {\it {Higgs-confinement phase
  transitions with fundamental representation matter}},  {\em Phys. Rev. D}
  {\bf 102} (2020), no.~10 105021, [\href{http://arxiv.org/abs/2007.08539}{{\tt
  arXiv:2007.08539}}].

\bibitem{Greensite:2021fyi}
J.~Greensite and K.~Matsuyama, {\it {Symmetry, Confinement, and the Higgs
  Phase}},  \href{http://arxiv.org/abs/2112.06421}{{\tt arXiv:2112.06421}}.

\bibitem{Gross:1980br}
D.~J. Gross, R.~D. Pisarski, and L.~G. Yaffe, {\it {QCD and Instantons at
  Finite Temperature}},  {\em Rev. Mod. Phys.} {\bf 53} (1981) 43.

\bibitem{Lee:1998bb}
K.-M. Lee and C.-h. Lu, {\it {SU(2) calorons and magnetic monopoles}},  {\em
  Phys. Rev. D} {\bf 58} (1998) 025011,
  [\href{http://arxiv.org/abs/hep-th/9802108}{{\tt hep-th/9802108}}].

\bibitem{tHooft:1981sps}
G.~'t~Hooft, {\it {Aspects of Quark Confinement}},  {\em Phys. Scripta} {\bf
  24} (1981) 841--846.

\bibitem{Lai:2021}
J.~Lai, {\it {to appear, 2021}}, .

\bibitem{Anber:2015wha}
M.~M. Anber and E.~Poppitz, {\it {On the global structure of deformed
  Yang-Mills theory and QCD(adj) on $ {\mathrm{\mathbb{R}}}^3\times
  {\mathbb{S}}^1 $}},  {\em JHEP} {\bf 10} (2015) 051,
  [\href{http://arxiv.org/abs/1508.00910}{{\tt arXiv:1508.00910}}].

\bibitem{Aitken:2017ayq}
K.~Aitken, A.~Cherman, E.~Poppitz, and L.~G. Yaffe, {\it {QCD on a small
  circle}},  {\em Phys. Rev. D} {\bf 96} (2017), no.~9 096022,
  [\href{http://arxiv.org/abs/1707.08971}{{\tt arXiv:1707.08971}}].

\bibitem{KorthalsAltes:1993ca}
C.~P. Korthals~Altes, {\it {Constrained effective potential in hot QCD}},  {\em
  Nucl. Phys. B} {\bf 420} (1994) 637--668,
  [\href{http://arxiv.org/abs/hep-th/9310195}{{\tt hep-th/9310195}}].

\bibitem{Hawking:1976ja}
S.~W. Hawking, {\it {Zeta Function Regularization of Path Integrals in Curved
  Space-Time}},  {\em Commun. Math. Phys.} {\bf 55} (1977) 133.

\bibitem{Weinberg:1995mt}
S.~Weinberg, {\em {The Quantum theory of fields. Vol. 1: Foundations}}.
\newblock Cambridge University Press, 6, 2005.

\bibitem{Zinn-Justin:2002ecy}
J.~Zinn-Justin, {\it {Quantum field theory and critical phenomena}},  {\em Int.
  Ser. Monogr. Phys.} {\bf 113} (2002) 1--1054.

\bibitem{Aitken:2018mbb}
K.~Aitken, A.~Cherman, and M.~\"Unsal, {\it {Vacuum structure of Yang-Mills
  theory as a function of $\theta$}},  {\em JHEP} {\bf 09} (2018) 030,
  [\href{http://arxiv.org/abs/1804.06848}{{\tt arXiv:1804.06848}}].

\bibitem{Myers:2007vc}
J.~C. Myers and M.~C. Ogilvie, {\it {New phases of SU(3) and SU(4) at finite
  temperature}},  {\em Phys. Rev. D} {\bf 77} (2008) 125030,
  [\href{http://arxiv.org/abs/0707.1869}{{\tt arXiv:0707.1869}}].

\bibitem{Myers:2009df}
J.~C. Myers and M.~C. Ogilvie, {\it {Phase diagrams of SU(N) gauge theories
  with fermions in various representations}},  {\em JHEP} {\bf 07} (2009) 095,
  [\href{http://arxiv.org/abs/0903.4638}{{\tt arXiv:0903.4638}}].

\bibitem{Bonati:2018rfg}
C.~Bonati, M.~Cardinali, and M.~D'Elia, {\it {$\theta$ dependence in trace
  deformed $SU(3)$ Yang-Mills theory: a lattice study}},  {\em Phys. Rev. D}
  {\bf 98} (2018), no.~5 054508, [\href{http://arxiv.org/abs/1807.06558}{{\tt
  arXiv:1807.06558}}].

\bibitem{Bonati:2019kmf}
C.~Bonati, M.~Cardinali, M.~D'Elia, and F.~Mazziotti, {\it {$\theta$-dependence
  and center symmetry in Yang-Mills theories}},  {\em Phys. Rev. D} {\bf 101}
  (2020), no.~3 034508, [\href{http://arxiv.org/abs/1912.02662}{{\tt
  arXiv:1912.02662}}].

\bibitem{Bhoonah:2014gpa}
A.~Bhoonah, E.~Thomas, and A.~R. Zhitnitsky, {\it {Metastable vacuum decay and
  $\theta$ dependence in gauge theory. Deformed QCD as a toy model}},  {\em
  Nucl. Phys. B} {\bf 890} (2014) 30--47,
  [\href{http://arxiv.org/abs/1407.5121}{{\tt arXiv:1407.5121}}].

\bibitem{Anber:2017rch}
M.~M. Anber and A.~R. Zhitnitsky, {\it {Oblique Confinement at $\theta\neq 0$
  in weakly coupled gauge theories with deformations}},  {\em Phys. Rev. D}
  {\bf 96} (2017), no.~7 074022, [\href{http://arxiv.org/abs/1708.07520}{{\tt
  arXiv:1708.07520}}].

\bibitem{Aitken:2018kky}
K.~Aitken, A.~Cherman, and M.~\"Unsal, {\it {Dihedral symmetry in $SU(N)$
  Yang-Mills theory}},  {\em Phys. Rev. D} {\bf 100} (2019), no.~8 085004,
  [\href{http://arxiv.org/abs/1804.05845}{{\tt arXiv:1804.05845}}].

\bibitem{DelDebbio:2006yuf}
L.~Del~Debbio, G.~M. Manca, H.~Panagopoulos, A.~Skouroupathis, and E.~Vicari,
  {\it {Theta-dependence of the spectrum of SU(N) gauge theories}},  {\em JHEP}
  {\bf 06} (2006) 005, [\href{http://arxiv.org/abs/hep-th/0603041}{{\tt
  hep-th/0603041}}].

\bibitem{tHooft:1973alw}
G.~'t~Hooft, {\it {A Planar Diagram Theory for Strong Interactions}},  {\em
  Nucl. Phys. B} {\bf 72} (1974) 461.

\bibitem{Witten:1979vv}
E.~Witten, {\it {Current Algebra Theorems for the U(1) Goldstone Boson}},  {\em
  Nucl. Phys. B} {\bf 156} (1979) 269--283.

\bibitem{Witten:1980sp}
E.~Witten, {\it {Large N Chiral Dynamics}},  {\em Annals Phys.} {\bf 128}
  (1980) 363.

\bibitem{Athenodorou:2020clr}
A.~Athenodorou, M.~Cardinali, and M.~D'Elia, {\it {Spectrum of Trace Deformed
  Yang-Mills Theories}},  \href{http://arxiv.org/abs/2010.03618}{{\tt
  arXiv:2010.03618}}.

\bibitem{tHooft:1979rtg}
G.~'t~Hooft, {\it {A Property of Electric and Magnetic Flux in Nonabelian Gauge
  Theories}},  {\em Nucl. Phys. B} {\bf 153} (1979) 141--160.

\bibitem{Cox:2021vsa}
A.~A. Cox, E.~Poppitz, and F.~D. Wandler, {\it {The mixed 0-form/1-form anomaly
  in Hilbert space: pouring the new wine into old bottles}},  {\em JHEP} {\bf
  10} (2021) 069, [\href{http://arxiv.org/abs/2106.11442}{{\tt
  arXiv:2106.11442}}].

\bibitem{Kitano:2021jho}
R.~Kitano, R.~Matsudo, N.~Yamada, and M.~Yamazaki, {\it {Peeking into the
  $\theta$ vacuum}},  \href{http://arxiv.org/abs/2102.08784}{{\tt
  arXiv:2102.08784}}.

\bibitem{Anber:2015kea}
M.~M. Anber, E.~Poppitz, and T.~Sulejmanpasic, {\it {Strings from domain walls
  in supersymmetric Yang-Mills theory and adjoint QCD}},  {\em Phys. Rev. D}
  {\bf 92} (2015), no.~2 021701, [\href{http://arxiv.org/abs/1501.06773}{{\tt
  arXiv:1501.06773}}].

\bibitem{Callan:1984sa}
C.~G. Callan, Jr. and J.~A. Harvey, {\it {Anomalies and Fermion Zero Modes on
  Strings and Domain Walls}},  {\em Nucl. Phys. B} {\bf 250} (1985) 427--436.

\bibitem{Komargodski:2017smk}
Z.~Komargodski, T.~Sulejmanpasic, and M.~\"Unsal, {\it {Walls, anomalies, and
  deconfinement in quantum antiferromagnets}},  {\em Phys. Rev. B} {\bf 97}
  (2018), no.~5 054418, [\href{http://arxiv.org/abs/1706.05731}{{\tt
  arXiv:1706.05731}}].

\bibitem{Hsin:2018vcg}
P.-S. Hsin, H.~T. Lam, and N.~Seiberg, {\it {Comments on One-Form Global
  Symmetries and Their Gauging in 3d and 4d}},  {\em SciPost Phys.} {\bf 6}
  (2019), no.~3 039, [\href{http://arxiv.org/abs/1812.04716}{{\tt
  arXiv:1812.04716}}].

\bibitem{Witten:1997ep}
E.~Witten, {\it {Branes and the dynamics of QCD}},  {\em Nucl. Phys.} {\bf
  B507} (1997) 658--690, [\href{http://arxiv.org/abs/hep-th/9706109}{{\tt
  hep-th/9706109}}].

\bibitem{SJRey:1998}
S.-J. Rey, {\it {unpublished, 1997}}, .

\bibitem{Armoni:2003ji}
A.~Armoni and M.~Shifman, {\it {On k string tensions and domain walls in N=1
  gluodynamics}},  {\em Nucl. Phys. B} {\bf 664} (2003) 233--246,
  [\href{http://arxiv.org/abs/hep-th/0304127}{{\tt hep-th/0304127}}].

\bibitem{Dunne:2000vp}
G.~V. Dunne, I.~I. Kogan, A.~Kovner, and B.~Tekin, {\it {Deconfining phase
  transition in (2+1)-dimensions: The Georgi-Glashow model}},  {\em JHEP} {\bf
  01} (2001) 032, [\href{http://arxiv.org/abs/hep-th/0010201}{{\tt
  hep-th/0010201}}].

\bibitem{Kovchegov:2002vi}
Y.~V. Kovchegov and D.~T. Son, {\it {Critical temperature of the deconfining
  phase transition in (2+1)-d Georgi-Glashow model}},  {\em JHEP} {\bf 01}
  (2003) 050, [\href{http://arxiv.org/abs/hep-th/0212230}{{\tt
  hep-th/0212230}}].

\bibitem{Simic:2010sv}
D.~Simic and M.~Unsal, {\it {Deconfinement in Yang-Mills theory through
  toroidal compactification with deformation}},  {\em Phys. Rev. D} {\bf 85}
  (2012) 105027, [\href{http://arxiv.org/abs/1010.5515}{{\tt
  arXiv:1010.5515}}].

\bibitem{Anber:2011gn}
M.~M. Anber, E.~Poppitz, and M.~Unsal, {\it {2d affine XY-spin model/4d gauge
  theory duality and deconfinement}},  {\em JHEP} {\bf 04} (2012) 040,
  [\href{http://arxiv.org/abs/1112.6389}{{\tt arXiv:1112.6389}}].

\bibitem{Anber:2012ig}
M.~M. Anber, S.~Collier, and E.~Poppitz, {\it {The SU(3)/$\mathbb{Z}_3$
  QCD(adj) deconfinement transition via the gauge theory/'affine' XY-model
  duality}},  {\em JHEP} {\bf 01} (2013) 126,
  [\href{http://arxiv.org/abs/1211.2824}{{\tt arXiv:1211.2824}}].

\bibitem{Anber:2013doa}
M.~M. Anber, S.~Collier, E.~Poppitz, S.~Strimas-Mackey, and B.~Teeple, {\it
  {Deconfinement in $\mathcal{N}=1$ super Yang-Mills theory on $\mathbb{R}^3
  \times \mathbb{S}^1$ via dual-Coulomb gas and ''affine'' XY-model}},  {\em
  JHEP} {\bf 11} (2013) 142, [\href{http://arxiv.org/abs/1310.3522}{{\tt
  arXiv:1310.3522}}].

\bibitem{Svetitsky:1982gs}
B.~Svetitsky and L.~G. Yaffe, {\it {Critical Behavior at Finite Temperature
  Confinement Transitions}},  {\em Nucl. Phys. B} {\bf 210} (1982) 423--447.

\bibitem{Yaffe:1982qf}
L.~G. Yaffe and B.~Svetitsky, {\it {First Order Phase Transition in the SU(3)
  Gauge Theory at Finite Temperature}},  {\em Phys. Rev. D} {\bf 26} (1982)
  963.

\bibitem{Shimizu:2017asf}
H.~Shimizu and K.~Yonekura, {\it {Anomaly constraints on deconfinement and
  chiral phase transition}},  {\em Phys. Rev. D} {\bf 97} (2018), no.~10
  105011, [\href{http://arxiv.org/abs/1706.06104}{{\tt arXiv:1706.06104}}].

\bibitem{Nelson:melt}
D.~R. Nelson, {\it {Study of melting in two dimensions}},  {\em Phys. Rev. B}
  {\bf 18} (178) 2318--2338.

\bibitem{Catterall:2008qk}
S.~Catterall, J.~Giedt, F.~Sannino, and J.~Schneible, {\it {Phase diagram of
  SU(2) with 2 flavors of dynamical adjoint quarks}},  {\em JHEP} {\bf 11}
  (2008) 009, [\href{http://arxiv.org/abs/0807.0792}{{\tt arXiv:0807.0792}}].

\bibitem{Hietanen:2008mr}
A.~J. Hietanen, J.~Rantaharju, K.~Rummukainen, and K.~Tuominen, {\it {Spectrum
  of SU(2) lattice gauge theory with two adjoint Dirac flavours}},  {\em JHEP}
  {\bf 05} (2009) 025, [\href{http://arxiv.org/abs/0812.1467}{{\tt
  arXiv:0812.1467}}].

\bibitem{DelDebbio:2009fd}
L.~Del~Debbio, B.~Lucini, A.~Patella, C.~Pica, and A.~Rago, {\it {Conformal
  versus confining scenario in SU(2) with adjoint fermions}},  {\em Phys. Rev.
  D} {\bf 80} (2009) 074507, [\href{http://arxiv.org/abs/0907.3896}{{\tt
  arXiv:0907.3896}}].

\bibitem{Athenodorou:2021wom}
A.~Athenodorou, Bennett, G.~Bergner, and B.~Lucini, {\it {Investigating the
  conformal behaviour of SU(2) with one adjoint Dirac flavor}},
  \href{http://arxiv.org/abs/2103.10485}{{\tt arXiv:2103.10485}}.

\bibitem{Bergner:2018unx}
G.~Bergner, S.~Piemonte, and M.~\"Unsal, {\it {Adiabatic continuity and
  confinement in supersymmetric Yang-Mills theory on the lattice}},  {\em JHEP}
  {\bf 11} (2018) 092, [\href{http://arxiv.org/abs/1806.10894}{{\tt
  arXiv:1806.10894}}].

\bibitem{vanBaal:1982ag}
P.~van Baal, {\it {Some Results for SU(N) Gauge Fields on the Hypertorus}},
  {\em Commun. Math. Phys.} {\bf 85} (1982) 529.

\bibitem{Nye:2000eg}
T.~M.~W. Nye and M.~A. Singer, {\it {An L**2 index theorem for Dirac operators
  on S**1 x R**3}},  \href{http://arxiv.org/abs/math/0009144}{{\tt
  math/0009144}}.

\bibitem{Poppitz:2008hr}
E.~Poppitz and M.~Unsal, {\it {Index theorem for topological excitations on
  R**3 x S**1 and Chern-Simons theory}},  {\em JHEP} {\bf 03} (2009) 027,
  [\href{http://arxiv.org/abs/0812.2085}{{\tt arXiv:0812.2085}}].

\bibitem{Callias:1977kg}
C.~Callias, {\it {Index Theorems on Open Spaces}},  {\em Commun. Math. Phys.}
  {\bf 62} (1978) 213--234.

\bibitem{Davies:1999uw}
N.~M. Davies, T.~J. Hollowood, V.~V. Khoze, and M.~P. Mattis, {\it {Gluino
  condensate and magnetic monopoles in supersymmetric gluodynamics}},  {\em
  Nucl. Phys. B} {\bf 559} (1999) 123--142,
  [\href{http://arxiv.org/abs/hep-th/9905015}{{\tt hep-th/9905015}}].

\bibitem{Poppitz:2020tto}
E.~Poppitz and F.~D. Wandler, {\it {Topological terms and anomaly matching in
  effective field theories on $\mathbb{R}^3\times \mathbb{S}^1$:. Part I.
  Abelian symmetries and intermediate scales}},  {\em JHEP} {\bf 01} (2021)
  091, [\href{http://arxiv.org/abs/2009.14667}{{\tt arXiv:2009.14667}}].

\bibitem{Poppitz:2009uq}
E.~Poppitz and M.~Unsal, {\it {Conformality or confinement: (IR)relevance of
  topological excitations}},  {\em JHEP} {\bf 09} (2009) 050,
  [\href{http://arxiv.org/abs/0906.5156}{{\tt arXiv:0906.5156}}].

\bibitem{Cordova:2018acb}
C.~C\'ordova and T.~T. Dumitrescu, {\it {Candidate Phases for SU(2) Adjoint
  QCD$_4$ with Two Flavors from $\mathcal{N}=2$ Supersymmetric Yang-Mills
  Theory}},  \href{http://arxiv.org/abs/1806.09592}{{\tt arXiv:1806.09592}}.

\bibitem{Nambu:1961tp}
Y.~Nambu and G.~Jona-Lasinio, {\it {Dynamical Model of Elementary Particles
  Based on an Analogy with Superconductivity. 1.}},  {\em Phys. Rev.} {\bf 122}
  (1961) 345--358.

\bibitem{Nambu:1961fr}
Y.~Nambu and G.~Jona-Lasinio, {\it {Dynamical model of elementary particles
  based on an analogy with superconductivity, II}},  {\em Phys. Rev.} {\bf 124}
  (1961) 246--254.

\bibitem{Hill:2002ap}
C.~T. Hill and E.~H. Simmons, {\it {Strong Dynamics and Electroweak Symmetry
  Breaking}},  {\em Phys. Rept.} {\bf 381} (2003) 235--402,
  [\href{http://arxiv.org/abs/hep-ph/0203079}{{\tt hep-ph/0203079}}]. [Erratum:
  Phys.Rept. 390, 553--554 (2004)].

\bibitem{Anber:2018iof}
M.~M. Anber and E.~Poppitz, {\it {Two-flavor adjoint QCD}},  {\em Phys. Rev. D}
  {\bf 98} (2018), no.~3 034026, [\href{http://arxiv.org/abs/1805.12290}{{\tt
  arXiv:1805.12290}}].

\bibitem{Poppitz:2019fnp}
E.~Poppitz and T.~A. Ryttov, {\it {Possible new phase for adjoint QCD}},  {\em
  Phys. Rev. D} {\bf 100} (2019), no.~9 091901,
  [\href{http://arxiv.org/abs/1904.11640}{{\tt arXiv:1904.11640}}].

\bibitem{Cordova:2019bsd}
C.~C\'ordova and K.~Ohmori, {\it {Anomaly Obstructions to Symmetry Preserving
  Gapped Phases}},  \href{http://arxiv.org/abs/1910.04962}{{\tt
  arXiv:1910.04962}}.

\bibitem{Cordova:2019jqi}
C.~C\'ordova and K.~Ohmori, {\it {Anomaly Constraints on Gapped Phases with
  Discrete Chiral Symmetry}},  {\em Phys. Rev. D} {\bf 102} (2020), no.~2
  025011, [\href{http://arxiv.org/abs/1912.13069}{{\tt arXiv:1912.13069}}].

\bibitem{Anber:2019nze}
M.~M. Anber and E.~Poppitz, {\it {On the baryon-color-flavor (BCF) anomaly in
  vector-like theories}},  {\em JHEP} {\bf 11} (2019) 063,
  [\href{http://arxiv.org/abs/1909.09027}{{\tt arXiv:1909.09027}}].

\bibitem{Anber:2020gig}
M.~M. Anber and E.~Poppitz, {\it {Generalized \textquoteright{}t Hooft
  anomalies on non-spin manifolds}},  {\em JHEP} {\bf 04} (2020) 097,
  [\href{http://arxiv.org/abs/2002.02037}{{\tt arXiv:2002.02037}}].

\bibitem{Manton:2004tk}
N.~S. Manton and P.~Sutcliffe, {\em {Topological solitons}}.
\newblock Cambridge Monographs on Mathematical Physics. Cambridge University
  Press, 2004.

\bibitem{Witten:1982df}
E.~Witten, {\it {Constraints on Supersymmetry Breaking}},  {\em Nucl. Phys. B}
  {\bf 202} (1982) 253.

\bibitem{Hori:2003ic}
K.~Hori, S.~Katz, A.~Klemm, R.~Pandharipande, R.~Thomas, C.~Vafa, R.~Vakil, and
  E.~Zaslow, {\em {Mirror symmetry}}, vol.~1 of {\em Clay mathematics
  monographs}.
\newblock AMS, Providence, USA, 2003.

\bibitem{Anber:2017ezt}
M.~M. Anber and E.~Poppitz, {\it {New nonperturbative scales and glueballs in
  confining supersymmetric gauge theories}},  {\em JHEP} {\bf 03} (2018) 052,
  [\href{http://arxiv.org/abs/1711.00027}{{\tt arXiv:1711.00027}}].

\bibitem{Seiberg:1996nz}
N.~Seiberg and E.~Witten, {\it {Gauge dynamics and compactification to
  three-dimensions}},  in {\em {Conference on the Mathematical Beauty of
  Physics (In Memory of C. Itzykson)}}, pp.~333--366, 6, 1996.
\newblock \href{http://arxiv.org/abs/hep-th/9607163}{{\tt hep-th/9607163}}.

\bibitem{Davies:2000nw}
N.~M. Davies, T.~J. Hollowood, and V.~V. Khoze, {\it {Monopoles, affine
  algebras and the gluino condensate}},  {\em J. Math. Phys.} {\bf 44} (2003)
  3640--3656, [\href{http://arxiv.org/abs/hep-th/0006011}{{\tt
  hep-th/0006011}}].

\bibitem{Bogomolny:1980ur}
E.~B. Bogomolny, {\it {Calculation of instanton-anti-instanton contributions in
  quantum mechanics}},  {\em Phys. Lett. B} {\bf 91} (1980) 431--435.

\bibitem{Zinn-Justin:1981qzi}
J.~Zinn-Justin, {\it {Multi - Instanton Contributions in Quantum Mechanics}},
  {\em Nucl. Phys. B} {\bf 192} (1981) 125--140.

\bibitem{Unsal:2021cch}
M.~\"Unsal, {\it {TQFT at work for IR-renormalons, resurgence and Lefschetz
  decomposition}},  \href{http://arxiv.org/abs/2106.14971}{{\tt
  arXiv:2106.14971}}.

\bibitem{Behtash:2015zha}
A.~Behtash, G.~V. Dunne, T.~Sch\"afer, T.~Sulejmanpasic, and M.~\"Unsal, {\it
  {Complexified path integrals, exact saddles and supersymmetry}},  {\em Phys.
  Rev. Lett.} {\bf 116} (2016), no.~1 011601,
  [\href{http://arxiv.org/abs/1510.00978}{{\tt arXiv:1510.00978}}].

\bibitem{Behtash:2015kva}
A.~Behtash, E.~Poppitz, T.~Sulejmanpasic, and M.~\"Unsal, {\it {The curious
  incident of multi-instantons and the necessity of Lefschetz thimbles}},  {\em
  JHEP} {\bf 11} (2015) 175, [\href{http://arxiv.org/abs/1507.04063}{{\tt
  arXiv:1507.04063}}].

\bibitem{Behtash:2015loa}
A.~Behtash, G.~V. Dunne, T.~Sch\"afer, T.~Sulejmanpasic, and M.~\"Unsal, {\it
  {Toward Picard\textendash{}Lefschetz theory of path integrals, complex
  saddles and resurgence}},  {\em Ann. Math. Sci. Appl.} {\bf 02} (2017)
  95--212, [\href{http://arxiv.org/abs/1510.03435}{{\tt arXiv:1510.03435}}].

\bibitem{Behtash:2018voa}
A.~Behtash, G.~V. Dunne, T.~Schaefer, T.~Sulejmanpasic, and M.~\"Unsal, {\it
  {Critical Points at Infinity, Non-Gaussian Saddles, and Bions}},  {\em JHEP}
  {\bf 06} (2018) 068, [\href{http://arxiv.org/abs/1803.11533}{{\tt
  arXiv:1803.11533}}].

\bibitem{Unsal:2020yeh}
M.~\"Unsal, {\it {Strongly coupled QFT dynamics via TQFT coupling}},
  \href{http://arxiv.org/abs/2007.03880}{{\tt arXiv:2007.03880}}.

\bibitem{Fujimori:2021oqg}
T.~Fujimori, M.~Honda, S.~Kamata, T.~Misumi, N.~Sakai, and T.~Yoda, {\it
  {Quantum phase transition and Resurgence: Lessons from 3d $\mathcal{N}=4$
  SQED}},  \href{http://arxiv.org/abs/2103.13654}{{\tt arXiv:2103.13654}}.

\bibitem{DiPietro:2021yxb}
L.~Di~Pietro, M.~Mari\~no, G.~Sberveglieri, and M.~Serone, {\it {Resurgence and
  $1/N$ Expansion in Integrable Field Theories}},
  \href{http://arxiv.org/abs/2108.02647}{{\tt arXiv:2108.02647}}.

\bibitem{Behtash:2015kna}
A.~Behtash, T.~Sulejmanpasic, T.~Sch\"afer, and M.~\"Unsal, {\it {Hidden
  topological angles and Lefschetz thimbles}},  {\em Phys. Rev. Lett.} {\bf
  115} (2015), no.~4 041601, [\href{http://arxiv.org/abs/1502.06624}{{\tt
  arXiv:1502.06624}}].

\bibitem{Poppitz:2012nz}
E.~Poppitz, T.~Sch\"afer, and M.~\"Unsal, {\it {Universal mechanism of
  (semi-classical) deconfinement and theta-dependence for all simple groups}},
  {\em JHEP} {\bf 03} (2013) 087, [\href{http://arxiv.org/abs/1212.1238}{{\tt
  arXiv:1212.1238}}].

\bibitem{Murayama:2021xfj}
H.~Murayama, {\it {Some Exact Results in QCD-like Theories}},  {\em Phys. Rev.
  Lett.} {\bf 126} (2021), no.~25 251601,
  [\href{http://arxiv.org/abs/2104.01179}{{\tt arXiv:2104.01179}}].

\bibitem{Holland:2003kg}
K.~Holland, M.~Pepe, and U.~J. Wiese, {\it {The Deconfinement phase transition
  of Sp(2) and Sp(3) Yang-Mills theories in (2+1)-dimensions and
  (3+1)-dimensions}},  {\em Nucl. Phys. B} {\bf 694} (2004) 35--58,
  [\href{http://arxiv.org/abs/hep-lat/0312022}{{\tt hep-lat/0312022}}].

\bibitem{Pepe:2006er}
M.~Pepe and U.~J. Wiese, {\it {Exceptional Deconfinement in G(2) Gauge
  Theory}},  {\em Nucl. Phys. B} {\bf 768} (2007) 21--37,
  [\href{http://arxiv.org/abs/hep-lat/0610076}{{\tt hep-lat/0610076}}].

\bibitem{DElia:2012pvq}
M.~D'Elia and F.~Negro, {\it {$\theta$ dependence of the deconfinement
  temperature in Yang-Mills theories}},  {\em Phys. Rev. Lett.} {\bf 109}
  (2012) 072001, [\href{http://arxiv.org/abs/1205.0538}{{\tt
  arXiv:1205.0538}}].

\bibitem{DElia:2013uaf}
M.~D'Elia and F.~Negro, {\it {Phase diagram of Yang-Mills theories in the
  presence of a $\theta$ term}},  {\em Phys. Rev. D} {\bf 88} (2013), no.~3
  034503, [\href{http://arxiv.org/abs/1306.2919}{{\tt arXiv:1306.2919}}].

\bibitem{Shuryak:2013tka}
E.~Shuryak and T.~Sulejmanpasic, {\it {Holonomy potential and confinement from
  a simple model of the gauge topology}},  {\em Phys. Lett. B} {\bf 726} (2013)
  257--261, [\href{http://arxiv.org/abs/1305.0796}{{\tt arXiv:1305.0796}}].

\bibitem{DeMartini:2021xkg}
D.~DeMartini and E.~Shuryak, {\it {Chiral Symmetry Breaking and Confinement
  from an Interacting Ensemble of Instanton-dyons in Two-flavor Massless QCD}},
   \href{http://arxiv.org/abs/2108.06353}{{\tt arXiv:2108.06353}}.

\bibitem{Shuryak:2018fjr}
E.~Shuryak, {\it {Lectures on nonperturbative QCD ( Nonperturbative Topological
  Phenomena in QCD and Related Theories)}},
  \href{http://arxiv.org/abs/1812.01509}{{\tt arXiv:1812.01509}}.

\bibitem{Shuryak:2021vnj}
E.~Shuryak, {\em {Nonperturbative Topological Phenomena in QCD and Related
  Theories}}, vol.~977 of {\em Lecture Notes in Physics}.
\newblock 3, 2021.

\bibitem{Chen:2020syd}
S.~Chen, K.~Fukushima, H.~Nishimura, and Y.~Tanizaki, {\it {Deconfinement and
  $\mathcal {CP}$ breaking at $\theta=\pi$ in Yang-Mills theories and a novel
  phase for SU(2)}},  {\em Phys. Rev. D} {\bf 102} (2020), no.~3 034020,
  [\href{http://arxiv.org/abs/2006.01487}{{\tt arXiv:2006.01487}}].

\bibitem{Iritani:2015ara}
T.~Iritani, E.~Itou, and T.~Misumi, {\it {Lattice study on QCD-like theory with
  exact center symmetry}},  {\em JHEP} {\bf 11} (2015) 159,
  [\href{http://arxiv.org/abs/1508.07132}{{\tt arXiv:1508.07132}}].

\bibitem{Cherman:2016hcd}
A.~Cherman, T.~Sch\"afer, and M.~\"Unsal, {\it {Chiral Lagrangian from Duality
  and Monopole Operators in Compactified QCD}},  {\em Phys. Rev. Lett.} {\bf
  117} (2016), no.~8 081601, [\href{http://arxiv.org/abs/1604.06108}{{\tt
  arXiv:1604.06108}}].

\bibitem{Cherman:2017tey}
A.~Cherman, S.~Sen, M.~Unsal, M.~L. Wagman, and L.~G. Yaffe, {\it {Order
  parameters and color-flavor center symmetry in QCD}},  {\em Phys. Rev. Lett.}
  {\bf 119} (2017), no.~22 222001, [\href{http://arxiv.org/abs/1706.05385}{{\tt
  arXiv:1706.05385}}].

\bibitem{Affleck:1982as}
I.~Affleck, J.~A. Harvey, and E.~Witten, {\it {Instantons and (Super)Symmetry
  Breaking in (2+1)-Dimensions}},  {\em Nucl. Phys. B} {\bf 206} (1982)
  413--439.

\bibitem{Poppitz:2009tw}
E.~Poppitz and M.~Unsal, {\it {Conformality or confinement (II): One-flavor
  CFTs and mixed-representation QCD}},  {\em JHEP} {\bf 12} (2009) 011,
  [\href{http://arxiv.org/abs/0910.1245}{{\tt arXiv:0910.1245}}].

\bibitem{Dunne:2018hog}
G.~V. Dunne, Y.~Tanizaki, and M.~\"Unsal, {\it {Quantum Distillation of Hilbert
  Spaces, Semi-classics and Anomaly Matching}},  {\em JHEP} {\bf 08} (2018)
  068, [\href{http://arxiv.org/abs/1803.02430}{{\tt arXiv:1803.02430}}].

\bibitem{Kanazawa:2019tnf}
T.~Kanazawa and M.~\"Unsal, {\it {Quantum distillation in QCD}},  {\em Phys.
  Rev. D} {\bf 102} (2020), no.~3 034013,
  [\href{http://arxiv.org/abs/1909.05222}{{\tt arXiv:1909.05222}}].

\bibitem{Unsal:2021xay}
M.~\"Unsal, {\it {Graded Hilbert spaces, quantum distillation and connecting
  SQCD to QCD}},  \href{http://arxiv.org/abs/2104.12352}{{\tt
  arXiv:2104.12352}}.

\bibitem{Sulejmanpasic:2016llc}
T.~Sulejmanpasic, {\it {Global Symmetries, Volume Independence, and Continuity
  in Quantum Field Theories}},  {\em Phys. Rev. Lett.} {\bf 118} (2017), no.~1
  011601, [\href{http://arxiv.org/abs/1610.04009}{{\tt arXiv:1610.04009}}].

\bibitem{Witten:1982fp}
E.~Witten, {\it {An SU(2) Anomaly}},  {\em Phys. Lett. B} {\bf 117} (1982)
  324--328.

\bibitem{Poppitz:2013zqa}
E.~Poppitz and T.~Sulejmanpasic, {\it {(S)QCD on $\mathbb{R}^{3} \times
  \mathbb{S}^{1}$: Screening of Polyakov loop by fundamental quarks and the
  demise of semi-classics}},  {\em JHEP} {\bf 09} (2013) 128,
  [\href{http://arxiv.org/abs/1307.1317}{{\tt arXiv:1307.1317}}].

\bibitem{Drach:2017btk}
V.~Drach, T.~Janowski, and C.~Pica, {\it {Update on SU(2) gauge theory with NF
  = 2 fundamental flavours}},  {\em EPJ Web Conf.} {\bf 175} (2018) 08020,
  [\href{http://arxiv.org/abs/1710.07218}{{\tt arXiv:1710.07218}}].

\bibitem{Davighi:2018xwn}
J.~Davighi and B.~Gripaios, {\it {Topological terms in Composite Higgs
  Models}},  {\em JHEP} {\bf 11} (2018) 169,
  [\href{http://arxiv.org/abs/1808.04154}{{\tt arXiv:1808.04154}}].

\bibitem{Kovner:1990nr}
A.~Kovner, B.~Rosenstein, and D.~Eliezer, {\it {Photon as Goldstone boson in
  (2+1)-dimensional Higgs model}},  {\em Mod. Phys. Lett. A} {\bf 5} (1990)
  2733--2740.

\bibitem{Kovner:1990pz}
A.~Kovner, B.~Rosenstein, and D.~Eliezer, {\it {Photon as a Goldstone boson in
  (2+1)-dimensional Abelian gauge theories}},  {\em Nucl. Phys. B} {\bf 350}
  (1991) 325--354.

\bibitem{Kovner:1992pu}
A.~Kovner and B.~Rosenstein, {\it {New look at QED in four-dimensions: The
  Photon as a Goldstone boson and the topological interpretation of electric
  charge}},  {\em Phys. Rev. D} {\bf 49} (1994) 5571--5581,
  [\href{http://arxiv.org/abs/hep-th/9210154}{{\tt hep-th/9210154}}].

\bibitem{Anber:2017tug}
M.~M. Anber and V.~Pellizzani, {\it {Representation dependence of k -strings in
  pure Yang-Mills theory via supersymmetry}},  {\em Phys. Rev. D} {\bf 96}
  (2017), no.~11 114015, [\href{http://arxiv.org/abs/1710.06509}{{\tt
  arXiv:1710.06509}}].

\bibitem{Ramond:2010zz}
P.~Ramond, {\em {Group theory: A physicist's survey}}.
\newblock 2010.

\bibitem{Anber:2018jdf}
M.~M. Anber and E.~Poppitz, {\it {Anomaly matching, (axial) Schwinger models,
  and high-T super Yang-Mills domain walls}},  {\em JHEP} {\bf 09} (2018) 076,
  [\href{http://arxiv.org/abs/1807.00093}{{\tt arXiv:1807.00093}}].

\bibitem{Anber:2018xek}
M.~M. Anber and E.~Poppitz, {\it {Domain walls in high-T SU(N) super Yang-Mills
  theory and QCD(adj)}},  {\em JHEP} {\bf 05} (2019) 151,
  [\href{http://arxiv.org/abs/1811.10642}{{\tt arXiv:1811.10642}}].

\bibitem{Tanizaki:2019rbk}
Y.~Tanizaki and M.~\"Unsal, {\it {Modified instanton sum in QCD and
  higher-groups}},  {\em JHEP} {\bf 03} (2020) 123,
  [\href{http://arxiv.org/abs/1912.01033}{{\tt arXiv:1912.01033}}].

\bibitem{Anber:2020xfk}
M.~M. Anber and E.~Poppitz, {\it {Deconfinement on axion domain walls}},  {\em
  JHEP} {\bf 03} (2020) 124, [\href{http://arxiv.org/abs/2001.03631}{{\tt
  arXiv:2001.03631}}].

\bibitem{Shifman:2008cx}
M.~Shifman and M.~Unsal, {\it {On Yang-Mills Theories with Chiral Matter at
  Strong Coupling}},  {\em Phys. Rev. D} {\bf 79} (2009) 105010,
  [\href{http://arxiv.org/abs/0808.2485}{{\tt arXiv:0808.2485}}].

\bibitem{Poppitz:2009kz}
E.~Poppitz and M.~Unsal, {\it {Chiral gauge dynamics and dynamical
  supersymmetry breaking}},  {\em JHEP} {\bf 07} (2009) 060,
  [\href{http://arxiv.org/abs/0905.0634}{{\tt arXiv:0905.0634}}].

\bibitem{ArabiArdehali:2019zac}
A.~Arabi~Ardehali, L.~Cassia, and Y.~L\"u, {\it {From Exact Results to Gauge
  Dynamics on $\mathbb{R}^3\times S^1$}},  {\em JHEP} {\bf 08} (2020) 053,
  [\href{http://arxiv.org/abs/1912.02732}{{\tt arXiv:1912.02732}}].

\bibitem{Anber:2017pak}
M.~M. Anber and L.~Vincent-Genod, {\it {Classification of compactified
  $su(N_c)$ gauge theories with fermions in all representations}},  {\em JHEP}
  {\bf 12} (2017) 028, [\href{http://arxiv.org/abs/1704.08277}{{\tt
  arXiv:1704.08277}}].

\bibitem{Anber:2019nfu}
M.~M. Anber, {\it {Self-conjugate QCD}},  {\em JHEP} {\bf 10} (2019) 042,
  [\href{http://arxiv.org/abs/1906.10315}{{\tt arXiv:1906.10315}}].

\bibitem{Anber:2021lzb}
M.~M. Anber, {\it {Condensates and anomaly cascade in vector-like theories}},
  {\em JHEP} {\bf 03} (2021) 191, [\href{http://arxiv.org/abs/2101.04132}{{\tt
  arXiv:2101.04132}}].

\bibitem{Golkar:2009aq}
S.~Golkar, {\it {Conformal windows of $SP(2N)$ and $SO(N)$ gauge theories from
  topological excitations on $R^3 \times S^1$}},  {\em JHEP} {\bf 11} (2009)
  076, [\href{http://arxiv.org/abs/0909.2838}{{\tt arXiv:0909.2838}}].

\bibitem{Dreiner:2008tw}
H.~K. Dreiner, H.~E. Haber, and S.~P. Martin, {\it {Two-component spinor
  techniques and Feynman rules for quantum field theory and supersymmetry}},
  {\em Phys. Rept.} {\bf 494} (2010) 1--196,
  [\href{http://arxiv.org/abs/0812.1594}{{\tt arXiv:0812.1594}}].

\bibitem{Birrell:1982ix}
N.~D. Birrell and P.~C.~W. Davies, {\em {Quantum Fields in Curved Space}}.
\newblock Cambridge Monographs on Mathematical Physics. Cambridge Univ. Press,
  Cambridge, UK, 2, 1984.

\bibitem{Schwartz:2014sze}
M.~D. Schwartz, {\em {Quantum Field Theory and the Standard Model}}.
\newblock Cambridge University Press, 3, 2014.

\bibitem{DiFrancesco:1997nk}
P.~Di~Francesco, P.~Mathieu, and D.~Senechal, {\em {Conformal Field Theory}}.
\newblock Graduate Texts in Contemporary Physics. Springer-Verlag, New York,
  1997.

\bibitem{Ponton:2001hq}
E.~Ponton and E.~Poppitz, {\it {Casimir energy and radius stabilization in
  five-dimensional orbifolds and six-dimensional orbifolds}},  {\em JHEP} {\bf
  06} (2001) 019, [\href{http://arxiv.org/abs/hep-ph/0105021}{{\tt
  hep-ph/0105021}}].

\bibitem{Weinberg:1979ma}
E.~J. Weinberg, {\it {Parameter Counting for Multi-Monopole Solutions}},  {\em
  Phys. Rev. D} {\bf 20} (1979) 936--944.

\end{thebibliography}\endgroup

\bibliographystyle{JHEP}

%=====================================

\end{document}